\documentclass[rmp,twocolumn,nofootinbib,endfloatsamsmath,amssymb,aps,10pt]{revtex4-1}
\pdfoutput=1
\usepackage{graphicx}
\usepackage{dcolumn}
\usepackage{bm}
\usepackage{bbm}
\usepackage[usenames]{color}
\usepackage{color} 
\usepackage{amsmath}
\usepackage{empheq}
\newcommand{\ket}[1]{\left|{#1}\right>}
\newcommand{\bra}[1]{\left<{#1}\right|}
\newcommand{\abs}[1]{\left|{#1}\right|}

\begin{document}

\title{Resonant Inelastic X-ray Scattering Studies of Elementary Excitations}

\author{Luuk J.P. Ament}
\affiliation{Institute-Lorentz for Theoretical Physics, Universiteit  Leiden - 2300 RA Leiden - The Netherlands}

\author{Michel van Veenendaal}
\affiliation{Advanced Photon Source - Argonne National Laboratory - Argonne - Illinois 60439 - USA}
\affiliation{Department of Physics - Northern Illinois University - De Kalb - Illinois 60115 - USA}

\author{Thomas P. Devereaux}
\affiliation{Stanford Institute for Materials and Energy Sciences, Stanford University and SLAC National Accelerator Laboratory - Menlo Park - CA 94025 - USA}

\author{John P. Hill}
\affiliation{Department of Condensed Matter Physics and Materials Science, Brookhaven National Laboratory - Upton - New York 11973 - USA}

\author{Jeroen van den Brink}
\affiliation{Institute for Theoretical Solid State Physics, IFW Dresden - 01069 Dresden - Germany}


\date{\today}

\begin{abstract}
In the past decade, Resonant Inelastic X-ray Scattering (RIXS) has made
remarkable progress as a spectroscopic technique. This is a direct result of the
availability of high-brilliance synchrotron X-ray radiation sources
and of advanced photon detection instrumentation. The technique's unique capability
to probe elementary excitations in complex materials by measuring their
energy-, momentum-, and polarization-dependence has brought RIXS to
the forefront of experimental photon science. We review both the
experimental and theoretical RIXS investigations of the past decade, focusing on those
determining the low-energy charge, spin, orbital and lattice
excitations of solids.  We present the fundamentals of RIXS as
an experimental method and then review the theoretical
state of affairs, its recent developments and discuss the different (approximate) methods to compute the dynamical RIXS response. The last decade's body of experimental RIXS data and its interpretation is surveyed, with an emphasis on RIXS studies of correlated electron systems, especially transition metal compounds.  Finally, we discuss the promise that RIXS holds for the near future, particularly in view of the advent of x-ray laser photon sources.
\end{abstract}

\maketitle

\tableofcontents

\section{Introduction}

Resonant Inelastic X-ray Scattering (RIXS) is a fast developing
experimental technique in which one scatters x-ray
photons inelastically off matter. It is a {\it photon-in} $-$ {\it
  photon-out}  spectroscopy for which one can, in principle, measure the energy, momentum,
and polarization change of the scattered photon. The change in
energy, momentum, and polarization of the photon are transferred to intrinsic
excitations of the material under study and thus RIXS provides
information about those excitations. RIXS is a resonant technique
in which the energy of the incident photon is chosen such that it
coincides with, and hence resonates with, one of the atomic x-ray
transitions of the system. The resonance can greatly enhance the inelastic scattering cross section, sometimes by many orders of magnitude, and offers a unique way to probe charge, magnetic, and orbital degrees of
freedom on selective atomic sites in a crystal. Early experimental work, and some more recent reviews include~\cite{Sparks1974,Bannett1975,Eisenberger1976a,Eisenberger1976b,Blume1985,Kotani2001,Schuelke2007}.

\begin{figure}
\begin{center}
\includegraphics[width=0.5\columnwidth]{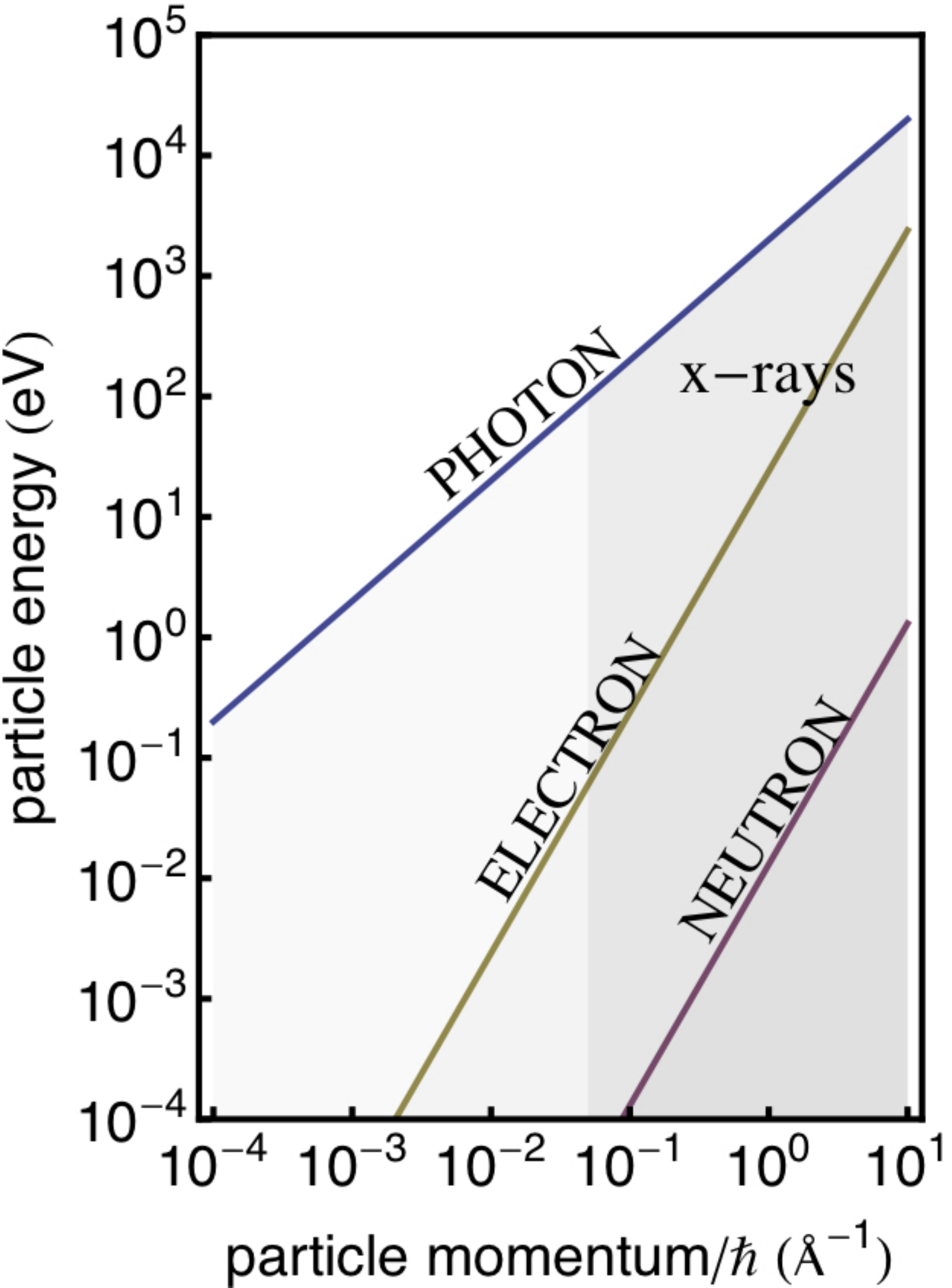}
\end{center}
\caption{(Kinetic) energy and momentum carried by the different elementary particles that are often used for inelastic scattering experiments. The scattering phase space --the range of energies and momenta that can be transferred in a scattering event-- of x-rays is indicated in blue, electrons in brown and neutrons in red. \label{fig:eplot}}
\end{figure}

\subsubsection{Features of RIXS as an experimental method}
Compared to other scattering techniques, RIXS has a number of unique features:  it covers a huge scattering phase-space, is polarization dependent, element and orbital specific, bulk sensitive and requires only small sample volumes. We briefly illustrate these features below and discuss them more extensively in the sections to follow.\\

1. RIXS exploits both the {\it energy and momentum} dependence of the
photon scattering cross-section. Comparing the energies of a neutron,
electron, and photon, each with a wavelength on the order of the
relevant length scale in a solid, {\it i.e.} the interatomic lattice
spacing, which is on the order of a few Angstroms, it is obvious that an x-ray photon has much more
energy than an equivalent neutron or electron, see Fig.~\ref{fig:eplot}. The
scattering phase space (the range of energies and momenta that can be
transferred in a scattering event) available to x-rays is therefore
correspondingly larger and is in fact without equal.
For instance, unlike photon scattering experiments with visible or infrared light, RIXS can probe the full dispersion of low energy excitations in solids.\\

2. RIXS can utilize the {\it polarization} of the photon: the nature
of the excitations created in the material can be disentangled through
polarization analysis of the incident and scattered photons, which
allows one, through the use of various selection rules, to characterize
the symmetry and nature of the excitations. To date, very few experimental facilities allow the polarization of the scattered photon to be measured~\cite{Braicovich2007c,Ishii2010}, though the incident photon polarization is frequently varied.
It is important to note that a polarization change of a
photon is necessarily related to an angular momentum change. Conservation of angular momentum means that any angular momentum lost by the
scattered photons has been transferred to elementary excitations in
the solid. \\

\begin{figure}
\begin{center}
\includegraphics[width=.8\columnwidth]{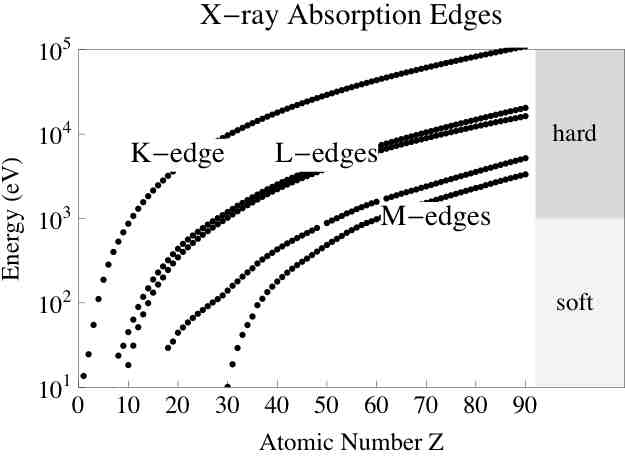}
\end{center}
\caption{Energy of $K$, $L_1$, $L_3$, $M_1$ and $M_5$ x-ray absorption edges as a function of atomic number $Z$. X-ray energies below 1 keV are referred to as soft, above as hard. \label{fig:plotedges}}
\end{figure}

3. RIXS is {\it element and orbital specific}: Chemical sensitivity
arises by tuning the incident photon energy to specific atomic transitions of the different types of
atoms in a material. Such  transitions are called  absorption edges. RIXS can even differentiate between the same
chemical element at sites with inequivalent chemical bondings, with
different valencies or at inequivalent crystallographic positions if the absorption edges in these cases are
distinguishable. In addition, the type of information that may be gleaned about the
electronic excitations can be varied by
tuning to different x-ray edges of the same
chemical element (e.g., $K$ edge for
  exciting $1s$ core electrons, $L$
edge for electrons in the $n=2$ shell or $M$
edge for $n=3$ electrons), since the photon excites different core-electrons into
different valence orbitals at each edge. The energies of these edges are shown in Fig.~\ref{fig:plotedges}. \\

4. RIXS is {\it bulk sensitive}: the penetration depth of resonant x-ray photons is material and scattering geometry-specific, but typically is on the order of a few $\mu$m for photons of 10 keV in the hard x-ray regime, and on the order of 0.1 $\mu$m for photons of 1 keV in the soft x-ray regime. \\

5. RIXS needs only {\it small sample volumes}: the photon-matter interaction is relatively strong, compared to for instance the neutron-matter interaction strength. In addition, photon sources deliver many orders of magnitude more particles per second, in a much smaller spot, than do neutron sources. These facts make RIXS possible on very small volume samples, thin films, surfaces and nano-objects, in addition to bulk single crystal or powder samples.\\

In principle RIXS can probe a very broad class of intrinsic excitations of the system under study -- as long as these excitations are overall charge neutral. This constraint arises from the fact that in RIXS the scattered photons do not add or remove charge from the system under study. In principle then, RIXS has a finite cross section for probing the energy, momentum and polarization dependence of, for instance, the electron-hole continuum and excitons in band metals and semiconductors, charge transfer and $dd$-excitations in strongly correlated materials, lattice excitations and so on. In addition magnetic excitations are also symmetry-allowed in RIXS, because the orbital angular momentum that the photons carry can in principle be transferred to the electron's spin angular moment.
This versatility of RIXS is an advantage but at the same time a complicating factor, because different types of excitations will generally be present in a single RIXS spectrum. An objective of this Review is to help disentangling these where needed.

The generic advantages of the RIXS technique listed above perhaps raise the question as to why this
spectroscopic technique is not as widely used as, say,
angle-resolved photoemission (ARPES) or neutron
  scattering. The main limitation is that the RIXS process is
  photon-hungry -- that is,  it requires a substantial incident photon flux to
  obtain enough scattered photons to collect spectra with a high enough
  resolution in energy and momentum in a reasonable time. With a required resolving power
  (defined as the incident photon energy divided by the energy
  resolution) of four orders of magnitude, RIXS has been a real challenge. Up until a few years ago this has limited
    RIXS experiments to measuring energy losses on the order of half
  an electron volt or greater. Thus neutron scattering and ARPES
  offered a more direct examination of the low energy excitations near the
  Fermi level. However, recent progress in RIXS instrumentation
  has been dramatic and this situation is now changing. One of the purposes
  of this review is to summarize this progress which is beginning to
  elevate RIXS into  an important condensed matter physics tool for probing elementary excitations in solids.

\begin{figure}
\begin{center}
\includegraphics[width=1.0\columnwidth]{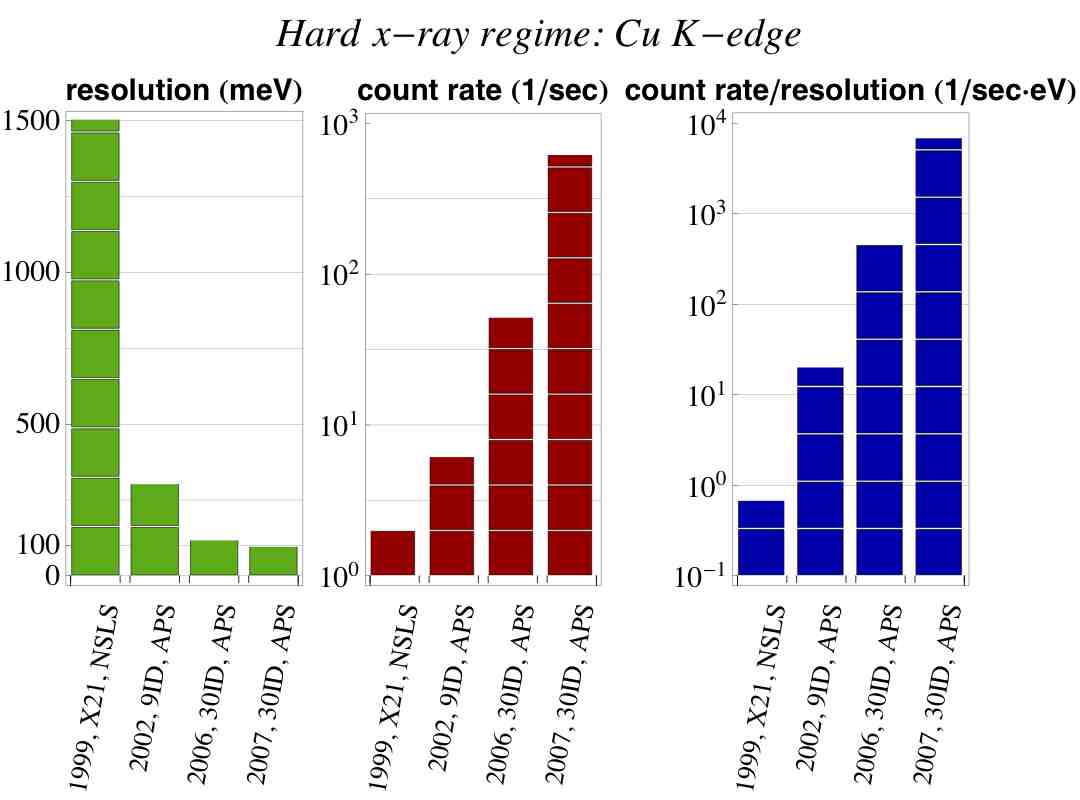}
\end{center}
\caption{Development of RIXS resolution, count rate and resulting orders of magnitude increase in the figure of merit (count rate/resolution) in the hard x-ray regime, as measured by the 6 eV feature in CuGeO$_3$ at the Cu K-edge (8990 eV). \label{fig:hard_RIXS_resolution}}
\end{figure}

\subsubsection{Progress of RIXS in the last decades}
As discussed above, the generic features of RIXS make it, in principle, an attractive
technique to study the intrinsic momentum dependent, low energy
response of a material. But there are of course practical limitations. The most critical of these is the energy
resolution, which is determined both by the availability of the instrumentation necessary to energy resolve the photons, and by the availability of tunable photon sources of sufficient intensity.

In order to tune the incident photon energy to a particular edge, a tunable x-ray photon source is essential. This can be achieved with synchrotron radiation sources and their increase in brilliance over the past decades has been many orders of magnitudes in the $10^3$-$10^4$ eV x-ray regime. The next generation photon sources include x-ray free electron lasers (FELs), which are coming on line at the time of writing. The peak brilliance of these sources is again orders of magnitude larger than that of the third generation synchrotrons and it is likely that these sources will provide further advances, particularly for time-resolved experiments. Such future developments will be touched upon in Section \ref{Summary}.

\begin{figure}
\begin{center}
\includegraphics[width=.45\columnwidth]{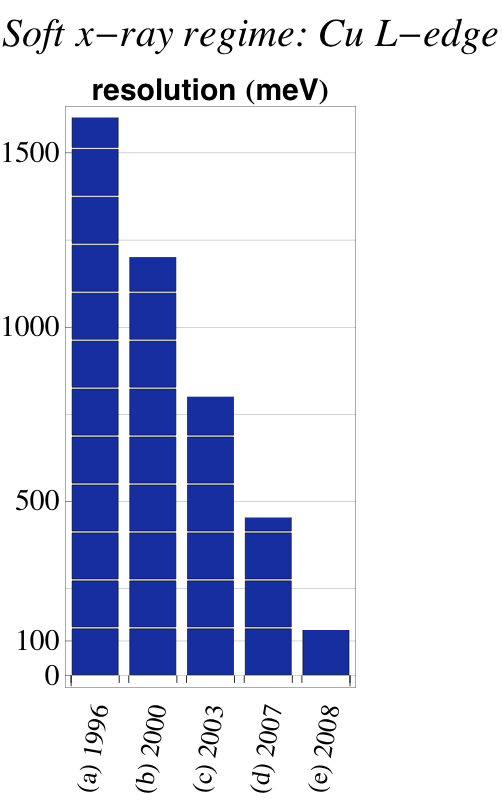}
\end{center}
\caption{Progress in soft x-ray RIXS resolution at the Cu L-edge at 931 eV  (a) \cite{Ichikawa1996}, BLBB @  Photon Factory (b) I511-3 @ MAX II \cite{Duda2000b}, (c) AXES @ ID08, ESRF\cite{Ghiringhelli2004} (d) AXES @ ID08, ESRF\cite{Braicovich2009a}, (e) SAXES @ SLS  \cite{Ghiringhelli}. Courtesy of G. Ghiringhelli and L. Braicovich. \label{fig:soft_RIXS_resolution}}
\end{figure}

This vast increase in photon flux has been matched by advances in the RIXS instrumentation: the monochromators, analyzers and spectrometers.  The resulting increase in resolution of RIXS experiments over time, as measured for instance at the hard x-ray Cu $K$-edges and soft x-ray Cu $L_3$, is shown in Figs.~\ref{fig:hard_RIXS_resolution} and~\ref{fig:soft_RIXS_resolution}, respectively. We will briefly discuss the current state-of-the-art in Sec.~\ref{sec:Experimental methods}

\begin{figure}
\begin{center}
\includegraphics[height=0.7\columnwidth]{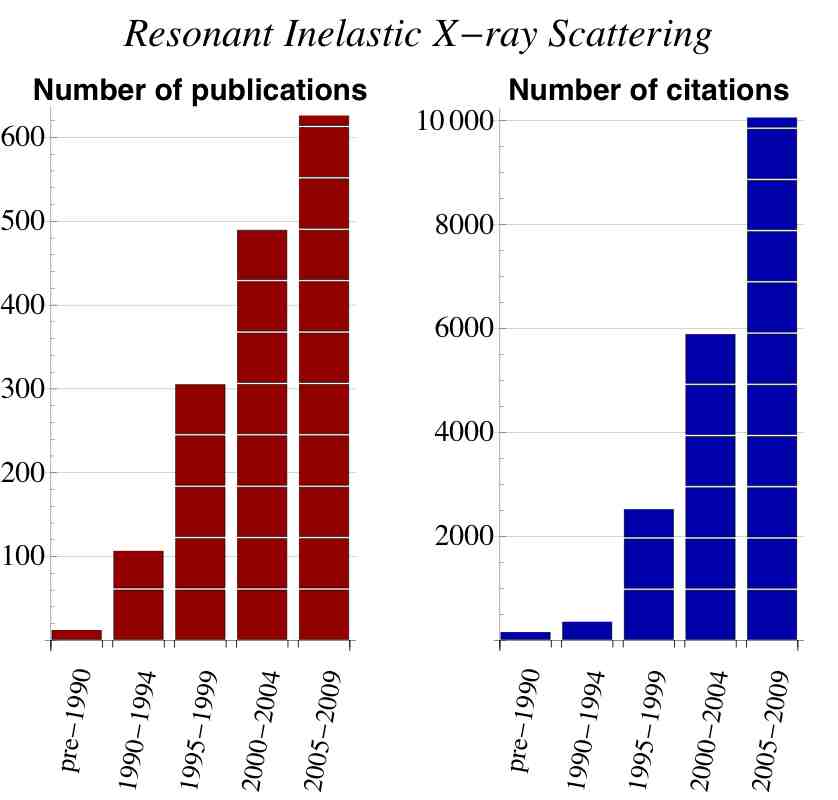}
\end{center}
\caption{Left: Number of publications on RIXS in five year periods. Right: Number of citations to RIXS publications in five year periods (Source: ISI Web of Knowledge). \label{fig:pubchart}}
\end{figure}

In concert with the great progress in the RIXS experiments over the past decades, there has been a similar rapid advance in the theoretical understanding of the scattering process and of the dynamic correlation functions that the technique probes.  Taken together, the theoretical and experimental advances have driven an enormous increase in the number of RIXS-related publications and citations to these publications, as shown in Fig.~\ref{fig:pubchart}.

It seems likely that this strong growth will continue. First, because of the ongoing push to better energy resolutions. Second, and perhaps more importantly, because there are a multitude of different x-ray absorption edges, in particular for the heavier elements in the periodic table, and each one of these can, in principle, be exploited for RIXS measurements. The bulk of RIXS data so far has been collected at transition metal and oxygen edges. This is motivated by the intense scientific interest in strongly correlated transition metal oxides such as the high-$T_c$ cuprate superconductors and the colossal magnetoresistance manganites. This focus on transition metal oxides is an accident of history. It has been very beneficial to the field, driving advances in instrumentation and theory at the relevant edges, but there is clearly a huge potential for growth as interest moves on to other materials and other fields.

\subsubsection{Probing Elementary Excitations with RIXS \label{sec:RIXS_excitations}}

The elementary excitations of a material determine many of its important physical properties, including transport properties and its response to external perturbations. Understanding the excitation spectrum of a system is understanding the system.

In this respect strongly correlated electron materials, e.g. transition metal oxides, are of special interest because the low energy electronic properties are determined by high energy electron-electron interactions (energies on the order of eV's). From these strong interactions and correlations a set of quantum many-body problems emerge, the understanding of which lies at the heart of present day condensed matter physics. Most often this many-body physics is captured in model Hamiltonians, the exact parameters of which must be determined experimentally. RIXS, along with other spectroscopic techniques, can play an important role there, even if it is as a spectroscopic technique applicable to many other materials and of course not limited to correlated systems.

In the following, we discuss the relevant excitation energy and momentum scale on which RIXS can probe the excitation spectrum of a solid. We then briefly introduce the kinds of elementary excitations that are accessible to RIXS.

\begin{figure}
\begin{center}
\includegraphics[width=.9\columnwidth]{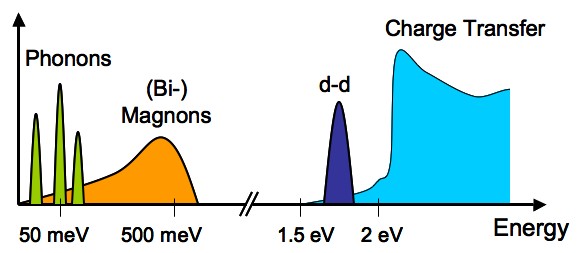}
\end{center}
\caption{Different elementary excitations in condensed matter systems and their approximate energy scales in strongly correlated electron materials such as transition metal oxides. \label{fig:elemenary_excitations}}
\end{figure}

\paragraph{Excitation Energy and Momentum Scale} As is shown in Fig.~\ref{fig:elemenary_excitations}, the elementary excitation spectrum in solids spans the range from plasmons and charge transfer excitations at a few eV, determining for instance optical properties, through excitons, $dd$ excitations and magnons down to phonons at the meV scale. In principle,  RIXS can measure the momentum-dependence of the excitation energy of all these modes -- i.e. their dispersion -- because the photon transfers momentum as well as energy to the material under study.

This is unusual if one is accustomed to optical light scattering, such as Raman scattering~\cite{Devereaux2007}. Photons in the visible range of the spectrum with an energy of a few eV carry negligible momentum compared to the quasi-momentum of the elementary excitations of a solid (fig. 1). A photon of 2 eV has a momentum of roughly $\hbar q= 10^{-27}$ kg~m/s, or a wavevector $q=10^{-3}$ \AA$^{-1}$  whereas elementary excitations in a crystal with a lattice constant of say 3 \AA \ have wavevectors up to $q= 2 \pi /3 \approx 2$ \AA$^{-1}$.  On this scale optical light scattering is in essence a zero momentum probe. To measure the dispersion of elementary excitations for momenta in a sizable portion of a typical Brillouin zone, x-rays with energy on the order of 1 keV or more are needed, corresponding to for instance the Cu $L$-edge.

\paragraph{Overview of elementary excitations} In this paragraph we briefly discuss the different elementary excitations accessible to RIXS -- a detailed review is provided in Chapter~\ref{sec:elementary}.

{\it Plasmons.}
Collective density oscillations of an electron gas are referred to as plasmons. They can be observed by inelastic x-ray scattering (IXS)  or by optical probes since they occur at finite energy for $q$=0. Plasmon-like excitations were also observed early on in RIXS~\cite{Isaacs1996}, but their resonant enhancement with respect to IXS is weak, and little work has been done since, see Section~\ref{sec:plasmons}.

{\it Charge-transfer excitations.}
Charge transport in a condensed matter system is determined by the energetics of moving electrons from one site to another. In a transition-metal oxide, there are two relevant energy scales for this process. The first is the energy associated with an electron hopping from a ligand site to a metal site. This is known as the charge transfer energy, $\Delta$, where $\Delta = E(d^{n+1}\underline{L})-E(d^n)$, and $\underline{L}$ represents a hole on the ligand site. The second energy scale is the energy, $U$, associated with moving a $d$ electron from one metal site to another where $U = E(d^{n+1})+E(d^{n-1}) -2E(d^n)$. Strongly correlated insulators may be classified by which of these two energies is the larger \cite{Zaanen1985}. If $U>\Delta$, then the gap is of the charge transfer type and the system is said to be a charge-transfer insulator. Conversely, if $U<\Delta$, then the gap is controlled by the $d-d$ Coulomb energy and the system is said to be a Mott-Hubbard insulator. 

The bulk of the interesting transition metal oxide compounds, including the cuprates, nickelates and manganites are all in the charge transfer limit. This means the lowest lying excitations across the optical gap are charge transfer excitations and therefore these are of central importance in these materials. Key questions include the size of the gap (typically on the order of a few eV) and the nature of the excitations - do they form bound exciton states? Are these localized or can they propagate through the lattice? What are their lifetimes, symmetries and temperature dependence, etc. While some studies have been performed using other techniques, notably EELS and optical conductivity measurements, RIXS offers a powerful probe for many of these questions and has been applied extensively. We review the experimental results and their theoretical interpretation in section~\ref{sec:CT}.

{\it Crystal Field and Orbital excitations.}
Many strongly correlated systems exhibit an orbital degree of freedom, that is, the valence electrons can occupy different sets of orbitals. Orbitally active ions are also magnetic: they have a partially filled outer shell. This orbital degree of freedom determines many physical properties of the solid, both directly, and also indirectly because the orbitals couple to other degrees of freedom. For instance, the orbital's charge distribution couples to the lattice, and according to the Goodenough-Kanamori rules for superexchange the orbital order also determines the spin-spin interactions. The nature of the orbital degree of freedom -- the orbital ground state and its excitations are an important aspect of strongly correlated systems.

In many Mott insulators this orbital physics is governed by the crystal field: the levels of the orbitally active ion are split and the orbital ground state is uniquely determined by local, single-ion considerations. The orbital excitations from this ground state are transitions between the crystal field levels. Crystal field transitions between different $d$ orbitals are called $dd$ excitations. Such excitations are currently routinely seen by RIXS and are now well understood.

In other cases the crystal field does not split the levels of the outer shell very much, leaving an orbital (quasi-)degeneracy in the ground state. This local low-energy degree of freedom can couple to orbital degrees of freedom on neighboring sites by superexchange processes, and in this way collective orbital excitations can emerge. The quanta of these collective modes are called orbitons, in analogy to spin waves and magnons. Definitive proof of the existence of orbitons remains elusive -- RIXS is contributing significantly to the search for orbitons. The distinctive advantages of RIXS over other spectroscopies when probing orbital excitations raise promising prospects, see Section~\ref{sec:orbitons}.

{\it Magnetic excitations.}
Magnetism and long-range magnetic ordering are arguably the best known and most studied consequences of the electron-electron interactions in solids. When usual magnetic order sets in, be it either of ferro-, ferri- or antiferromagnetic type, the global spin rotation symmetry in the material is broken. As a result characteristic collective magnetic excitations emerge. The resulting low energy quasiparticles, the magnons, and the interactions between them, determine all low temperature magnetic properties. Magnon energies can
extend up to $\sim 0.3$ eV (e.g. in cuprates) and their momenta up to
$\sim 1$ \AA$^{-1}$. Recently magnon dispersions have been measured for the first time at the Cu $L$-edge on thin films of La$_2$CuO$_4$~\cite{Braicovich2009b}. In $K$-edge RIXS bi-magnon excitations and their dispersions have also been observed~\cite{Hill2008}.

A melting of the long-range ordering, for instance through an increase in quantum fluctuations as a result of the introduction of mobile charge carriers in a localized spin system, or by the frustration of magnetic interactions between the spins, can result in the formation of spin-liquid ground states. Spin liquids potentially have elusive properties such as high-temperature superconductivity or topological order, which one is only beginning to explore and understand. Some of the more exotic magnetic excitations that emerge from these ground states, such as spinons and triplons can also be observed by RIXS~\cite{Schlappa2009}. In Section~\ref{sec:magnetic}, we review both the experimental and theoretical status of magnetic RIXS in detail.

\subsection{The RIXS Process}
\begin{figure}
\begin{center}
\includegraphics[height=.65\columnwidth]{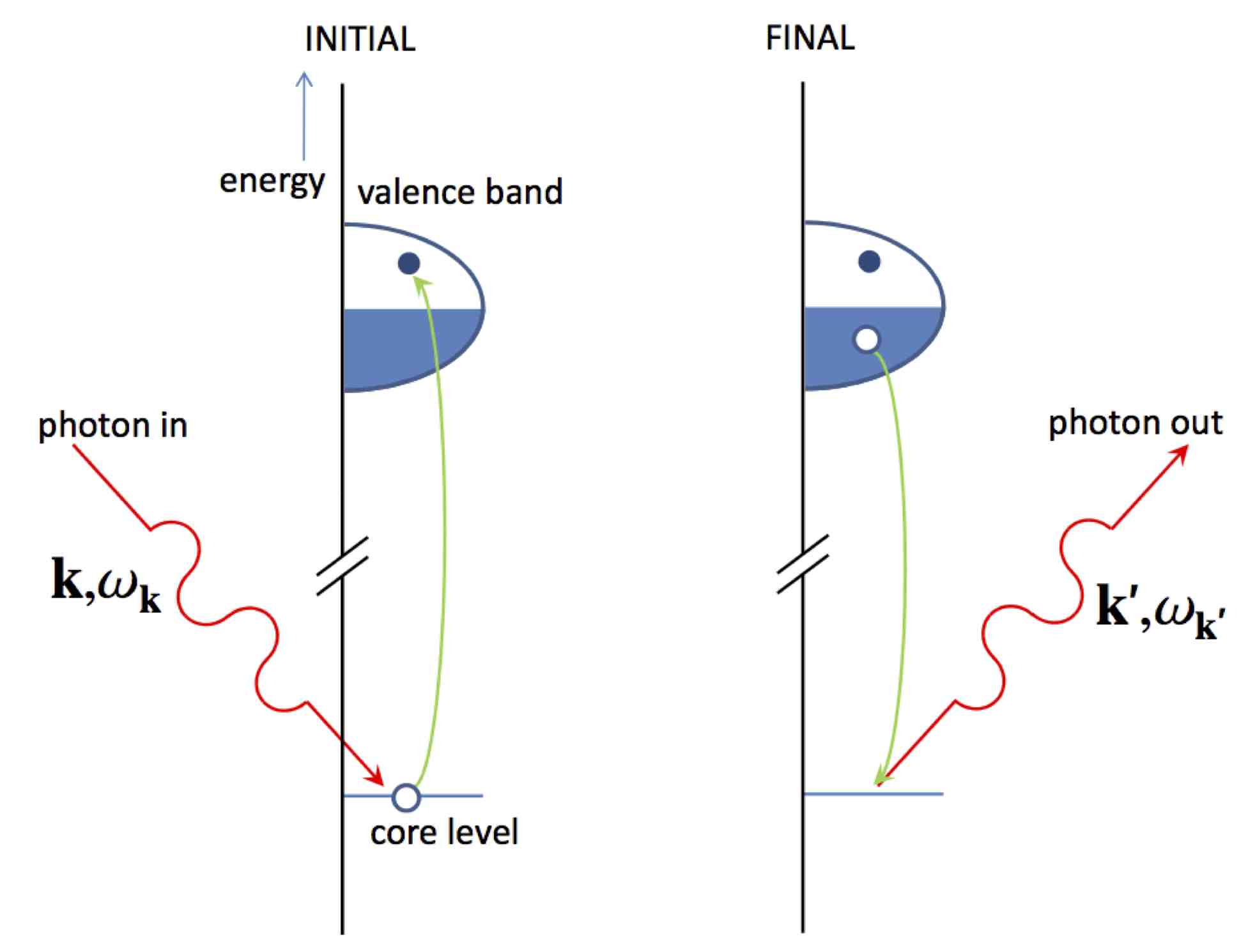}
\end{center}
\caption{In a direct RIXS process the incoming x-rays excite an
  electron from a deep-lying core level into the empty valence. The
  empty core state is then filled by an electron from the occupied
  states under the emission of an x-ray. This RIXS process creates a
  valence excitation with momentum $\hbar {\bf k}'-\hbar {\bf k}$ and
  energy $\hbar \omega_{\bf k}-\hbar \omega_{{\bf k}'}$. \label{fig:direct}}
\end{figure}

The microscopic picture of the resonant inelastic x-ray scattering process is most easily explained in terms of an example. We will choose a copper-oxide material as a typical example, but it should be stressed once more that the focus of RIXS on transition metal oxides is rather an accident of history.
 In a copper-oxide material, one can tune the incoming photon energy to resonate with the copper $K$, $L$ or $M$ absorption edges, where in each case the incident photon promotes a different type of core electron into an empty valence shell, see Figs. ~\ref{fig:direct} and ~\ref{fig:indirect}. The electronic configuration of Cu$^{2+}$ is $1s^22s^22p^63s^23p^63d^9$, with the partially filled $3d$ valence shell characteristic of transition metal ions. The copper $K$-edge transition $1s \rightarrow 4p$, is around 9000 eV and in the {\it hard x-ray} regime. The $L_{2,3}$-edge $2p \rightarrow 3d$ ($\sim$ 900 eV) and $M_{2,3}$-edge $3p \rightarrow 3d$ ($\sim$ 80 eV) are {\it soft x-ray} transitions. Alternatively, by tuning to the Oxygen $K$-edge, one can choose to promote an O $1s$ to an empty $2p$ valence state, which takes $\sim$ 500 eV.

After absorbing a soft or hard x-ray photon, the system is in a highly energetic, unstable state: a hole deep in the electronic core is present. The system quickly decays from this intermediate state, typically within 1-2 femtoseconds. Decay is possible in a number of ways, for instance via an Auger process, where an electron fills the core-hole while simultaneously emitting another electron. This non-radiative decay channel is not relevant for RIXS, which instead is governed by fluorescent decay, in which the empty core-state is filled by an electron and at the same time a photon is emitted.

There are two different scattering mechanisms by which the energy and momentum of the emitted photon can change from the incident one. These are known as {\it direct} and {\it indirect} RIXS. The distinction between these two is discussed below.

\subsubsection{Direct and Indirect RIXS}
\label{sec:Direct and Indirect RIXS}
Resonant inelastic x-ray scattering processes are classified as either
direct or indirect~\cite{Brink2005,Brink2006}. This distinction is
useful because the cross-sections for each are quite different. When
direct scattering is allowed, it is the dominant inelastic scattering channel, with indirect processes contributing only in higher order. In contrast, for the large class of experiments for which direct scattering is forbidden, RIXS relies exclusively on indirect scattering channels.

\paragraph{Direct RIXS}
For direct RIXS, the incoming photon promotes a core-electron to an empty valence band state, see Fig.~\ref{fig:direct}. Subsequently an electron from a {\it different} state in the valence band decays and annihilates the core-hole.

The net result is a final state with an electron-hole excitation, since
an electron was created in an empty valence band state and a hole in the
filled valence band, see Fig.~\ref{fig:direct}. The electron-hole
excitation can propagate through the material, carrying momentum
$\hbar {\bf q}$ and energy $\hbar \omega$. Momentum and energy
conservation require that ${\bf q}={\bf k^\prime}-{\bf k}$ and
$\omega=\omega_{\bf k^\prime}-\omega_{\bf k}$, where  $\hbar {\bf k}$ ($\hbar {\bf k^\prime}$) and  $\hbar \omega_{\bf k}$ ($\hbar \omega_{\bf k^\prime}$)  are the momentum and energy of the incoming (outgoing) photon, respectively.

For direct RIXS to occur, both photoelectric transitions -- the
initial one from core to valence state and succeeding one from
conduction state to fill the core hole -- must be
possible. These transitions can for instance be an
initial dipolar transition of $1s \rightarrow 2p$ followed by the
decay of another electron in the $2p$ band from $2p \rightarrow 1s$,
in for example wide-band gap insulators. This happens for instance at the
$K$-edge of oxygen, carbon and silicon. At transition metal
$L$-edges, dipole transitions give rise to direct RIXS via  $2p \rightarrow 3d$ absorption and subsequent  $3d \rightarrow 2p$ decay. In all these cases, RIXS probes the valence and conduction states directly. Although the direct transitions into the valence shell dominate the spectral line shape, the spectral weight can be affected by interactions in the intermediate-state driven by, for example, the strong core-hole potential.

\begin{figure}
\begin{center}
\includegraphics[height=.65\columnwidth]{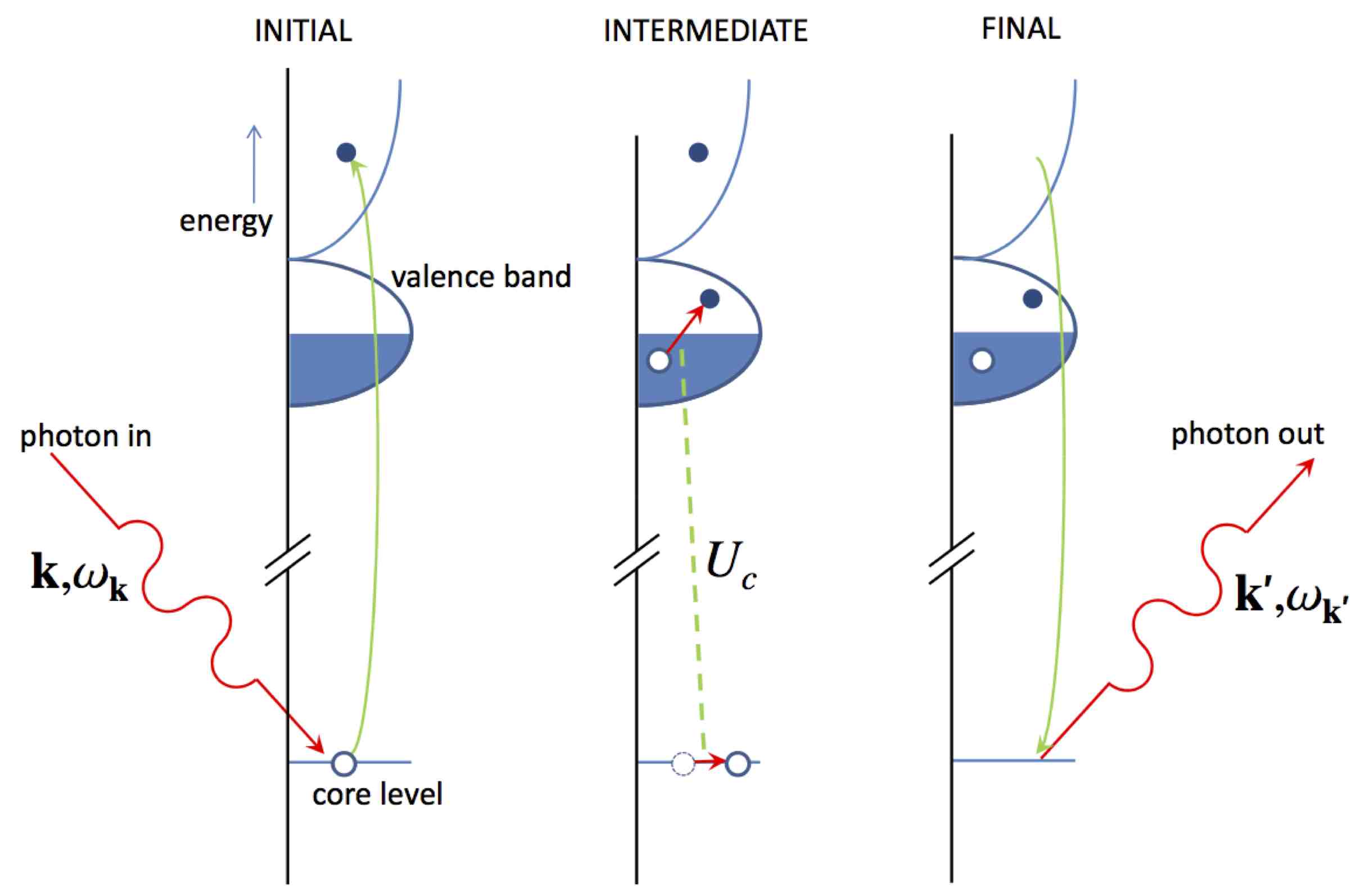}
\end{center}
\caption{In an indirect RIXS process, an electron is excited from a deep-lying core level into the valence shell. Excitations are created through
the Coulomb interaction $U_c$ between the core hole (and in some cases the excited electron) and the valence electrons.
\label{fig:indirect}}
\end{figure}

\paragraph{Indirect RIXS}
The indirect RIXS process is slightly more complicated. For pure indirect RIXS to occur, photoelectric transitions from the core-state to conduction-band states must be weak. Instead, the incoming photon promotes a core-electron into an empty state several electron volts above the Fermi level. Subsequently the electron from this same state decays again to fill the core-hole, see Fig. \ref{fig:indirect}. The most studied example is RIXS at the transition-metal $K$-edges ($1s\rightarrow 4p$). Obviously, in the absence of any additional interaction, no inelastic scattering would be observed. But in the intermediate state a core-hole is present, which exerts a strong potential on the $3d$ valence electrons, that therefore tend to screen the core-hole. The core-hole potential scatters these valence electrons, thereby creating electron-hole excitations in the valence band. After the $4p \rightarrow 1s$ decay, the electron-hole excitations are then left behind in the system.

Indirect RIXS is thus due to shakeup excitations created by the intermediate state core-hole. The fact that close to the absorption edge the $1s$ core-hole and $4p$ electron bind together to form an exciton does not change this picture conceptually -- in this case, one may think of the valence electrons as scattering off this exciton.
\subsection{Relation to other spectroscopies}

\subsubsection{RIXS, (N)IXS, XAS, XES and EELS}

{\it Resonant inelastic x-ray scattering} is sometimes also referred to as {\it resonant x-ray emission} spectroscopy, {\it resonant x-ray fluorescence} spectroscopy, and {\it resonant Raman} spectroscopy. These different names all indicate essentially the same process. Throughout this review, we will only use the term resonant inelastic x-ray scattering (RIXS).

RIXS is a two-step process involving a photoelectric transition from a deep-lying core level into the valence shell, with the material absorbing an x-ray photon. After evolving for a very short period in this highly-excited state, the system decays under the  emission of an x-ray photon.  

The term inelastic x-ray scattering (IXS) is reserved for processes where the photon scatters inelastically by interacting with the charge density of the system. IXS measures the dynamic charge structure factor $S_{\bf q}(\omega)$. To emphasize the difference with RIXS, IXS is sometimes also referred to as {\it non-resonant IXS} (NIXS).

{\it X-ray absorption spectroscopy} (XAS) is the first step of the RIXS process and thus RIXS is related to it. A common way to measure XAS is to study the decay products of the core hole that the x-ray has created, either by measuring the electron yield from a variety of Auger and higher-order processes (known as electron yield) or by measuring the radiative decay (fluorescence yield). The total fluorescence yield corresponds approximately to the integration of all possible RIXS processes.

The term {\it X-ray emission spectroscopy} (XES) is generally used for the situation where one studies the radiative decay of a core hole created by X-rays or electrons with an incident energy far away from an absorption edge. However, even with the incident energy not tuned to a resonance, its theoretical description is much closer to RIXS than to IXS, with the modification that the initial excitation is not into the valence shell, but into the vacuum continuum at much higher energy.

{\it Electron-energy-loss spectroscopy} (EELS), is comparable to RIXS and NIXS in the sense that in the inelastic electron scattering process charge neutral excitations are created. The EELS cross section is determined by the charge structure factor $S_{\bf q}(\omega)$ of the system under study \cite{Schnatterly1979,Schuster2009} and is, therefore, closely related to NIXS. When dealing with excitations in the valence shell,  the RIXS cross section can, within certain theoretical approximations, also be related to the structure factor, see Section~\ref{sec:effindirect}. It should be stressed, however, that to what extent, and under what circumstances, the RIXS response is proportional to $S_{\bf q}(\omega)$ is still an open question and an active subject of theoretical and experimental investigations.

The EELS intensity is limited due to space-charge effects in the beam. It has the advantage that it is very sensitive for lower momentum transfers, around $q=0$ but its intensity rapidly decreases for large $q$. Further, at large $q$, multiple scattering effects become increasingly important, making it hard to interpret the spectra above $q \sim$ 0.5 - 1.0 \AA$^{-1}$. Since EELS is measured in transmission it requires thin samples and measurements in the presence of magnetic or electric fields are not possible due to their detrimental effect on the electron beam. There are no such restrictions for x-ray scattering and EELS therefore provides spectroscopic information that is complementary to IXS/RIXS. Comparing spectra measured by these different techniques can be a fruitful method  both the physical properties of the materials investigated and to deepen our understanding of the techniques themselves.

\subsubsection{Raman and Emission peaks in RIXS spectra}

\begin{figure}
  \begin{center}
  \includegraphics[width=0.8\columnwidth]{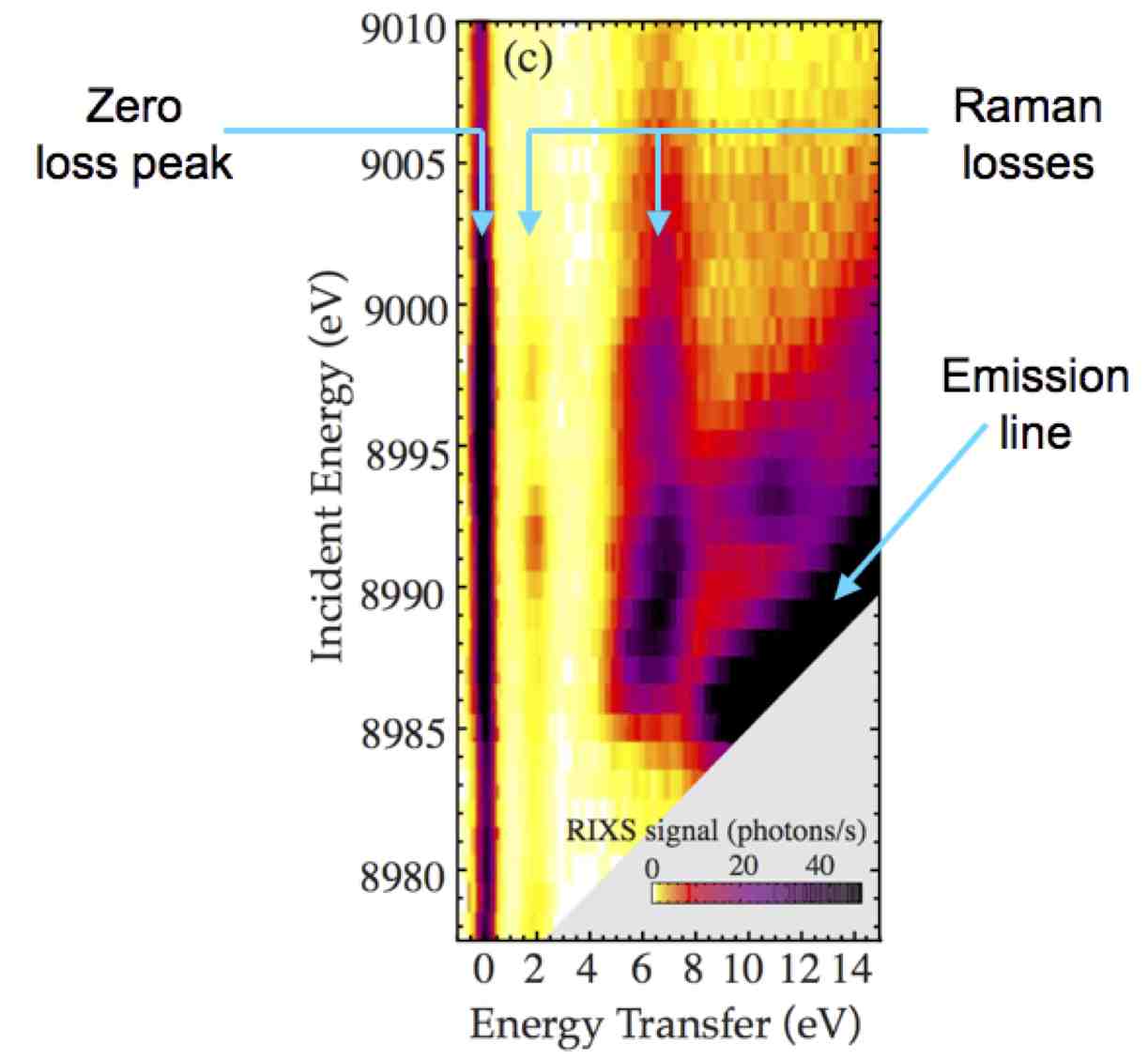}
  \end{center}
   \caption{Two classes of inelastic energy-loss features in the RIXS spectrum of CuB$_2$O$_4$~\cite{Hancock2009}. The RIXS intensity is represented on a color scale versus transferred energy (energy loss) $\hbar \omega$ and incident energy $\hbar \omega_{\bf k}$. The zero-loss line is the vertical line at zero transferred energy. X-ray Raman features are parallel to that line and tend to resonate strongly at a specific incident energy, for instance the $\hbar \omega=$2 eV loss feature resonating at $\hbar \omega_{\bf k}=$8992 eV. Emission lines appear as diagonal features in this  $\hbar \omega$ - $\hbar \omega_{\bf k}$ plot because in this case energy of the emitted photon $\hbar \omega_{{\bf k}'}$ is roughly constant. \label{fig:Hancock2009_2}}
\end{figure}

Two classes of inelastic energy-loss features are distinguished in RIXS, depending on how their energy $\hbar \omega$ changes upon changing the incident energy $\hbar \omega_{\bf k}$. For one set of RIXS features, the energy $\hbar \omega$ lost by the photon and transferred to the material is independent of  the incoming photon energy $\hbar \omega_{\bf k}$ and they appear as vertical lines in a plot of transferred energy $\hbar\omega$ versus the incoming energy $\hbar\omega_{\bf k}$, see  Fig.~\ref{fig:Hancock2009_2}. This implies that when the incident photon energy is increased, the energy of the outgoing photons $\hbar \omega_{{\bf k}'}$ increases by the same amount. This is similar to Stokes phonon losses measured in optical Raman scattering. Using this analogy one can refer to these energy-loss peaks as {\it resonant x-ray Raman} features.

The other class of loss features is characterized by the energy of the outgoing photons $\hbar \omega_{{\bf k}'}$ being independent of the incident energy  $\hbar \omega_{\bf k}$. This happens when the absorption process promotes a deep core-electron to the vacuum continuum far above the Fermi level and subsequently an electron from, for example, a shallow core-level decays radiatively, filling the deep core and emitting a photon. The energy of the x-ray photon emitted in such a core-core transition is constant, i.e. independent of the incident energy $\hbar \omega_{\bf k}$. Thus in this situation, when the incident energy $\hbar \omega_{\bf k}$ increases,  the energy lost by the photon $\hbar \omega$ increases by the same amount. As the presence of such loss features does not depend on the incident energy being close to an absorption edge or a resonance, one can refer to such peaks in the RIXS spectrum as {\it x-ray emission} or {x-ray fluorescence} features.

In a plot of the transferred energy $\hbar\omega$ versus the incoming energy $\hbar\omega_{\bf k}$, x-ray emission features move along the diagonal, see Fig. \ref{fig:Hancock2009_2}.  Note that both Raman and emission features can appear close in energy. One should also consider that for incident photon energies {\it below} the edge, emission lines occur at constant energy loss. Thus, a full description of the RIXS cross section should include both phenomena.
\subsection{Aim of this review}

Here, we review the field of resonant inelastic x-ray scattering from elementary excitations in solids, with a strong focus on strongly correlated materials. This implies we do not include RIXS on gas phase atoms or molecules, reviewed in~\textcite{Salek2003,Rubensson2000}, liquids, or non-crystalline and biologically relevant transition metal complexes~\cite{Glatzel2005}. Elementary excitations here are understood as excitations of electronic, lattice, orbital or magnetic origin with an energy of up to a few electron volts above the Fermi level. Such low-energy excitations have become accessible experimentally in the past decade due to the progress in instrumentation and x-ray sources. We discuss both these experimental developments and the simultaneous theoretical developments. Pre-2001 RIXS work has been reviewed by~\textcite{Kotani2001}. Other reference material includes the books of~\textcite{Schuelke2007} and~\textcite{Groot2008}.

\subsection{Structure of this Review}

In the following section, we  survey the experimental methods and foundations of RIXS, including the practicalities of the method, and instrumentation used in soft and hard x-ray RIXS. Sections~\ref{sec:Theory_fundamentals} and \ref{sec:Calculating_crosssection}  provide a self-contained review of the theory of RIXS, including effective theories and approximate descriptions that are used to calculate the complex RIXS process.

Readers not interested in the experimental methods or theoretical background can directly proceed to Section~\ref{sec:elementary}, where the experimental and theoretical investigations of the past decade on the different elementary excitations measured by RIXS are reviewed side by side. These include plasmons, inter-band and charge transfer excitations, charge density wave, crystal field, orbital excitations, orbitons and finally magnetic excitations. A summary and outlook is presented in Section~\ref{Summary}. An appendix on spherical tensors and a table of symbols are included at the end of this review.

\section{Experimental methods and foundations}
\label{sec:Experimental methods}

\subsection{Practicalities of the method}

\subsubsection{Choice of Absorption Edge}
The first choice for an experimentalist interested in doing a RIXS experiment is which absorption edge to work at. The most common edges for transition-metal compounds are the transition-metal $L$- and $K$-edges; less work has been done on transition-metal $M$-edges and ligand edges, such as the oxygen $K$ edge. Although at first, one might assume that there should be little difference, it  turns out that different valence excitations have significantly different spectral weights at different absorption edges.

The transition-metal $L$- ($2p\rightarrow 3d$) and $M$- ($3p\rightarrow 3d$) edges are in many ways comparable. The dipolar transitions directly involve the  $3d$ orbitals. This allows access to all $3d$ orbitals providing the experimentalist with the opportunity to study $dd$ excitations and orbital effects. In addition, since spin is not a good quantum number in the intermediate state, one is also able to create single-magnon excitations.  Since the excited electron effectively screens the core hole, there is relatively little shake up apart from the excitations made by the dipolar transitions.

At the transition-metal $K$-edge ($1s\rightarrow 3d/4p$), direct transitions into the $3d$ are quadrupolar and therefore relatively weak. The major RIXS intensity is related to the $1s\rightarrow 4p$ transitions. Since the excited $4p$ electron poorly screens the core hole, significant shake-up processes occur due to the transient presence of a Coulomb potential related to the $1s$ hole and $4p$ electron. These can include transitions across the charge-transfer gap. Since x-rays in the $K$-edge absorption region carry significant momentum, these charge excitations can be studied as a function of both energy and momentum. Since no spin-orbit coupling is present in the intermediate state, spin is a good quantum and one cannot observe single magnon excitations, although spin-conserving excitations, such as bi-magnons are certainly possible. The ligand edge is used less often and provides complementary information to $L$- and $M$-edge RIXS.

\subsubsection{Choice of Incident Energy}
\begin{figure}
  \begin{center}
  \includegraphics[width=1.0\columnwidth]{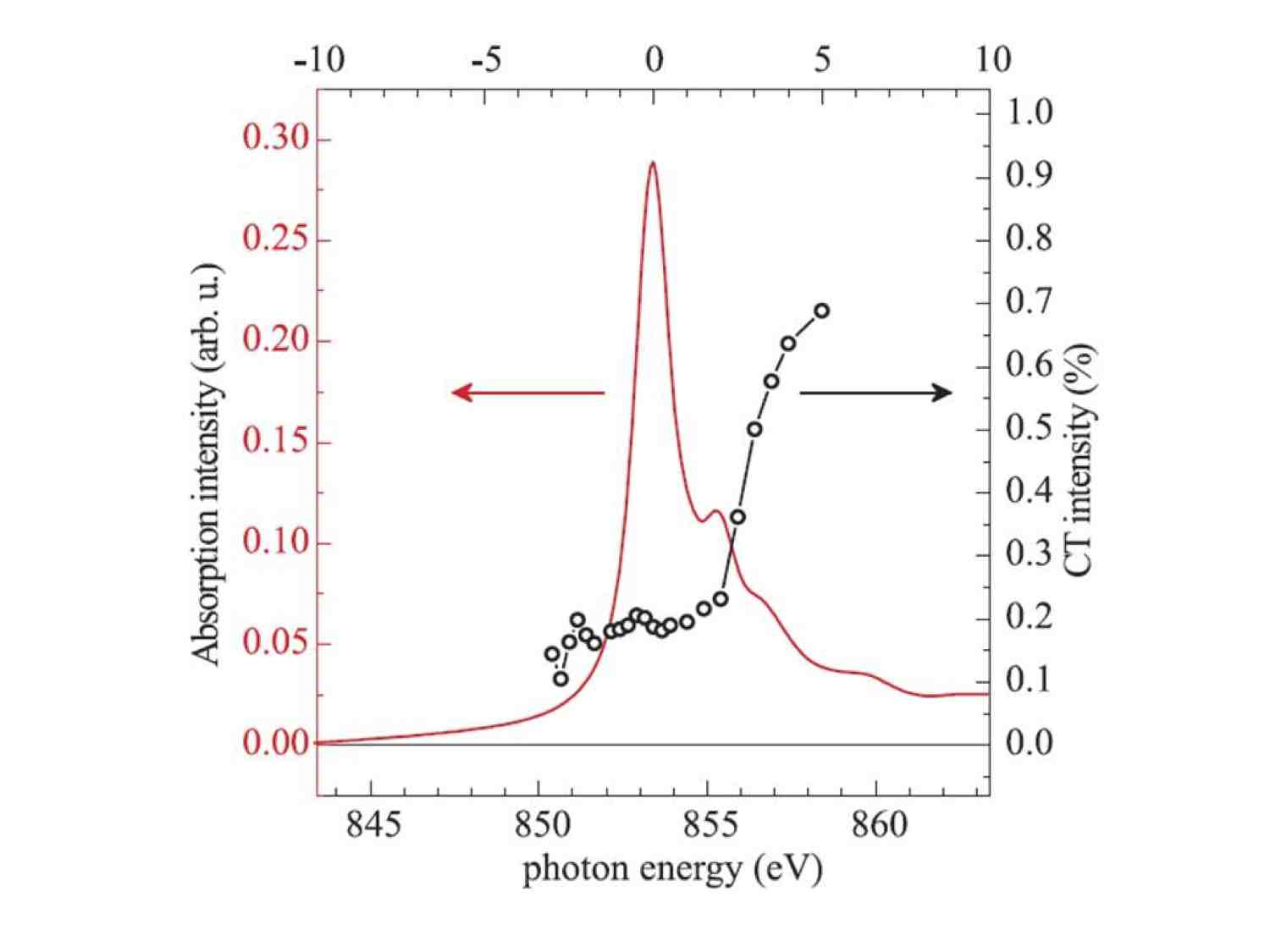}
  \end{center}
    \caption{Relative intensity of charge transfer spectral weight in NiO (defined as all intensity between 5 eV and 20 eV energy loss) compared to the whole RIXS spectrum. The charge transfer processes dominate for incident energies more than 4 eV above the peak in the absorption. Figure taken from Ghiringhelli {\em et al.} \cite{Ghiringhelli2005}}
 \label{fig:LedgeCT_IntDep}
  \end{figure}

After deciding upon an excitation edge, one has the choice of incident energy. In general, in an experimental run, scans are made for a range of different energies. The intensities of the features can strongly depend on the incoming photon energy. At an $L$-edge, probing the main absorption line usually leads to transitions that are metal centered. When using energies in the satellite region above the main line, the charge transfer features are emphasized. Figure \ref{fig:LedgeCT_IntDep} shows this for NiO - a few eV above the absorption maximum, the inelastic processes are dominated by CT excitations. At the $K$ edge, one can distinguish several regions. In the weak pre-edge region, direct excitations are made into the $3d$ shell, making it comparable to RIXS at the $L$- and $M$-edges. Here, $dd$ excitations are emphasized. For the main edge, the RIXS spectrum is dominated by charge-transfer excitations. The most efficient screening mechanisms appear at the lowest energies. For example, for cuprates, the most efficient screening mechanisms are excitations across the charge-transfer gap. Scanning further above the edge into regions where the core hole is not well screened, increases the intensities of the unscreened final states at higher energies.

\subsubsection{Relative Intensity of Elastic Line}

One of the most striking features of RIXS  is that the size of the elastic line relative to the inelastic scattering is very different at the $K$-, $L$- or $M$-edges . For $K$- and $M$-edge experiments, the elastic scattering is so large that it is typically the tails of this peak that determine whether a specific low energy feature can be seen, and not the resolution Full Width at Half Maximum (FWHM) -- although a better resolution narrows the elastic peak. In contrast, for $L$-edge experiments, the inelastic scattering typically dominates the elastic. It is therefore important to understand the elastic scattering, and to the extent possible seek to minimize it in order to carry out a successful experiment.

There are several contributions to the elastic scattering. The first of these are phonons. While strictly inelastic processes, scattering from phonons is generally unresolved and therefore contributes to the apparent $\hbar \omega = 0$ signal. For single crystals of high quality, this may contribute as much as 50\% of the elastic intensity at room temperature and may be minimized by cooling to as low a temperature as possible.  The overall intensity of the phonon scattering varies as $(\bf{q} \cdot \bm{\delta})^2$, where ${\bm \delta}$ is the displacement vector for the atomic motion, and $\bf{q}$ is the total momentum transfer. As such it increases rapidly at the higher momentum transfers where $K$-edge data tends to be taken. In contrast, the smaller ${\bf q}$ at $L$- and $M$-edges may be one factor why the elastic response is smaller in these experiments.

Other contributions to the elastic scattering include diffuse scattering due to strain or defects in the crystal. Diffuse scattering is spread throughout reciprocal space, peaking in the vicinity of Bragg peaks in characteristic patterns and is commonly known as Huang diffuse scattering. This scattering also scales with $q^2$ and so becomes an increasing problem at higher momentum transfers. The contribution from such scattering is thus minimized by choosing crystals of as high a crystalline quality as possible and by working at low $q$.

A similar source of elastic scattering is the crystalline surface quality. A rough surface will produce diffuse, static scattering spread throughout reciprocal space. For rough surfaces, the integrated intensity of this scattering  drops off rapidly away from Bragg peaks. Very smooth surfaces will give rise to rods of scattering connecting the Bragg peaks along directions normal to the surface, known as crystal truncation rods. These may be avoided by moving away in reciprocal space in a direction parallel to the surface.

All the contributions listed above are seen in the non-resonant scattering channel, and as such have ${\bm \epsilon} \cdot {\bm \epsilon}^{'}$ polarization dependence, where ${\bm \epsilon}  ({\bm \epsilon}^{'})$ is the incident (outgoing) photon polarization. As such, these contributions may be minimized by working with a scattering geometry for which the incident photon is polarized in the plane of scattering ($\pi$ incident) and the scattering angle is 90$^{\circ}$. Such scattering geometries are used routinely in soft x-ray experiments working at the $L$- and $M$-edges, but until recently have been less commonly used for $K$-edges experiments, and this is another reason why the $K$-edge elastic intensity can appear large. Of course such requirements are at odds with the desire to move around in reciprocal space and investigate the dispersion of a given excitation. There are certain cases - for example, one or two dimensional materials, where the dispersion in one or two directions may be negligible and one may take advantage of this fact to maintain the 2$\theta$ = 90$^{\circ}$ condition whilst still varying the  momentum transfer in a relevant direction. This has proved of great utility in studying such low dimensional materials at both $L$- and $K$-edges.

Finally, when considering the relative strength of the elastic and inelastic contributions, one must consider the size of the resonant enhancement for the inelastic channel. For the $L$-edge experiments, it is typically, though not always, a direct RIXS process, and as such the inelastic scattering is relatively strong. Further, the resonances are large. This latter fact is evidenced by the strong ``white line'' features seen at $L$-edges in absorption data and is contrast with the very weak peaks, or absence of peaks in $M$- and $K$-edge XAS spectra. As a result the resonant enhancements are significantly larger for the $L$-edge experiments.  The net result of these two effects is that the inelastic scattering is much stronger at an $L$-edge than a $M$- or $K$-edge.

\subsubsection{Momentum Dependence}

Generally, RIXS spectra on elementary electronic excitations show a dependence on the momentum that the photon transfers to the solid. This experimental observation concurs with the theoretical observations of Platzman and Isaacs~\cite{Platzman1998} who argued that while the matrix elements involved in the resonant scattering process are all highly local, the coherent sum of such transitions on different sites would lead to overall crystal momentum conservation, and further, that momentum-breaking processes such as the creation of phonons would be small (by a factor ${\omega_D}/{\Gamma}$, where $\omega_D$ is the Debye frequency and $\Gamma$ the core-hole lifetime).

In exploring the momentum dependence of the inelastic scattering in a $K$-edge RIXS experiment, one has a choice of which Brillouin zone in which to take the data (for TM $L$-edge and $M$-edge RIXS experiments, the energy is such that the photon momentum is within the first Brillouin zone).  In studies of La$_2$CuO$_4$, \textcite{Kim2007} showed that to a first approximation the spectra did not depend on which Brillouin zone they were measured in.
They therefore concluded that the momentum dependence observed in RIXS experiments arises from the crystal lattice, i.e from electronic states obeying Bloch's theorem and
respecting the periodicity ${\bf q}={\bm \kappa} +  n{\bf G}$, where ${\bf q}$ is the total momentum, ${\bm \kappa}$ the reduced wave-vector and ${\bf G}$ a reciprocal lattice vector\footnote{Note that contrary to the present convention many papers use ${\bf Q}$ and ${\bf q}$ as the total and reduced momentum, respectively.}. The reason that the observed RIXS spectra depend on ${\bm \kappa}$ but not on $n{\bf G}$ is that the local core-hole interaction is the main excitation mechanism, and this has the translational symmetry of the lattice. This point will be discussed in detail in Section \ref{sec:effindirect}.

A detailed study of the Brillouin Zone dependence of the zone center scattering in Nd$_2$CuO$_4$ was carried out by \textcite{Chabot-Couture2010}. After careful normalization of their data, the authors conclude that there was a small Brillouin zone dependent component of about 10\% of the inelastic signal. Its intensity is best described by an empirical form; $\cos^2(\theta - \eta)$, where $\theta$ is half the scattering angle and $\eta$ the deviation of the scattering vector from the surface normal (in this case along the c-axis of the crystal). \textcite{Chabot-Couture2010} argue that the functional form originates from a photon polarization dependence of the cross-section -- not the momentum-dependence of the excitation spectrum of the solid. This implies it is a correction to the {\it 4p as a spectator} approximation to $K$-edge RIXS discussed in section~\ref{sec:effindirect}.

\begin{figure}
  \begin{center}
  \includegraphics[width=0.6\columnwidth]{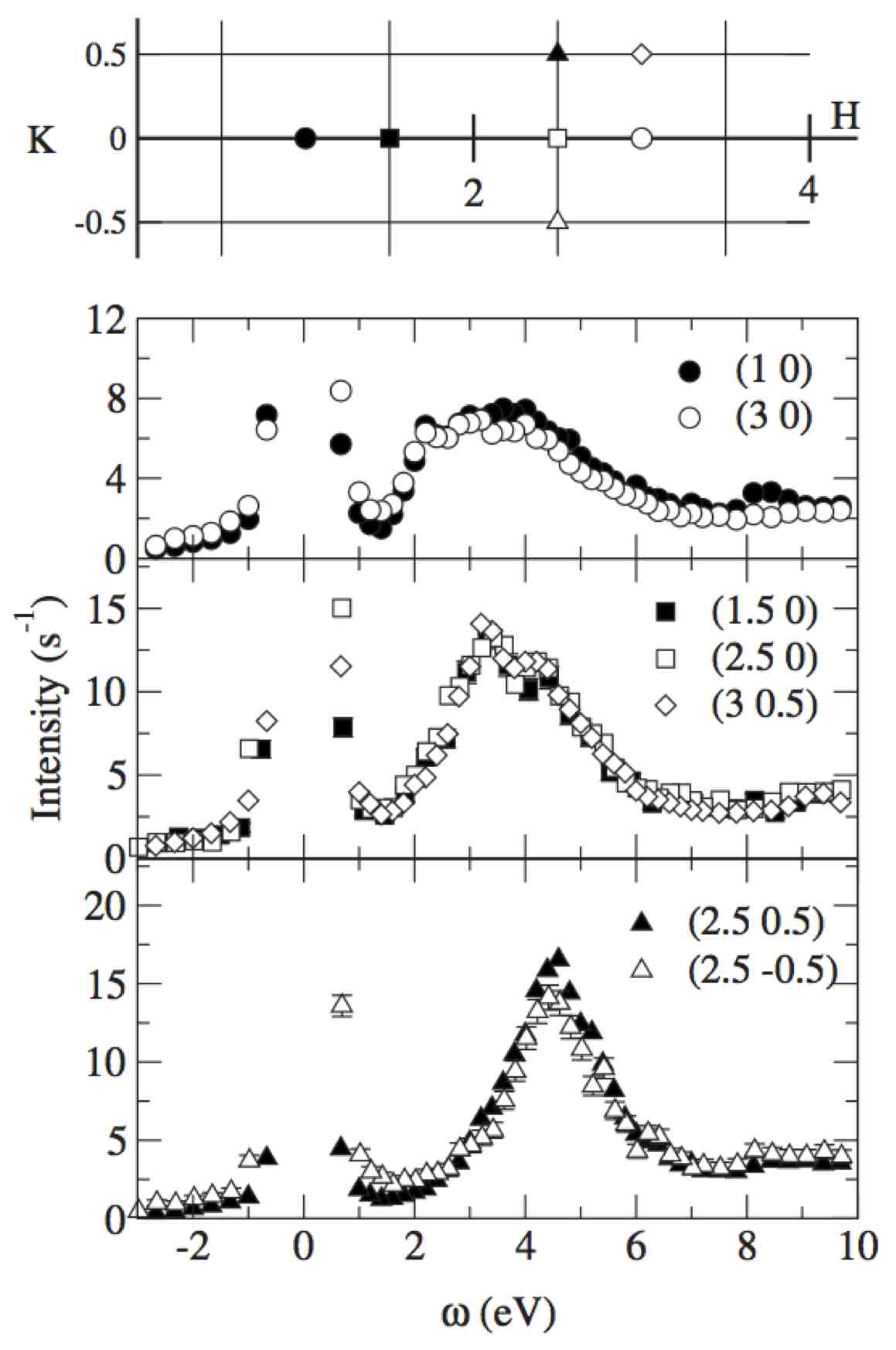}
  \end{center}
  \caption{Comparison of RIXS spectra on La$_2$CuO$_4$ taken at different total q positions but with an equivalent reduced wave vector $\kappa$ for three different q's corresponding to ${\boldsymbol \kappa} =(0,0)$, $(\pi,0)$ and $(\pi,\pi)$, from top to bottom. The two-dimensional reciprocal space net is shown in the top panel. The same symbols are used both to denote the q position in reciprocal space and to show the spectra. All data were obtained with $\hbar \omega_{\bf k}$=8991 eV \cite{Kim2007}. }
\label{fig:Kim2007_2}
\end{figure}

\subsubsection{Normalization of data}

During the RIXS scattering process, the incident and outgoing photons are attenuated by Auger emission and core-level fluorescence as they propagate through the sample. This  attenuation, called self absorption, depends strongly on the scattering geometry used, and determines the illuminated sample volume. Correcting for this extrinsic effect is thus important in order to compare spectra collected in different scattering geometries. In Ref.~\cite{Chabot-Couture2010}, the authors describe how an extrinsic 20-25 \% attenuation (or enhancement) of the measured Cu $K$-edge RIXS spectra can be corrected for by using the fluorescence signal of a known emission line (measured in the same scattering geometry) and Cu $K$-edge X-ray absorption spectra. Similar considerations also apply for $L$- and $M$-edge experiments and in general care should be taken to properly account for self absorption.
\subsection{Instrumentation}

A RIXS instrument may be divided into two parts, the {\it beamline}, which is required to deliver a well-collimated, highly monochromatic, and often focused, beam onto the sample, and the {\it spectrometer}, which must provide the sample environment, manipulate the scattering angles to vary the momentum transfer, collect the scattered radiation over an appropriate solid angle, and energy analyze the scattered radiation. All of these are highly non-trivial tasks and it has taken a great deal of work performed over a large number of years by a large number of people to reach the current state-of-the-art. In parallel to the instrumentation development there has been a tremendous development of synchrotron sources in terms of ever improving brightness in the relevant energy ranges. The combined effect is that there has been dramatic improvement in energy resolution that can be achieved in momentum-resolved RIXS instruments and still have sufficient count rates for experiments to be performed in a finite amount of time.

Hard x-ray instruments are invariably built around Bragg crystal optics to achieve the very high energy resolution required, ultimately benefitting from the very highly ordered crystalline lattice of (typically) Si or Ge that is well matched to the x-ray wavelengths being utilized. Such schemes do not work well in the soft x-ray regime, where the long wavelengths prohibit the use of crystal optics. Instead artificial periodic structures are utilized - diffraction gratings.  This difference means a hard x-ray RIXS spectrometer is fundamentally different from a soft RIXS instrument. In the following two sections, an example of each is presented.

\subsubsection{A soft RIXS instrument}

\begin{figure}
  \begin{center}
  \includegraphics[width=0.8\columnwidth]{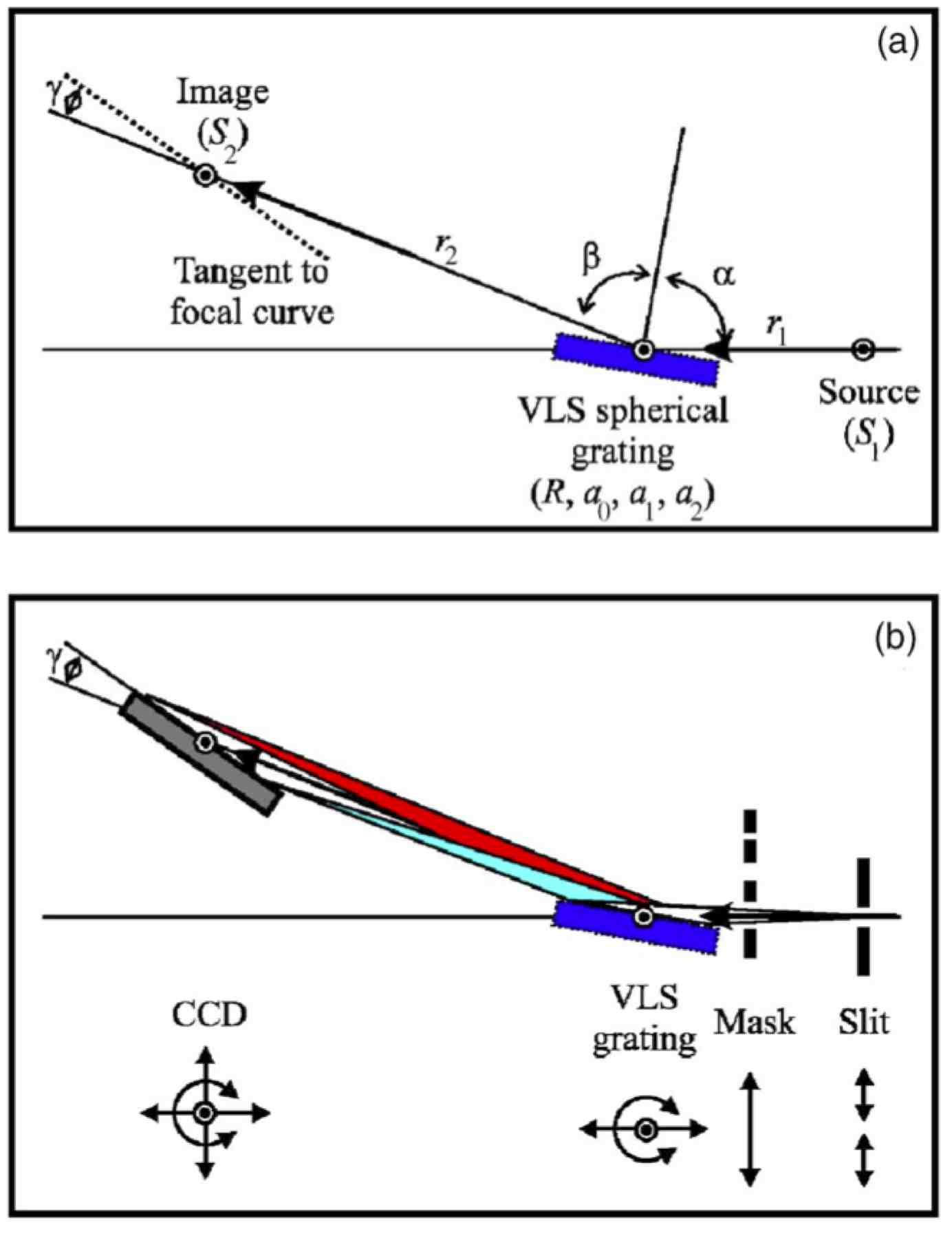}
  \end{center}
    \caption{ (Color online) The schematic layout of the SAXES spectrometer, at the Swiss Light Source. It is based on a variable-line-spacing spherical grating and a two dimensional position sensitive CCD detector. Panel A: the main optical parameter symbols are defined. Panel B: the main mechanical components are sketched together with their degrees of freedom. Figure from Ghiringhelli {\em et al.} \cite{Ghiringhelli2006a} \label{fig:SAXES_layout}}
  \end{figure}

As mentioned above, diffraction gratings lie at the heart of soft x-ray beamlines and spectrometers. These are lined gratings that are typically operated in grazing incidence. The lines act as a series of slits and the light diffracts from these. The resolution is determined by the number of lines per mm. These are not particularly efficient optics, and offer small collecting angles when used as analyzers. Therefore bright x-ray sources, synchrotrons and in the coming years soft x-ray free-electron lasers, are a pre-requisite for high-resolution RIXS. Pioneers in the construction of such instruments were~\textcite{Callcott1986} and~\textcite{Nordgren1989} and others~\cite{Shin1995,Dallera1996,Cocco2001,Cocco2004,Hague2005,Chuang2005,Tokushima2006}.

As an example, we discuss here the instrument at the Swiss Light Source. It consists of a beamline, known as ADRESS for ``ADvanced RESonant Spectrscopies", and a spectrometer, known as SAXES for ``superadvanced x-ray emission spectrometer". The beamline is based on a design that utilizes a plane grating monochromator (PGM) diffraction grating with a line spacing of up to 4200 mm$^{-1}$, together with collimating mirrors upstream of the PGM and focusing mirrors downstream of the PGM. (Another common choice is to combine these latter two optical elements into a single element -- a spherical grating). Operating in the energy range 400-1800 eV, ADRESS has a number of key features that make it well optimized for RIXS experiments. In particular, it delivers a high flux: $\sim 10^{13}$ ph/s/0.01\% BW at 1 keV, with a variable incident polarization in a small spot (8.5 $\mu$m x 100 $\mu$m). This latter is essential in allowing high energy resolution in the spectrometer, as we shall see below. Further, the energy resolution is extremely good, with a resolving power of over 30,000 at 1 keV (i.e. an energy resolution of 30 meV)~\cite{ADRESSurl2010}.

\begin{figure}
  \begin{center}
  \includegraphics[width=1.0\columnwidth]{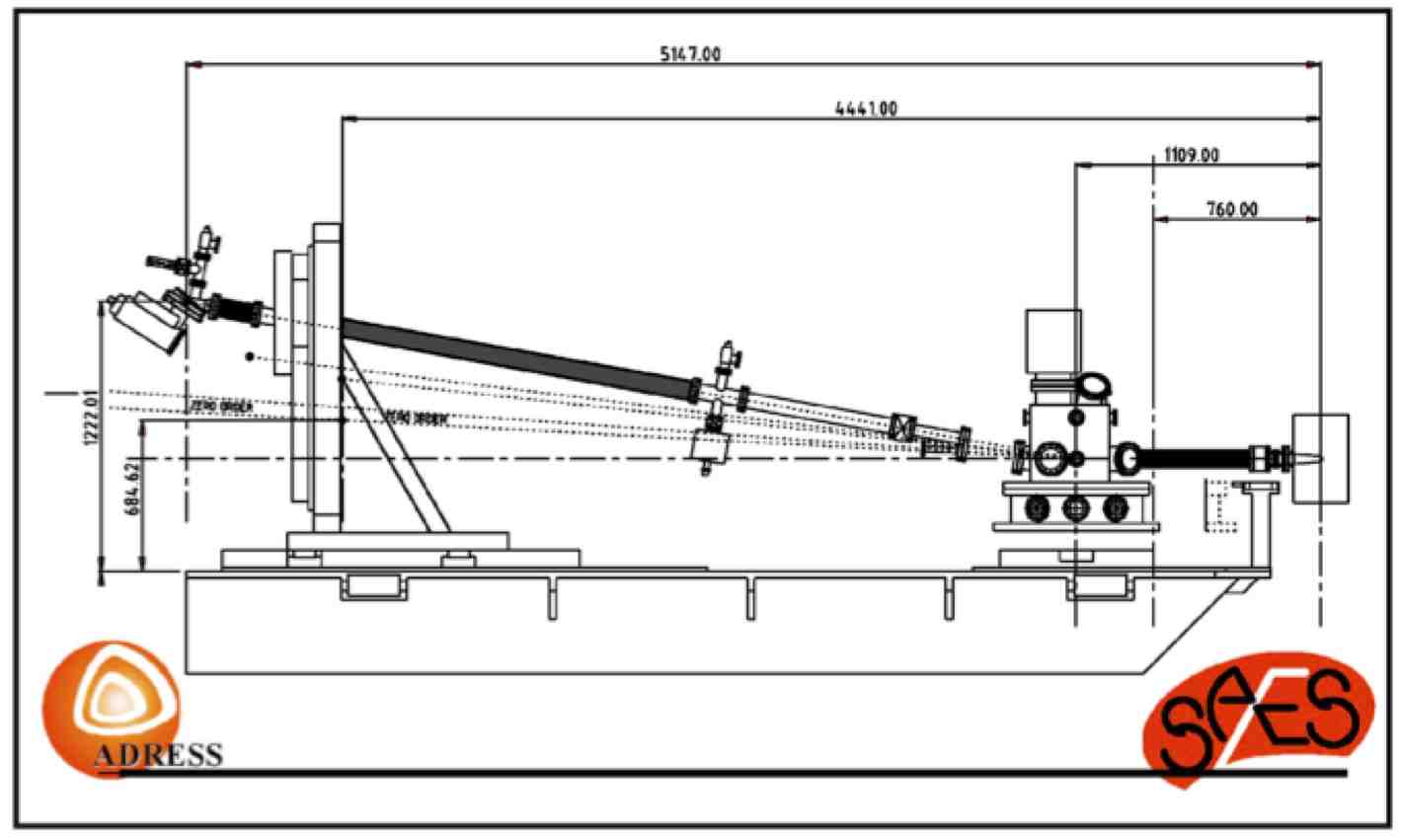}
  \end{center}
    \caption{The mechanical design of SAXES. The sample is placed at the far right. Diffraction grating is in the vacuum chamber approximately 1m from the sample. The CCD is inclined at an angle of 20$^{\circ}$ and is 5.1 m from the sample. The whole spectrometer is supported by a single steel girder and can rotate around a vertical axis about the sample on air cushions. This allows the momentum transfer to be varied. Figure from~\textcite{Ghiringhelli2006a} \label{fig:MechDesignSAXES}}
\end{figure}

The spectrometer, SAXES, must collect the scattered radiation from the sample and provide both momentum and energy resolution. Its design and performance is described in detail in~\textcite{Ghiringhelli2006a}. The basic design is shown in figure~\ref{fig:SAXES_layout}.

The spectrometer is designed around a single optical element, a spherical variable line space (VLS) grating which collects the scattered radiation and disperses it onto a CCD camera. The choice of a single grating - as opposed to a plane grating and a concave collection mirror - simplifies the minimization of higher order optical aberrations at the price of reduced collection angle. The overall energy resolution of the instrument is largely determined by the line spacing of the grating and the spatial resolution of the CCD camera. In recent years there has been steady improvement in the manufacturing technology for making these gratings. The grating in SAXES has an average line spacing $a_0$=3200 mm$^{-1}$ and a fixed radius of curvature of $R = 58.55$ m. The grating angle and position, $\alpha$ and $r_1$ respectively, are optimized at each energy to minimize coma aberration. Slits 5 mm downstream of the sample determine the illumination of the grating by the scattered beam and may be apertured down to improve the resolution. The grating is estimated to be 3-4\% efficient, operating at an incident angle of $88.3^{\circ}$.

\begin{figure}
  \begin{center}
  \includegraphics[width=0.6\columnwidth]{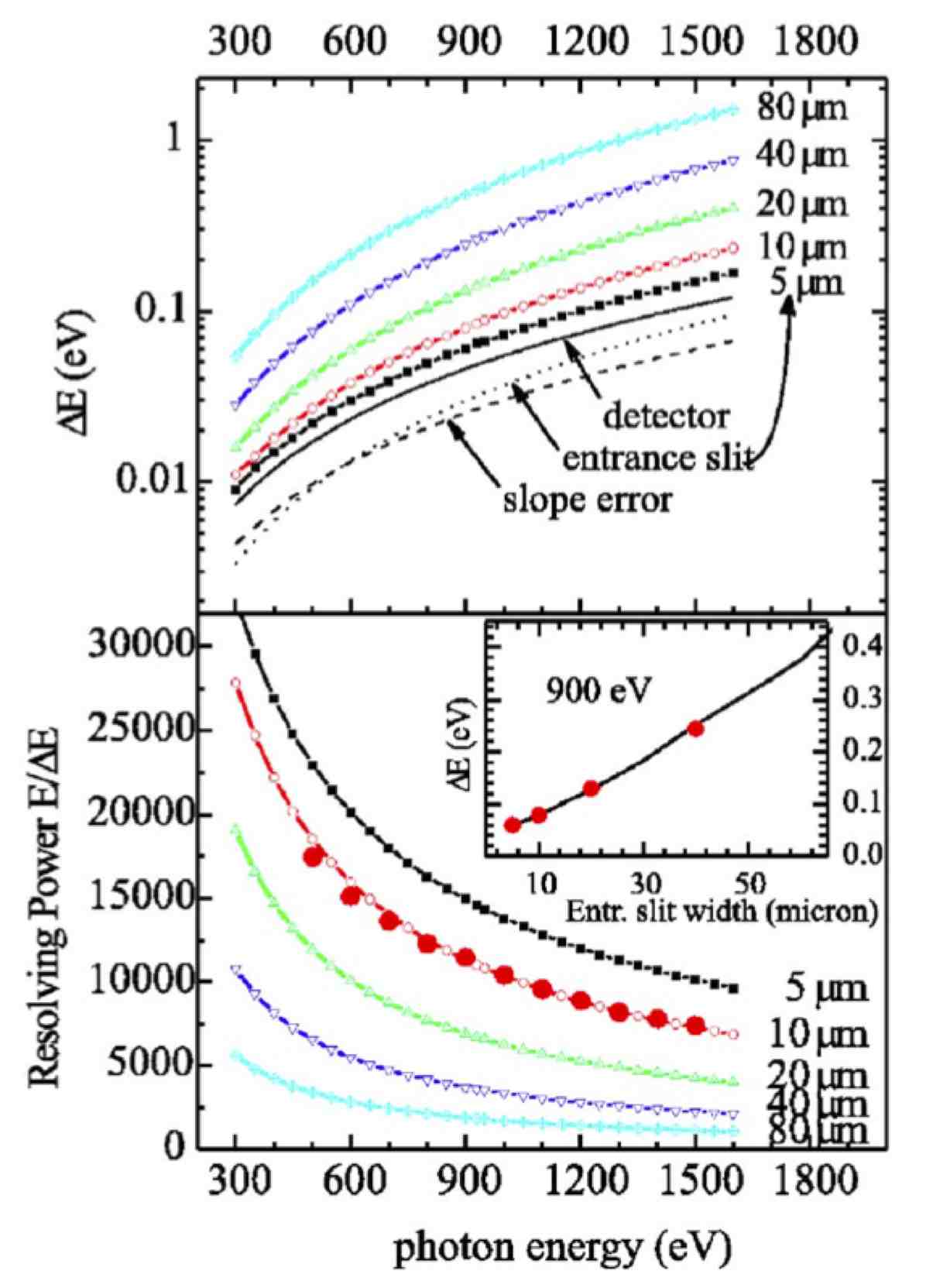}
  \end{center}
  \caption{The calculated resolving power and energy resolution of SAXES. Top panel: The expected energy resolution for various openings of the entrance slit.  Bottom panel: The corresponding values of the resolving power, E/$\Delta$E.  Figure from~\textcite{Ghiringhelli2006a} \label{fig:SAXES_RP}}
\end{figure}

The dispersed light is then allowed to propagate in vacuum until it is incident on the CCD camera. In principle, the energy resolution is
then determined by the distance the light is propagated ($r_2 = 4.03$ m, figure \ref{fig:SAXES_layout}) and the pixel size on the camera. In fact, the physics of the CCD means that while the actual pixel size is 13.5 $\mu$m x 13.5 $\mu$m, a single photon will appear to be observed
by several pixels due to charge leakage between neighboring pixels~\cite{Ghiringhelli2006a}. The camera is therefore operated at a
grazing incidence (20$^\circ$) to match the effective pixel size to the charge distribution caused by a single photon. Finally, it is cooled by liquid
nitrogen to $-100\ ^{\circ}$C to reduce the dark current to negligible levels.

The whole spectrometer (grating plus CCD) is then mounted on a large (5 m) girder as shown in figure~\ref{fig:MechDesignSAXES}. This allows control of the vertical and transverse position of the spectrometer together with its roll and yaw, to correctly position the spectrometer about the scattered beam. Finally, the whole girder is able to rotate in the horizontal plane about the sample position -- moving on air cushions over a granite substrate -- to control the scattered momentum transfer.

The calculated performance of this instrument is shown in figure~\ref{fig:SAXES_RP} as a function of photon energy and for various openings of the entrance slit (i.e. the slit just downstream of the sample). At the smallest slit settings (5 $\mu$m), it is seen that a resolving power of better than 10,000 is expected throughout the whole energy range. One can see that the effective pixel size from the detector is the largest single contribution to the energy resolution suggesting that further improvements in such spectrometers will require associated advances in detector technology. Overall, this instrument is, when coupled with the ADRESS beamline, capable of sub-100 meV resolutions. As shall be seen in later sections, this has led to dramatic advances in the experimental results.

\subsubsection{A hard RIXS instrument}

RIXS instruments in the hard x-ray regime are built around Bragg crystal optics and as such, are fundamentally different from soft RIXS instruments, such as the one discussed above. Here, we describe the so-called MERIX instrument (for Medium Energy Resolution Inelastic Xray) which operates on the 30 ID beamline at the Advanced Photon Source, USA. Other, related instruments include an earlier version of MERIX designed by the same team and in operation at 9ID, APS~\cite{Hill2007}.

\begin{figure}
  \begin{center}
  \includegraphics[width=1.0\columnwidth]{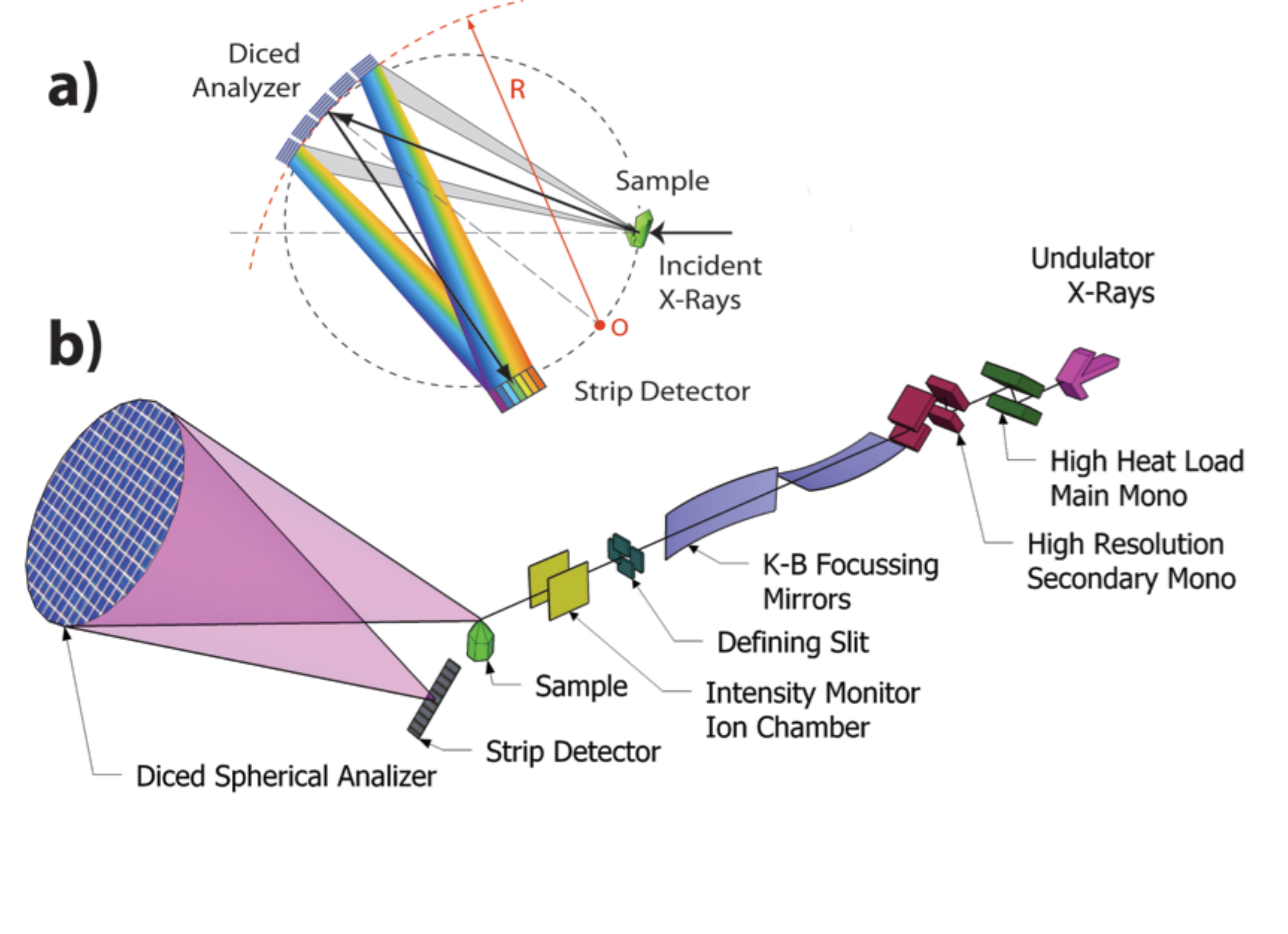}
  \end{center}
  \caption{Schematic of optical components in the 30 IDB inelastic x-ray scattering beamline at the Advanced Photon Source, USA.  Figure courtesy of D. Casa (APS). \label{fig:30ID_beamline_layout}}
\end{figure}

The beamline is designed to provide high incident flux at medium energy resolutions ($\sim$ 70 meV) for incident photon energies from 4.9 keV to 15 keV. It uses two linearly polarized 2.5 m long 30 mm period devices as a radiation source. The primary monochromator comprises two water-cooled diamond (111) crystals in a fixed exit geometry. This is capable of handling the heat load from the undulators and provides an incident band pass of $\sim$0.6 eV at 10 keV. This beam is then incident on a secondary monochromator comprised of 4 asymmetrically cut Si(400) crystals. These are arranged in two pairs to provide a fixed output beam with zero offset from the low resolution beam. This arrangement allows the experimenter to quickly switch from the high-band pass, high flux beam from the diamond monochromator to the narrow-band pass, lower flux configuration, a feature that is frequently of great value. The design of the Si(400) high-resolution monochromator is such that it provides a more or less constant band pass of $\sim$70 meV over the 5-15 keV range of the beamline, with an average spectral efficiency (the number of incident photons that lie within the required band pass that are passed by the optic) of 70 \%. Following the high-resolution monochromator, there are two plane mirrors arranged in a Kirkpatrick-Baez configuration to provide horizontal and vertical focusing of the monochromatic beam. The focused spot size is 45  $\times 6 \mu$m$^2$ (H x V)  and the incident flux on the sample is $2 \times 10^{12}$ photons/s/100meV at 9 keV.  A schematic of the beamline layout is shown in figure~\ref{fig:30ID_beamline_layout}.

 \begin{figure}
  \begin{center}
  \includegraphics[width=1.0\columnwidth]{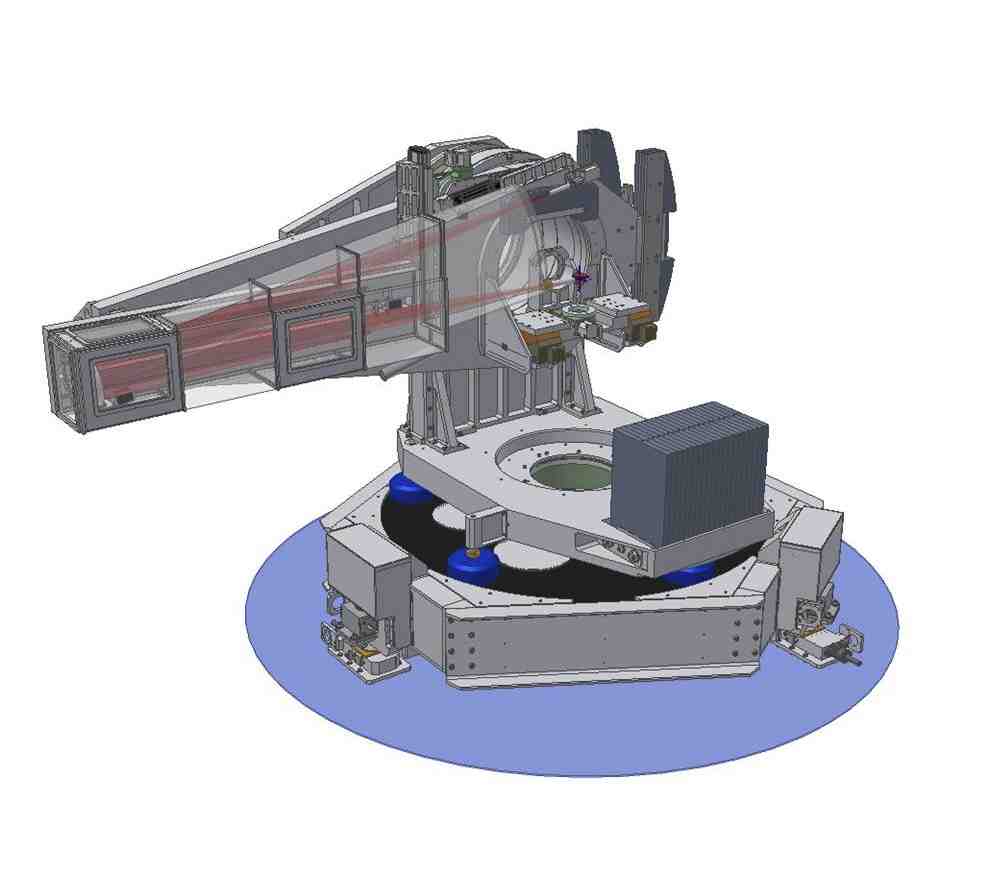}
  \end{center}
    \caption{The MERIX spectrometer for hard x-ray inelastic x-ray scattering at 30 IDB, Advanced Photon Source, USA. The sample is located at the center of the instrument and the beam is incident from the right hand side of the figure. The red cones show the path of the scattered light onto the analyzer and back onto the detector. There are three degrees of freedom on the sample and the detector arm may move in both the vertical and horizontal plane. The detector is typically a position sensitive detector. Figure courtesy of S. Coburn (BNL).}
 \label{fig:MERIX}
\end{figure}

 The MERIX spectrometer itself is shown in figure \ref{fig:MERIX}. It features a 2 m detector arm which is able to move in both the vertical and horizontal plane. This allows the experimenter to vary the incident polarization with respect to the momentum transfer ($\sigma$ incident or $\pi$ incident geometries). Further, fixing the scattering angle at 90$^{\circ}$ in the horizontal plane with horizontally polarized incident light minimizes the intensity of the elastic scattering due to the ${\bm \epsilon} \cdot{\bm \epsilon}^{'}$ dependence of most sources of such scattering. The analyzer is situated inside a tank filled with He gas. This minimizes absorption and background scattering due to windows that would otherwise have to be present. The analyzer may be moved along a linear track to ensure it is placed on the Rowland circle. Common distances used are 1 and 2 m.

The analyzer is the heart of the instrument. In hard RIXS instruments, these are again based on Bragg optics. A single crystal is positioned to intercept the scattered radiation, and Bragg reflect it onto a detector. The scattered energy is then scanned by changing the Bragg angle of the crystal. In order to maximize the scattered intensity (at the expense of momentum space resolution) a spherically-bent analyzer is used to intercept a large solid angle. At MERIX, analyzers are typically 10 cm in diameter and intercept $\sim$10 mSr at 1 m. These are operated in the so-called Rowland circle geometry, such that the sample, analyzer and detector are all mounted on a circle of diameter R. When this radius is equal to radius of curvature of the analyzer, one has the so-called Johan geometry, which is an approximation to the ideal figure (Johansson geometry). This is shown in figure~\ref{fig:30ID_beamline_layout}a. For the very center of the analyzer, positioned right on the Rowland circle, all appropriate scattering conditions are met. For the parts of the analyzer away from this point, aberrations are introduced.

The energy resolution of such analyzers has four contributions; i) the intrinsic Darwin width of the Bragg reflection being utilized  - this is a function of the Debye-Waller factor, the structure factor and the Bragg angle itself - and the strain introduced by the bending which broadens this resolution according to the equations of dynamical diffraction theory. ii) The contribution in the vertical and horizontal directions, due to the Johan error discussed above. This increases quadratically with $r/R$, where $r$ is the radius of the analyzer, and decreases with increasing Bragg angle. iii) The source size resulting from the finite beam footprint on the sample. This increases the energy resolution, linearly increasing with the angular width of the source as seen by a point on the analyzer and decreasing with increasing Bragg angle. 

The introduction of strain due to the bending of the analyzer causes the largest contribution to the energy resolution and therefore in recent years, diced crystals have been used (following their highly successful use for very high-resolution, meV non-resonant spectroscopy~\cite{Verbeni2005}). The dicing however, leads to the fourth contribution to the energy resolution; iv) The finite block size, due to the angular width of that block as seen by the source point. This is linear in block size and decreases with increasing Bragg angle. Because the finite size contributions, and the Johan corrections all scale with $\cot \theta_B$, these are minimized when at back-scattering, $\theta_B=90^{\circ}$, thus diced analyzers work extremely well in these conditions.

Unfortunately, in a RIXS experiment, and in contrast to non-resonant IXS, the incident energy is chosen by the resonance process. One must then find a set of Bragg planes in a suitable analyzer material for which the Bragg angle at this energy is close to $\theta_B=90^{\circ}$. This is by no means a trivial task. It turns out that at the Cu $K$-edge, Ge(733) has $\theta_B=86.8^{\circ}$ and this is sufficiently close to 90$^{\circ}$ that relatively good energy resolutions have been achieved. However, for other transition metal $K$-edges, neither Si nor Ge has any appropriate reflections and resolutions are compromised. Further, even at the Cu $K$-edge, dicing still limits the ultimate resolution achievable. Therefore in the very latest generation instruments, the traditional point detector is replaced with a position sensitive detector~\cite{Huotari2005,Huotari2006}.

As shown in figure~\ref{fig:30ID_beamline_layout}(a), and as first emphasized by~\textcite{Huotari2005,Huotari2006}, the scattered radiation is dispersed from the Bragg reflection from a single block with an energy-angle correlation given by Bragg's law. With a point detector, this spread is integrated over and gives rise to the broadened energy resolution (point iv) above). However, with a position senstive detector, this information can be recovered, removing this broadening effect. Further, there is the added advantage that these energies - corresponding to different energy transfers, may be counted simultaneously, greatly increasing the throughput of the instrument.  The block size no longer contributes to the energy resolution, but rather now determines the range of energy transfers that may be measured simultaneously. Such dispersive schemes, echoing those of the soft RIXS instrument discussed above, are the latest instrumentation development, improving both resolution and count rate (see figure~\ref{fig:hard_RIXS_resolution}). With such a scheme, MERIX has achieved 83 meV FWHM at the Cu $K$-edge (overall resolution), of which the biggest contribution is the monochromator. Further, good resolutions ($\leq$ 100 meV) may be achieved at the other transition metal $K$-edges using this scheme and Si or Ge reflections. In addition, groups are now exploring alternate analyzer materials, including sapphire (Al$_2$O$_3$), quartz (SiO$_2$) and LiNbO$_3$, which offer many more reflections and the chance to move closer to back scattering and thus improved resolution. While a great deal of development will be required for each of these, it seems likely that the resolution of hard RIXS will continue to improve in the coming years.

\section{Theory of RIXS: I. Fundamentals \label{sec:Theory_fundamentals}}
In RIXS, the solid is taken from a ground state with energy $E_g$, to a final-state with excitations and an energy $E_f$. The energy and momentum of the excitation is determined by the difference in photon energy $\hbar \omega_{\bf k}-\hbar \omega_{{\bf k}'}$  and momentum $ \hbar {\bf k}'-\hbar {\bf k}$, respectively. The RIXS intensity can in general be written in terms of a scattering amplitude as
\definecolor{myblue}{rgb}{.8, .8, 1}
\newcommand*\mybluebox[1]{%
\colorbox{myblue}{\hspace{1em}#1\hspace{1em}}}
\begin{empheq} [box=\mybluebox]{align}
I(\omega,{\bf k},{\bf k}',{\boldsymbol \epsilon},{\boldsymbol \epsilon}') = \sum_f   &|{\cal F}_{fg}({\bf k}, {\bf k}^\prime, {\boldsymbol \epsilon}, {\boldsymbol \epsilon}^\prime,\omega_{\bf k})|^2 \nonumber \\
 & \times \delta(E_f+\hbar \omega_{{\bf k}^\prime}-E_g-\hbar \omega_{{\bf k}}) \nonumber
\end{empheq}
where the delta function enforces energy conservation and the amplitude ${\cal F}_{fg}({\bf k}, {\bf k}^\prime, {\boldsymbol \epsilon}, {\boldsymbol \epsilon}^\prime, \omega_{\bf k})$ reflects which excitations are probed and how, for instance, the spectral weights of final state excitations depend on the polarization vectors, ${\boldsymbol \epsilon}$ and ${\boldsymbol \epsilon}^\prime $ of the incoming and outgoing x-rays, respectively. The following sections derive the RIXS scattering amplitude and demonstrate how it can be broken down into separate pieces.

\begin{center}
\begin{table*}[ht]
\caption{ \label{scatteringamplitude}
In the theory of RIXS, the  scattering amplitude ${\cal F}_{fg}$ occurring in the Kramers-Heisenberg equation is separated into several pieces. One  can split off the angular and polarization dependence ${\bf T}^x$, leaving fundamental scattering amplitudes ${\bf F}^{x}$. Several  approximation schemes then break down these scattering amplitudes into a resonance function $P$ and effective transition operators $W^x$. }
\centering
\begin{tabular}{|c|c|c|c|} \hline
\multicolumn{4}{|c|}{Factorizing the Kramers-Heisenberg Scattering Amplitude ${\cal F}_{fg}$}\\ \hline
& \multicolumn{3}{c|}{~~~scattering amplitude~~~ = ~~~~~fundamental scattering amplitudes $\times$ angular and polarization dependence }\\
exact {\phantom{\Large{$\downarrow$}}}& \multicolumn{3}{l|}{{~~~${\cal F}_{fg}$}($\bf k$, $\bf k^\prime$, $\boldsymbol \epsilon$, $\boldsymbol \epsilon^\prime$, $\omega_{\bf k}$) ~~ = ~~$\sum_x ~~~~~~~~~~{\bf F}^{x}(k,k',\omega_{\bf k}$)
 $~~~~~~~~~~~~~~~~\cdot~~~~~~~~~~~~~~~{\bf T}^x$($\bf \hat{k}$, $\bf \hat{k}^\prime$, $\boldsymbol \epsilon$, $\boldsymbol \epsilon^\prime$) } \\
\hline
\multicolumn{4}{c}{\vspace{-0.75em}} \\
 \multicolumn{2}{c}{} & \multicolumn{2}{r}{\Huge $\Downarrow$~~~~~~~~~~~~~~~~~~~~~~~~~~~~~~~~~}  \\
 \multicolumn{4}{c}{\vspace{-0.75em}} \\  \cline{3-4}
 \multicolumn{2}{c|}{~~~~~~~~~~~~~~~~~~~~~~~~~~~~~~~~~~~~~~~~~} & \multicolumn{1}{c|}{} & \multicolumn{1}{l|}{~~~~~fundamental ~~~~~~~~~~~~~~~~~~~~~~~~~~~~~~~~~~~~~~~~~~~~~ effective}\\
  \multicolumn{2}{c|}{~~~~~~~~~~~~~~~~~~~~~~~~~~~~~~~~~~~~~~~~~} & \multicolumn{1}{c|}{} & \multicolumn{1}{l|}{scattering amplitude~~~=~resonance function $\times$  transition operators}\\
 \multicolumn{2}{l|}{}& approximate & \multicolumn{1}{l|}{$~~~~~{\bf F}^{x}(k,k',\omega_{\bf k})~~~~~~~$ =$ ~~~~~~P(\omega_{\bf k},\omega_{{\bf k}'})$
 $~~~~\times~~~~~~\langle f|W^x |g\rangle $
 }
\\      \cline{3-4}
\end{tabular}
\end{table*}
\end{center}

First, we need to derive a general expression for the RIXS scattering
amplitude. Section \ref{sec:basictheory} looks at the interaction
between photons and matter. RIXS refers to the process where the
material first absorbs a photon. The systems is then in a short-lived
intermediate state, from which it relaxes radiatively. In an
experiment, one studies the x-rays emitted in this decay process. This
two-step process cannot be described simply by using Fermi's Golden Rule,
but requires a higher-order treatment, known as the Kramers-Heisenberg
equation~\cite{Kramers1925}. Since the absorption and emission are single-photon
processes, the interactions between the x-rays and the material are
dominated by the terms in the cross-section proportional to ${\bf p}\cdot{\bf A}$, where ${\bf p}$ is the
momentum of the electrons in the material and ${\bf A}$ is the vector
potential of the photon. The interaction between the x-rays and the material depends on external quantities, such as
wavevector ${\bf k}$ and polarization vectors ${\boldsymbol \epsilon}$
of the x-rays, and operators, such as ${\bf p}$ and ${\bf r}$. As a
result the electronic transitions are intermingled. Sections III.A.2 and III.A.3 describe how the scattering amplitude can be split into an angular and polarization dependence ${\bf T}^x({\hat{\bf k}}, {\hat{\bf k}}^\prime, {\boldsymbol \epsilon}, \boldsymbol \epsilon^\prime)$ related to the experimental geometry and  spectral functions  $ {\bf F}^{x}(k,k',\omega_{\bf k})$ that measure the properties of the material, see Table \ref{scatteringamplitude}. This separation can be done exactly. It is important to note that there are only a finite number of fundamental scattering amplitudes ${\bf F}^{x}(k,k',\omega_{\bf k})$ and that the RIXS scattering amplitude is a linear combination of these fundamental scattering amplitudes weighted by the angular functions ${\bf T}^x({\hat{\bf k}}, {\hat{\bf k}}^\prime, {\boldsymbol \epsilon}, \boldsymbol \epsilon^\prime)$.

The next step is to understand the fundamental scattering amplitudes. This is done in Section \ref{sec:Calculating_crosssection}. This can be done numerically, as is outlined in Section \ref{sec:numerical}. In addition, several authors have used approximation schemes in order to provide more insight into the scattering amplitude. Generally, the approximations involve the propagation of the system in the time between the absorption and emission processes. The schemes generally allow the separation of the fundamental scattering amplitudes into a resonance function $P(\omega_{\bf k},\omega_{{\bf k}'})$ and an effective transition between ground and final states $\langle f|W^x |g\rangle$, see Table \ref{scatteringamplitude}.  The resonance function gives the strength of the fundamental scattering amplitude which is a combination of radial matrix elements of the transition operators and energy denominators that describe the resonant effect as a function of $\omega_{\bf k}$. The effective transition operators create excitations in the valence shell similar to an optical excitation. In certain cases, these operators can also be related to correlation functions such as the dynamic structure factor. The  approximations
depend on the RIXS process. Direct RIXS is approximated by using a fast-collision approximation, see Section \ref{sec:effdirect}. Indirect RIXS can be approached via perturbative methods or an ultra-short core-hole lifetime expansion, see Section \ref{sec:effindirect}.
\subsection{Interaction of light and matter \label{sec:basictheory}}

To develop the theory of RIXS, we first need to derive the Hamiltonian
that describes the interaction of the incident X-ray beam with the
electrons in the sample. The interaction terms in this Hamiltonian are
small, controlled by the dimensionless fine structure constant $\alpha
= e^2/4\pi \epsilon_0 \hbar c \approx 1/137$, with $e=|e|$ the
magnitude of the elementary charge and $\epsilon_0$ the permittivity
of free space. Therefore they can be treated as a perturbation to the
terms in the Hamiltonian that describe the system under study. To
second order in such a perturbation theory, we obtain the
Kramers-Heisenberg formula, which describes RIXS very well. We need to
go to second order because two interactions are needed: one to create
the core hole, and one for the subsequent radiative de-excitation.

\subsubsection{Kramers-Heisenberg cross section}
The incident X-rays are described by an electromagnetic field with
vector potential ${\bf
  A}({\bf r},t)$. The coupling between such a field and electrons is
given by the theory of quantum electrodynamics. It is common to start
from the exactly solvable case of a single electron without
potentials (${\bf A}$ and electric potential $\phi ({\bf r},t)$). Then, the potentials are (perturbatively) introduced and
one takes two limits. The first of these is that the electrons travel at speeds, $v$, small compared
to the speed of light. This is a good approximation even for, for
instance, copper $1s$ core electrons, where we estimate $v \sim \hbar Z/m
a_0 \approx 0.21 c$ with $Z$ the atomic number for copper and $a_0$
the Bohr radius. At first glance, $v/c$ might not appear small in this latter case,
but $\gamma = 1/\sqrt{1-v^2/c^2} \approx 1.02$ and relativistic
effects are still small. The second limit is that the potentials
related to both the electrons and photons in the system are small
compared to twice the mass of the electron: $e\phi /2 mc^2,\
e\abs{{\bf A}} /2 mc \ll 1$ ($m$ is the electron mass). Although the intrinsic
potentials of materials diverge close to the nuclei, they may be treated consistently within the whole procedure for $Z \ll 137$ (see \textcite{Messiah1962}, page 948). Photon potentials at existing x-ray sources satisfy these limits. However, in future very strongly focussed X-ray Free Electron Lasers, the electric field of the photon is projected to exceed
$10^{16}$ V/m, which gives $e\abs{{\bf A}}\sim 2mc$ at a photon energy
of $\sim 8$ keV and these approximations are no longer valid. However,
for the purposes of this review, such effects are neglected and the
formalism is developed for non-relativistic electrons in small
potentials.

In these limits, one obtains for a system with $N$ electrons, in SI units (see pages 944-947 in \textcite{Messiah1962}, pages 85-88 in \textcite{Sakurai1967})
\begin{eqnarray}
H &=& \sum_{i=1}^{N} \biggl[ \frac{\left({\bf p}_i + e {\bf A}({\bf r}_i) \right)^2}{2m} + \frac{e\hbar}{2m} {\boldsymbol \sigma}_i \cdot {\bf B}({\bf r}_i) + \frac{e\hbar}{2(2mc)^2} \times  \nonumber \\
	&{\boldsymbol \sigma}_i& \cdot \biggl( {\bf E}({\bf r}_i) \times \left( {\bf p}_i + e{\bf A}({\bf r}_i) \right) - \left( {\bf p}_i + e{\bf A}({\bf r}_i) \right) \times {\bf E}({\bf r}_i) \biggr) \biggr] \nonumber \\
	&+& \frac{e\hbar^2 \rho({\bf r}_i)}{8(mc)^2 \epsilon_0} + H_{\rm Coulomb} + \sum_{{\boldsymbol \kappa},{\boldsymbol \varepsilon}} \hbar \omega^{\phantom{\dag}}_{{\boldsymbol \kappa}} \left( a^{\dag}_{{\boldsymbol \kappa}{\boldsymbol \varepsilon}} a^{\phantom{\dag}}_{{\boldsymbol \kappa}{\boldsymbol \varepsilon}} + \frac{1}{2} \right)
	\label{Ham}
\end{eqnarray}
where ${\bf p}_i$, ${\bf r}_i$ and ${\boldsymbol \sigma}_i$ are, respectively, the momentum and position operators and the Pauli matrices acting on electron $i$. ${\bf A}({\bf r})$ is the vector potential, ${\bf E}({\bf r}) = -\nabla \phi - \partial {\bf A} /\partial t$, the electric field, and ${\bf B}({\bf r}) = \nabla \times {\bf A}$, the magnetic field. $a^{(\dag)}_{{\boldsymbol \kappa}{\boldsymbol \epsilon}}$ annihilates (creates) a photon in the mode with wave vector ${\boldsymbol \kappa}$ and polarization vector ${\boldsymbol \varepsilon}$. The second term yields the Zeeman splitting, and the third includes spin-orbit coupling. The interaction of electrons with an external electric potential and with other electrons and nuclei in the sample (including the Darwin term) are all described by $H_{\rm Coulomb}$. The vector potential can be expanded in plane waves as
\begin{equation}
	{\bf A}({\bf r}) = \sum_{{\boldsymbol \kappa},{\boldsymbol \varepsilon}}\sqrt{\frac{\hbar}{2 {\cal V} \epsilon_0 \omega_{\boldsymbol \kappa}}} \left( {\boldsymbol \varepsilon}\ a^{\phantom{\dag}}_{{\boldsymbol \kappa}{\boldsymbol \varepsilon}} e^{i {\boldsymbol \kappa}\cdot {\bf r}} + {\boldsymbol \varepsilon}^* a^{\dag}_{{\boldsymbol \kappa}{\boldsymbol \varepsilon}} e^{-i {\boldsymbol \kappa}\cdot {\bf r}} \right),
	\label{vectorpotential}
\end{equation}
where $\cal V$ is the volume of the system.

In order to derive the photon scattering cross section one splits the Hamiltonian $H$ into an electron-photon interaction part, $H'$, and the remaining terms, $H_0$, which describe the electron and photon dynamics in the absence of electron-photon interactions. $H'$ is then treated as a perturbation to $H_0$. To calculate the RIXS cross section in this perturbation scheme, it is assumed that there is a single photon in the initial state with momentum $\hbar {\bf k}$, energy $\hbar \omega_{\bf k}$ and polarization ${\boldsymbol \epsilon}$ that is scattered to $(\hbar {\bf k}', \hbar \omega_{{\bf k}'}, {\boldsymbol \epsilon}')$ in the final state. Photon scattering then induces a change in the material from ground state $\ket{g}$ to final state $\ket{f}$, with energies $E_g$ and $E_f$ respectively. In the process, the photon loses momentum $\hbar {\bf q} = \hbar {\bf k} - \hbar {\bf k}'$ and energy $\hbar \omega = \hbar \omega_{\bf k} - \hbar \omega_{{\bf k}'}$ to the sample. Fermi's Golden Rule to second order gives the transition rate for this process:
\begin{eqnarray}
	w &=& \frac{2\pi}{\hbar} \sum_{\mathtt f} \biggl| \bra{{\mathtt f}} H' \ket{{\mathtt g}} \nonumber \\
	&+& \sum_n \frac{\bra{{\mathtt f}}H'\ket{n} \bra{n}H' \ket{{\mathtt g}}}{E_{\mathtt g}-E_n} \biggr|^2 \delta (E_{\mathtt f}-E_{\mathtt g})
\label{FermiGoldenRule}
\end{eqnarray}
where the initial state $\ket{{\mathtt g}} = \ket{g;{\bf k} {\boldsymbol \epsilon}}$, intermediate state $\ket{n}$ and the final state $\ket{{\mathtt f}} = \ket{f;{\bf k}'  {\boldsymbol \epsilon}'}$ are eigenstates of $H_0$ with energies $E_{\mathtt g} = E_g + \hbar \omega_{\bf k}$, $E_n$ and $E_{\mathtt f} = E_f + \hbar \omega_{{\bf k}'}$, respectively. The first order amplitude in general dominates the second order, but when the incoming X-rays are in resonance with a specific transition in the material ($E_{\mathtt g} \approx E_n$), then the second order terms become large. The second order amplitude causes resonant scattering, while the first order yields non-resonant scattering.  

In order to derive $H'$ it is useful to classify the terms of
Eq.~(\ref{Ham}) by powers of ${\bf A}$. Terms of $H$ that are quadratic in ${\bf A}$ are the only ones to contribute to the first order amplitude, because they contain terms proportional to $a^{\dag}_{{\bf k}' {\boldsymbol \epsilon}'} a^{\phantom{\dag}}_{{\bf k} {\boldsymbol \epsilon}}$ and $a^{\phantom{\dag}}_{{\bf k} {\boldsymbol \epsilon}} a^{\dag}_{{\bf k}' {\boldsymbol \epsilon}'}$. To be specific, the quadratic contribution from the first term of $H$ gives rise to non-resonant scattering, while the third term of $H$ yields magnetic non-resonant scattering. Although both appear in the first order scattering amplitude, they in principle also contribute to the second order, but we neglect these processes because they are of order $\alpha^{3/2}$.

The interaction terms linear in ${\bf A}$ do not contribute to the
first order amplitude, but do contribute to the second order. They
thus give rise to resonant processes. In the following, we neglect such contributions that come from the third term of Eq.~(\ref{Ham}), because they are of second order in two separate expansions. Firstly, this term of $H$ is of second order in the limits discussed above, and secondly, it appears in the second order of the scattering amplitude.

Finally, all terms in Eq.~(\ref{Ham}) that are independent of ${\bf A}$ are included in $H_0$.
The relevant remaining terms are
\begin{eqnarray}
	H' &=& \sum_{i=1}^N \left[ \frac{e}{m} {\bf A}({\bf r}_i)\cdot{\bf p}_i + \frac{e^2}{2m} {\bf A}^2({\bf r}_i) + \frac{e\hbar}{2m} {\boldsymbol \sigma}_i \cdot \nabla \times {\bf A}({\bf r}_i) \right. \nonumber \\
	&-& \left. \frac{e^2\hbar}{(2mc)^2} {\boldsymbol \sigma}_i \cdot \frac{\partial {\bf A} ({\bf r}_i)}{\partial t} \times {\bf A}({\bf r}_i) \right],
	\label{Hprime}
\end{eqnarray}
where the gauge was fixed by choosing $\nabla \cdot {\bf A}({\bf r}) = 0$ so that ${\bf A}\cdot {\bf p} = {\bf p} \cdot {\bf A}$.

The two terms of $H'$ that contribute to the first order amplitude are the one
proportional to ${\bf A}^2$ and the ${\boldsymbol \sigma}\cdot
(\partial {\bf A}/\partial t) \times {\bf A}$ term. The latter is
smaller than the former by a factor $\hbar \omega_{{\bf k}(')}/mc^2
\ll 1$, and is therefore neglected. The first order term in Eq.~(\ref{FermiGoldenRule}) then becomes
\begin{equation}
	\frac{e^2}{2m} \bra{{\mathtt f}} \sum_i {\bf A}^2 ({\bf r}_i) \ket{{\mathtt g}} = \frac{\hbar e^2}{2m {\cal V} \epsilon_0} \frac{{\boldsymbol \epsilon}'^* \cdot {\boldsymbol \epsilon}}{\sqrt{\omega_{\bf k}\omega_{{\bf k}'}}} \bra{f} \sum_i e^{i {\bf q}\cdot {\bf r}_i} \ket{g}.
\end{equation}
When the incident energy $\hbar \omega_{\bf k}$ is much larger than any
resonance of the material, the scattering amplitude is dominated by
this channel, which is called Thompson scattering. In scattering
from a crystal at zero energy transfer, this term contributes amongst others to the Bragg peaks. It also gives rise to non-resonant inelastic scattering. In practice, RIXS spectra show a strong resonance behavior, demonstrating that for these processes, it is the second order term that dominates the first order term. We therefore omit the ${\bf A}^2$ contribution in the following. More details on non-resonant inelastic X-ray scattering can be found in, for instance, \textcite{Schuelke2007,Rueff2010}.

The second order amplitude in Eq.~(\ref{FermiGoldenRule}) becomes
large when $\hbar \omega_{\bf k}$ matches a resonance energy of the
system, and the incoming photon is absorbed first in the intermediate
state, creating a core hole. The denominator $E_g+\hbar \omega_{\bf
  k}-E_n$ is then small, greatly enhancing the second order scattering amplitude.  We neglect the other, off-resonant processes here, though they do give an important contribution to non-resonant scattering \cite{Blume1985}. The resonant part of the second order amplitude is
\begin{eqnarray}
	&\ & \frac{e^2 \hbar}{2m^2 {\cal V}\epsilon_0 \sqrt{\omega_{\bf k} \omega_{{\bf k}'}}} \sum_n  \sum_{i,j=1}^N \nonumber \\
	&\times& \frac{\bra{f} e^{-i{\bf k}' \cdot {\bf r}_i} \left({\boldsymbol \epsilon}'^* \cdot {\bf p}_i - \frac{i\hbar}{2} {\boldsymbol \sigma}_i \cdot {\bf k}' \times {\boldsymbol \epsilon}'^* \right) \ket{n}}{E_g+\hbar \omega_{\bf k}-E_n+i\Gamma_n} \nonumber \\
	&\times& \bra{n} e^{i{\bf k}\cdot {\bf r}_j} \left({\boldsymbol \epsilon}\cdot {\bf p}_j + \frac{i\hbar}{2} {\boldsymbol \sigma}_j \cdot {\bf k} \times {\boldsymbol \epsilon} \right) \ket{g},
\end{eqnarray}
where a lifetime broadening $\Gamma_n$ is introduced for the intermediate states. This accounts for the many non-radiative interaction terms that are not included in $H'$ (for example Auger decay), which make the intermediate states very short lived.

Resonant scattering can thus occur via a magnetic and a non-magnetic
term. An estimate shows that the latter dominates.
The size of localized $1s$ copper core orbitals is roughly $a_0/Z \approx 0.018 $ \AA\ so that for $10$ keV photons the
exponential $e^{i{\bf k}\cdot {\bf r}}$ is close to unity and can be
expanded. The non-magnetic term can induce a dipole transition of
order $\abs{{\bf p}} \sim \hbar Z/a_0 \sim 5.9 \cdot 10^{-23}$ kg m/s,
whereas the magnetic term gives a dipole transition of order $({\bf
  k}\cdot {\bf r}) \hbar \abs{{\bf k}}/2 \sim 2.5 \cdot 10^{-25}$ kg
m/s.
We thus ignore the magnetic term here, and the relevant transition operator for the RIXS cross section is
\begin{empheq} [box=\mybluebox]{align}
 {\cal D}= \frac{1}{im\omega_{\bf k}}\sum_{i=1}^N   e^{i{\bf k}\cdot {\bf r}_i}{\boldsymbol \epsilon}\cdot{\bf p}_i,
\label{operator}
\end{empheq}
where a prefactor has been introduced for convenience in the following expressions.

The double-differential cross section $I(\omega,{\bf k},{\bf k}',{\bm \epsilon},{\bm \epsilon}')$ is now obtained by multiplying by the density of photon states in the solid angle $d\Omega$ $(= {\cal V} {\bf k}'^2 d|{\bf k}'| d\Omega /(2\pi)^3)$, and dividing by the incident flux $c/{\cal V}$ \cite{Sakurai1967,Blume1985,Schuelke2007}:
\begin{eqnarray}
I(\omega,{\bf k},{\bf k}',{\bm \epsilon},{\bm \epsilon}') &=& r_e^2m^2\omega_{{\bf k}'}^3\omega_{\bf k} \sum_{\tt f}|{\cal F}_{fg}({\bf k}, {\bf k}^\prime, {\boldsymbol \epsilon}, {\boldsymbol \epsilon}^\prime, \omega_{\bf k}, \omega_{{\bf k}^\prime})|^2  \nonumber \\
&& ~~~\times  \delta (E_g-E_f+\hbar \omega) ,
\label{KramersH}
\end{eqnarray}
 where the classical electron radius $r_e = \frac{1}{4\pi \epsilon_0} \frac{e^2}{mc^2}$. The scattering amplitude at zero temperature is given by
\begin{eqnarray}
&& {\cal F}_{fg}({\bf k}, {\bf k}^\prime, {\boldsymbol \epsilon}, {\boldsymbol \epsilon}^\prime, \omega_{\bf k}, \omega_{{\bf k}^\prime}) =  \sum_n
	 \frac{\bra{f} {\cal D}'^\dagger  \ket{n} \bra{n} {\cal D} \ket{g}}{E_g+\hbar \omega_{\bf k}-E_n+i\Gamma_n} ,
\label{eq:K-H}
\end{eqnarray}
where the prime in ${\cal D}'$ indicates it refers to transitions related to the outgoing X-rays. Eqs. (\ref{KramersH}) and (\ref{eq:K-H}) are referred to as the Kramers-Heisenberg equations, which are generally used to calculate the RIXS cross section.

Alternatively, we can rewrite the denominator for the intermediate-states in terms of a Green's function, which is also referred to as the intermediate-state propagator, which describes the system in the presence of a core hole:
\begin{empheq} [box=\mybluebox]{align}
G(z_{\bf k})=\frac{1}{z_{\bf k}-H} =\sum_n \frac{|n\rangle \langle n|}{z_{\bf k}-E_n},
\end{empheq}
where $|n\rangle$ forms a complete basis set and
\begin{empheq} [box=\mybluebox]{align}
z_{\bf k}=E_g+\hbar \omega_{\bf k}+i\Gamma,
\end{empheq}
where $\Gamma$ is taken to be independent of the intermediate
states. The quantity $z_{\bf k}$ is the energy of the initial state
combined with the finite lifetime of the core hole. In the following we will often suppress the explicit label $\bf k$ of $z_{\bf k}$ and denote it simply by $z$, with an implicit incident energy dependence. With the core-hole propagator $G$ and transition operators $D$ in place, the RIXS scattering amplitude ${\cal F}_{fg}$ finally reduces to the elegant expression
\begin{empheq} [box=\mybluebox]{align}
 {\cal F}_{fg} = \bra{f} {\cal D}'^\dagger G(z_{\bf k}){\cal D} \ket{g}.
 \label{eq:scatteringampl}
\end{empheq}

\subsubsection{Scattering amplitude in dipole approximation}
In the previous section, Eqs. (\ref{KramersH}) and (\ref{eq:K-H}) give the Kramers-Heisenberg expression for RIXS. The next step is to separate the part pertaining to the geometry of the experiment from the fundamental scattering amplitudes that relate to the physical properties of the system, see  Table \ref{scatteringamplitude}. In addition, better-defined transition operators will be obtained. Due to the complexity of the multipole expansion, we first give a derivation in the dipole limit allowing the reader to better follow the arguments. In the next section, we present the higher order transitions.

In the dipole limit, one asumes that $e^{i{\bf k} \cdot {\bf r}_i}\cong e^{i{\bf k} \cdot {\bf R}_i}$ where ${\bf R}_i$ indicates the position of the ion to which the electron $i$ is bound. Note that ${\bf R}_i$ is not an operator. This has as a result that the electronic transitions are due to the momentum operator ${\bf p}$ and Eq. \ref{operator} becomes
\begin{equation}
  {\cal D}={\boldsymbol \epsilon} \cdot {\bf D} ~~~{\rm with} ~~~~
 {\bf D}= \frac{1}{im\omega_{\bf k}}\sum_{i=1}^N e^{i{\bf k} \cdot {\bf R}_i} {\bf p}_i,
 \label{transoper}
\end{equation}
Generally, the matrix elements are expressed in terms of the position operator ${\bf r}$. For example, in the absorption step, one can write
\begin{align}
     \bra{n} {\bf D} \ket{g} &=
\sum_{i=1}^N \frac{e^{i{\bf k} \cdot {\bf R}_i}}{im\omega_{\bf k}}\bra{n}{\bf p}_i\ket{g} =\sum_{i=1}^N\frac{e^{i{\bf k} \cdot {\bf R}_i}}{\hbar \omega_{\bf k}}
\bra{n} [\frac{{\bf p}_i^2}{2m},{\bf r}_i] \ket{g} \nonumber \\
    &\cong \sum_{i=1}^N\frac{e^{i{\bf k} \cdot {\bf R}_i}}{\hbar \omega_{\bf k}} (E_n-E_g)\bra{n} {\bf r}_i\ket{g} \nonumber \\
    &\cong \sum_{i=1}^N e^{i{\bf k} \cdot {\bf R}_i} \bra{n} {\bf r}_i\ket{g}
\nonumber
\end{align}
where $\hbar \omega_{\bf k} \cong E_n-E_g$. The operator thus reduces to
the well-known dipole operator ${\bf D}=\sum_{i=1}^N e^{i{\bf k} \cdot {\bf R}_i} {\bf r}_i$ that
causes electronic transitions.

The next step is to separate the part that pertains to the geometry of
the experiment (the polarization vectors ${\boldsymbol \epsilon}'$ and
${\boldsymbol \epsilon}$) from the physical properties of the
system. Ultimately, our interest lies in the spectral functions of a
material. The geometry is chosen in an optimal way to measure
them. Using spherical-tensor algebra, see Eq. (\ref{recoupling}) in
the Appendix, we can rewrite the scattering amplitude,
Eq.~(\ref{eq:scatteringampl}), remaining in the dipole limit,
Eq.~(\ref{transoper}), as
\begin{equation*}
{\cal F}_{fg} = \sum_{x=0}^2  [x]n_{11x}^2 \bra{f} [{\boldsymbol \epsilon}'^*,{\boldsymbol \epsilon}]^x\cdot [{\bf D}^\dagger ,G(z_{\bf k})  {\bf D}]^x \ket{g},
\end{equation*}
using the shorthand $[l_1\cdots l_n]=(2l_1+1)\cdots (2l_n+1)$; $n_{11x}$ is a normalization constant, and $[~ ,~]^x$ is a tensor product, see Appendix. Since the tensor product couples tensors of rank 1 (the polarization vectors and the position vector ${\bf r}$), the rank  $x$ of the tensor products  can assume the values 0, 1, and 2. The fundamental scattering amplitudes are given by
\begin{equation}
{\bf F}^{x}(z_{\bf k})= \bra{f}  [{\bf D}^\dagger ,G(z_{\bf k})  {\bf D}]^x \ket{g},
\end{equation}
For each value of $x$, there are $2x+1$ components $F^{x}_q$ with
$q=-x,-x+1,\cdots,x$.  Note that, whereas there is an infinite number
of different scattering amplitudes, for dipole transitions, there are
only nine fundamental ones ( $3\times 3=1+3+5=9$) .  All the other possible scattering amplitudes are combinations of these fundamental scattering amplitudes with a weighting determined by the angular dependence;
\begin{equation}
{\bf T}^{x} ({\boldsymbol \epsilon},{\boldsymbol \epsilon}')
= [x]n_{11x}^2 [ {\boldsymbol \epsilon}'^*,{\boldsymbol \epsilon}]^x,
\label{T1x}
\end{equation}
which again has nine components $T_q^{x} ({\boldsymbol
  \epsilon},{\boldsymbol \epsilon}')$. For $x=0,1$, the angular
dependence is given by the inner product, $T_0^{0} ({\boldsymbol
  \epsilon},{\boldsymbol \epsilon}')=\frac{1}{3}{\boldsymbol
  \epsilon}'^*\cdot{\boldsymbol \epsilon}$,  and the outer product, $T_\alpha^{1} ({\boldsymbol \epsilon},{\boldsymbol \epsilon}')=\frac{1}{2}({\boldsymbol \epsilon}'^*\times {\boldsymbol \epsilon})_\alpha$ of the polarization vectors, respectively.  The total scattering amplitude in the dipole limit can now be written as
\begin{equation}
 {\cal F}_{fg}({\boldsymbol \epsilon},{\boldsymbol \epsilon}',\omega_{\bf k}) = \sum_{x=0}^2  {\bf T}^{x} ({\boldsymbol \epsilon},{\boldsymbol \epsilon}')\cdot {\bf F}^{x}(z_{\bf k}).
\end{equation}

The spectra for different $x$ and $q$ are combinations of the spectra for different polarizations. Usually, the scattering amplitudes are calculated in terms of  the components $D_{\alpha}$  of the dipole operator, where $\alpha=1,0,-1$ in spherical symmetry or $\alpha=x,y,z$ in Cartesian coordinates. The spectra for different polarizations are then combined to form the fundamental scattering amplitudes. This can be compared with x-ray absorption. The circular dichroic spectrum (the $x=1$ fundamental spectrum for x-ray absorption) is usually calculated by subtracting the spectra for left and right circularly polarized light ($\alpha=\pm 1$). The scattering amplitudes in terms of the components of the dipole operator are given by
\begin{align}
F_{\alpha'\alpha} &=  \bra{f} D_{\alpha'}^\dagger G(z_{\bf k})  D_{\alpha} \ket{g} \nonumber \\
&=\sum_n    \frac{\bra{f} D_{\alpha'}^\dagger\ket{n}
\bra{n} D_{\alpha}\ket{g}}{\hbar \omega_{\bf k}+E_g-E_n +i\Gamma}.
\label{eq:Fqq}
\end{align}
Note that again there are only nine spectra and this is just a representation of the nine fundamental spectra in a different basis. The simplest scattering amplitude is the isotropic one given by $x=0$. The tensor containing the isotropic scattering amplitudes ${\bf F}^0$ has only one component $F^0_0$,
\begin{equation}
   F^{0}_0 = F_{00} +F_{11}+F_{-1,-1} =F_{xx} +F_{yy}+F_{zz},
\end{equation}
which is just a sum of all the different polarization components.  For the expressions in spherical symmetry, note that, since $r_{i\alpha'}^\dagger=(-1)^{\alpha'} r_{i,-\alpha'}$, there is no net transfer of angular momentum to the system for the isotropic scattering amplitude. Since the angular dependence is given by
${\bf T}^{0}={\boldsymbol \epsilon}'^* \cdot{\boldsymbol \epsilon}$,
the isotropic contribution to the spectral line shape is removed in
many experiments by the use of  a $90^\circ$ scattering condition with
the incoming polarization vector in the scattering plane ($\pi$
polarized). This makes the incoming polarization vector perpendicular
to both possible outgoing polarization vectors and therefore
${\boldsymbol \epsilon}'^*\cdot {\boldsymbol \epsilon}=0$. In
addition, this has the advantage that it strongly reduces the
non-resonant ${\bf A}^2$ term from the experimental RIXS data (which
has the same polarization dependence). This contributes mostly to the elastic line and is frequently the major experimental impediment to measuring low-energy excitations.

Tensors of rank $x=1$ have  three components. For example, the $q=0$ component is given by
\begin{equation}
F^{1}_{0} =F_{11}- F_{-1,-1}=F_{xy}-F_{yx}.
\end{equation}
For resonant elastic x-ray scattering, the $F^{1}_{0}$ scattering
amplitude is the one that gives rise to, amongst others, magnetic scattering. The angular dependence for $x=1$, is given by an outer product ${\bf T}^{1}={\boldsymbol \epsilon}'^*\times {\boldsymbol \epsilon}$.

At this point it is useful to make a comparison with X-ray absorption (XAS) and resonant X-ray scattering (RXS), which are determined by the scattering amplitude $ {\cal F}_{gg}$ where
\begin{eqnarray}
&I_{\rm XAS}({\boldsymbol \epsilon},\omega_{\bf k}) = -\frac{1}{\pi} {\rm Im} [{\cal F}_{gg}({\boldsymbol \epsilon},{\boldsymbol \epsilon},\omega_{\bf k})] \\
&I_{\rm RXS}({\boldsymbol \epsilon},{\boldsymbol \epsilon}',\omega_{\bf k}) = |{\cal F}_{gg}({\boldsymbol \epsilon},{\boldsymbol \epsilon}',\omega_{\bf k})|^2 ,
\end{eqnarray}
where for X-ray absorption, there is only a polarization vector for the incident X-rays, and ${\boldsymbol \epsilon}'\equiv {\boldsymbol \epsilon}$.  Since  for XAS and RXS the "final" state is equivalent to the ground state in the scattering amplitude ($|f\rangle= |g\rangle$), an additional restriction is imposed on the scattering. In many symmetries, this means that only the $q=0$ component contributes, reducing the scattering amplitude determining  XAS and RXS to
\begin{equation}
{\cal F}_{gg}(\omega_{\bf k})=
\sum_{x=0}^2 T^{x}_0 ({\boldsymbol \epsilon},{\boldsymbol \epsilon}')  F^{x}_0(z_{\bf k}).
\end{equation}
This implies that of the $3\times 3=9$  components in the full scattering amplitude, only 3 components, corresponding to $x=0,1,2$ and $q=0$, remain. For X-ray absorption, these correspond to the well-known isotropic, circular dichroic, and linear dichoic spectra, respectively.

\subsubsection{Scattering amplitude for a multipole expansion}

We next generalize the ideas from the previous section to include the different types of multipoles arising from the ${\bf p}\cdot{\bf A}$  interaction in Eq. (\ref{Hprime}). Since the dipolar and quadrupolar transitions in RIXS are predominantly excitations from a localized core hole into the valence states, the common approach is to expand the plane wave in the vector potential, see Eq. (\ref{vectorpotential}), around the site where the absorption takes place. Essentially, one is using an approximation of the type $e^{i{\bf k}\cdot {\bf r}_i}\cong 1+i{\bf k}\cdot {\bf r}_i$ but in spherical harmonics. In the previous section, we treated the case that $e^{i{\bf k}\cdot {\bf r}_i}\cong 1$.
The plane wave can be expanded in terms of spherical harmonics $Y_{lm}
(\theta,\varphi)$ and spherical Bessel functions $j_l$ \cite{Varshalovich1988}
\begin{equation*}
e^{i{\bf k}\cdot {\bf r}_i}
=4\pi \sum_{l=0}^\infty \sum_{m=-l}^l i^l j_l(kr_i) Y^*_{lm} (\theta_{\hat {\bf k}},\varphi_{\hat {\bf k}})  \cdot Y_{lm} (\theta_{{\hat {\bf r}}_i},\varphi_{{\hat {\bf r}}_i}).
\end{equation*}
 In order to arrive at the standard transition operators, it makes  sense,  at this point, to rewrite the above equation in terms of spherical tensors, see Appendix. A common set of tensors are the normalized spherical harmonics, which we write as the tensor ${\hat {\bf r}}^{(l)}$ with components ${\hat  r}^{(l)}_m= \sqrt{4\pi/[l]} Y_{lm} (\theta_{\hat {\bf r}},\varphi_{\hat {\bf r}})$. In addition, ${\bf r}^{(l)}=r^l {\hat {\bf r}}^{(l)}$. Note that ${\bf r}^{(0)}=1$. For spherical harmonic tensors of order $l=1$, the
 superscript is dropped ${\bf r}={\bf r}^{(l)}$. This allows us to rewrite the expansion as
\begin{eqnarray}
   e^{i{\bf k}\cdot {\bf r}_i} =\sum_{l=0}^\infty [l] i^l j_l(kr) {\hat {\bf k}}^{(l)} \cdot {\hat {\bf r}_i}^{(l)}.
\end{eqnarray}
As in the previous section, we want to separate the momentum and polarization vectors of the x-rays (the geometry of the experiment) from the transitions in the material under consideration (the fundamental spectra). This can be done by recoupling the different tensors. Recoupling of the tensors using Eq. (\ref{recoupling}) \cite{Varshalovich1988,Brink1962,Yutsis1988} leads to
\begin{eqnarray}
{\cal D}&=&\frac{1}{im\omega}\sum_{i=1}^N e^{i{\bf k}\cdot {\bf r}_i} {\boldsymbol \epsilon}\cdot {\bf p}_i \nonumber \\
&=&\frac{1}{im\omega} \sum_{i=1}^N\sum_{lL} \frac{[lL] i^l}{[l]!!} n_{1lL}^2 k^l
[{\bf p}_i,{\bf r}_i^{(l)}]^L \cdot [{\boldsymbol \epsilon},
{\hat {\bf k}}^{(l)} ]^L
\label{expansion}
\end{eqnarray}
where  the approximation $j_l(kr)\cong (kr)^l/[l]!!$  for $kr\ll 1$  has been used, with the double factorial $l!!=l(l-2)\cdots$. Note that the operators acting on the electrons, namely the momentum ${\bf p}_i$ and position ${\bf r}_i$, are coupled together to form an effective operator ${\bf D}^{lL}$ of rank $L=l-1,l,l+1$. The quantities related to the photons, namely, the wavevector ${\bf k}$ and polarization ${\boldsymbol \epsilon}$ also form a tensor of rank $L$.  Let us first consider the transition operators in Eq.  (\ref{expansion}) by introducing the transition operators \cite{Veenendaal1998}
\begin{equation}
{\bf D}^{lL} =\frac{p_{lL}(k)}{im\omega} \sum_{i=1}^N [{\bf p}_i,{\bf r}_i^{(l)}]^L,
\end{equation}
with $p_{lL}(k)= [lL]  n_{1lL}^2 k^l/[l]!!=1,k/2,k/6$ for $lL=01,11,12$, respectively. The values of $l$ and $L$ give rise to the  usual dipolar ($lL=01$), magnetic dipolar ($11$), and quadrupolar ($12$) transition operators. For $l=0$, one has ${\bf r}_i^{(0)}=1$ and the operator simplifies to ${\bf D}^{01}=\sum_i {\bf p}_i/im\omega$, which is equivalent to the dipole operator in Eq. (\ref{transoper}) in the previous Section.
 In cartesian coordinates, the operator
$ [{\bf r}_i,{\bf p}_i]^1= {\bf L}_i$ and ${\bf D}^{11}=\frac{\alpha a_0}{2}\sum_i  {\bf L}_i$, with $a_0$ the Bohr radius and the angular momentum given in $\hbar$. The orbital moment forms, together with the Zeeman term in Eq. (\ref{Ham}),  the magnetic dipole transition.  Since magnetic dipole transitions are of the order of $\alpha^2=1/137^2$, i.e.  about five orders of magnitude, smaller than the electric dipole transitions of the same wavelength, they will be neglected in the remainder of the paper. The next operator is ${\bf D}^{12}=\frac{k}{6}\sum_{i=1}^N{\bf r}^{(2)}_i$ which is the electric quadrupole operator.

In the remainder, we limit ourselves to the electric $L$-pole transitions, and we can drop the $l=L-1$ from the expressions, i.e. ${\bf D}^{L-1,L}\rightarrow {\bf D}^L$. The transition operators are then ${\cal D} \sim [{\boldsymbol \epsilon},{\bf k}^{(L-1)} ]^L\cdot {\bf D}^L$ with ${\bf D}^L= p_L(k) \sum_{i=1}^N {\bf r}_i^{(L)}$ with $L=1,2$ for dipolar and quadrupolar transitions, respectively. The relative strengths of the components of the multipole transition operators $r^{(L)}_M$ depend on the direction of polarization and wavevector through $[{\boldsymbol \epsilon},{\bf k}^{(L-1)} ]^L$.
These  reduce to ${\boldsymbol \epsilon}$ and $[{\boldsymbol \epsilon},{\bf k} ]^2$, for the electric dipolar and quadrupolar transitions, respectively. As discussed in the previous section,  the part that depends on the geometry of the experiment and the fundamental spectra (9 and 25 for dipolar and quadrupolar transitions, respectively) that describe the physical properties of the system can be separated exactly. This can again be achieved by applying a recoupling using Eq. (\ref{recoupling}) on the scattering amplitude, which can then be rewritten as
\begin{equation*}
F_{fg}({\bf k},{\bf k}',{\boldsymbol \epsilon},{\boldsymbol \epsilon}',
\omega_{\bf k})=\sum_{x=0}^{2L} {\bf T}^{Lx} ({\hat {\bf k}},{\hat {\bf k}}',{\boldsymbol \epsilon},
{\boldsymbol \epsilon}') \cdot {\bf F}^{Lx}(k,k',\omega_{\bf k}).
\end{equation*}
Neglecting interference effects between different multipoles, the scattering amplitude for a particular multipole is given by \cite{Veenendaal1998}
\begin{equation*}
{\bf F}^{Lx}(k,k',\omega_{\bf k})
= \sum_n  \frac{[\bra{f} ({\bf D}^L)^\dagger\ket{n},
\bra{n} {\bf D}^L\ket{g}]^x}{\hbar \omega_{\bf k}+E_g-E_n +i\Gamma},
\end{equation*}
 which has angular dependence
\begin{equation*}
{\bf T}^{Lx} ({\hat {\bf k}},{\hat {\bf k}'},{\boldsymbol \epsilon},{\boldsymbol \epsilon}')
= [x]n_{LLx}^2 [ [{\boldsymbol \epsilon}'^*,{\hat {\bf k}}'^{(L-1)} ]^L,[{\boldsymbol \epsilon},{\hat {\bf k}}^{(L-1)} ]^L]^x.
\end{equation*}
The above equations give an exact separation of the Kramers-Heisenberg expression for RIXS into an angular dependence and fundamental scattering amplitude achieving the first step in Table \ref{scatteringamplitude}.
\subsection{Direct and Indirect RIXS}
\begin{center}
\begin{table*}
\centering\begin{tabular}{|c|c|c|c|}
\hline
\multicolumn{4}{|c|}{Intermediate State Propagator}\\
\hline
& \multicolumn{3}{c|} {RIXS amplitude ${\cal F}_{\tt fg}=\langle {\tt f} | {\cal D}^{\dagger} G {\cal D} | {\tt g}\rangle$}\\
Definition& \multicolumn{3}{c|}{  Intermediate State Propagator $G(z_{\bf k})=\frac{1}{z_{\bf k}-H}$; $H=H_0 + H_C$}\\
& \multicolumn{3}{c|} { $H_0$ ground state, $H_C$ core-hole Hamiltonian}\\   \hline
& \multicolumn{3}{c|}{} \\
& \multicolumn{3}{c|}{$G=G_0+G_0 H_C G$~~~~~~~~~~~~~~~~~~~~~~~~~~~~~~~~~~~~ } \\
Exact Propagator  & \multicolumn{3}{c|}{\Large $\swarrow$ ~~~~~~ $\searrow$~~~~~~~~~~~~~~~~~~~~~~~} \\
& \multicolumn{3}{c|}{\phantom{\tiny .}} \\
\cline{2-4}
& {\it Direct RIXS} & \multicolumn{2}{c|}{\it Indirect RIXS}\\
& ~~~~~~~~~~{${\cal F}^{direct}_{\tt fg}=\langle {\tt f} | {\cal D}^{\dagger} G_0 {\cal D} | {\tt g}\rangle$} ~~~~~~~~~~& \multicolumn{2}{c|}{~~${\cal F}^{indirect}_{\tt fg}=\langle {\tt f} | {\cal D}^{\dagger} G_0 H_C G {\cal D}| {\tt g}\rangle $ }\\ \hline
\multicolumn{4}{c}{} \vspace{-0.75em} \\
\multicolumn{1}{c}{} & \multicolumn{1}{c}{\Large $\downarrow$  } & \multicolumn{2}{c}{~~~~~~~~~~~~~~~~~~\Large $\swarrow \searrow$ ~~~~~~~~~~~~~~~~~~} \\
\multicolumn{4}{c}{} \vspace{-0.75em} \\  \hline
Approximation & Fast Collision$^{(a)}$ & Perturbation Expansion$^{(a-g)}$ & Ultrashort Core-hole Life-time$^{(h-j)}$ \\
to propagator & $G(z_{\bf k}) \rightarrow 1/z_{\bf k}$ &  $G_0 H_C G \rightarrow G_0 H_C G_0$ & $G_0 H_C G \rightarrow G_0 H_C G_C  $ \\   \hline
Valence excitation &  caused by transition operators $D$ &  \multicolumn{2}{c|}{ caused by core-hole interaction $H_C$}\\  \hline
\end{tabular}
\caption{Theoretical approach to the intermediate state propagator, classifying direct and indirect RIXS processes and common approximations to the propagator\footnote{$^a$\cite{Luo1993}, $^b$\cite{Platzman1998}, $^c$\cite{Doring2004}, $^d$\cite{Abbamonte1999a}, $^e$\cite{Nomura2004}, $^f$\cite{Nomura2005}, $^g$\cite{Markiewicz2006}, $^h$\cite{Brink2005}, $^i$\cite{Brink2006}, $^j$\cite{Ament2007}.}.
}
\label{tab:propagator}
\end{table*}
\end{center}

In the previous section the Kramers-Heisenberg expression for the RIXS scattering amplitude ${\cal F}_{\tt fg}$ , Eq.~\ref{eq:K-H}, was derived and re-expressed as a product of a photon absorption operator $\cal D$, the intermediate state propagator $G$ and a photon emission operator  $\cal D^{\dagger}$, sandwiched between the RIXS final and ground state:
\begin{eqnarray}
  {\cal F}_{\tt fg}=\langle {\tt f} | {\cal D}^{\dagger} G {\cal D} | {\tt g}\rangle.
\label{eq:propagator}
\end{eqnarray}
The presence of the intermediate state propagator is what makes the theory of RIXS complicated -- and  interesting. The propagator $G$ is defined in terms of the inverse of the total Hamiltonian $H$ of the system, $G(z_{\bf k})=(z_{\bf k}-H)^{-1}$, where the operator $H$ naturally divides into the ground state Hamiltonian $H_0$ (governing the quantum system without a core-hole) and the core-hole Hamiltonian $H_C$ perturbing the system after photon absorption: $H=H_0+H_C$. It should be noted that even if one commonly refers to $H_C$ as the core-hole Hamiltonian, it also includes the interaction between the electron excited into the conduction band and the rest of the material. As core-hole and excited electron together form an exciton, their separate effects on the system cannot, in principle, be disentangled .

At this point it is useful to separate the full propagator $G$ into an unperturbed one $G_0 = (z_{\bf k}-H_0)^{-1}$ and a term that contains the core-hole Hamiltonian $H_C$, using the identity $G=G_0+G_0 H_C G$. This also separates  the RIXS amplitude into two parts, which define {\it direct} and {\it indirect} RIXS~\cite{Brink2006}:
\begin{empheq} [box=\mybluebox]{align}
{\cal F}^{direct}_{\tt fg}=\langle {\tt f} | {\cal D}^{\dagger} G_0 {\cal D} | {\tt g}\rangle
\label{eq:direct}
\end{empheq}
and
\begin{empheq} [box=\mybluebox]{align}
{\cal F}^{indirect}_{\tt fg}=\langle {\tt f} | {\cal D}^{\dagger} G_0 H_C G {\cal D}| {\tt g}\rangle.
\label{eq:indirect}
\end{empheq}
Note that this definition of direct/indirect RIXS, based on the Kramers-Heisenberg expression, is exact.

For the direct RIXS amplitude, the core-hole does not play a role --
the photon absorption and emission matrix elements determine which
electronic transitions are allowed. The physical picture that arises
for direct RIXS is that an incoming photon promotes a core-electron to
an empty valence state and subsequently an electron from a different state in the valence band decays, annihilating the core-hole, see Fig.~\ref{fig:direct}. Thus for direct RIXS to occur, both photoelectric transitions -- the initial one from core to valence state and the succeeding one from valence state to fill the core hole -- must be possible. These transitions can, for example, be an initial dipolar transition of $1s \rightarrow 2p$ followed by the decay of another electron in the $2p$ band from $2p \rightarrow 1s$. This happens at the $K$-edge of oxygen, carbon and silicon. In addition, at transition metal (TM) $L$-edges, dipole transitions causing direct RIXS are possible via  $2p \rightarrow 3d$ and $3d \rightarrow 2p$ dipolar transitions. In all these cases RIXS probes the valence and conduction states directly.

For indirect RIXS, the scattering amplitude depends critically  on the perturbing core-hole Hamiltonian -- without it the indirect scattering amplitude vanishes. In general the scattering, ${\cal F}^{indirect}_{\tt fg}$, arises from the combined influence of $H_C$ and transition matrix elements $\cal D$. Most often for indirect RIXS, $\cal D$/$\cal D^\dagger$ create/annihilate an electron in the same state, far above the Fermi level. For instance at the TM $K$-edge, the $1s \leftrightarrow 4p$ process creates/annihilates an electron in $4p$ states electonvolts above the TM $3d$ valence shell. The delocalized $4p$ electron can then be approximated as being a {\it spectator} because (Coulomb) interactions involving the localized core-hole are usually much stronger and dominate the scattering cross section.

It should be noted that if scattering is direct, as for instance at TM $L$-edges,  indirect processes can also contribute to the total scattering amplitude. However, as indirect scattering arises in this case as a higher order process, it is normally weaker than the leading order direct scattering amplitude. Conversely, in case of indirect RIXS, direct processes are absent by definition.

\section{Theory of RIXS: II. Calculating the cross-section\label{sec:Calculating_crosssection}}

Computing the exact RIXS response of a material using the full propagator is in general a daunting if not impossible task. This section describes the different approximate methods used to evaluate the Kramers-Heisenberg scattering amplitude for a specific system.

The most straightforward approach is to evaluate the exact scattering amplitude Eq.~(\ref{eq:scatteringampl}) and intensity Eq.~(\ref{KramersH}) numerically for a simplified system.  One can for instance neglect or simplify correlation and core-hole interaction effects and resort to an effective single-particle description, for instance with band structure methods. Another possibility is to keep the correlation effects but reduce the number of degrees of freedom by considering instead of the full macroscopic system only a small cluster of atoms~\cite{Kotani2001}. We will briefly discuss such cluster methods in Section \ref{sec:numerical}.

A complementary approach is to approximate the intermediate-state core-hole propagator. At first, this appears less appealing than a direct evaluation of the RIXS cross section. However, one should note that, in particular for strongly correlated systems, numerical calculations can only be done for small systems, such as an ion, a cluster of atoms, or an impurity model. Often, one is also interested in the coupling of RIXS to collective excitations, not present in these small systems. Therefore, effective methods that project out the intermediate states are of crucial importance for understanding the RIXS cross section. These approaches provide effective response functions for RIXS which can be related to response functions for other spectroscopic techniques. As a next step one can then compute, in cases where it is feasible, the effective RIXS response function exactly. Depending on whether one considers direct or indirect RIXS, one uses Eqs. (\ref{eq:direct}) or (\ref{eq:indirect}), as a starting point to derive an effective expression for the RIXS cross section.

For {\it direct} RIXS the transition operators are key to the scattering process. One can highlight this physics by neglecting the dependence on $H_C$ of the full intermediate state propagator, so that $G \rightarrow 1/z_{\bf k}$ -- this amounts to neglecting the {\it indirect} contributions to the scattering process. This is known as the {\it fast collision approximation}~\cite{Luo1993,Veenendaal2006}. This approximation allows the reduction of the full direct RIXS amplitude to a rather elementary response function of effective scattering operators. This is a great advantage and simplification, even if evaluating this response function for a real (strongly correlated) material still presents a challenge in itself -- as is evaluating any other response function for a correlated electron system. These effective theories for direct RIXS are reviewed in Section~\ref{sec:effdirect}.

In the case of indirect RIXS the scattering is due to the intermediate state core-hole potential. The approximation closest at hand is to treat this perturbing potential as being weak, expand $G$ in a perturbation series and retain only the leading terms \cite{Platzman1998,Doring2004,Abbamonte1999a,Nomura2004,Nomura2005,Takahashi2007,Markiewicz2006}. To deal with the strong core-hole interaction, the Ultrashort Core-hole Lifetime (UCL) expansion of the core-hole propagator was developed~\cite{Brink2005,Brink2006,Ament2007}, which treats the core-hole potential as the dominant energy scale. Section~\ref{sec:effindirect} presents a survey of the different effective theories for indirect RIXS.
\subsection{Numerical Evaluation of Kramers-Heisenberg \label{sec:numerical}}

There are several numerical approaches used to evaluate the Kramers-Heisenberg equation directly. Although there are many different methods to calculating the electronic structure, the general approach is always the same. First, one finds the eigenstates and eigenvalues for ground, intermediate, and final states. Subsequently, one calculates the relevant electric-multipole matrix elements between the core levels and the valence shell for the appropriate scattering condition, and evaluates Eq. (\ref{eq:Fqq}).

\subsubsection{Single particle approach}
A common approach in the description of the electronic structure of solids is the use of an independent-particle approximation. After applying a Local-Density Approximation, the electronic structure can be calculated {\it ab initio}. Numerical methods based on this approximation can be successfully applied to systems where the interaction between the electrons can be described by an effective potential.  For RIXS, this approach is particularly useful in the study of semiconductors and wide-band insulators \cite{Johnson1994,Carlisle1995,Jia1996,Denlinger2002,Strocov2004,Strocov2005,Kokko2003}.

Here, the absorption process excites an electron from a deep-lying core level into the bands above the Fermi level. Varying the energy of the incoming photon energy over the absorption edge essentially probes the unoccupied density of states. The de-excitation removes an electron from the states below the Fermi level. The final states have an excited electron-hole pair. Since the transferred energy and momentum are known, the RIXS spectrum is essentially a momentum-resolved convolution of the projected occupied and unoccupied density of states. For relatively shallow core levels such as the $1s$ level of Carbon, the momentum transfer is very small and the created electron-hole pairs have $q\cong 0$.

This approach entirely neglects any interaction that can occur in the intermediate state between the core-hole and valence-shell electrons. It is well known that for semiconductors or large band-gap insulators, the attractive Coulomb interaction between the core hole and the valence electrons gives rise to an the intermediate state with an exciton consisting of the core hole and the electron excited into the conduction band. Since this is a two-particle problem, it can be solved exactly using Bethe-Salpeter type equations~\cite{Shirley1998,Veenendaal1997}. In a metallic system, the absence of a gap leads to the creation of low-energy electron-hole pair excitations that screen the core-hole potential. This response leads to X-ray edge  singularities in the RIXS spectral line shape~\cite{Nozieres1974,Ahn2009}.

\subsubsection{Exact diagonalization and other cluster methods}
In recent years, the majority of RIXS experiments have been performed on materials that display strong Coulomb interactions between the electrons. For these systems, {\it ab initio} methods based on the assumption of independent particles are not applicable and one generally resorts to model Hamiltonians. In these calculations one defines an appropriate basis set. Subsequently one sets up the Hamiltonian using parameters obtained from {\it ab initio} methods, Hartree-Fock type calculations, or in a more heuristic fashion. The Hamiltonian is then diagonalized for ground, initial, and final states and the RIXS cross section is calculated using Eq. (\ref{eq:Fqq}).

Exact diagonalization methods have the advantage of being able to treat many-body effects since the eigenstates are not limited to a single Slater determinant. The major disadvantage is that the basis set increases very rapidly since all possible permutations of electron in the basis set need to be considered. Despite these limitations, these approaches have been successful in describing the RIXS cross section, in particular for the $L$ and $M$ edges in transition-metal compounds. This has been described extensively in Refs.~\cite{Harada2000,Agui2009,Veenendaal2004,Ghiringhelli2006a,Ghiringhelli2008,Agui2005,Magnuson2002,Chiuzbaian2008,Groot1998,Magnuson2002a,Matsubara2005,Chiuzbaian2005,Ghiringhelli2005,Huotari2008,Vernay2009,Ferriani2004} and reviewed in Refs. \cite{Groot2008} and \cite{Kotani2001}.

Since the many-body basis set that one can handle numerically is limited, it is essential to decide which states, interactions and electronic configurations are to be taken into account, and conversely, which ones can safely be neglected.

In direct RIXS, after a $np\rightarrow 3d$ (with $n=2,3$) dipolar transition, the excited electron is strongly bound to the $np$ core hole. Therefore the most important configurations are the metal-centered ones. The process can then be described as $np^6 3d^m\rightarrow np^5 3d^{m+1}\rightarrow np^6 (3d^m)^*$, where $m$ are the number of $3d$ electrons and $(3d^m)^*$ indicates an excited $3d$ state. Since $L$ and $M$ RIXS in transition-metal compounds is mainly a direct process, many excited states can be accessed.

In order to interpret the $dd$ type excitations that are typically observed at these edges, it is important that the Hamiltonian include a Coulomb interaction that includes higher-order terms. This interaction splits the $3d$ states into Coulomb multiplets that can be separated by tenths of eV to several eV \cite{Griffith1961}. A well-known effect of these higher-order terms in the Coulomb interaction is Hund's rule that determines the ground state for a free ion. To mimic the effect of a solid, a crystal field can be introduced, which further splits the Coulomb multiplets. An additional interaction that needs to be included is the spin-orbit coupling, in particular since at the $L$ edges in transition-metal compounds, the $2p$ spin-orbit coupling is by far the strongest interaction in the intermediate state, splitting the absorption edge into two distinct features. Since the direct RIXS process creates excitations involving all $3d$ states, it is in addition very important to properly include dipole/quadrupole transitions, since polarization effects are crucial in the interpretation of the RIXS at the $L$ and $M$ edges in transition-metal compounds.

A next step in including solid-state effects is to take into account the ligands, such as oxygen, surrounding the ion where the scattering takes place explicitly in the calculation. Usually, a metal ion is surrounded by 4-6 ligands. The Hamiltonian then needs to include the hybridization between the metal and the ligands and the on-site energy of the ligands. An important parameter in this respect is the charge-transfer energy $\Delta=E(3d^{m+1}{\underline L})-E(3d^m)$, where ${\underline L}$ indicates a hole on the ligands,  that gives the energy needed to transfer an electron from the ligands to the metal site. This approach is known as (small) cluster calculations, referring to the cluster of atoms taken into account in the exact diagonalization. Instead of including only the surrounding ligands, one can also include the bands corresponding to the ligands. This is known as a single-ion Anderson impurity model. However, often it is more important to include additional metal sites in the small cluster calculations, especially when considering strongly covalent and/or doped materials.

For indirect RIXS, other considerations are often more important in limiting the size of the basis set. Here, valence excitations are shake-up structures related to the screening of the strong transient core-hole potential in the intermediate state. Since the screening processes often only involve the $3d$ states closest to the Fermi level, not all $3d$ orbitals are necessarily relevant. On the other hand, the spatial extent of the cluster becomes increasingly important, since screening charge is pulled in, not only from neighboring ligands, but also from neighboring metal sites. Furthermore, since the excited electron can usually be considered a spectator, the transfer of angular momentum to the valence band is much smaller and polarization effects are less important.

Therefore, the appropriate reduction of the basis set depends strongly on the type of final state excitations that one considers. Let us illustrate this with an example:  consider a divalent copper compound, which has holes in the $3d_{x^2-y^2}$ orbital. In describing the $L$ and $M$ edge RIXS it is important to include all the $3d$ orbitals and their appropriate crystal fields, since the final states consist primarily of local $dd$ transitions. In addition, spin flips can occur due to the strong core-hole spin-orbit coupling in the intermediate state. In addition, since this is a direct process, polarization effects are important in reproducing the correct RIXS intensities. At the $K$ edge, on the other hand, the screening dynamics in the intermediate state primarily involve the $3d_{x^2-y^2}$ orbitals. If the calculation is restricted to a small cluster containing only one copper ion \cite{Hill1998}, only high-energy bonding-antibonding between the $3d^m$ and $3d^{m+1}{\underline L}$ configurations on a local copper plaquette are included. This misses out the important charge-transfer excitations between different copper plaquettes. In order to describe these and obtain insight into their dispersion additional simplifications can be made by reducing the copper-oxygen plaquette to an effective single site system. This leads to the well-known single-band Hubbard model, where a number of exact diagonalization studies for RIXS have been carried out \cite{Tsutsui1999}.
\subsection{Effective theory for direct RIXS \label{sec:effdirect}}

The assumption in direct RIXS is that the excitations are predominantly created by the radiative transitions. Essentially, this approach was followed in papers on systems with wide bands such as boron nitride \cite{Jia1996}, diamond \cite{Ma1992}, and graphite \cite{Carlisle1995}. In these systems, the intermediate state propagator (or, equivalently, the x-ray absorption spectrum) can be described as the ground state plus an  electron excited into the unoccupied states. The RIXS spectrum is then interpreted as the creation of electron-hole pairs and effectively a (projected) joint density of states due to the two consecutive radiative transitions. The polarization dependence of the direct RIXS process shows up nicely for graphite \cite{Carlisle1995} in the change in intensity as a function of the angle of the polarization of the incoming light with respect to the graphite planes. However, for strongly correlated systems, such as transition-metal compounds, such an intuitive approach to the intermediate states often fails due to intermediate state interactions. However, at the transition-metal $L$ and $M$ edges, the assumption that the excitations are, to first order, due to the dipolar transitions is still valid. In order to understand what valence excitations can be created by the radiative process, one can apply the UCL approximation to lowest order (also known as the fast-collision approximation). In this limit, the intermediate-state propagator is replaced by a resonance factor. 

\subsubsection{Fast collision approximation and effective transition operators \label{sec:fastcollision}}
To obtain effective valence operators for the $L$ and $M$ edge, the dipole operator is expressed in second quantization \cite{Judd1967}. For a transition between a core level and the valence shell with angular momentum $c$ and $l$, respectively, we have
\begin{equation}
{\hat D}^L_M=\sqrt{[cl]} P_{n_c c,nl}{\hat w}^{clL}_M
\label{transition2nd}
\end{equation}
where the $\sqrt{[cl]}$ has been kept separate for use in the derivation of sum rules and effective operators. The reduced matrix elements are contained in
\begin{align}
P_{nl,n_cc}^L &=  p_{L-1,L}(k)(-1)^{l} \left (
 \begin{array}{ccc}
 l & L & c \\
0 & 0& 0
 \end{array}
\right ) \nonumber \\
 &\;\;\;\; \times n_{clL} 
 \int_0^\infty dr  r^{2+L} R_{nl}(r) R_{n_cl_c}(r) ,
\label{reduced}
\end{align}
where the coefficient $n_{ll'x}$ is defined in the Appendix; The term in parentheses is a $3j$ symbol, which is proportional to the Clebsch-Gordan coefficients. The integral is over the radial parts of the wavefunctions. Although it is important in determining the strength of the transition, in general, it only scales the spectrum and we can often take $P_{nl,n_cc}\equiv 1$. The essential physics is contained in the operator
\begin{equation}
{\hat w}^{clL}_M=\sum_{mm'\sigma}  w^{clL}_{mm'M} l^\dagger_{m'\sigma} c_{m\sigma},
\label{tensorop}
\end{equation}
where $c_{m\sigma}$ annihilates an electron in the core level with orbital component $m$ and spin $\sigma$ and $l^\dagger_{m'\sigma}$ creates an electron in the valence shell with quantum number $m'$ and the same spin. The coefficients are given by 
\begin{equation}
w^{ll'L}_{mm'M}=
(-1)^{l'-m'}
\left (
 \begin{array}{ccc}
 l' & L & l \\
-m' & M& m
 \end{array}
 \right )
n_{ll'L}^{-1} .
\label{coefoper}
\end{equation}
The $3j$ symbol imposes the well-known selection rule $m'=m+M$ since the sum of the elements in the lower row has to equal zero.

In direct RIXS, the spectral line shape is mainly determined by the transition operators, see Eq. (\ref{eq:direct}). In order to obtain effective operators that give the combined action of the absorption and emission process, we need to  approximate the intermediate-state propagator. In order to show the procedure, let us first consider a single edge, before considering the situation where the edge is split by the core-hole spin-orbit coupling. This is usually done using a fast-collision approximation \cite{Luo1993,Veenendaal2006}
\begin{equation*}
F^x_{fg,q}(z) = \bra{f} [{\hat {\bf D}}^\dagger,G(z_{\bf k}){\hat {\bf D}}]^x_q \ket{g} \cong {\bar G}(z_{\bf k}) \bra{f} [{\hat {\bf D}}^\dagger,{\hat {\bf D}}]^x_q \ket{g},
\end{equation*}
where  ${\bar G}(z_{\bf k})=(z_{\bf k}-{\bar E})^{-1}$. This approximate intermediate state propagator is then effectively the resonance function $P(\omega_{\bf k})$, see Table \ref{scatteringamplitude}, describing the strength of the scattering amplitude as a function of the incoming photon energy.
 The dipole operator can be written as  ${\hat D}^L_M=\sqrt{[cl]}{\hat
   w}^{clL}_M$, where the reduced matrix element, which only scales
 the spectrum, has been taken unity, $P_{n_c c,nl}\equiv 1$. Since
 neither the ground state nor the final states contain a core hole, we can effectively remove the creation and annihilation operators related to the core states \cite{Luo1993,Veenendaal2006}. Using spherical tensor algebra, it can be shown  \cite{Luo1993,Veenendaal2006}
\begin{empheq} [box=\mybluebox]{align}
F^x_{fg,q}(z)&= {\bar G}(z_{\bf k})[cl] \langle f| [({\hat {\bf w}}^{clL})^\dagger,{\hat {\bf w}}^{clL}]^x_q |g\rangle\nonumber   \\
&={\bar G}(z_{\bf k}) [c] \langle f| {\hat {\underline w}}^{llx}_q |g\rangle,
\label{scatspinindependent}
\end{empheq}
where the usual situation $l=c+L$ has been taken; ${\hat {\bf w}}^{clL}$ is a tensor containing the operators ${\hat w}^{clL}_M$ with $M=L,L-1,\cdots,-L$. Equation (\ref{scatspinindependent}) is the central expression for the fundamental scattering amplitude in the absence of core-hole spin-orbit coupling. Note that the combined action of the two operators ${\hat {\bf w}}^{clL}$ representing the absorption and emission between the core level and the valence shell has been replaced by one effective hole operator 
\begin{eqnarray}
{\hat {\underline w}}^{llx}_q=\sum_{mm'\sigma}   w^{llL}_{mm'M} l_{m'\sigma} l^\dagger_{m\sigma}
\label{wllx}
\end{eqnarray}
 acting only on the valence shell. The coefficients are again given by Eq. (\ref{coefoper}). For $x=0$, the operator ${\hat {\underline w}}^{ll0}_0=n_h$ is the number of holes operator. Within an ionic framework, this only contributes to the elastic line, although it can give rise to inelastic satellite structures. For $x=1,2$, the operator is the angular momentum  ${\hat {\underline {\bf w}}}^{ll1}={\bf L}/l$ and the quadrupolar moment, respectively. The transitions made by these operators will be discussed in the Section on $dd$ excitations.

In many cases, the core-hole spin-orbit coupling cannot be neglected. For example at the $L_{23}$-edges ($2p\rightarrow 3d$) in transition-metal compounds, one has two distinct edges corresponding to a total angular momentum $j^\pm=c\pm \frac{1}{2}=\frac{3}{2},\frac{1}{2}$ of the core hole with angular momentum $c=1$. The same evaluation of the combined action of the transition operators for absorption and emission can be done at a particular spin-orbit split edge. Restricting ourselves to the common situation where $l=c+Q$, the expressions can be reduced to
\begin{empheq} [box=\mybluebox]{align}
F^z_{fg,q}(j^\pm)&= 
 {\bar G}_j(z_{\bf k})[cl] \langle f| [({\hat {\bf w}}^{jlL})^\dagger,{\hat {\bf w}}^{jlL}]^z_q |g\rangle \nonumber \\
&= {\bar G}_j(z_{\bf k}) \langle f| {\hat {\underline V}}^{llz}_q (j^\pm) |g\rangle ,
\label{operspindependent}
\end{empheq}
where the fast-collision approximation is now done for a particular spin-orbit split edge ${\bar G}_j(z_{\bf k})=(z_{\bf k}-{\bar E}_j)^{-1}$, where ${\bar E}_j$ is an average energy for each edge with $j=l\pm \frac{1}{2}$.  The rank of the total tensor is now $z$ (reserving the $x$ for the orbital part of the operators). The operator is now 
\begin{align}
 {\hat {\underline V}}^{llz}_q (j^\pm) 
 &= (j^\pm +\frac{1}{2}){\hat {\underline v}}^{z0z}_q \nonumber \\ &\mp \frac{c}{[z]} 
 \left [ 
 {\hat {\underline v}}^{z-1,1z}_q -
\frac{(z\mathopen{+}1)(z\mathopen{+}2)}{2}
 {\hat {\underline v}}^{z+1,1z}_q
 \right ], 
 \label{Vllz}
\end{align}
with $j^\pm=c\pm \frac{1}{2}$. Since the spin is no longer a good quantum number at a particular spin-orbit split edge, the operators are now spin dependent
\begin{equation}
{\hat v}^{xyz}_q = 2[{\hat w}^{llx},{\bf S}^y]^z_q ,
\end{equation}
where ${\bf S}^0=1$ and ${\bf S}^1={\bf S}$. The $ll$ in the
superscript has been assumed implicitly to avoid crowding of
indices. The rank $z=0,\cdots,2L$ is still related to the rank of the
transition operators; $y$ indicates the spin dependence and can assume
the values 0 and 1, corresponding to spin-independent and
spin-dependent operators respectively; $x=z-1,1,z+1$ relates to the
orbital part of the operator. First, note that for the
spin-independent operators ${\hat v}^{z0z}_q=w^{llx}_q$ and that when
adding the transition operators from the different edges, ${\hat
  {\underline V}}^{llz}_q (j^+) +{\hat {\underline V}}^{llz}_q (j^-)
=[c]{\hat {\underline v}}^{z0z}_q $, the result in the absence of
spin-orbit coupling is reproduced. Therefore, the first term on the
right-hand side in Eq. (\ref{Vllz}) gives, at each edge, the transition as if there was no spin-orbit interaction. Note that the intensity at each edge is proportional to the multiplicity of the core level $j^\pm+\frac{1}{2}$. The additional operators are spin-dependent. Most notable are the spin ${\bf v}^{011}=2{\bf S}$ and the spin-orbit coupling $v^{110}_0=\frac{2}{l} {\bf L}\cdot{\bf S}$. The effect of the spin-dependent operators will be discussed in the Section on magnetic excitations.

\begin{table}[t]
\begin{ruledtabular}
\caption{Orbitals for the $s$, $p$, and $d$ orbitals with $l=0,1,2$, respectively. The value of $\mu=l,l-1,\cdots,-l$.  The symmetry notations in $O_h$ (octahedral) and $D_{4h}$ are also given. }
\begin{tabular}{crlllc}
 orbital & $l$ &  $\mu$  & $O_h$ & $D_{4h}$  \\
\hline  \\
$s$ &0  & 0 &  $A_1$ & $A_1$ \\
$p_x$ &1 & 1 &  $T_1$ & $E$ \\
$p_z$ & 1 & 0 & $T_1$ & $A_2$ \\
$p_y$ &1 & $-1$ &  $T_1$ & $E$ \\
$d_{x^2-y^2}$ &2 & 2 &  $E$ & $B_1$ \\
$d_{zx}$ &2 & 1  & $T_2$ & $E$ \\
$d_{3z^2-r^2}$ & 2 & 0 & $E$ & $A_1$ \\
$d_{yz}$ & 2 & $-1$ & $T_2$ & $E$ \\
$d_{xy}$ & 2 & $-2$ &  $T_2$ & $B_2$ \\
\end{tabular}
\nopagebreak
\label{tesseralT}
\end{ruledtabular}
\end{table}

\subsubsection{Selection rules}
Equation (\ref{scatspinindependent}) shows that, within the
fast-collision approximation, the combined action of the transition
operators for absorption and decay can be written as an effective
transition operator in the valence shell. The rank of the transition
operator is $x$. For dipole operators, we have $x=0,1,2$. The presence
of the $x=1$ term might seem surprising since this is comparable to an
effective dipole operator (due to ${\bf r}$, a tensor of rank 1) in
the valence shell, and it is well known that dipole transitions within
one shell are not allowed. However, such transitions arise from the
reduced matrix element in Eq. (\ref{reduced}) and not from the
transition operator in Eq. (\ref{coefoper}). Since the reduced matrix
element in RIXS is the product of the reduced matrix elements for
absorption and emission, the $x=1$ term is
allowed for transition-metals at the transition-metal $L$ and $M$ edges.  For the components, we have the well-known selection rule
\begin{equation}
   m'=m+q.
\end{equation}
When performing RIXS at a particular spin-orbit split edge the effective operators in the valence shell can be written in terms of spin-dependent operators, see Eq. (\ref{operspindependent}). The selection rule is then effectively modified to \cite{Veenendaal2006}
\begin{empheq} [box=\mybluebox]{align}
m'+\sigma'=m+\sigma+q.
\label{selection}
\end{empheq}
When $x=0$, there is no net  transferred angular momentum to the system ($q=0$). In the absence of spin flips, the effective operator is the number of holes ${\underline v}^{000}_0=n_h$ which only contributes to the elastic line. However, when considering spin-dependent operators, inelastic features can be observed. The spin-dependent $x=0$ operator is the spin-orbit coupling ${\underline v}^{110}_0=(2/l){\bf L}\cdot{\bf S}$. Inelastic scattering in spherical symmetry is due to the $L_\pm S_\mp$ term, where the change in angular momentum is compensated by the change in spin.

Many materials are more appropiately described in a lower symmetry than spherical. Often, the basis set used for octahedral symmetry is used for transition-metal compounds. Unfortunately, the standard notation is not very conducive for the definition of selection rules. A more appropiate notation is given by the use of tesseral harmonics
\begin{equation}
	Z_{l\mu}({\hat {\bf r}})= A_\mu \sqrt{\frac{4 pi}{[l]}} Y^l_\mu ({\hat {\bf r}}) +B_\mu \sqrt{\frac{4 pi}{[l]}} Y^l_{-\mu} ({\hat {\bf r}}) , 
	\label{tesseral}
\end{equation}
where, with $s_\mu=\pm 1$ for $\mu\ge 0,\mu<0$,  $A_\mu=1,\frac{1}{\sqrt{2}}s_\mu^\mu i^{(1+s)/2}$ and $B_\mu=0,\frac{1}{\sqrt{2}} s_\mu^{\mu+1}  i^{(1-s)/2}$, for $\mu=0$ and $|\mu|>0$, respectively. Roman and Greek letters are used for the components of the angular momentum in for spherical and tesseral harmonics, respectively. The tesseral harmonics  correspond to the well-known orbitals, see Table \ref{tesseralT}. The operators $w^{llx}_q$ can be transformed to $W^{llx}_\xi$ in a tesseral basis via a unitary transformation, see Appendix. The selection rule for transition between shells $l$ and $l'$ in Eq. (\ref{selection}) then becomes \cite{Veenendaal2010}
\begin{empheq} [box=\mybluebox]{align}
\mu'= s_\mu s_\xi t_{l+l'+x} |\mu\pm \xi |
\label{selectionmu}
\end{empheq}
 respectively, and $t_l=\pm 1$ for $l$ is even/odd. In addition,  the transition to $\mu'=0$ is not allowed when $s_\mu s_\xi t_{l+l'+x}<0$. A notable example of the difference between is the transition operator ${\underline w}^{101}=-L_0/l=-L_z/l$. In spherical symmetry, this gives a transition  $m'=m$, i.e. the orbital does not change. However, for tesseral harmonics, one has $\mu'=-\mu$, and therefore scattering $x^2-y^2 \leftrightarrow xy$ and  $zx\leftrightarrow yz$. Obviously, the selection rules should give the same results as those that can be obtained from general group-theoretical considerations, see Table \ref{tesseralT}. Examples of the use 
 of these selection rules are given in Section \ref{sec:crystal_field_excitations}.

\subsubsection{Sum rules}
Sum rules relating the integrated spectral weight to ground-state
expectation values of such quantities as the orbital and spin moment
\cite{Thole1993,Carra1993} have played a major role in x-ray
absorption and x-ray magnetic circular dichroism studies. The
derivation of such rules is significantly more complex for
higher-order spectroscopies, due to the presence of the
intermediate-state propagator, and approximations are necessary. Using
the fast-collision approximation, sum rules can be derived in the
situation where the decay is between two core levels \cite{Carra1994,Veenendaal1996a}. Since the decay process can then be decoupled from the excitation process, the sum rules relate the integrated RIXS intensities to expectation values of one-particle operators \cite{Carra1994,Veenendaal1996a}. When the decay involves the valence shell, the RIXS sum rules involve two-particle operators which are difficult to interpret. When the polarization of the outgoing x-rays is essentially isotropic, numerical evaluation of the integrated intensities shows \cite{Veenendaal1996b} that the sum rules are close to those for x-ray absorption  \cite{Thole1993,Carra1993}.
\subsection{Effective theory for indirect RIXS \label{sec:effindirect}}

In the previous section, we have seen that for direct RIXS the lowest term in the UCL expansion (also known as the fast-collision approximation), can be written in terms of effective transition operators corresponding to the states into which the absorption takes places. However, often for indirect RIXS these terms do not contribute to the RIXS cross section. For example, $K$-edge RIXS is dominated by excitations into the transition-metal $4p$ states. Since the $4p$ states are usually almost completely empty, the effective operators for direct RIXS only contribute to the elastic line, where the effective transition operator creates an electron in the valence shell in the excitation step and annihilates it again in the emission process.

Experimentally, however, RIXS is observed at the $K$-edge. Particularly prominent are the charge-transfer type excitations  that will be discussed in Section~\ref{sec:CT}. Also the excitation of $dd$ transitions and magnons
have been observed, see Sections~\ref{sec:crystal_field_excitations} and~\ref{sec:magnetic}.
The general consensus is that these excitations are created through the interaction between the valence shell and the $1s$-$4p$ excitation created in the absorption process. Most work has focused on the interaction with the potential of the $1s$ core hole, which is known to be of the order of 6-8 eV. This potential can be written as
\begin{eqnarray}
H_C=\sum_{{\bf k}{\bf k}'{\bf q}\mu\sigma\sigma'} U_{1s,3d} d^\dagger_{{\bf k}+{\bf q},\mu\sigma}s^\dagger_{{\bf k}'-{\bf q},\sigma'}
s_{{\bf k}'\sigma'}d_{{\bf k}\mu\sigma},
\label{eq:coreH}
\end{eqnarray}
where $\mu$ sums over the different orbitals. The potential can in principle contain exchange terms, but these are negligible at the $K$-edge. The transient presence of this potential in the intermediate state leads to strong screening dynamics in the valence shell giving rise to the final-state excitations. This Section discusses some of the methods used to describe the excitations created by interactions in the intermediate state.

\subsubsection{Momentum dependence for indirect RIXS}
Recognizing that for indirect RIXS the core-hole dominates the scattering process has an important consequence for the momentum dependence. In the hard X-ray regime photons have a momentum $\bf q$ that can span several Brillouin zones because it is larger than the reciprocal lattice vectors ${\bf G}$.  The photon momentum reduced to the first Brillouin zone is, by definition, ${\bm \kappa} = {\bf q}-n{\bf G}$.  The translational invariance and localized nature of the core potential in Eq.~(\ref{eq:coreH}) imply that the momentum dependence of RIXS is determined by the {\it reduced} momentum ${\bm \kappa}$. It will only weakly depend on $n{\bf G}$ as in reality a finite, but small, length-scale is associated with the core potential. RIXS spectra will therefore appear practically identical in different Brillouin zones. This is confirmed experimentally by~\textcite{Kim2007}. The weak variations found in~\textcite{Chabot-Couture2010}, are attributed by the authors to polarization effects.

This is remarkable because in IXS the {\it total} momentum ${\bf q}$ determines the scattering amplitude. The reason for this is that in IXS ${\bf q}$ enters directly into the transition matrix elements, which in RIXS are dominated by dipolar transitions for which $e^{i{\bf q}\cdot{\bf r}}\cong 1$ and are therefore independent of ${\bf q}$.

In the following, we review how in certain limits the indirect RIXS amplitude can be related to the dynamic electronic structure factor $S_{\bf k} (\omega)$, which is directly measured by IXS. The important difference is thus that IXS measures $S_{\bf q} (\omega)$ and RIXS is, in these cases, related to $S_{\bm \kappa} (\omega)$.

\subsubsection{Perturbative approach}

\begin{figure}
   \begin{center}
   \includegraphics[width=0.5\columnwidth]{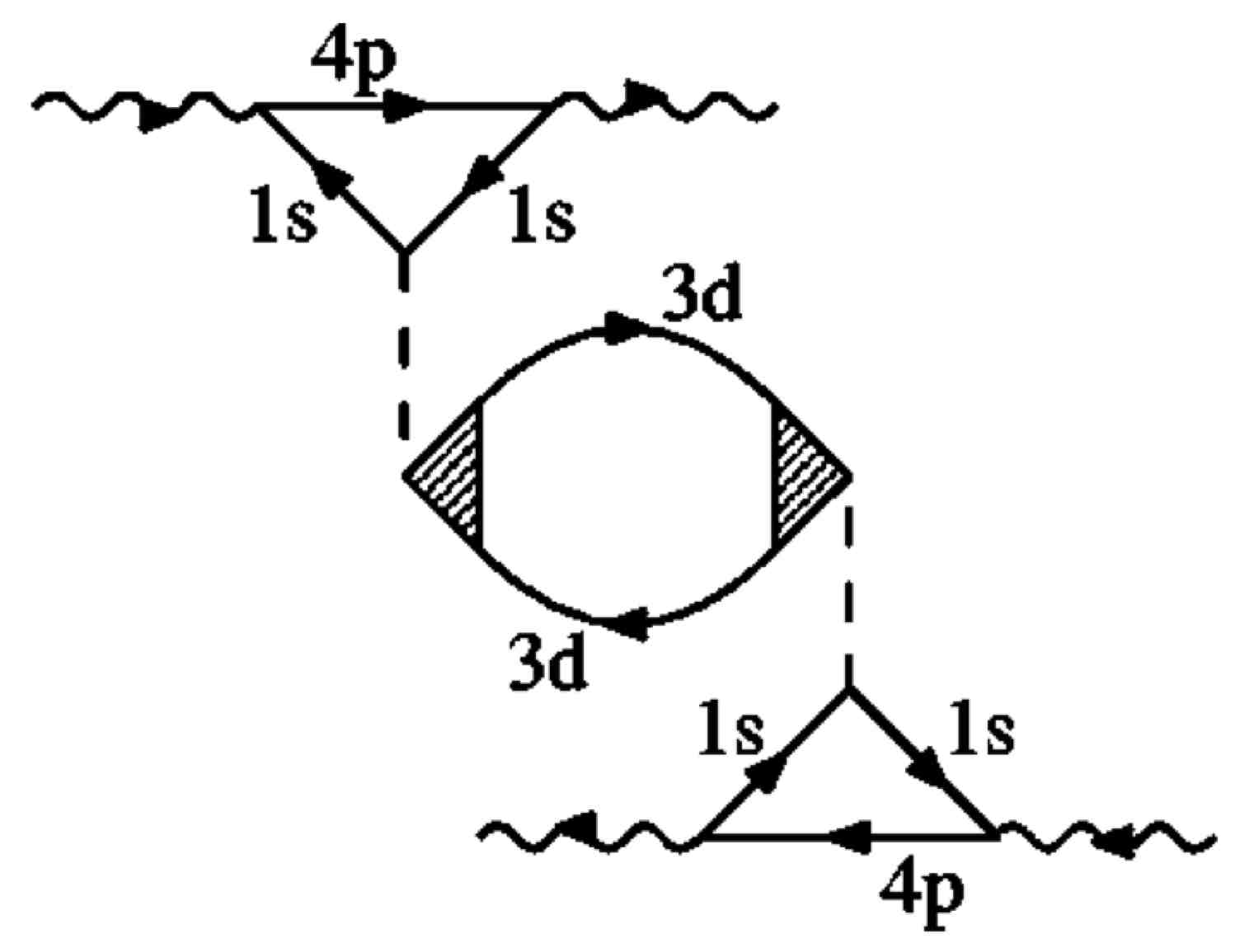}
   \end{center}
   \caption{Feynman diagram for the transition probability in an indirect RIXS process in the Born approximation. GreenÕs functions for Cu $1s$, $4p$, and $3d$ electrons correspond to the solid lines labeled $1s$, $4p$, and $3d$, respectively. The wavy and broken lines represent the photon propagator and core hole potential $U_{1s,3d}$, respectively. The shaded triangle is the effective scattering vertex renormalized interactions between the valence electrons in the $3d$-shell~\cite{Nomura2005}.
\label{fig:Nomura2005_2}}
\end{figure}

The most straight-forward approach to include effects of the interaction $H_C$ between the core hole and the valence shell is the use of perturbation theory. This amounts to replacing $G$ by $G_0$ in Eq.~(\ref{eq:indirect}) \cite{Platzman1998,Abbamonte1999a,Doring2004,Nomura2005,Nomura2004,Semba2008}, so that
\begin{eqnarray}
   {\cal F}^{indirect}_{ fg}&=&\langle f | {\cal D}^{\dagger} G_0 H_C G_0 {\cal D}| g\rangle,
\label{directindirect}
\end{eqnarray}
which is also referred to as the Born approximation and shown in terms of Feynman diagram expansion in Fig.~\ref{fig:Nomura2005_2}. For dipolar $1s\rightarrow 4p$ transitions at the $K$-edge, we have ${\cal D}=\sqrt{3}P_{1s,4p}^{1}\sum_{{\boldsymbol \kappa}{\bf    k}\alpha}\varepsilon_\alpha p^\dagger_{{\boldsymbol \kappa}+{\bf    k},\alpha\sigma} s_{{\boldsymbol \kappa}\alpha \sigma}$ with $\alpha=x,y,z$, and $P_{1s,4p}^{1}$ the reduced matrix element containing the integral over the radial parts of the wavefunction, see Eq. (\ref{reduced}).

In indirect RIXS, one considers the case where the $1s$-$4p$ exciton created in the absorption step is annihilated in the emission process. Since there is a momentum transfer ${\bf q}$ from the photons to the system, this implies that the momentum of the $1s$-$4p$ exciton must have changed in the intermediate state. This can only be a result of interactions of the $1s$-$4p$ exciton with the valence shell. If the dominant interaction is the Coulomb interaction of the core hole with the valence shell, then the isotropic scattering amplitude can be rewritten as~\cite{Doring2004,Schuelke2007}
\begin{empheq} [box=\mybluebox]{align}
   {\cal F}^{indirect}_{ fg}({\bf q},\omega)=P(\omega_{\bf k},\omega_{{\bf k}'})T({\boldsymbol \epsilon},{\boldsymbol \epsilon}') \langle f |   \rho_{\bf q} | g\rangle,
\end{empheq}
Note that all the operators involving the $1s$ and $4p$ states have been removed from the expression. The density operator is
\begin{eqnarray}
   \rho_{\bf q}=\sum_{{\bf k}\sigma}  d^\dagger_{{\bf k}+{\bf q},\sigma}d_{{\bf k}\sigma}.
\end{eqnarray}
The resonance behavior is determined by the resonant function
\begin{eqnarray}
   P(\omega_{\bf k},\omega_{{\bf k}'})=3(P_{1s,4p}^{1})^2\frac{U_{1s,3d}}{(z_{\bf k}-\hbar\omega)z_{\bf k}} ,
\label{resonancefunctionPT}
\end{eqnarray}
using Eq. (\ref{reduced}) and the fact that $\hbar\omega=\hbar\omega_{\bf k}-\hbar \omega_{{\bf k}'}=E_f$. The resonant function is more complex than for direct RIXS reflecting the fact that this is a higher-order excitation.
The polarization dependence requires some careful consideration. In the situation, where the $4p$ electron is a spectator, an electron is excited into the $4p$ band with momentum ${\bf k}$ and band index $n$ and subsequently removed from the same state,
\begin{eqnarray}
T({\boldsymbol \epsilon},{\boldsymbol \epsilon}')=\frac{1}{N}\sum_{{\bf k}\alpha \alpha'}\epsilon'^*_{\alpha'}\epsilon_\alpha
\langle 0|p_{{\bf k}\alpha'}|E_{{\bf k}n}^{4p}\rangle \langle E_{{\bf k}n}^{4p}|p^\dagger_{{\bf k}\alpha} |0\rangle
\end{eqnarray}
with $\alpha=x,y,z$. In the atomic limit, this expression reduces to ${\boldsymbol \epsilon}'^*\cdot{\boldsymbol \epsilon}$, since the orbital of the $4p$ electron is unchanged in the intermediate state. In the absence of band effects, a change in the polarization therefore implies that the angular momentum of the valence electrons has changed. However, the $4p$ states form wide bands that are mixtures of the different $4p$ orbitals, and these local-symmetry arguments only apply at the $\Gamma$-point. Therefore, the use of a scattering condition where the incoming polarization vector is perpendicular to the outgoing polarization vectors does not necessarily imply that a symmetry change has to occur for the valence electrons.

The essential physics of the material is contained in the fundamental scattering amplitude
\begin{eqnarray}
   F_{fg}({\bf q})=\langle f |   \rho_{\bf q} | g\rangle.
\label{eq:fundindirect}
\end{eqnarray}
This quantity is directly related to the dynamic structure factor, through
\begin{eqnarray}
   S_{\bf q} (\omega)&=&\sum_f |F_{fg}|^2 \delta(E_f-\hbar\omega) \nonumber  \\
&=&-\frac{1}{\pi} {\rm Im} \langle g|\rho_{-{\bf q}}\frac{1}{\hbar \omega-H+i0^+}\rho_{\bf q}|g\rangle,
\label{eq:sq}
\end{eqnarray}
which corresponds to the bubble in the Feynman diagram in Fig.~\ref{fig:Nomura2005_2}. It should be noted that RIXS measures a projected $S_{\bf q}(\omega)$, meaning that $\rho_{\bf q}$ contains only $d^\dagger_{{\bf k}+{\bf q},\sigma}d_{{\bf k}\sigma}$ terms. This is a direct result of the fact that the core-hole Coulomb interaction does not scatter between different orbitals. This is different from IXS, where in principle the photon can induce a direct transition from the $d$ states to the ligands. This does not imply that RIXS does not create charge-transfer excitations, since the charge-transfer states also have $d$ character.
In \cite{Igarashi2006} multiple scattering corrections to the Born approximation are also considered on the basis of a Keldysh Green's function formalism. It was found that multiple scattering effects lead to small modifications in the shape of the RIXS spectrum, which partly justifies the Born approximation for wide gap insulators such as La$_2$CuO$_4$ and NiO~\cite{Takahashi2007}.

The scattering amplitudes for direct and indirect RIXS in Eqs. ~(\ref{eq:fundindirect}) and~(\ref{scatspinindependent}) look similar. For direct RIXS, the detailed dependence on the polarization is given in the fundamental scattering amplitude $F^x_{fg,q}$. When including the polarization dependence for indirect RIXS, one obtains a similar fundamental scattering amplitude. The quantity $F_{fg}$ in Eq.  (\ref{eq:fundindirect}) then reduces to $F^0_{fg,0}$ and corresponds to the isotropic term. In the absence of interactions causing a transfer of angular momentum between the $4p$ and the valence shell, the indirect RIXS amplitude is simply proportional to $\rho_{\bf q}$. In terms of tensors,  $\rho_{\bf  q}=w^{dd0}_0({\bf q})$.

A significant difference between the two processes is that when direct excitations are made into the valence shell ({\it e.g.} $2p/3p\rightarrow 3d$), the effect of the operator ${\underline w}^{dd0}_0=n_h$ is relatively small, since the excited electron screens the $2p/3p$ core-hole potential very well. The isotropic contribution then mainly contributes to the elastic line. In indirect RIXS, the excited delocalized $4p$ electron does not screen the $1s$ core-hole potential very well. This produces instead in the intermediate state appreciable screening dynamics of the valence electrons -- the reason why the $\rho_{\bf q}$ response generates significant inelastic scattering intensity for indirect RIXS.

\subsubsection{Ultrashort Core-hole Lifetime expansion \label{sec:UCL}}

The potential that the core-hole exerts on the valence electrons is strong, the attraction $U_{1s,3d}$ between a $1s$ core-hole and $3d$ electron is typically $\sim 6-8$ eV, which is of the same order as the $d-d$ Coulomb interaction $U_{3d,3d}$ that appears in Hubbard-like models. Treating such a strong interaction as a weak perturbation renders a perturbation expansion uncontrolled. To deal with the strong core-hole interaction, the Ultrashort Core-hole Lifetime (UCL) expansion was developed in~\cite{Brink2005,Brink2006,Ament2007}, which treats the core-hole potential as the dominating energy scale.

The UCL relies on three observations. First for most RIXS intermediate
states, the core hole lifetime broadening is quite large: typically
$\Gamma$ is of the order of $1$ eV. This yields a time scale $\tau =
1/2\Gamma = 4$ fs. Only during this ultrashort time, the system is
perturbed by the core hole. Many elementary excitations have an
intrinsic time scale that is much larger than $4$ fs. This intrinsic timescale is the fundamental oscillation period, related to the inverse frequency $\omega$ of an excitation with energy $\hbar \omega$. For example,
phonons have a typical energy scale up to 100 meV, and magnons up to
250 meV, thus corresponding to timescales almost an order of magnitude
larger. Even low energy electronic valence band excitations can be within this range.

The resulting physical picture of a RIXS process involving low-energy
excitations is therefore that the dynamics in the intermediate state
are limited because of lack of time, provided that the excitation time
scale is not decreased significantly by the core hole. The second
observation is that the core-hole potential can to good approximation
be treated as a local potential, i.e. its dominating effect is to
perturb electrons on the same atom where the core hole resides. Finally,
the core-hole is considered to be immobile, which is a reliable
assumption for the deep core-states such as Cu
$1s$.

The calculation of the indirect RIXS amplitude within the UCL
expansion by~\cite{Brink2005,Brink2006,Ament2007} is based on a series
expansion of the Kramers-Heisenberg equation, (Eq.~\ref{eq:K-H}). But a
Green's function approach is equally viable, which then starts by inserting in Eq.~(\ref{eq:indirect}) the identity $G = G_C+G_C H_0 G$
\begin{eqnarray}
{\cal F}^{indirect}_{\tt fg}&=&\langle {\tt f} | {\cal D}^{\dagger} G_0 H_C G_C(1+ H_0 G) {\cal D}| {\tt g}\rangle,
\label{eq:indirect1}
\end{eqnarray}
where  the Green functions, $G_0=(z_{\bf k}-H_0)^{-1}$, $G_C=(z_{\bf k}-H_C)^{-1}$,
$G=(z_{\bf k}-H)^{-1}$, correspond to the Hamiltonian of the unperturbed
system $H_0$, the valence electron - core-hole interaction $H_C$, and
the total Hamiltonian $H=H_0+H_C$. The UCL is best illustrated by
considering the core-hole Hamiltonian $H_C = U_C \sum_i
\rho_i^{s}\rho_i^{d}$, where $U_C=U_{1s,3d}$ and $\rho_i^{1s}$
($\rho_i^{3d}$) are the density operators counting the number of $1s$
core-holes ($3d$ electrons) at site $i$. The simplest system one can
consider is where one allows the $3d$ states to be only occupied by
either $0$ or by $1$ electron, for instance due to strong correlation
effects in the $3d$ shell. As there is only one localized core-hole
present in the intermediate state, $H_C$ then has the interesting
property $H_C^l = U^{l-1}_C H_C$ for any integer
$l>0$~\cite{Brink2005,Brink2006}, which implies that $H_C$ is either
$0$ or $U_C$.
This directly implies the relation
$H_C G_C=H_C(z_{\bf k}-U_C)^{-1}$. One now obtains for the indirect RIXS amplitude
\begin{eqnarray}
   {\cal F}^{indirect}_{\tt fg}&=& \bra{{\tt f}} {\cal D}^{\dagger} G_0 \frac{H_C}{z_{\bf k}-U_C} (1+ H_0 G){\cal D} \ket{{\tt g}},
\label{eq:ucl}
\end{eqnarray}
Note that this expression is exact, but of course specific for the
present form of the core-hole potential; generalized forms are given
in~\cite{Brink2005,Brink2006,Ament2007}, which include the spin and
possible orbital degrees of freedom of the $3d$ electrons.

In the leading order of the UCL expansion one retains in Eq.~(\ref{eq:ucl}) the first order term in $H_C$ so that
\begin{equation}
   {\cal F}^{indirect}_{\tt fg} = \frac{\bra{{\tt f}} {\cal D}^{\dagger} H_C {\cal D} \ket{{\tt g}}}{(z_{\bf k}-\omega) (z_{\bf k}-U_C)}=P(\omega_{\bf k},\omega_{{\bf k}'}) \bra{{\tt f}} \rho^d_{\bf q} \ket{{\tt g}},
\label{eq:indirect2}
\end{equation}
where the resonance function
\begin{equation}
  P(\omega_{\bf k},\omega_{{\bf k}'})= (P^1_{1s,4p})^2 U_C ((z_{\bf
    k}-\omega) (z_{\bf k}-U_C))^{-1} \label{resonancefunctionUCL}
\end{equation}
was introduced, where $P^1_{1s,4p}$ is the $1s \rightarrow 4p$ dipole transition amplitude. The generic shape of the resonance function depends on the form of the core-hole potential. It is remarkable that the RIXS amplitude found in leading order of the strong coupling UCL is directly related to the dynamic structure factor $S_{\bf q} (\omega)$ of Eq.~(\ref{eq:sq}), which is a situation very similar to the weak coupling perturbative approach, see Eq.~(\ref{eq:fundindirect}). In fact for $U_C \rightarrow 0$ the strong coupling UCL resonance function reduces to the perturbative one. This result has important implications for the interpretation of RIXS spectra since this approach then suggests that with proper handling of the prefactor, RIXS can be considered as a weak probe that measures $S_{\bf q}(\omega)$.

The sub-leading contributions to the indirect UCL scattering amplitude of Eq.~(\ref{eq:ucl}) are
of the type $H_C H_0 H_C$. Such terms {\it a priori} cannot be reduced to a response of $\rho_{\bf q}$ because $H_0$ and $H_C$ do not commute. Physically this term corresponds to an electron (or hole) hopping onto the core-hole site in the intermediate state. Denoting the hopping amplitude as $t$, these contributions to the scattering amplitude is down by a factor $t/(z_{\bf k}-U_C)$ with respect to the leading term. When going off-resonance corrections to the UCL expansion thus become progressively smaller. On resonance these terms constitute contributions to the RIXS intensity of the order of $(t/\Gamma)^2$, which are thus governed by $U_C$ and the inverse core-hole life-time $\Gamma$. Corrections to the UCL are thus smaller for shorter-lived core-holes. In cuprates, for instance the effective $3d$ valence bandwidth $t \approx 0.4$ eV and such corrections are expected to be moderate. Using for a specific system the commutation relation $H_0$ and $H_C$ such a higher order term can be calculated explicitly and again be cast in the form of a product of a resonance function and a generalized charge response function.

The observation that within the UCL the RIXS cross section can be factored into a resonant pre-factor and the dynamic structure factor, $S_{\bf q}(\omega)$ was tested experimentally~\cite{Kim2009a}, see also Section~\ref{sec:incident_energy_dependence}. \textcite{Kim2009a} reporting an empirical comparison of Cu $K$-edge indirect RIXS spectra, taken at the
Brillouin-zone center, with optical dielectric loss functions measured in a number of copper oxides:  Bi$_2$CuO$_4$, CuGeO$_3$, Sr$_2$Cu$_3$O$_4$Cl$_2$, La$_2$CuO$_4$, and Sr$_2$CuO$_2$Cl$_2$. Analyzing both incident and scattered-photon resonances  \textcite{Kim2009a} extracted an incident-energy-independent response function. The overall spectral features of the indirect resonant inelastic x-ray scattering response function were found to be in a reasonable agreement with the optical dielectric loss function over a wide energy range. In the case of Bi$_2$CuO$_4$ and CuGeO$_3$ \cite{Kim2009a} observe that the incident-energy-independent response function, $S_{{\bf q}=0} (\omega )$, matches very well with the dielectric loss function, $-{\rm Im} (1/\epsilon(\omega))$  measured with spectroscopic ellipsometry, suggesting that the local core-hole approximation treatment of the UCL works well in these more localized electron systems. Corner-sharing two-dimensional copper oxides exhibit more complex excitation features to those observed in the dielectric loss functions, likely related to non-local core-hole screening effects.

The UCL expansion describes the RIXS cross section in the limits of small and large core-hole potential, respectively. In the intermediate region, one has to resort to numerical calculations \cite{Ahn2009}. In the dynamics structure factor, excitations are created via $\rho_{\bf q}$, implying that electron and holes are excited in an equivalent fashion. When dynamical effects are strong in the intermediate state, this can change and an asymmetry in the excitation of electron and holes can occur \cite{Ahn2009}. Since the screening electron is strongly bound to the core hole in the intermediate state, it is more likely to be scattered to higher lying states. The hole excitations on the other hand can delocalize and have a tendency to be closer to the Fermi level.

Besides charge excitations also magnetic and orbital excitations were
studied with the UCL. Theoretically the two-magnon response of
antiferromagnetic La$_2$CuO$_4$ was calculated within the UCL
\cite{Brink2007,Forte2008a},  agreeing nicely with experiment
\cite{Hill2008}. Collective orbital excitations were investigated
theoretically for LaMnO$_3$ \cite{Forte2008b} and for YTiO$_3$
\cite{Ament2009b} and compared to experiments on titanates
\cite{Ulrich2009}. These indirect RIXS investigations on magnetic and
orbital excitations will be reviewed in Sections~\ref{sec:magnetic} and ~\ref{sec:orbitons}, respectively.

\section{Elementary Excitations Measured by RIXS \label{sec:elementary}}
In this Section, the elementary excitations, as outlined briefly in the introduction, are discussed in detail. The Sections start with high-energy excitations, such as plasmons, interband and charge-transfer transitions, via intermediate-energy transitions, such as $dd$ excitations, to low-energy excitations, such as orbitons, magnons, and phonons.
\subsection{Plasmons, Interband Excitations \label{sec:plasmons}}

A significant portion of the early RIXS work concerned direct RIXS in wide-band insulators, such as Si \cite{Rubensson1990}, diamond \cite{Ma1992,Johnson1994}, graphite \cite{Carlisle1995}, BN \cite{Jia1996}, B$_2$O$_3$ \cite{OBrien1993}, hexaborides \cite{Denlinger2002}, MgB$_2$ \cite{Zhang2003}, and GaN \cite{Strocov2004,Strocov2005}. In these systems, the RIXS process is very intuitive, as described in these papers and more conceptually in Refs. \cite{Ma1994,Ma1995}. The absorption step excites an electron into the empty conduction band and the deexcitation removes an electron from the occupied valence band. The final excitation is an electron-hole pair created across the gap. One of the important conclusions of these experiments was to show the feasibility of momentum resolution. Generally, these RIXS experiments are rather well understood within an independent-particle framework. However, on closer inspection, the presence of intermediate-state relaxation can be observed. In graphite, it was demonstrated \cite{Veenendaal1997} that the Coulomb interaction between the $1s$ core hole and the excited electron creates an intermediate-state exciton. The result is that, although the final states are equivalent, the spectral weight is redistributed as a function of the incoming photon's energy. Scattering of electron-hole pairs by long-range Coulomb interactions leads to collective excitations known as plasmons. Plasmonlike excitations were also observed early on in RIXS \cite{Isaacs1996}. Whereas the measurements of RIXS on wide-band insulators is important, and covered in more detail by Kotani and Shin \cite{Kotani2001}, RIXS experiments in recent years have focused more on strongly-correlated electron systems.
\subsection{Charge Transfer Excitations\label{sec:CT}}

\begin{figure}
  \begin{center}
  \includegraphics[width=.9\columnwidth]{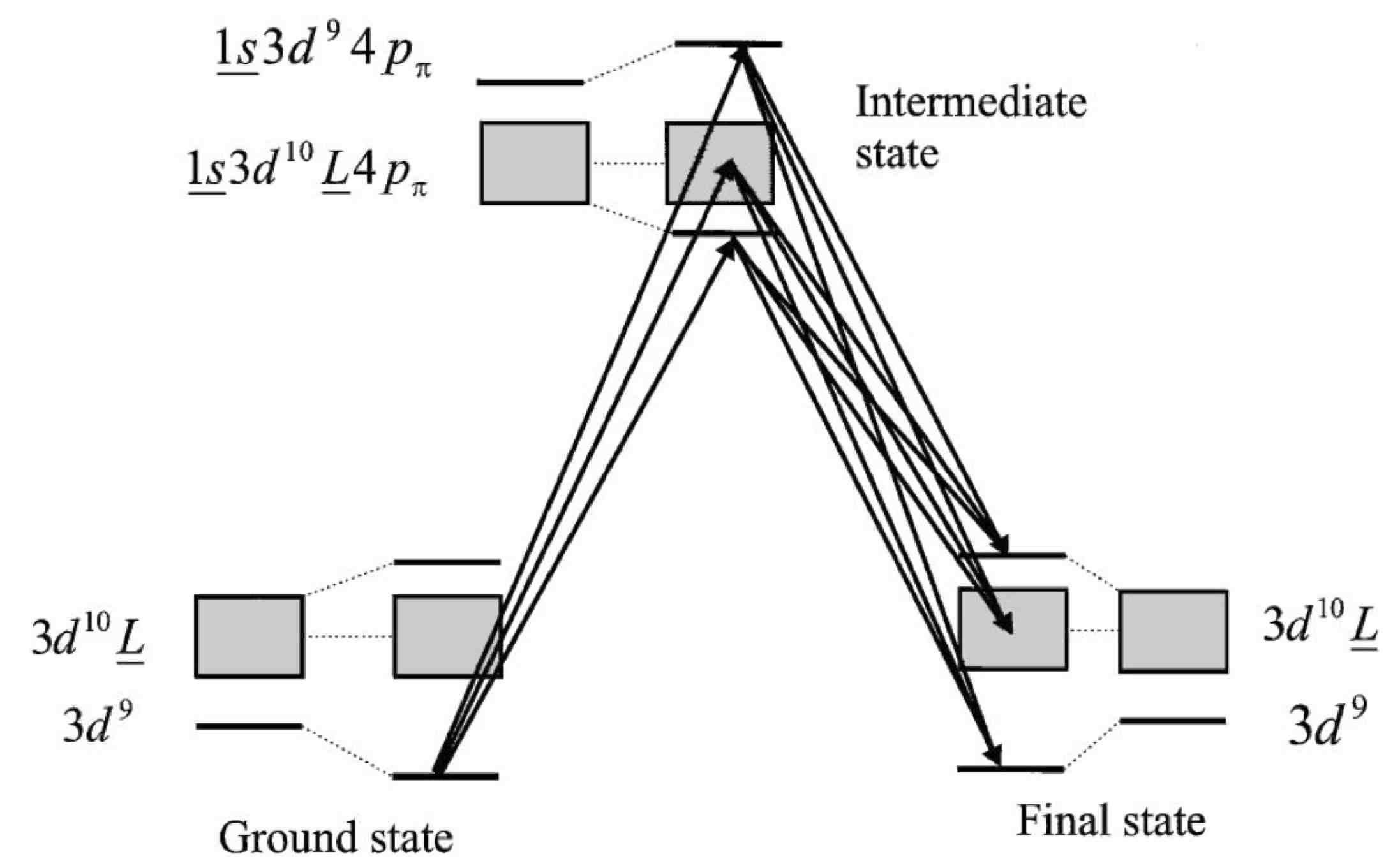}
  \end{center}
    \caption{Indirect RIXS scattering process for observing charge transfer excitations at a metal $K$-edge, illustrated for the case of a copper oxide . Figure from \cite{Hill1998}}
 \label{fig:KedgeCT RIXS}
  \end{figure}

To date, charge transfer excitations have been studied at metal $K$-  $L$-
and $M$- edges and oxygen $K$-edges. For the first three cases, a
core-hole is created on the metal site, and charge moves in from the ligand site to screen the core hole, which then decays. Following the decay, there is a finite probability that some rearrangement of the charge remains - creating a charge transfer excitation. Thus, this is an indirect RIXS process. This case is illustrated in Fig.~\ref{fig:KedgeCT RIXS} drawn for the Cu $K$-edge. Similar diagrams may be drawn for the $L$ and $M$ -edges, where the core hole is respectively in the $2p$ or $3p$ shell. At the oxygen $K$-edge, the core hole is created on the oxygen site and the photoelectron annihilates an oxygen $2p$ hole, which is strongly hybridized with a metal $d$ orbital. Then a different O $2p$ electron decays, filling the O $1s$ core hole and leaving the system in an excited state. Thus this is a direct RIXS process.

In the study of charge transfer excitations, RIXS offers the ability to measure their momentum dependence and to carry out studies under extreme conditions, such as pressure~\cite{Rueff2010}. These advantages, combined with the fact that the energy scales are well matched to the achievable experimental resolutions mean that a large amount of experimental work has been performed utilizing RIXS. Here we review that work, focusing on what has been learned about the character of these excitations in the various compounds studied to date.

\subsubsection{Early work}

We first briefly mention the early work in this field, defined here as work carried out before 2000, and typically with lower resolution than the later work, to be discussed in more detail below.

\paragraph{Metal K-edge}
The first work on charge transfer excitations at a metal $K$-edge was the seminal work of Kao and co-workers \cite{Kao1996} who studied them in NiO. This was notable for being the first $K$-edge RIXS experiment and was largely responsible for opening up the field to higher energy x-rays. Platzman and Isaacs followed this work with a study of NiSi$_{1.5}$Se$_{0.5}$ \cite{Platzman1998}.

Following this work, attention immediately switched to the cuprates and there were a series of studies of increasing sophistication  on  a variety of cuprates. The first of these was work on Nd$_2$CuO$_4$ \cite{Hill1998} in which the 6 eV antibonding charge transfer excitation was observed and explained with a simple Anderson impurity model.  Abbamonte {\em et al.} then carried out work on La$_2$CuO$_4$ and Sr$_2$CuO$_2$Cl$_2$ \cite{Abbamonte1999a}. This higher-resolution work revealed the charge transfer gap at 2eV and hinted at its q-dependence.

The 6 eV feature is a bonding-antibonding transition that can be understood within the limit of a small cluster. For a transition-metal ion surrounded by ligands, the ground state for $n$ electrons (with $n=9$ for divalent copper compounds) is given approximately by $\alpha |d^n\rangle +\beta|d^{n+1}\underline{L}\rangle$, where $\underline{L}$ stands for a hole on the ligand. For positive charge-transfer energies, one has $\alpha>\beta$. In the presence of a core hole $\underline{c}$, the configuration is modified to $\alpha'|\underline{c}d^{n+1}\underline{L}\rangle+\beta' |\underline{c}d^n\rangle$, when considering a particular spin-orbit split edge. After emission of a photon, the system can return to the ground state. However, there is also a finite probability that the antibonding state $\beta |d^n\rangle -\alpha|d^{n+1}\underline{L}\rangle$ is reached. The energy difference between these two states is about 6-8 eV. This is called a local screening process since the screening charge comes from the ligands surrounding the site with the core hole \cite{Veenendaal1993a,Veenendaal1993b}.

However, there is an alternative longer-range screening process, which is often more effective, in particular in cuprates that have a small charge-transfer energy. In this  nonlocal screening process \cite{Veenendaal1993a,Veenendaal1993b}, a screening electron comes from a ligand surrounding neighboring copper sites. The intermediate state is $|\underline{c}d^{n+1}\rangle$ on the site with the core hole. An electron is then removed from a neighboring site. For cuprates, this creates a two-hole state known as a Zhang-Rice singlet (ZRS) \cite{Zhang1988}. In the final state, one is then left, with a transition across the charge-transfer gap, which is about 2-3 eV in cuprates. The filled $3d^{10}$ state and the ZRS can move independently giving rise to dispersive features. This latter fact was more fully explored in Ca$_2$CuO$_2$Cl$_2$ \cite{Hasan2000} who observed strong dispersion of the 2 eV feature along the ($\pi,\pi$) direction and weaker dispersion along the $(\pi, 0)$ direction. This was modeled with a $ t,t^{'},t^{''}-U$ model calculated on a 4x4 cluster.

Early works exploring the polarization dependence of the charge-transfer states included Hamalainen {\em et al.} \cite{Hamalainen2000} and Abbamonte {\em et al.} \cite{Abbamonte1999b}.
		
\begin{figure}
  \begin{center}
  \includegraphics[width=1.0\columnwidth]{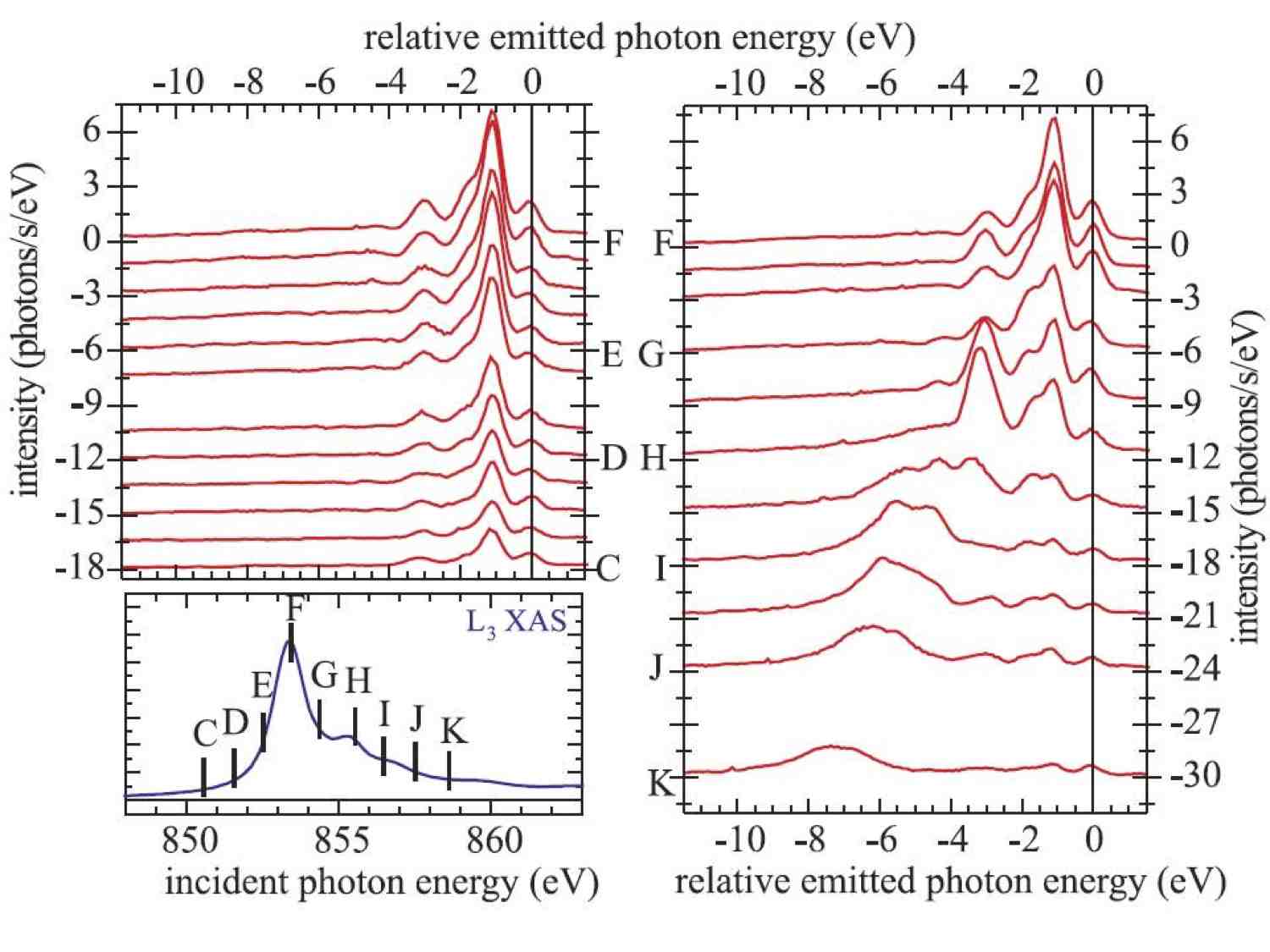}
  \end{center}
    \caption{Experimental Ni L$_3$ RIXS spectra measured with 750 meV combined resolution. The
lower left panel shows the Ni L$_3$ absorption spectrum measured with the same incident energy
resolution and the labels of the excitation energies used for the RIXS spectra. The polarization of the incident photons was V-pol, i.e. perpendicular to the scattering plane. Figure taken from \cite{Ghiringhelli2005}}
 \label{fig:LedgeNiO}
  \end{figure}

\paragraph{Metal L-edge}
Charge-transfer peaks are generally relatively weak at the $L$ edges. The main absorption feature has a configuration $\alpha'' |\underline{c}d^{n+1}\rangle+\beta''|\underline{c}d^{n+2}\underline{L}\rangle$. Note that the electron excited into the $3d$ shell directly screens the core hole and there is no need for screening electrons from neighboring transition-metal sites. Therefore, nonlocal screening effects are expected to be small and one expects mainly transitions between bonding and antibonding states of the transition-metal site and its surrounding ligands.

These bonding-antibonding charge-transfer features become more prominent by tuning the incoming photon energy to the weak satellite features above the main absorption edge. These satellites have wavefunctions of the form $\alpha''|\underline{c}d^{n+2}\underline{L}\rangle -\beta''|\underline{c}d^{n+1}\rangle$ with a higher ligand character and therefore the overlap with final states that have a higher ligand character is larger. The satellites were already observed early on in FeTiO3 and KMnO4 \cite{Butorin1997} with approximately 1 eV resolution. Higher resolution (400 meV) experiments were done on TiO2 \cite{Harada2000}, see Fig. \ref{fig:Harada2000_6}.
  \begin{figure}
  \begin{center}
  \includegraphics[width=1.0\columnwidth]{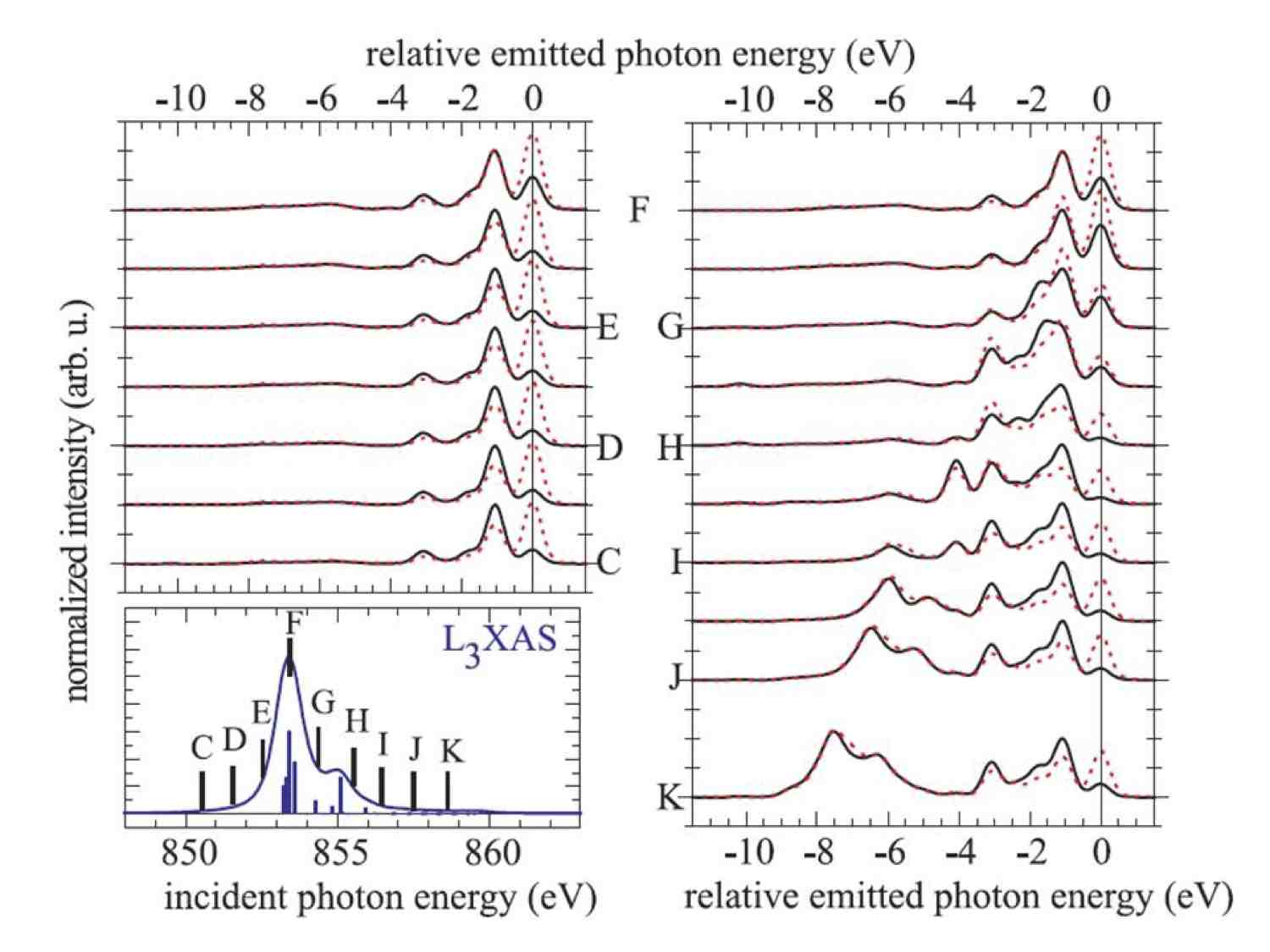}
  \end{center}
    \caption{The RIXS and absorption spectra calculated with the Anderson impurity model at the same excitation energies used in figure \ref{fig:LedgeNiO}. Colors and symbols: red dashed lines are V-pol, black solid lines are H-pol. The spectra were broadened with a 500 meV FWHM Guassian in the final state. Figure taken from \cite{Ghiringhelli2005}}
 \label{fig:NiO Ledge calcs}
  \end{figure}\noindent
The antibonding states are at 14 eV energy loss. The antibonding states are usually at  higher energies for early transition-metal compounds reflecting the strong coupling between the ligands and the mainly empty $3d$ states. Other prominent ligand features, often described as non-bonding states, are observed around  5-10 eV. For TiO$_2$, the intermediate states are given by $\alpha'' |\underline{c}d_\mu\rangle+\beta''|\underline{c}d_\mu d_{\mu'}\underline{L}_{\mu'}\rangle$, where the symmetry $\mu$ of the $3d$ orbitals is introduced. When the electron is removed from the orbital $d_{\mu}$, the system returns again to the ground state symmetry. However, when the deexcitation removes the electron from the $d_{\mu'}$ orbital, one is left with a configuraton $|\underline{c}d_\mu\underline{L}_{\mu'}\rangle$. Note that the hole on the ligands no longer couples to the $3d$ states, since the $d_\mu$ electron has a different symmetry. Therefore, since there is no coupling to the $d^0$ state, the features between 5-10 eV are nonbonding states \cite{Harada2000}.

Another model strongly-correlated system is NiO, which has a $3d^8$
ground state. Several experiments have been performed that show $dd$
excitations and the presence of bonding-antibonding charge-transfer
features around 6-8 eV, in particular above the main absorption edge,
see Fig.~\ref{fig:LedgeNiO}
\cite{Ishii2001,Magnuson2002,Ghiringhelli2005}. These RIXS data were
also modelled using a NiO$_6$ cluster. Note, the presence of
polarization dependence, not only in the $dd$ excitations, but also in
the satellites. The good agreement between theory and experiment, see Fig.~\ref{fig:NiO Ledge calcs}, demonstrates the current understanding of the RIXS process at transition-metal $L$ edges is in good shape.

\paragraph{Metal M-edge}

Amongst the very first RIXS work on charge transfer excitations, Butorin {\em et al.} studied CeO$_2$ and UO$_3$ at the Ce and U $M$-edges \cite{Butorin1996}. Using the Anderson Impurity model, these authors showed that this technique was sensitive to charge transfer excitations and that reasonable parameters could reproduce the experimental results.

\paragraph{Oxygen K-edge}
The oxygen $K$ edge has been used only sporadically. In 2002, Harada {\em et al.} used O $K$-edge RIXS to study Sr$_2$CuO$_2$Cl$_2$ \cite{Harada2002}. These authors used the polarization dependence of the observed scattering, together with theoretical expectations for the polarization of the $dd$ excitations to conclude that the feature at 2 eV was a ZRS CT excitation. In addition, they observe double magnon features around 0.5 eV. \textcite{Duda2006} measured the oxygen $K$ edge of NiO. They observed the presence of features related to $dd$ excitations ($\sim$1 eV), nonlocal screening \cite{Veenendaal1993a} ($\sim$4-5 eV), and local bonding-antibonding transitions (above 5 eV). In addition, they claim the presence of double-singlet excitations, where the two holes are transferred between two neighboring nickel sites creating two singlet excitations, which are about $2J_H$, with $J_H$ the Hund's exchange splitting, higher in energy than the ground state that consists of two triplet states. The possibility of measuring momentum dependence at the Oxygen $K$ edge has also been suggested \cite{Okada2007}. \textcite{Learmonth2009} used oxygen $K$-edge RIXS to probe the electronic structure of the band insulator $\alpha$-MoO$_3$ and observe that when the incident energy is resonantly tuned to the $\pi^\star$ states at the Fermi level, momentum selectivity allows the observation of states only near certain symmetry points reflecting the highly anisotropic electronic structure of $\alpha$-MoO$_3$.

\subsubsection{Momentum Dependence}

\paragraph{0D Cuprates}

Materials in which the CuO$_4$ plaquettes are isolated from each other by virtue of the crystal structure have been termed zero dimensional. Such systems are interesting since they represent perhaps the simplest incarnation of the transition metal oxide physics - and one that is particularly amenable to cluster model-type calculations. In the context of RIXS experiments, they can also therefore provide insight into the resonant cross-section.

Kim et al., studied Bi$_2$CuO$_4$ at the Cu $K$-edge \cite{Kim2009a}. In this paper, the authors analyzed their data by dividing out a resonant prefactor, following the prescription of Ref. \cite{Ament2007} with $\omega_{res}$ defined by the absorption spectrum and $\Gamma$ and $|U|$, the core-hole lifetime and core-hole potential, respectively, viewed as fitting parameters. In practice, such a scaling approach is very similar to that first taken by Abbamonte {\em et al.} \cite{Abbamonte1999a}, and by D\"oring {\em et al.} \cite{Doring2004} with $|U|$=0, or Lu {\em et al.} who took $|U|$=1 eV. \cite{Lu2006}.
Kim {\em et al.} found that for $\Gamma=$ 3.2 eV and $|U|$=5.7 eV, the data taken at different incident energies for Bi$_2$CuO$_4$ collapsed onto a single curve, and further that this curve closely resembled the loss function as independently obtained from ellipsometry measurements on the same sample. This demonstrates empirically, the utility of the UCL approach, discussed above, at least for zero dimensional systems. In the same paper, these authors look at a number of other cuprates. They find similar agreement between the underlying RIXS response (i.e. without the resonant prefactor) and the loss function in most cases, though this agreement is better in the 1D edge-sharing cuprate CuGeO$_3$, than in two-dimensional cuprates. Such edge sharing cuprates are relatively localized systems  because of the very small overlap of the oxygen orbitals in neighboring plaquettes as a result of the 90$^{\circ}$ bonds and this suggests that the UCL works best in localized systems. This is to be expected because of the local approximation to the core hole in the UCL. For the more two dimensional compounds studied in the same paper (La$_2$CuO$_4$, Sr$_2$Cu$_3$O$_4$Cl$_2$ and Sr$_2$CuO$_2$Cl$_2$) the agreement was less good and additional features were seen in the reduced RIXS spectrum that were not present in the loss function, suggesting that non-local features in the intermediate state were important in determining the lineshape and that were not captured in the UCL.

Hancock {\em et al.}  studied a different zero-dimensional cuprate,
CuB$_2$O$_4$ \cite{Hancock2009,Hancock2010}. These authors observed
several features in the Cu $K$-edge RIXS spectrum for this material,
specifically peaks at 1.9 eV, 6.8 eV and 10.9 eV. They observed no
momentum dependence in any of these features, consistent with the zero
dimensionality of the system. They argued that the 1.9 eV feature,
which is resolution limited in energy, was not a charge-transfer
excitation, but rather a $d-d$ excitation. The higher dimensional
cuprates do see a charge transfer excitation in this same energy range
($\sim$2 eV), but in those cases it is attributed to a non-local CT
excitation (i.e. where the ligand hole is shared between neighboring
plaquettes and is not on the same site as the filled $d$ shell). Such
an excitation cannot exist in these zero dimensional cuprates, and
only the Molecular Orbital (MO) antibonding excitation is seen - where
the ligand hole and filled Cu $d$ shell are on the same
plaquette. This occurs at 6.8 eV in this system. The authors suggest
that the dd excitation channel could arise from symmetry breaking
interaction with the 4p photoelectron, or the participation of
phonons. They further note that the lineshape (Gaussian) and incident
energy dependence of the energy loss and the width of the MO
excitation are difficult to understand in the context of an electron
only model for the RIXS cross-section
\cite{Hancock2009,Hancock2010}. They argue that the origin of these
features lies in electron-phonon coupling during and after the scattering process and construct a simple Frank-Condon model; as the charge is rearranged in the intermediate and final states, the potential landscape for the lattice is changed and the lattice is abruptly out of equilibrium, and thus contains many phonons.  Their model reproduces the salient features qualitatively, though not quantitatively. They find that the coupling shifts the MO excitation energy upwards, relative to the bare energy, by an amount $\hbar\omega_{ph}\gamma^2_{g-f}$, where $\omega_{ph}$ is a typical phonon energy (taken to be 75 meV in this work), and $\gamma_{g-f}$ is a measure of the electron-phonon coupling \cite{Hancock2010}. They further argue that this shift may explain the anomalous dependence of the MO energy on the Cu-O bond length observed by Kim {\em al.} \cite{Kim2004d}.

\paragraph{1D Cuprates}

The one dimensional cuprates have attracted attention again because of their relationship to the two dimensional superconducting cuprates, but also because interesting new behavior is expected in one dimension and because theoretical approaches tend to be more reliable in one dimension where, for example, larger clusters may be calculated. The one dimensional cuprates studied to date may be classified according to the connectivity of the CuO$_4$ plaquettes. Either these are connected along their edges, so-called edge-sharing networks, in which neighboring plaquettes share two oxygens, or at their corners (neighboring plaquettes share a single oxygen), so-called corner-sharing networks.

\subparagraph{Edge-sharing 1D cuprates}

The two prototypical edge-sharing 1D cuprates are Li$_2$CuO$_2$ and CuGeO$_3$. The charge transfer excitations in these systems are of interest because of the contrasts one may draw with the corner-sharing 2D networks of the high-T$_c$ cuprates. In these latter systems, the ZRS excitation is the basic building block of many effective theories of the electronic structure. It is stabilized by the antiferromagnetic exchange between the Cu and the O and, as discussed, can propagate to neighboring plaquettes. In the edge-sharing network the Cu-O-Cu bond angle is close to 90$^{\circ}$. According to the Goodenough-Kanamori-Anderson rules this should make the nearest-neighbor copper-copper exchange slightly ferromagnetic. Thus the dynamics of this excitation should be quite different in this geometry.

Li$_2$CuO$_2$ and CuGeO$_3$ differ slightly in their the Cu-O-Cu bond angle. It is 94$^{\circ}$ in  Li$_2$CuO$_2$ and 98$^{\circ}$ in CuGeO$_3$. This is significant because higher order processes such as Cu-O-O-Cu hopping contribute antiferromagnetic terms to the magnetic exchange and this makes the net exchange sensitive to slight variations in this angle. Both have been studied extensively using RIXS. Cu $K$-edge RIXS has been carried out on both Li$_2$CuO$_2$ \cite{Kim2004b,Qian2005} and CuGeO$_3$ \cite{Hasan2003,Suga2005,Qian2005,Vernay2008} and for Li$_2$CuO$_2$, both O $K$-edge \cite{Duda2000a,Bondino2007,Learmonth2007}and Cu $L$-edge RIXS \cite{Learmonth2007}has been performed. The gross features are similar in all these studies across these two materials. Specifically, an excitation is seen around 1.7 (2.2) eV energy loss, another around 3.8 (5.4) eV and a third at 6.5 (7.6) eV, where the numbers refer to CuGeO$_3$ (Li$_2$CuO$_2$). There is in fact a variety of interpretations of these features.

The 2 eV feature is attributed to a $dd$ \cite{Kim2004b,Suga2005,Learmonth2007,Duda2004}, while others attribute it to a d-intersite ZRS excitation \cite{Qian2005,Vernay2008}. There is general agreement that the feature near 3.8 eV in CuGeO$_3$ is due to a non-local charge transfer excitation with the hole weight on the neighboring plaquettes and that the 6.8 eV feature is a local CT excitation \cite{Suga2005,Bondino2007,Vernay2008}.
In Li$_2$CuO$_2$, Kim {\em et al.} attribute the 5.5 eV feature to a local CT excitation on the basis of an agreement with a calculation by Weht and Pickett \cite{Weht1998}, an assignment agreed to by Vernay {\em et al.} on the basis of a Cu$_3$O$_8$ cluster calculation \cite{Vernay2008}. The feature at 7.6 eV in this material, which resonates at the poorly screened intermediate state is also believed to be a CT excitation \cite{Kim2004b} although this has not been reproduced to date in a calculation.

A number of authors looked at the momentum dependence of these various features with Cu $K$-edge RIXS \cite{Kim2004b,Qian2005,Suga2005} with the momentum transfer along the chain direction. These studies observed very little, if any, dispersion - consistent with the expected isolated nature of the plaquettes. Kim {\em et al.} did observe some small dispersion of the 5 eV and 7.5 eV feature in the direction perpendicular to the CuO chains in Li$_2$CuO$_2$ with a bandwidth of $\sim$100 meV, suggesting that there is significant interchain coupling \cite{Kim2004b}.

\subparagraph{Corner-sharing 1D cuprates}

In corner-sharing 1D cuprates, neighboring CuO$_4$ plaquettes share a single oxygen and the Cu-O-Cu bond is 180$^{\circ}$. As a result the hopping along the chain is relatively large and quite different behavior is expected, and observed, in the charge transfer excitations of such materials. Two materials have been commonly studied in this class of materials, SrCuO$_2$\cite{Kim2004a,Qian2005,Wray2007} and Sr$_2$Cu$_3$ \cite{Hasan2002,Suga2005,Seo2006}. In the latter, there is a single CuO chain per unit cell, in the former there are two essentially independent such chains, connected to each other in an edge sharing geometry.

\begin{figure}
  \begin{center}
  \includegraphics[width=0.7\columnwidth]{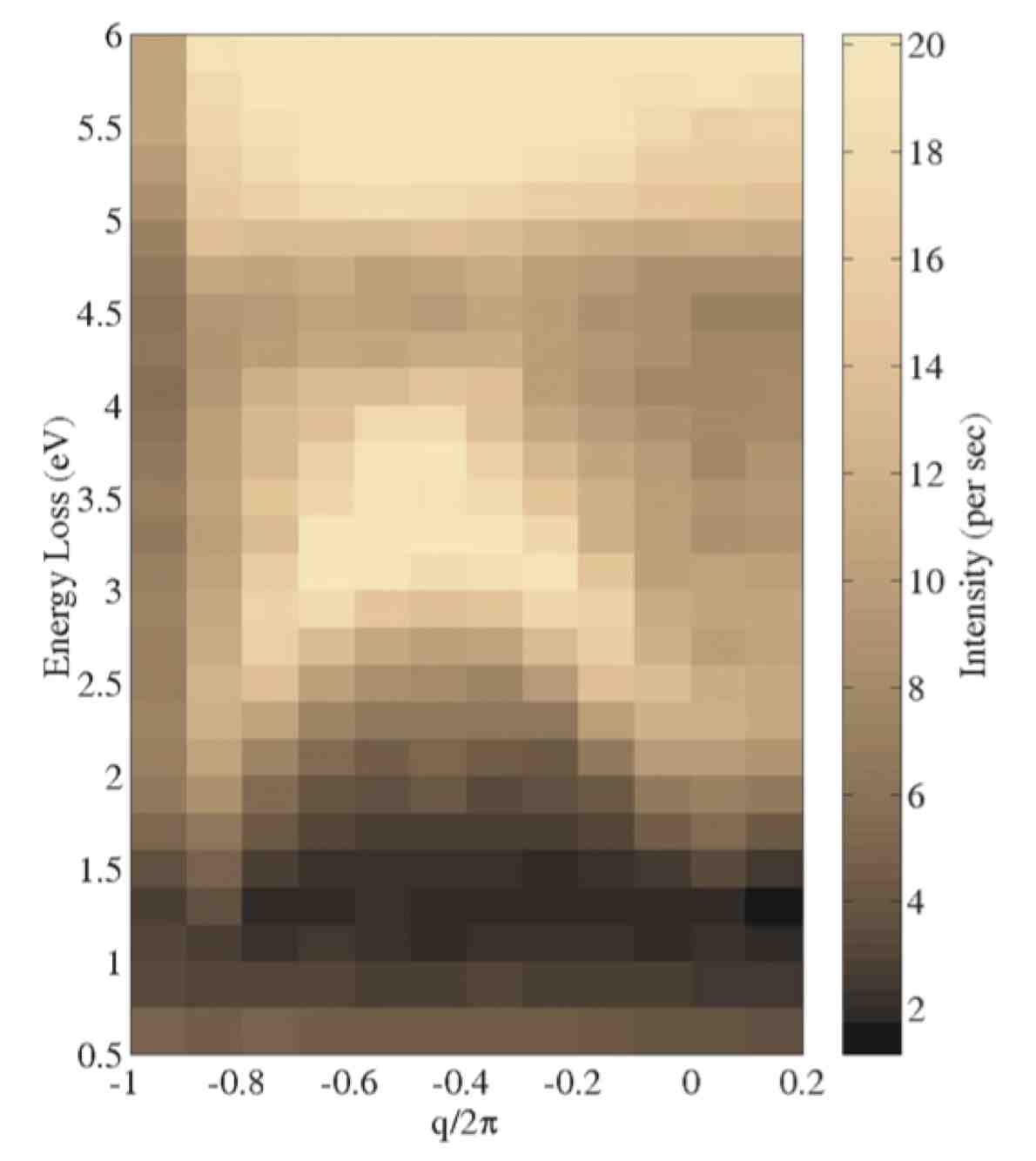}
  \end{center}
    \caption{(color online). Pseudocolor plot of Cu $K$-edge RIXS intensity as a function
of momentum and energy transfer ($\hbar\omega_{\bf k}$=8982 eV)as measured in the 1D corner-sharing cuprate, SrCuO$_2$. The highly dispersive feature has been interpreted as a holon-anti-holon continuum. Figure taken from~\cite{Kim2004a}}
 \label{fig:Kim2004a}
  \end{figure}

In general, the results from the different experiments for these two materials are in broad agreement. A sinusoidally dispersing feature is observed between $\sim$2 and 4 eV, with a maximum energy around reduced momentum transfer q=$\pi$ and a higher energy, non-dispersive feature at $\sim$6 eV, see Fig. \ref{fig:Kim2004a}. This latter is attributed to a local charge transfer (bonding to anti-bonding) excitation. The interpretation of the dispersive feature, however, is more interesting. In 1D, $S=\frac{1}{2}$ chains, the low energy excitations are expected to be in the form of spinons (which carry spin and no charge) and holons (which carry charge and no spin).

Hasan {\em et al.}, noting that the dispersion was larger than in the 2D cuprates, and on the basis of comparison with exact diagonalization calculations of the single band Hubbard model \cite{Tsutsui2000}, identified the dispersion of the onset of the excitation as representing the dispersion of holons in this 1D material \cite{Hasan2002}. Kim {\em et al.} noted that the continuum starts at a much lower energy and has a weaker dispersion than that of the peak \cite{Kim2004a}. They carried out dynamical density matrix renormalization group calculations of the extended single band Hubbard model for a 128 site cluster. Parameters were obtained by fitting to optical conductivity data on the same material. They found $t$ = 0.435 eV, $V/t$ = 1.3, and $U/t$ = 7.8. The results of these calculations were consistent with the observed RIXS spectra, implying that the dispersive peak represents a continuum of spin-less charge $\pm$e collective modes (holons and antiholons). In contrast to previous calculations \cite{Tsutsui2000,Stephan1996},
no bound holon-antiholon exciton was observed, even at reduced $q=\pi$. The onset dispersion is weaker and out of phase with the dispsersion of the peak \cite{Kim2004a} and this is consistent with low energy field theory predictions  which say that it should follow the single spinon dispersion $\frac{\pi}{2}J|\sin q|+\Delta$ \cite{Controzzi2002}.

In contrast to these somewhat exotic pictures of the low energy excitations, Suga {\em et al.} interpret the same feature in Sr$_2$CuO$_3$ as an excitation from the ZRS to Upper Hubbard Band (UHB), with a dispersion that reflects the dispersion of these two bands - as calculated in a Hartree Fock approximation, with an RPA treatment of correlations \cite{Nomura2004,Suga2005}. With the parameter set; $t_{x,dp}=t_{y,dp} = -1.4$ eV, $t^{'}_{pp}=-0.7$ eV and $U_{dd}$ = 11 eV, they were able to reproduce the energy position and dispersion of this feature, though their calculations predict a sharpening in energy of the feature at reduced q=$(2n+1)\pi$ not seen in the experiments. Other calculations on  corner-sharing 1D cuprates were performed by  Okada and Kotani using exact diagonalization \cite{Okada2006}.

 A third 1D corner-sharing compound, namely the so-called ``telephone number compound'' Sr$_{14}$Cu$_{24}$O$_{41}$, and variants, has also been studied \cite{Wray2007,Ishii2007b,Higashiya2008}. This material is ``self-doped'' that is the parent compound already has holes in it, with a nominal Cu$^{2.25+}$ valence. This material has a complicated crystal structure that may be thought of as being made up of two corner-sharing chains (the "ladder") and an edge-sharing chain, each with different periodicities (which are incommensurate with each other). This fact allows the identification of spectral features with the two structural components based on their momentum dependence. In general, all three works see very similar phenomenology.

 Wray {\em et al.} studied  Sr$_{14}$Cu$_{24}$O$_{41}$ and compared the response to 1D SrCuO$_2$ and 2D Nd$_2$CuO$_4$. In the 14-24-41 compound, they found a broad dispersive peak below about 4 eV, which they fit to two broadened Lorentzians, and a non-dispersive higher energy peak at 5.3 eV. The found the two low energy peaks to disperse with a periodicity of that of the ladder, exhibiting a maximum energy at  $\pi_{ladder}$. They noted that this dispersion was intermediate between the 1D SrCuO$_2$ and the 2D Nd$_2$CuO$_4$, suggesting that this implied a continuous evolution of the dynamics on varying the dimensionality of the corner-sharing network from 1D to 2D. These authors modeled this dispersion based on a single band Hubbard model in which the dispersion of the quasi-particles is given by $ 2t_L\cos (kc_L) \pm t_{\perp}$ in the ladder and $ 2t_{ch}\cos (kc_{ch})$, where the subscripts refer to the ladder and chain and $t_{\perp}$ is the hopping across ladder rungs. This gives a continuum of excitations and the authors associated the observed peaks with the edges of this continuum where intensity is expected to become large.

 Higashiya {\em et al.} also found two excitations in Sr$_{14}$Cu$_{24}$O$_{41}$ with the same momentum dependence as observed by Wray {\em et al.}, one at around 4eV, which they treated as a single excitation and associated it with excitations from the ZRS band into the UHB on the chain, and  one around 5.5 eV associated with the local charge transfer excitation which appears to reside on both the ladder and the chain \cite{Higashiya2008}. (With the incident photon energy set to the pre-edge, they also saw a feature at 2 eV which they associated with d-d transitions). They found that the 4 eV feature exhibited significant dispersion, with a bandwidth of approximately 1 eV and a periodicity commensurate with the ladder, exhibiting a maximum energy at $\pi_{ladder}$, consistent with Wray {\em et al.}.  The 5.5 eV feature is dispersion-less. These authors also carried out a cluster calculation on a Cu$_6$O$_{18}$ corner-sharing 1D cluster, with electron correlation treated exactly \cite{Okada1997,Okada2006}. They found that in order to model the data, it was important to include a Cu 3d - O 2p intersite interaction, $U_{pd}$. Without this interaction, the calculated energy losses were too low by about 1.5 eV (i.e. by $U_{pd}$). With this addition, they were able to reproduce the incident energy dependence and momentum dependence observed in the data for the 4 eV and 5.5 eV features. The full parameter set in their model was  $\Delta$ = 3.5 eV, ($dp\sigma$) = -1.5 eV, ($pp\sigma$)-($pp\pi$) = 0.65 eV, $U_{dd}$ = 9.0 eV, $U_{pp}$ = 4.0 eV, $U_{pd}$ = 1.5 eV and $Q$ = 7.0 eV \cite{Higashiya2008}.

Ishii {\em et al.} studied Sr$_{14}$Cu$_{24}$O$_{41}$,
La$_5$Sr$_9$Cu$_{24}$O$_{41}$, and Sr$_{2.5}$Ca$_{11.5}$Cu$_{24}$O$_{41}$. The hole concentration
in La$_5$Sr$_9$Cu$_{24}$O$_{41}$ is very small in both the ladder and the chain,
while for Sr$_{2.5}$Ca$_{11.5}$Cu$_{24}$O$_{41}$ the doping in the ladder is high enough that the material becomes a superconductor under high pressure \cite{Ishii2007b}. These authors also identify the excitation between 2-4 eV as being associated with the ladder structural unit. In this case, in addition to the momentum dependence, they also looked at Ca$_2$Y$_2$Cu$_5$O$_{10}$ which lacks the ladders, but is otherwise similar. The Cu $K$-edge RIXS spectrum on this latter compound lacked the 2-4 eV feature. They attribute this feature to excitation from the ZRB to the UHB of the ladder \cite{Ishii2007b}. As discussed above, this feature is highly dispersive. Ishii {\em et al.} found that the peak position shifted to higher energies as the doping increased, but the dispersion remained roughly the same. In addition, they looked at the dispersion along the ''rungs'' of the ladder (i.e normal to the ladder direction) as well as along the ladder itself, finding the gap increased to higher energies in this direction. They modeled their data with exact diagonalization calculations for a single band Hubbard model on a ladder. This model reproduced the dispersion in both directions \cite{Ishii2007b}. They attribute the insensitivity of the dispersion to doping to the fact that in these ladder compounds the antiferromagnetic correlations are left largely unchanged by doping - in contrast to the 2D cuprates, for which the magnetic correlations are strongly dependent and this reduction in AF correlation has been argued to reduce the dispersion of this excitation as seen by RIXS \cite{Tsutsui2003}.

\paragraph{2D Cuprates}

In this section, we review the work on charge transfer excitations in the 2D cuprates. In doing so, we will focus on the features between 2 eV and 4 eV. The other prominent CT feature is that around 6 eV, first seen in Nd$_2$CuO$_4$~\cite{Hill1998} and La$_2$CuO$_4$~\cite{Abbamonte1999a} and subsequently in all systems featuring CuO$_4$ plaquettes. As has already been discussed above, this feature is attributed to a local CT excitation within the plaquette from the bonding to antibonding orbitals. It's dependence on the Cu-O bond length is discussed in \cite{Kim2004d}. It is dispersion-less and is correctly predicted by, for example, single ion Anderson models. As such, it is relatively uncontroversial.  There has been much more discussion on the complex starting around 2 eV, coincident with the optical gap in these materials~\cite{Uchida1991}.

In this context, 2D planar cuprates studied to date with $K$-edge RIXS include La$_{2-x}$Sr$_x$CuO$_4$~\cite{Kim2002,Lu2006,Wakimoto2005b,Collart2006,Ellis2008},
Nd$_{2-x}$Ce$_x$CuO$_4$ \cite{Hill1998,Ishii2005b}, Ca$_{2}$CuO$_2$Cl$_2$ \cite{Hasan2000}, Sr$_{2}$CuO$_2$Cl$_2$~\cite{Abbamonte1999a}, YBa$_2$Cu$_3$O$_{7-\delta}$ \cite{Ishii2005a}, and HgBa$_2$CuO$_{4+\delta}$ \cite{Lu2005}. At the same time, soft x-ray work has been done at the O $K$-edge on Sr$_{2}$CuO$_2$Cl$_2$ \cite{Harada2002} and at the Cu $L$-edges on La$_{2-x}$Sr$_x$CuO$_4$,~\cite{Ghiringhelli2004, Ghiringhelli2007} and Sr$_2$CuO$_2$Cl$_2$, Bi$_2$Sr$_2$CaCu$_2$O$_8$, and Nd$_{2-x}$Ce$_x$CuO$_4$~\cite{Ghiringhelli2004}.

The picture of the charge transfer complex, and its momentum and incident energy dependence has slowly evolved as the energy resolution of the instruments has improved. Early experiments saw this as a single feature dispersing \cite{Abbamonte1999a,Hasan2000}. Subsequently, with an energy resolution of 300 meV, it was seen as two features, one around 2 eV, the other around 4 eV, with different dispersions in-plane and no dispersion out of the CuO plane~\cite{Kim2002}. The authors speculated that the lower energy feature was a highly-dispersive, almost-bound exciton predicted by Zhang and Ng~\cite{Zhang1998}. Highly dispersive features were also seen in the hole-doped cuprate, Nd$_{1.85}$Ce$_{0.15}$CuO$_4$ \cite{Ishii2005b}, though one should note that here they are intraband transitions. The 2 eV feature is only seen around the origin. This fact, and the dispersion of the complex was explained by an exact diagonalization calculation of a 4 $\times$ 4 cluster \cite{Ishii2005b}.

\begin{figure}
  \begin{center}
  \includegraphics[width=.6\columnwidth]{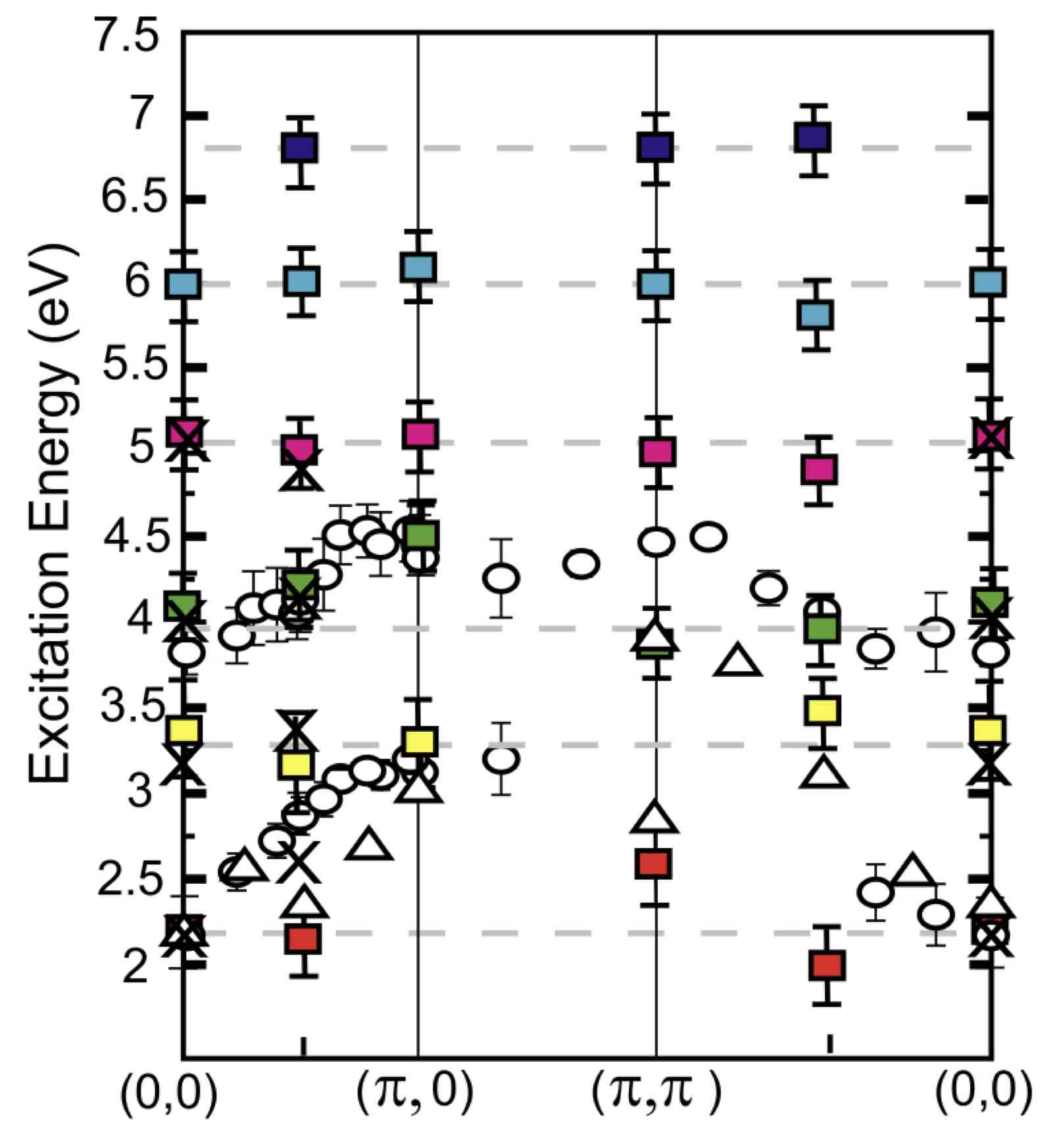}
  \end{center}
    \caption{(color online). Dispersion of CT excitations for optimally doped Hg1201 (squares), compared with previous results below 6 eV for Ca$_2$CuO$_2$Cl$_2$ (triangles)~\cite{Hasan2000} and La$_2$CuO$_4$ (circles)~\cite{Kim2002}, and new data for La$_2$CuO$_4$ (crosses). Figure from \textcite{Lu2005}.}
 \label{fig:Lu2005_3}
\end{figure}
However, in later work, initially on an optimally doped superconductor HgBa$_2$CuO$_{4+\delta}$ \cite{Lu2005}, it was pointed out that as a function of incident photon energy, a multiplet of discrete peaks were observed at approximately 2, 4, 5, and 6 eV and that each of these was more or less dispersion-less in position (with an upper bound of 0.5 eV dispersion for each feature), though the relative spectral weight varied with reduced momentum transfer in the 2D Brillouin zone. This is shown in figure \ref{fig:Lu2005_3}. This multiplet of charge transfer excitations was then seen in La$_2$CuO$_4$ suggesting that such a picture is common to the 2D cuprates \cite{Lu2006}. They argue against the bound exciton picture on the basis of the very small dispersion of the 2 eV feature. Still later work, at still better resolution (130 meV) mapped out this variation in La$_2$CuO$_4$ \cite{Ellis2008}. The authors speculate that the lowest energy feature, around 2 eV may still be a bound exciton, but that its dispersion is significantly reduced from the predictions of \textcite{Zhang1998} because coupling to the phonons gives it a larger effective mass \cite{Ellis2008}. Excitonic states in this energy region are also found in an exact diagonalization study, even if for small cluster sizes the intensity of these features tends to be overestimated~\cite{Chen2010}.

As might be expected, the full range of theoretical options discussed in Section \ref{sec:Calculating_crosssection} have been applied to calculating the RIXS cross-section in 2D cuprates. Amongst the first of these was exact diagonalization of small clusters \cite{Hill1998,Tsutsui1999,Hamalainen2000,Hasan2000,Tsutsui2003,Ishii2005b}
These calculations were able to reproduce (and predict) the general features of the dispersive excitations, assigning them to transitions from the effective ZRS band to the UHB.

In addition, in work led by Igarashi and co-workers \cite{Nomura2005,Takahashi2008,Takahashi2009,Nomura2006}
use realistic band structures, include correlation effects at the level of RPA and multiple scattering effects from the core-hole (not a large correction to the Born approximation) to calculate the interband transitions for a three band Hubbard model. They find the interband transitions, plus correlations, can explain, broadly, the features of the lower resolution data, i.e. the two peak dispersing structure in La$_2$CuO$_4$, without the need to appeal to bound excitons. The fine structure observed in later experiments is not seen in these calculations. A similar approach is taken by Markiewicz and Bansil \cite{Markiewicz2006,Markiewicz2008} who calculated the RIXS response for both hole doped and electron doped cuprates. These authors found 5 ``RIXS bands'' in the undoped case, from 2 eV to 6 eV and mapped out their dispersion. They found that for electron doping, the lowest RIXS band was shifted down to nearly zero energy at the $\Gamma$ point, in agreement with the experimental results of \cite{Ishii2005b}.

\paragraph{3D Cuprates}
\textcite{Doring2004} studied the charge transfer excitations in the 3D cuprate, CuO, in order to investigate their polarization and momentum dependence. In this material, the CuO$_4$ plaquettes are close to rectangular and arranged in two chains \cite{Doring2004}. They found no momentum dependence to the 5.4 eV feature, when either the direction or the magnitude of the momentum transfer was varied. This is consistent with their assignment of this feature to a strictly local charge transfer excitation \cite{Doring2004}. No lower energy CT excitations were observed, though the energy resolution was only 0.8 eV. Ghiringhelli {\em et al.} looked at this same compound using Cu $L$-edge RIXS \cite{Ghiringhelli2004} and saw weak CT excitations out to 5 eV.

\textcite{Kim2009b} looked at a different 3D copper compound, namely Cu$_2$O ("cuprite") . In contrast to most of the rest of the materials in this review, this material is a weakly correlated system and in fact is a direct gap semiconductor with well characterized exciton lines, of the Mott-Wannier type. The exciton shows up clearly in the $K$-edge RIXS data (taken with an energy resolution of 130 meV) and shows strong sinusoidal dispersion with reduced momentum transfer, along the $\Gamma - R$ direction, with a bandwidth of $\sim$0.5 eV. The authors note that if one assumes that in a weakly correlated system, RIXS measures the joint density of states, then this dispersion should  reflect the lower bound of possible transitions and follow the dispersion of the valence band. This latter has been measured by ARPES in this material to be 0.6 eV, in good agreement with this interpretation of the RIXS result. This same material has also been looked at with Cu $L$-edge RIXS. \cite{Hu2008}.

\paragraph{Nickel-based compounds}

\subparagraph{NiO}
NiO is one of the prototypical transition
metal oxides and has thus received a great deal of attention over the years. Although simple band theory predicts it to be a
good conductor, it is in fact an antiferromagnetic insulator at room temperature with a relatively large band gap (determined
to be about 4 eV by optical measurements).  Following the early work discussed above, this compound has been returned to repeatedly for studies at the O $K$-edge \cite{Duda2006}, the Ni $L$-edge \cite{Ishii2001,Magnuson2002,Matsubara2005,Ghiringhelli2005,Kotani2006} and the Ni $K$-edge \cite{Kao1996,Shukla2003,Huotari2008}

At the Ni $L$-edges, one sees a number of features. With the incident photon energy tuned to the peak of the absorption, features are seen at around 1.1, 1.6 and 3.0 eV energy loss. These are assigned to $d-d$ excitations (including spin-flip processes and non-local pair singlet creation) \cite{Ishii2001,Magnuson2002,Matsubara2005,Ghiringhelli2005}, and are discussed in section \ref{sec:crystal_field_excitations}. If, however, the incident photon energy is tuned to features in the absorption spectrum corresponding to a $d^9{\underline L}$ configuration, then the charge transfer excitations are emphasized, as shown in figure \ref{fig:LedgeCT_IntDep}.
Figure \ref{fig:LedgeNiO} shows the evolution of the charge transfer complex as a function of incident energy \cite{Ghiringhelli2005}. There is broad consensus in understanding such data,  based on a Single Ion Anderson Model (SIAM) type approach, that is one Ni site surrounded by 6 oxygens, as discussed above in the context of the cuprates.  Specifically, the model includes parameters for the charge transfer energy, $\Delta$, defined here as the splitting between the 3$d^8$ and the 3$d^{9}{\underline L}$ configurations in the absence of hybridization, and a finite bandwidth, $W$, for the oxygen 2$p$ states (approximated with a number of evenly spaced discrete levels \cite{Matsubara2005}). Spin-orbit parameters and Slater integrals are taken from Hartree-Fock calculations, scaled by an ad-hoc scaling of typically 0.85 and the full 3d multiplet is included. In contrast to the cuprates, three interaction configurations are included, {\em viz}: 3$d^8$, 3$d^{9}{\underline L}$ and 3$d^{10}{\underline L}^2$. The 3$d$ wavefunctions are allowed to expand (contract) due to the addition of one 3$d$ electron (presence of the core-hole). The RIXS process is then calculated using the Kramers-Heisenberg formula. Using parameters that also work for photoemission and optical spectroscopy produces broad agreement with the experimental data, see figure \ref{fig:NiO Ledge calcs}.

The limitation of the SIAM approach is that it is only able to describe strictly local processes. In interpreting their O $K$-edge data, Duda et al \cite{Duda2006} utilized a Ni$_6$O$_{19}$ cluster, which they found was the minimum size that reproduced the excitation at 4.5 eV, arguing that this was therefore a non-local charge transfer (NLCT) excitation, as it is in the cuprates. (They also found that the 1.9 eV feature is a non-local double singlet creation excitation, see section \ref{sec:crystal_field_excitations})

This cluster is too large to carry out a full exact diagonalization calculation, but by disregarding the $d-d$ excitations and the next-nearest neighbor Ni-to-Ni CT paths, and restricting the basis set to O 2$p_{x,y,z}$ and Ni 3$d_{x^2-y^2,3z^2-r^2}$, then the calculation could be carried out. Parameters very similar to the Ni $L$-edge work above were obtained. In particular, it was found that $\Delta = 3.8 eV$ and $U_{dd}$=7. 2 eV \cite{Duda2006} (c.f. $\Delta = 3.5 eV$ and $U_{dd}$=7. 2 eV from the $L$-edge work \cite{Ghiringhelli2005}). This consistency is reassuring and furthermore demonstrates that NiO is indeed in the charge transfer limit.

\begin{figure}
  \begin{center}
  \includegraphics[width=.8\columnwidth]{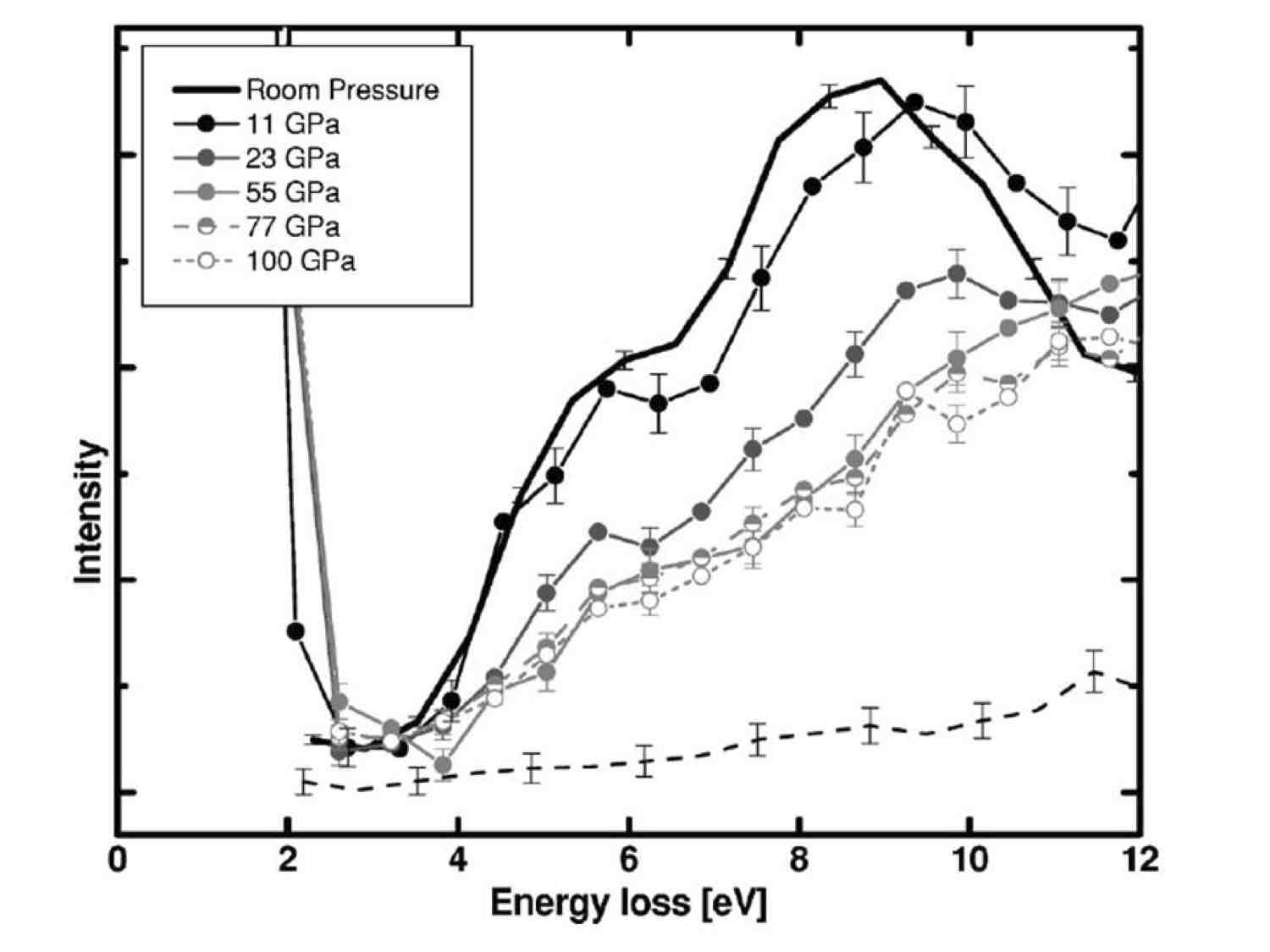}
  \end{center}
    \caption{Charge transfer excitations in NiO as a function of pressure. Data were taken in a powder sample at q = 6 \AA$^{-1}$ with the incident energy tuned to the pre-edge resonance. Bottom dashed line: nonresonant background. As the pressure is increased the resonant intensity decreases and the 5.3 and 8.5 eV structures are seen to smear out due to the increasing electronic bandwidth. Figure taken from Shukla {\em et al.} \cite{Shukla2003}}
 \label{fig:KedgeNiO_underPressure}
  \end{figure}

A holy grail of the field of strongly correlated electron systems is to understand the development of the strong correlation as one tunes from the well-understood limit of good metals to that of the strongly correlated bad metals. A step towards this understanding can be taken by studying how NiO evolves out of the charge-transfer limit as pressure is applied, and the bandwidths driven up. The hard x-ray nature of the Ni $K$-edge allows the use of diamond anvil pressure cells, and thus the pressure dependence of the charge transfer excitations may be studied with this technique. Shukla {\em et al.} carried out such studies on NiO powder \cite{Shukla2003}. Working at room temperature and utilizing a horizontal scattering geometry and a scattering angle of 2$\theta =90^{\circ}$ to minimize the elastic scattering from the sample and from the diamond, these workers were able to observe the pressure dependence of the charge transfer complex up to 100 GPa, figure \ref{fig:KedgeNiO_underPressure}. They found that the position of the features (for example at 5.3 V and 8.5 eV), shifted very little as a function of pressure, but that the peaks became smeared out as the bands broadened, suggesting that there would ultimately be a bandwidth-driven metal-insulator transition in this material.

\cite{Takahashi2007} calculated the $K$-edge RIXS spectra for NiO, treating the correlations in NiO in an Hartree-Fock approach on a multi-orbital tight-binding model, augmented with dynamic correlation effects within the Random Phase Approximation. The core-hole - valence interaction is treated perturbatively, within the Born approximation, which is controlled by the very large gap of NiO. Two prominent peaks are found around 4-6 and 8 eV in the continuum spectra which originate from the 3d $e_g$ electron densities of states.

\begin{figure}
  \begin{center}
  \includegraphics[width=.9\columnwidth]{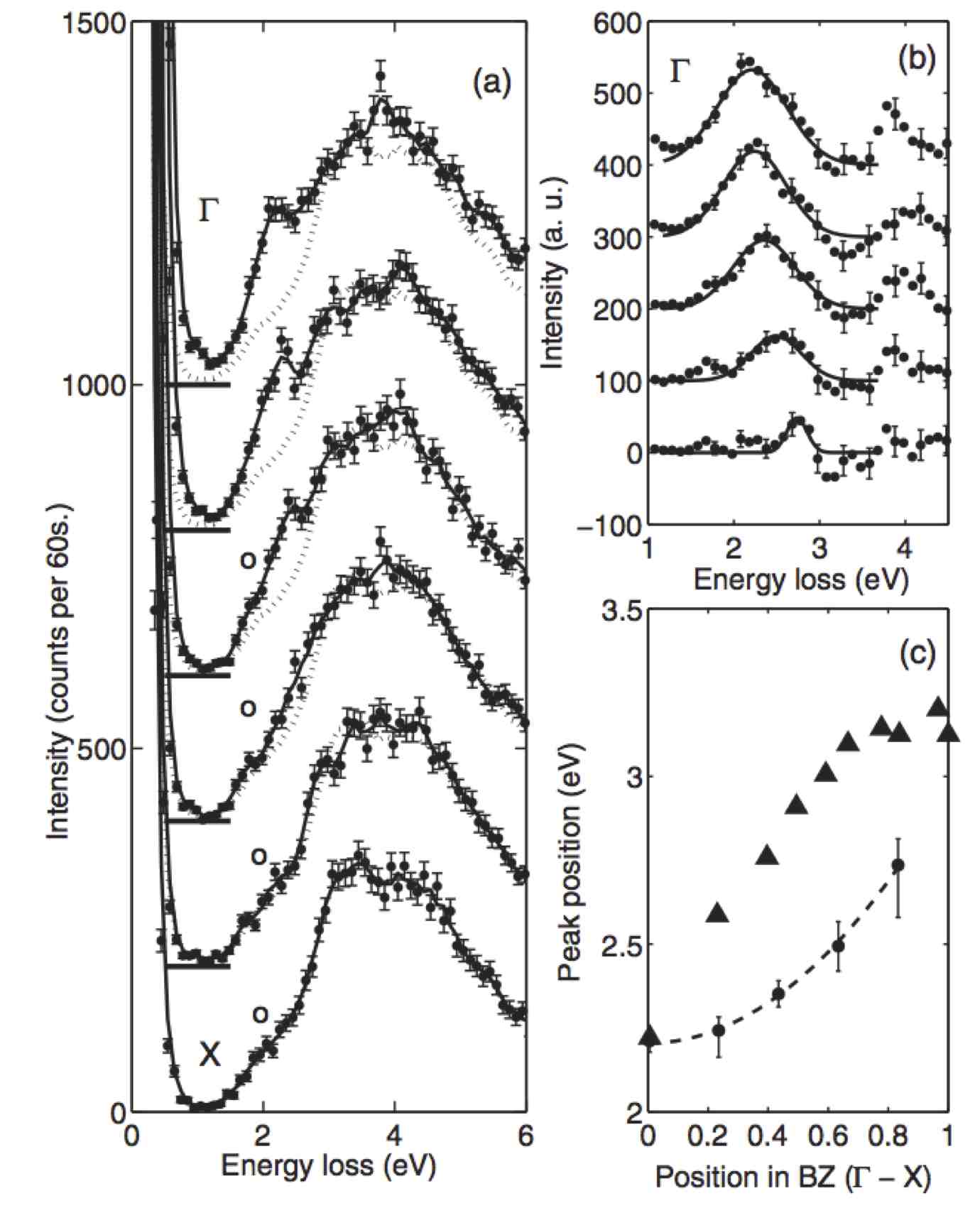}
  \end{center}
    \caption{ (a) RIXS dispersion in La$_2$NiO$_4$ along the [100]=[010]
direction. The spectra have been shifted vertically, the baseline is
indicated for each spectrum. Solid line: three-point smoothed
guide for the eye; dotted line: BZ edge spectrum; open circles
label crystal field excitations. (b) Differences between smoothed
spectra at various points in the BZ and the BZ edge spectrum
from (a). (c) Dispersion of the exciton. Figure taken from Collart {\em et al.} \cite{Collart2006}}
 \label{fig:Nickelate exciton}
  \end{figure}

\subparagraph{Nickelates}
A natural extension to the NiO work is to look at La$_2$NiO$_4$, which is the structural analog of the 2D cuprates discussed above. As revealed by Ni $K$-edge work \cite{Collart2006}, the charge transfer excitations behave quite differently in the nickelate than in the cuprate. The charge transfer gap is seen around 4 eV, in good agreement with optical measurements on this system. A shoulder is seen on the CT complex, which is interpreted as a low lying CT exciton at 3.6 eV. In contrast to the case of the lowest lying CT exciton observed in La$_2$CuO$_4$, this exciton is strongly localized in the nickelate, see figure \ref{fig:Nickelate exciton}. The authors explain this difference by noting that the lowest lying excitation in the cuprate can form a S=0 singlet bound exciton, involving two sites, both of which are spinless and can therefore propagate freely through  lattice without disturbing the the S=1/2 antiferromagnetic background. In contrast in the nickelate system, while there can be a S=0 singlet exciton, $\ket{d^9{\underline L}}$, this is on a single plaquette and cannot propagate through the S=1 antiferromagnetic background without disturbing that background. The non-local CT excitation discussed above for NiO, $\ket{d^8{\underline L};d^9}$ may also form an excitonic state, but this is not spinless. Thus on general grounds, a different dispersion is expected, and in fact observed.

\paragraph{Manganese-based compounds}

\begin{figure}
  \begin{center}
  \includegraphics[width=.6\columnwidth]{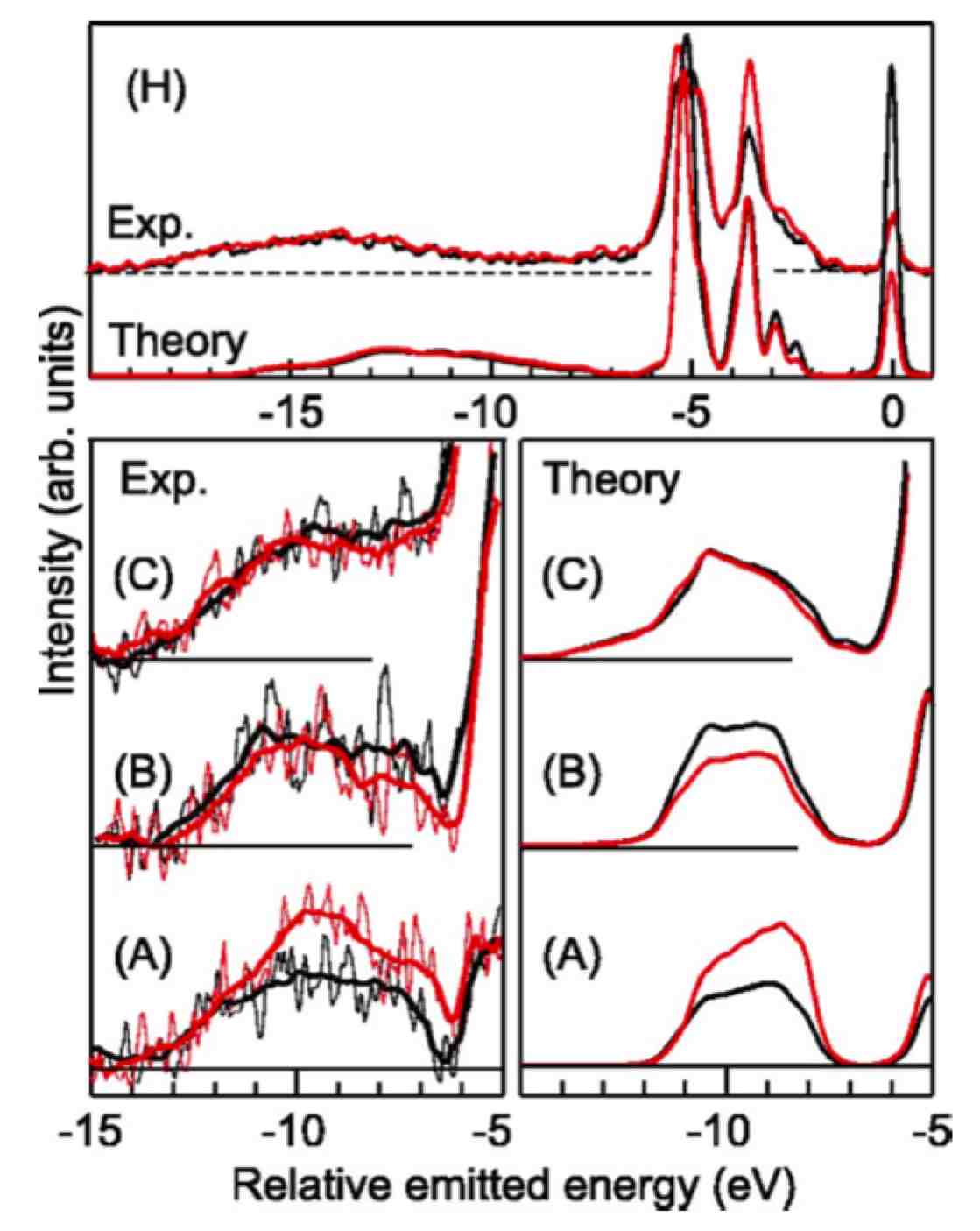}
  \end{center}
    \caption{(Color online) Comparison between the calculated Anderson model results and the experimen  in the charge transfer excitation region as  measured by $L$-edge RIXS in MnO. Labels A-C refer to incident energies in the vicinity of the peak of the L$_3$ resonance. H is the peak of the L$_2$ resonance. Figure taken from Ref \cite{Ghiringhelli2006}}
 \label{fig:MnOGhiringhelli}
  \end{figure}

\subparagraph{MnO}
Early work on MnO was carried out by Butorin {\em et al.} who saw both $d-d$ and CT excitations at the Mn $L_{2,3}$ edges with about 1 eV energy resolution \cite{Butorin1996b}. These authors, however, did not attempt to model the charge transfer excitations. Ghiringhelli {\em et al.} continued the study of MnO with a much improved resolution of 320 meV \cite{Ghiringhelli2006}. With this resolution, the authors found a wealth of features in the inelastic spectrum. In particular, a number of sharp excitations were observed out to 6 eV, attributed to local $d-d$ excitations, and a broad excitation between 7 eV and 14 eV attributed to charge transfer excitations. The authors modeled the spectra using the SIAM, and a local crystal field model. Both reproduced the $d-d$ excitations well, but as expected, only the SIAM approach reproduced the charge transfer excitations,  as shown in figure \ref{fig:MnOGhiringhelli}. The parameters used in the SIAM model were the charge-transfer energy, $\Delta$=6.5 eV, the on-site Coulomb interaction, $U_{dd}$=7.2 eV, the intra-atomic core-hole potential, $U_{dc}$=8.0 eV, the hybridization strengths, $V(e_g)=-2V(t_{2g})$=1.8 eV, the crystal field splitting $10Dq$=0.5 eV and the O 2p bandwidth, W=3.0 eV. This latter was modeled as 6 discrete states. Spin orbit coupling and Slater integrals were calculated using atomic Hartree-Fock codes with the latter then being scaled by 80\%. The results were then broadened to account for lifetime effects, experimental resolution and the discrete O 2p states. As is shown in the figure, in general the agreement is reasonably good, reproducing the position, width, intensity and polarization dependence.

\subparagraph{Manganites}

Agui {\em et al.} utilized Mn $L$-edge RIXS to study the charge transfer excitations in the doped, bilayered manganite, La$_{1.2}$Sr$_{1.8}$Mn$_2$O$_7$, for which the Mn ions have a nominal valence of $d^{3.6+}$ \cite{Agui2005}. Data were taken at room temperature with the incident and outgoing wavevectors were in the plane, such that the contributions from the $3d_{x^2-y^2}$ were emphasized. They observed CT features from 5 eV energy loss to 10 eV energy loss with a tail extending even higher. The 5 eV feature was interpreted as the $3d^{5}\underline{L}$ charge transfer excitations. These authors obtained good agreement with cluster calculations for the Mn$^{3+}$ ion when the full multiplet and JT distortion were taken into account. They found no need to include any contribution from Mn$^{4+}$ sites to explain the   spectra and concluded that the electronic organization was one of $d^{3+} + d^{4+}$, rather than $d^{3.6+}$

Kuepper {\em et al.} studied the single layer parent manganite LaSrMnO$_4$, working at the Mn L$_{2,3}$- and O K- edges with 0.6 eV energy resolution. At the Mn $L$-edge, they were able to observe local Mn $dd$ excitations around 2.2 eV and charge transfer excitations in the range 6-9 eV energy loss which were well reproduced by multiplet charge transfer calculations, including the details of the fine structure. \cite{Kuepper2006}. At the O $K$-edge the energy loss spectra reflects the site specific local partial density of states (LPDOS). By tuning the incident energy through the oxygen $K$-edge, the contributions from the in-plane (O1) and out-of-plane (O2) inequivalent oxygens may be emphasized. While good agreement was seen existing LDA+U band structure calculations \cite{Park2001} for the out-of-plane oxygens, less good agreement was found for the in-plane oxygens, suggesting that further work is required to understand the electronic structure in these materials.

\subparagraph{Cubic Manganites}

Inami et al \cite{Inami2003} carried out the first $K$-edge RIXS study of LaMnO$_3$. They observed peaks at 2.5, 8, and 11 eV attributing the 2.5-eV peak to a transition from effective LHB to UHB across the Mott gap, while the 8- and 11-eV peaks were assigned to excitations from the O 2p bands to the empty Mn 3d and 4s/4p bands, respectively. These authors also measured the momentum and polarization dependence of the features. No significant dispersion was observed in any of the features, while the 2.5 eV feature showed a two-fold periodicity in intensity under rotations around the scattering vector \cite{Inami2003}. Following an approach detailed in \cite{Kondo2001}, these authors calculated the inelastic response in the orbitally ordered state and found that it had the same weak dispersion and characteristic polarization dependence of the 2.5-eV peak.

It should be noted that this picture is not uncontroversial. \textcite{Semba2008} also calculated the $K$-edge RIXS spectrum for LaMnO$_3$. Describing the RIXS intensity as a product of an incident-photon-dependent factor and a density-density correlation function in the 3d states, they calculated the incident photon dependence from an {\em ab initio} band structure calculation of the 4p density of states. The correlation function was evaluated using a multiorbital tight-binding model calculated within the Hartree-Fock approximation. Correlation effects were included via the random-phase approximation, which were shown to strongly modify the spectra. Reasonable agreement was then obtained with the extant experimental data and the authors concluded that the features arose from band-to-band transitions to screen the core-hole potential - rather than orbital excitations created by the off-diagonal part of the 3d-4p Coulomb interaction as proposed by Inami {\em et al} \cite{Inami2003}. This debate echoes a similar one concerning the origin of the resonant enhancement of {\em elastic} scattering at the Mn $K$-edge when looking at orbital wave-vectors. Here it was eventually concluded that the 3d-4p interaction was too small to give rise to the observed effects \cite{Shen2006}.

\begin{figure}
  \begin{center}
  \includegraphics[width=.5\columnwidth]{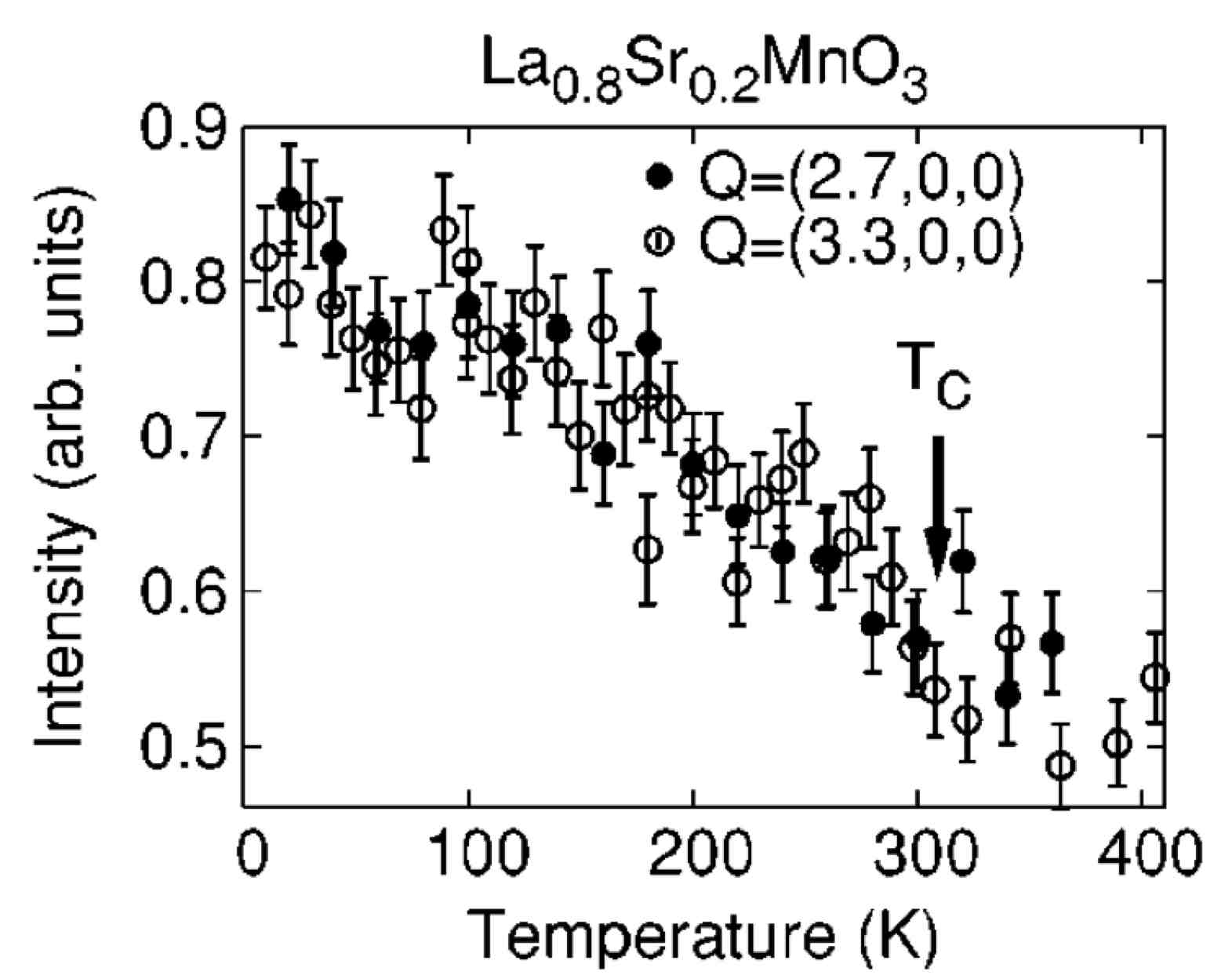}
  \end{center}
    \caption{Temperature dependence of the $K$-edge RIXS intensity at fixed energy transfer of 2.25 eV in La$_{0.8}$Sr$_{0.2}$MnO$_3$. The arrow indicates the insulator to ferromagnetic metal transition temperature (T$_C$)(Figure taken from Ref \cite{Ishii2004}}
 \label{fig:Ishii_tdep}
  \end{figure}

Other studies followed this first experimental work, notably Ishii {\em et al.} \cite{Ishii2004} who studied the doped systems La$_{2-x}$Sr$_x$MnO$_3$ for x=0.2 and 0.4 and found that the gap was partially filled by doped holes,but that the 2 eV feature remained. They also studied the temperature dependence of these features and found that the 2 eV increased in energy at lower temperatures and that the temperature dependence is only observed along the (100) direction, and not the (110) direction - using $P_{bnm}$. The authors speculated that the origin for this anisotropy lay in the magnetic interactions and the relative alignment of spins in the different directions on crossing the insulator to ferromagnetic metal transition. Indeed the RIXS intensity tracks the ferromagnetic order parameter, figure \ref{fig:Ishii_tdep}.

This early temperature dependence study foreshadowed more detailed work by Grenier {\em et al.} who studied a range of cubic manganites \cite{Grenier2005}. Specifically, these authors studied the temperature dependence of the inelastic scattering at (0,2.3,0) in La$_{0.875}$Sr$_{0.125}$MnO$_3$, Pr$_{0.6}$Ca$_{0.4}$MnO$_3$, La$_{0.7}$Ca$_{0.3}$MnO$_3$ and Nd$_{0.5}$Sr$_{0.5}$MnO$_3$. These were chosen because they exhibit a variety of different ground states, respectively: Ferromagnetic insulator, ferromagnetic metal, antiferromagnetic insulator, and antiferrimagnetic insulator. In some cases, exhibiting more than one such phase as a function of temperature. What these authors found is that the integrated spectral weight from 2eV to 5 eV, tracked the magnetic correlations, increasing on entering a ferromagnetic phase and decreasing on entering an antiferromagnetic phase, regardless of the metallicity (or otherwise) of that phase.

In order to understand these results, the authors turned to LDA+U calculations of LaMnO$_3$ \cite{Grenier2005}. They argued that the observed temperature dependence arose because, in the small-q region, the low-energy features in the RIXS spectra are dominated by intersite d-d transitions between Mn atoms (mediated via the O-2p). Ignoring improbable spin-flip excitations, such intersite transitions can occur only between neighbors with the same spin alignment, since no states exist at low-energy in the opposite spin channel. Thus the more pairs of ferromagnetically aligned neighbors in the system, the stronger the low-energy spectra will be, consistent with the experiments.

Working at the Mn $L$-edge, Butorin {\em et al.} showed that the AIM and full multiplet theory could explain the RIXS profiles in thin films of La$_{0.5}$Ca$_{0.5}$MnO$_3$ as long as the calculations were performed for Mn$^{3+}$ ions and not Mn$^{4+}$ \cite{Butorin2002}. Related work in cubic manganites includes that in La$_{1-x}$Na$_x$MnO$_3$ and La$_{1-x}$Ba$_x$Mn$_{1-y}$M$_y$O$_3$, with TM=Co,Ni \cite{Kuepper2000,Bondino2004}

\paragraph{Other Compounds}

Charge transfer excitations have also been studied in a range of other oxides in recent years. Examples include CoO  at the Co $L$-edges \cite{Magnuson2002} and L and $M$-edges \cite{Chiuzbaian2008}. Here as in the materials discussed above, the SIAM was found to describe the data well. In ref. \cite{Chiuzbaian2008}it was further shown that the same set of parameters could describe both data sets providing reassurance as to their validity. Charge transfer excitations, at 6 eV, have also been studied in NaV$_2$O$_5$ at the V $L$-edges, see e.g. ref \cite{Zhang2002,Zhang2006}. In this latter work, the angular dependence was also probed. It was found that the CT excitation had very little angular dependence, though the ratio of the CT to $d-d$ excitations changed, with the $d-d$ becoming substantially larger as the incidence angle was increased. This was attributed to the change in the incident polarization varying the excitation into the different $d$ orbitals \cite{Zhang2006}.

Agui {\em et al.} studied FeTiO$_3$ ("Ilmenite") at the Ti $L$ edge~\cite{Agui2009}. Here, Fe is divalent Fe$^{2+}$, d$^6$ and Ti is tetravalent Ti$^{4+}$ ,d$^0$. Both are octahedrally coordinated and occupy neighboring sites in the basal plane. Importantly, the neighboring octahedra are face-sharing. In addition to the usual charge transfer features seen between 5-10 eV energy loss, new features were seen at 2.5 eV and 4.5 eV that were absent in TiO$_2$ \cite{Matsubara2000} - and also absent in single cluster (TiO$_6$) calculations. These authors therefore carried out calculations for a double cluster (FeTiO$_9$), i.e. the central Ti, surrounded by its oxygen neighbors and a neigboring Fe and its ligands. Such a double cluster was able to reproduce the new features in the spectra and they were therefore attributed to intermetallic charge transfer excitations (i.e. between  the Fe and the Ti). The authors state that these excitations are not mediated by the oxygen, but are rather ``direct'' intermetallic charge transfer excitations.

\subsubsection{Incident Energy Dependence}
\label{sec:incident_energy_dependence}

By its very nature, RIXS is characterized by a very strong incident energy dependence to the inelastic response. Here, by incident energy dependence, we are not simply referring to which edge is being utilized, but also to the position within the edge. As discussed in section \ref{sec:Experimental methods}, the incident energy determines not only the overall size of the signal, but also the relative weight of the various excitations that are observed. This makes it hard {\em a priori} to determine the response function of the, for example, $3d$ electrons from a series of measurements when that response function is apparently strongly dependent on the energy at which it was measured.

\begin{figure}
  \begin{center}
  \includegraphics[width=.6\columnwidth]{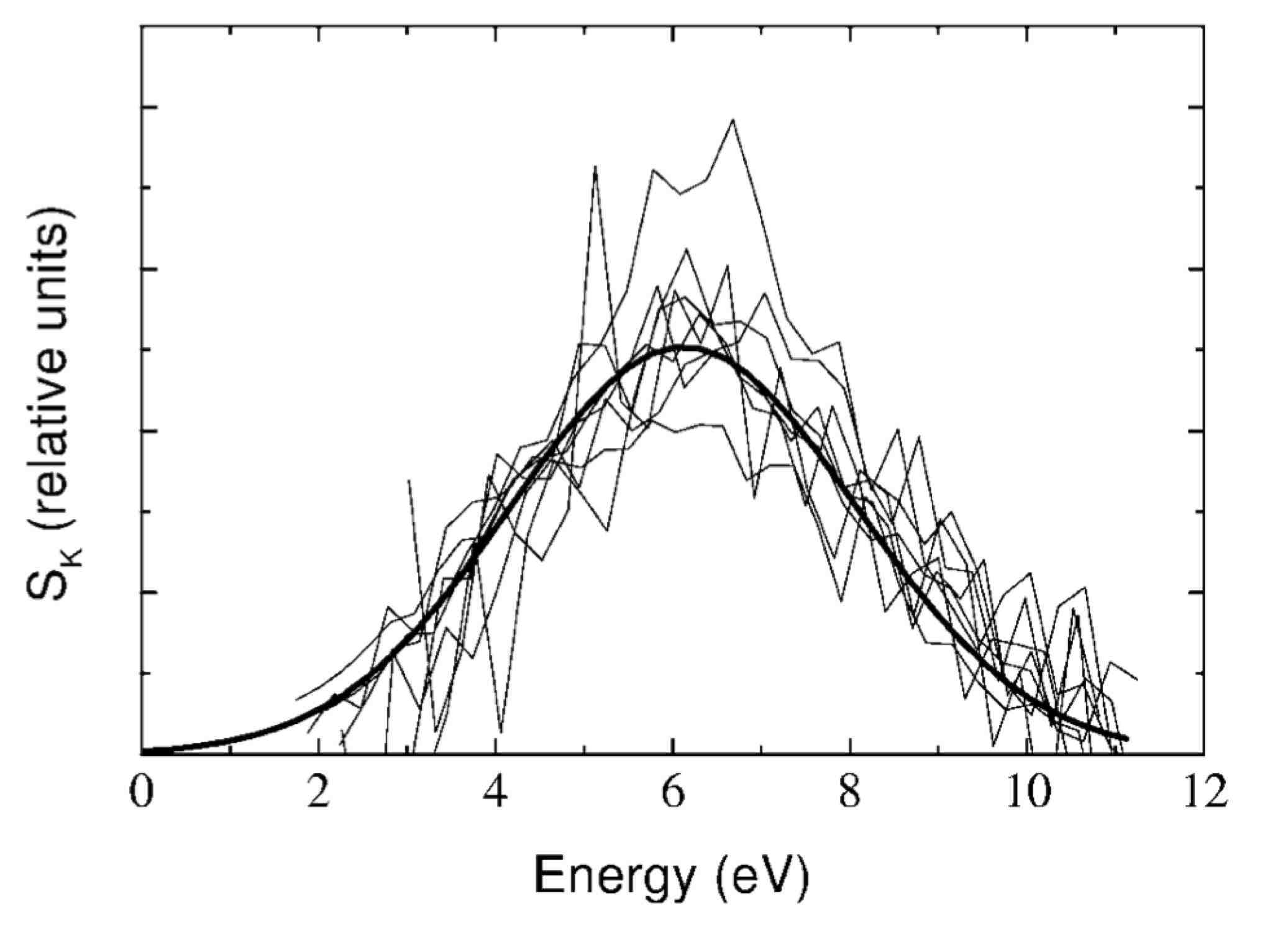}
  \end{center}
    \caption{The result of dividing the energy denominator, equation \ref{eqn:Expt_prefactor}, out from a series of incident energy spectra for the molecular orbital charge transfer excitation at 6 eV in La$_2$CuO$_4$. The thick line is a Gaussian fit~\cite{Abbamonte1999a}.}
 \label{fig:Abbamonte1999a_2}
 \end{figure}

Focusing here on indirect RIXS measured at the Cu $K$-edge, there have been a number of attempts to "divide out'' the incident energy dependence and experimentally derive something that might represent a characteristic response function, see Section~\ref{sec:effindirect}. Fundamentally, these attempts are based on the assumption -- motivated by various theoretical approaches -- that the RIXS response may be factored into a resonant prefactor and a reduced response function. The resonant prefactor is generally taken to be a variant of

\begin{eqnarray}
P(\omega,\omega_i) &=& \frac{1}{(\omega_{k^\prime}-\omega_{res})^2+{\Gamma_{k'}}^2} \times  \nonumber \\
 &&\frac{1}{(\omega_k-\omega_{res}-|U_C|)^2+{\Gamma_k}^2}.
\label{eqn:Expt_prefactor}
\end{eqnarray}

Where $\omega_{res}$, $\Gamma_{k,k'}$ and $U_C$ are potentially experimentally determined parameters representing the resonance energy, the core-hole lifetimes (in principle allowed to be different) and the core-hole potential, respectively. Expressions similar to this have been derived a number of times from different perspectives (see e.g. \cite{Abbamonte1999a,Doring2004,Brink2006,Ament2007})

The first attempt to recover the response function was due to~\textcite{Abbamonte1999a}, who motivated such an expression (albeit with $U_C$=0 and $\Gamma =\Gamma_{k}=\Gamma_{k'}$) based on a third-order shake-up description of RIXS~\cite{Abbamonte1999a,Platzman1998}. They then took $\omega_{res}$, $\Gamma$ as fitting parameters, fitting the data to find the best fit values. Dividing each individual spectrum by the appropriate resonant prefactor then gave the results shown in figure~\ref{fig:Abbamonte1999a_2}, in which all the data (in this case for the molecular orbital charge transfer excitation in La$_2$CuO$_4$), approximately collapse onto a single curve. They argued that the resulting function was a projected response function, that is the valence part of the dynamic structure factor, $S_{\bf q}(\omega)$, weighted by the form factor of the core state.

D\"oring {\em et al.} took a similar approach, deriving equation \ref{eqn:Expt_prefactor} (again with $U_C=0$) again from third order perturbation theory for shakeup processes \cite{Doring2004}. In looking at CuO they observed that the dispersion of the shakeup features with incident energy required that two intermediate state resonances were involved, with distinct excitation energies, labeled A and B. Interestingly, they found $\Gamma_{k}\neq \Gamma_{k'}$ for resonance A at 8981.4 eV, but $\Gamma_{k}= \Gamma_{k'}$ for resonance B, at 8995.8. With these parameters, they could simulate the incident energy dependence of their features, both in terms of intensity and energy position \cite{Doring2004}

 \begin{figure}
  \begin{center}
  \includegraphics[width=.95\columnwidth]{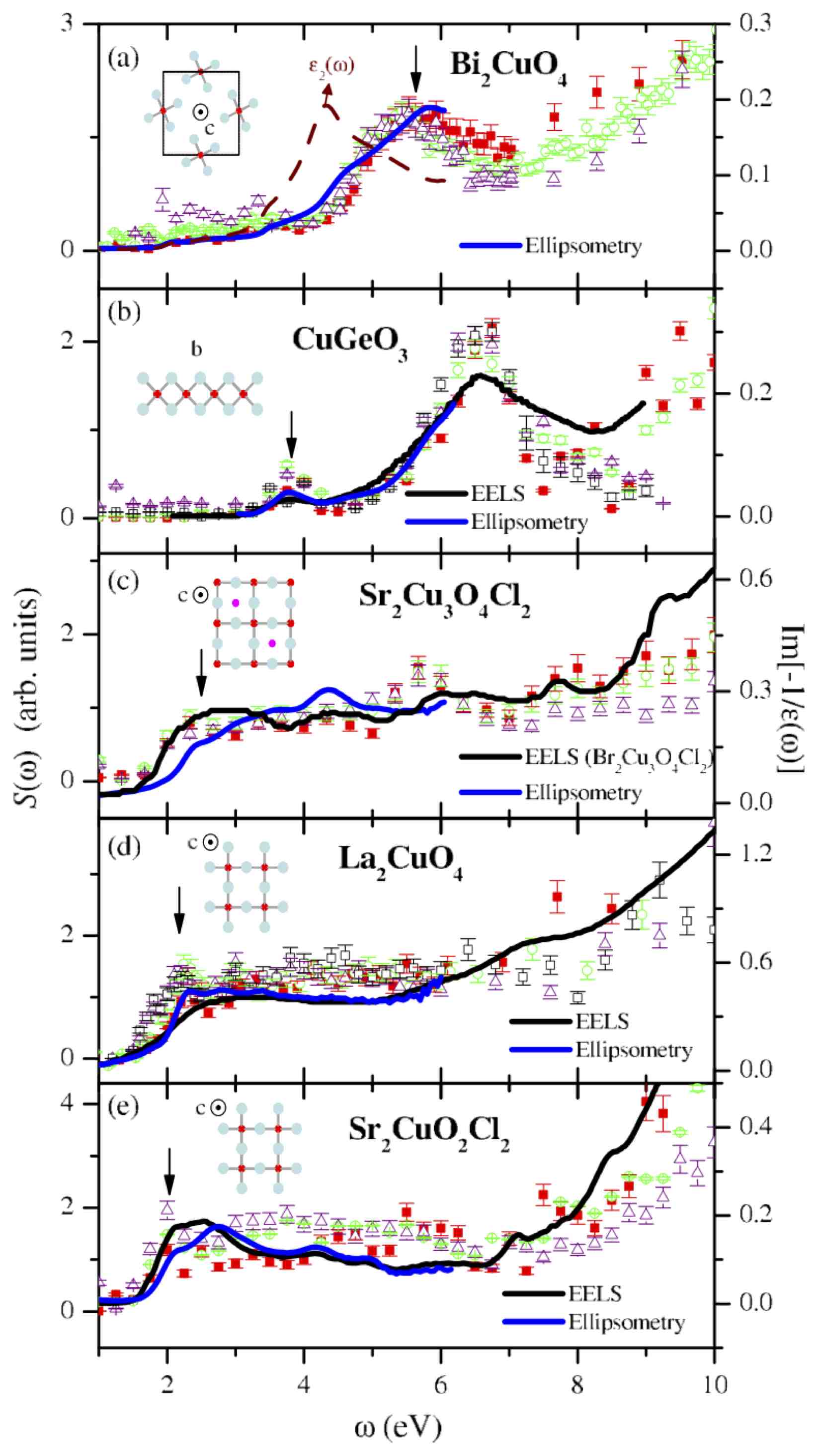}
  \end{center}
    \caption{(Color online) Comparison of S($\omega$) with the
      measured $Im(-1/\epsilon(\omega))$ for various
      cuprates. Different symbols are used to denote RIXS spectra
      obtained with different incident energies. EELS data from the
      literature are also shown for comparison. These incident photon
      energies (in eV) are (a)  8997.5 ($\blacksquare$), 8998.5
      ($\bigcirc$), 8999.5  ($\triangle$); (b) 8988 ($\blacksquare$),
      8990 ($\bigcirc$), 8992  ($\triangle$), 8994  ($\square$); (c)
      8992 ($\blacksquare$), 8993 ($\bigcirc$), 8994   ($\square$);
      (d) 8991 ($\blacksquare$), 8992 ($\bigcirc$), 8993
      ($\triangle$), 8994 ($\bigcirc$); and (e) 8993 ($\blacksquare$),
      8994  ($\bigcirc$), 8995 ($\square$). The EELS data for CuGeO3,
      La2CuO4, and Sr2CuO2Cl2 are taken from
      Refs.~\cite{Atzkern2001,Terauchi1999,Moskvin2002},
      respectively. Although there is no EELS spectrum for
      Sr$_2$Cu$_3$O$_4$Cl$_2$ available in the literature, that of
      Ba$_2$Cu$_3$O$_4$Cl$_2$ is shown~\cite{Knauff1996}. Figure from~\textcite{Kim2009a}.
    }
 \label{fig:Kim2009a_2}
  \end{figure} 

\textcite{Lu2006} followed the same approach to model their data in La$_2$CuO$_4$. They found 4 energy loss features at 2.3, 3, 4 and 7.2 eV, respectively and showed that their experimental data could be modeled with the incident energy dependence of Eq.~(\ref{eqn:Expt_prefactor}). They found $\Gamma_{k}= \Gamma_{k'}$ and $U_C=0$, for the 3 lower excitations and $\Gamma_{k}=1.7$ eV, $ \Gamma_{k'}=1.5$ eV and $U_C=1$ eV for the high energy, 7.2 eV feature. Similar results were obtained with the incident polarization in the CuO plane or normal to it, apart from an overall 4 eV shift in the incident energies \cite{Lu2006}.

 Following this, and with a view to testing the prediction that the underlying response function is closely related to $S_{\bf q} (\omega)$~\cite{Brink2006,Ament2007}, Kim {\em et al.} carried out a similar procedure, dividing the resonant prefactor out for a range of cuprates in the hopes of obtaining the material dependent response function. A range of cuprates were studied; Bi$_2$CuO$_4$, CuGeO$_3$, Sr$_2$Cu$_3$O$_4$Cl$_2$, La$_2$CuO$_4$ and Sr$_2$CuO$_2$Cl$_2$ \cite{Kim2009a}. Taking the peak in the XAS spectrum for each material, $\omega_{XAS}$ as being the incident energy resonance (i.e. $\omega_{XAS} = \omega_{res}+ |U_C|$) , and forcing $\Gamma =\Gamma_{k}=\Gamma_{k'}$, there are three fitting parameters for each compound. These were chosen to optimize the data collapse. Values for $\Gamma$ ranged from 2.5 - 3.5 eV for these materials and $U_C$ from 2.6 - 5.7 eV~\cite{Kim2009a}. The results of this procedure are shown in figure \ref{fig:Kim2009a_2}. The data collapse is reasonable, though not perfect. The collapsed data are then compared with measurements of the loss function from EELS and ellipsometry experiments on the same materials. Again the agreement is reasonable, for example, the leading edge is more or less reproduced for all materials, though, it is not perfect. Notably, it is better for the zero dimensional Bi$_2$CuO$_4$ and edge-sharing CuGeO$_3$ cuprates than for corner-sharing geometries. This suggests that the ultra-short core-hole lifetime expansion \cite{Brink2006,Ament2007} works better in more localized systems.

\subsection{Charge-density waves}

Recent work by Wakimoto {\it et al.} \cite{Wakimoto2009} has indicated the possibility of RIXS to provide information on charge-density waves. Charge excitations  were studied in stripe-ordered 214 compounds La$_{5/3}$Sr$_{1/3}$NiO$_4$ and 1/8-doped La$_{2-x}$(Ba or Sr)$_x$ CuO$_4$. They studied the momentum-dependence of RIXS at the Ni and Cu $K$ edges. At a momentum transfer corresponding to the charge stripe spatial period, an increase in inelastic spectral weight around transferred energies of 1 eV is observed for both the diagonal (nickelate) and parallel (cuprates) stripes.
These features, which are below the regular charge-transfer gap, are interpreted as collective stripe excitations or anomalous softening of the charge excitonic modes of the in-gap states. These modes are related to fluctuations of the charged stripes, i.e. collective charge stripe motion in the form of meandering and compression modes. Another plausible explanation is that the observed anomaly is due to the in-gap state of the striped system, in particular, because the excitonic mode of the in-gap state may have anomalous softening.

\subsection{Crystal field and Orbital Excitations; Orbitons \label{sec:crystal_field_excitations}}

The RIXS spectra show the presence of local $dd$ transitions at the
$K$, $L$, and $M$-edges. In optical spectroscopy local $dd$
transitions are forbidden because they are not dipole allowed. Due to
an additional coupling to the lattice can merely become weakly
allowed. However, the transition matrix element in $L$- and $M$-edge
RIXS is very different since it is the product of the reduced matrix
elements for the absorption and emission steps, which are both
allowed. The main body of RIXS work on $dd$ excitations is therefore
done at the transition metal $L$ and $M$-edges, to be surveyed in Section~\ref{sec:dd_direct}.

At the transition-metal $K$ edge weak $dd$ excitations are observed in the pre-edge region governed by $1s\rightarrow 3d$ quadrupolar transitions. The presence of $dd$ excitations in the $1s\rightarrow 4p$ region in transition-metal compounds is less easily understood since it is not {\it a priori} clear how the intermediate state exciton exactly couples to $dd$ excitations, see Section~\ref{sec:dd_indirect}.

At the $L$ and $M$-edges the $dd$ excitations are well understood. In general they can be described in a relatively local approach consisting of a  single ion in a crystal field, or an ion hybridizing with its surrounding ligands~\cite{Griffith1961,Sugano1970}. The observed spectral lines are related to crystal-field excitations, {\it e.g.}, $t_{2g}\rightarrow e_g$ in transition-metal compounds in octahedral symmetry and multiplet terms due to  higher-order effects in the Coulomb interaction. In order to describe the $dd$ excitations, a many-body approach is essential.  

A detailed description of Coulomb multiplet effects is beyond the scope of this review and is given in several books~\cite{Griffith1961,Sugano1970}. Let us instead consider the prototypical example of a system with just two spins. The underlying reason that independent-particle models fail to describe Coulomb multiplet is that they are single-Slater determinant approaches. 

First consider the spin part. In an independent-particle picture, two spins can only interact via $-J\sum_{ij}S_{iz} S_{jz}$, which only makes a distinction between  parallel $(\ket{\uparrow\uparrow}$ and $\ket{\downarrow\downarrow})$ and antiparallel ($\ket{\uparrow\downarrow} $ and $\ket{\downarrow\uparrow}$) spins. The energies are $\mp \frac{1}{4}J$, respectively. In a many-body framework, however, spins can interact via a Heisenberg exchange $-J\sum_{ij}{\bf S}_i\cdot {\bf S}_j$ which has as eigenstates a triplet with $S=1$ ($\ket{\uparrow\uparrow}$, $\frac{1}{\sqrt{2}}(\ket{\uparrow\downarrow} +\ket{\downarrow\uparrow})$, $\ket{\downarrow\downarrow})$ and a singlet with $S=0$ ($\frac{1}{\sqrt{2}}(\ket{\uparrow\downarrow} +\ket{\downarrow\uparrow})$), with energies $-\frac{1}{4}J$ and $\frac{3}{4}J$, respectively. Note that the $S=0$ and 1 states both have an $M_S=0$ component, in which the spins are antiparallel. These terms are described as a linear combination of Slater determinants. 

In addition to the dependence on the spin, the Coulomb interaction has a strong orbital component, leading to the well-known $LS$ coupling schemes to describe the eigenstates of the Coulomb interaction. For example, an atom/ion in spherical symmetry with two $3d$ electrons or two $3d$ holes has five distinct multiplets: $^3F$, $^3P$, $^1G$, $^1D$, and $^1S$. Hund's rule indicates that the $^3F$ is the lowest term. Within an independent-particle framework without orbital polarization effects, one is only able to describe two different energies: spins parallel and spin antiparallel. When lowering the symmetry, the different Coulomb  multiplet split. For example, the $^3F$ term branches into $^3T_1$, $^3T_2$, and $^3A_2$, where for two electrons the $^3T_1(t_{2g}^2)$ is the lowest in energy and for $3d^8$, the $^3A_2$ with two $e_g$ holes is the lowest. 

The $dd$ excitations are often rather well described with numerical codes using a single ion, small cluster, or Anderson-impurity type calculations~\cite{Groot2008,Kotani2001}. Despite an often lower resolution, the advantage of RIXS over optical spectroscopy is that the $dd$ transition are allowed and the relative intensities can be described in a straightforward fashion. This allows a better identification of the $dd$ excitations.

\subsubsection{dd excitations with direct
  RIXS \label{sec:dd_direct}}
The creation of $dd$ excitations with direct RIXS is best understood in an ionic framework. The absorption of an x-ray photon starting from $3d^n$ ground state
produces a $\underline{c}3d^{n+1}$ intermediate state, where $\underline{c}$ denotes a hole in a core level. The excited electron is strongly bound to the core hole, forming a local exciton. It therefore does not easily delocalize to neighboring sites. 

The subsequent deexcitation can produce an excited $3d^n$ configuration. RIXS at transition-metal L and M edges has been measured for $3d^0$ systems~\cite{Butorin1997,Harada2000,Agui2009}, early-transition-metal compounds~\cite{Kurmaev2003,Ulrich2008,Ulrich2009,Zhang2002,Zhang2006,Braicovich2007a} via manganese compounds~\cite{Agui2005,Ghiringhelli2006a,Ghiringhelli2008} to late-transition-metal systems, such as CoO~\cite{Magnuson2002,Chiuzbaian2008}, NiO~\cite{Ishii2001,Magnuson2002a,Matsubara2005,Chiuzbaian2005,Ghiringhelli2005} and copper compounds~\cite{Butorin1995,Kuiper1998,Duda2000a,Ghiringhelli2004,Ghiringhelli2007}.

\begin{figure}
   \begin{center}
   \includegraphics[width=0.7\columnwidth]{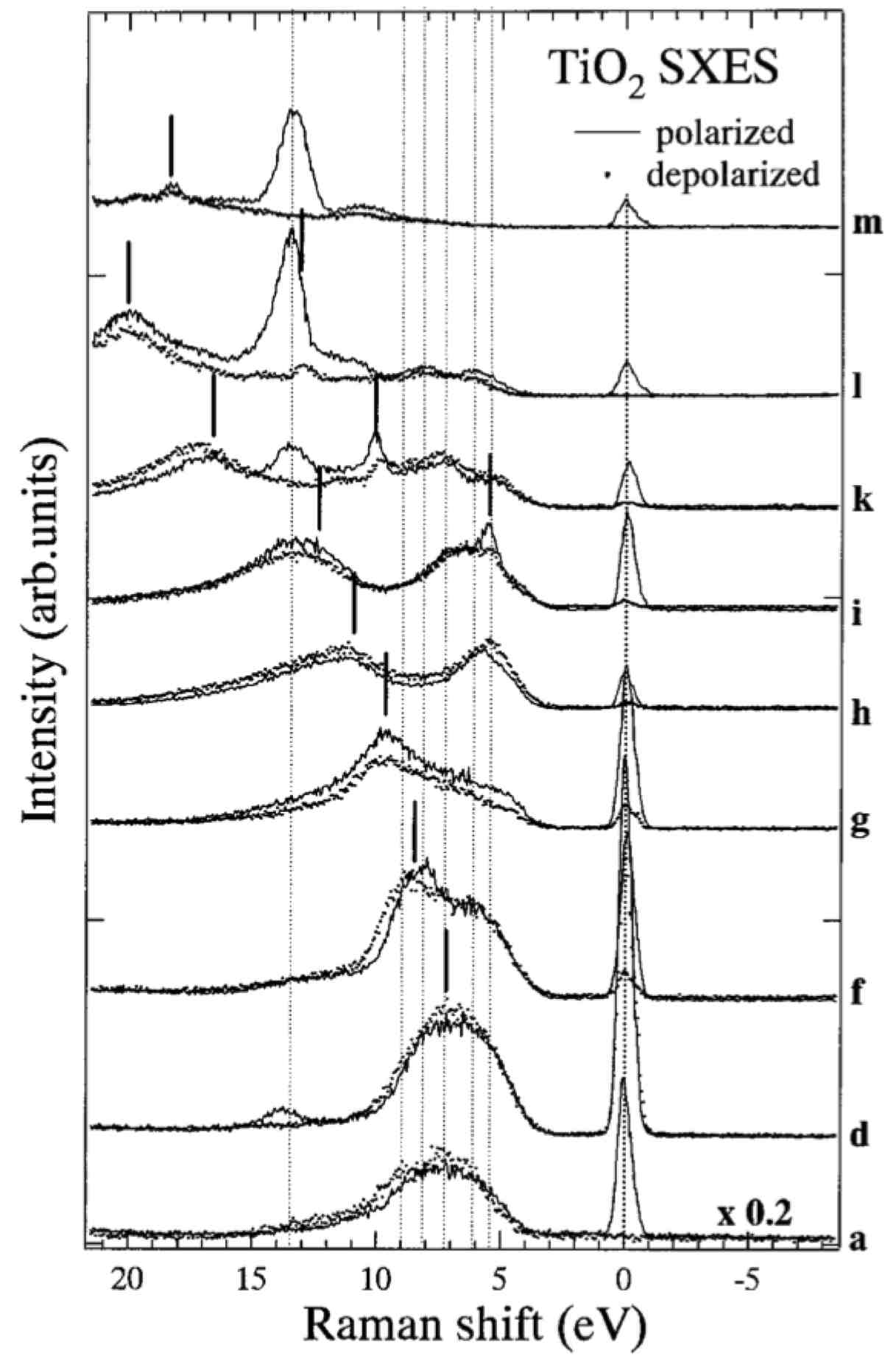}
   \end{center}
   \caption{Polarization dependent soft-x-ray Raman scattering spectra
     of TiO$_2$ plotted relative to the elastic peaks located at
     zero. Five vertical dotted lines between 5 eV and 10 eV indicate
     Raman scattering to the multiple splitting $3d^1 L^{-1}$
     non-bonding states, whereas a higher energy shift line around 14
     eV indicates Raman scattering to the antibonding state. Solid
     bars on each spectrum indicate peak positions of the spectrum
     labeled n fluorescence~\cite{Harada2000}.
    \label{fig:Harada2000_6}}
\end{figure}

The RIXS spectral line shapes often show a strong dependence on the polarization of the incoming and outgoing X-rays. The angular dependence can be understood by looking at the angular dependence of the effective transition operators for direct RIXS as described in Section~\ref{sec:effdirect}. Let us first consider the simplest effective operator with $x=0$, which equal the number operator for holes, ${\underline w}^{dd0}_0=n_h$, see Eqs. (\ref{scatspinindependent}) and (\ref{wllx}). The angular dependence is given by $\bm{\epsilon}'^*\cdot \bm{\epsilon}$, see Eq. (\ref{T1x}). In an atomic model, this term only gives rise to elastic scattering. Obviously, this scattering is strong for early transition-metal compounds that have many empty $3d$ states. The dependence on the polarization vectors can be most clearly seen is a system such as TiO$_2$ ($3d^0$)~\cite{Harada2000}, although it is still present for late transition-metal systems, such as the cuprates~\cite{Ghiringhelli2004}. 

In RIXS experiments, one often employs a (close to) 90$^\circ$ scattering condition. When the outgoing polarization is not measured, the outgoing beam can be described by two polarization vectors $\bm{\epsilon}'_1$ and $\bm{\epsilon}'_2$,  in and perpendicular to the scattering plane, respectively. When the incoming light is linearly polarized, the direction of the polarization vector can be rotated around the incoming beam. When the incoming polarization vector is in the scattering plane ($\pi$-polarized), it is perpendicular to both outgoing polarization vectors at this 90$^\circ$ scattering angle and $\bm{\epsilon}'^*_i\cdot \bm{\epsilon}=0$ for $i=1,2$. This means that there is no contribution of the scattering operator ${\underline w}^{dd0}_0$. 

This is clearly seen in for instance TiO$_2$, where for $\pi$-polarization, the elastic line is close to zero, see Fig.~\ref{fig:Harada2000_6}. When the incoming polarization vector is perpendicular to the scattering plane ($\sigma$-polarized), it is still perpendicular to $\bm{\epsilon}'_1$, but parallel to $\bm{\epsilon}'_2$. Hence, one should get a contribution from ${\underline w}^{dd0}_0$, and indeed for TiO$_2$ a large elastic line is observed, see Fig.~\ref{fig:Harada2000_6}~\cite{Harada2000}. 

\begin{figure}
   \begin{center}
   \includegraphics[width=0.7\columnwidth]{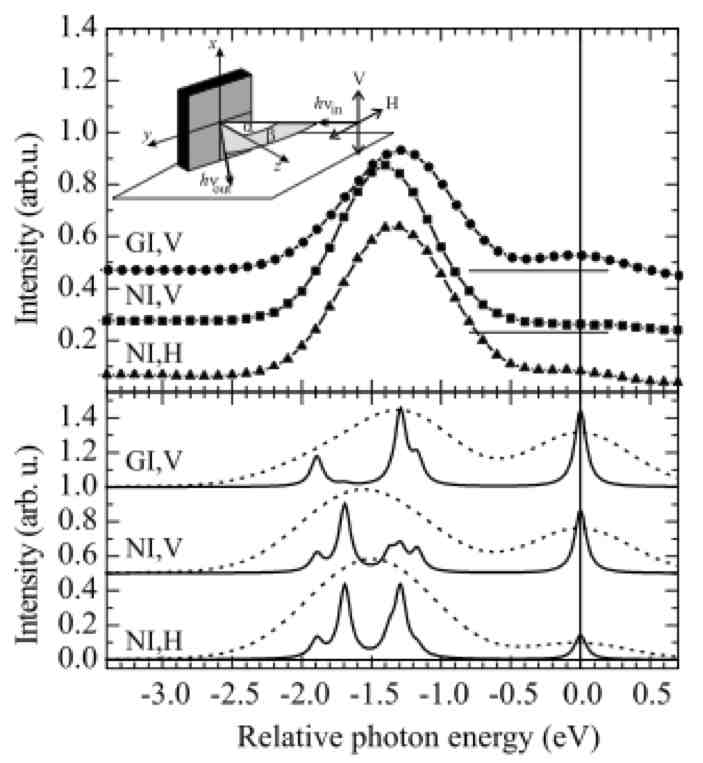}
   \end{center}
   \caption{Upper panel: RIXS spectra of Sr$_2$CuO$_2$Cl$_2$ measured at grazing and normal incidence with V and H polarizations ([GI,V], [NI,V], [NI,H]). Bottom panel: calculated spectra for the same scattering geometries. Upper panel: RIXS spectra of Sr$_2$CuO$_2$Cl$_2$ measured~\cite{Ghiringhelli2004}. 
   \label{fig:Ghiringhelli2004_3}}
\end{figure}

\begin{figure}
   \begin{center}
   \includegraphics[width=.9\columnwidth]{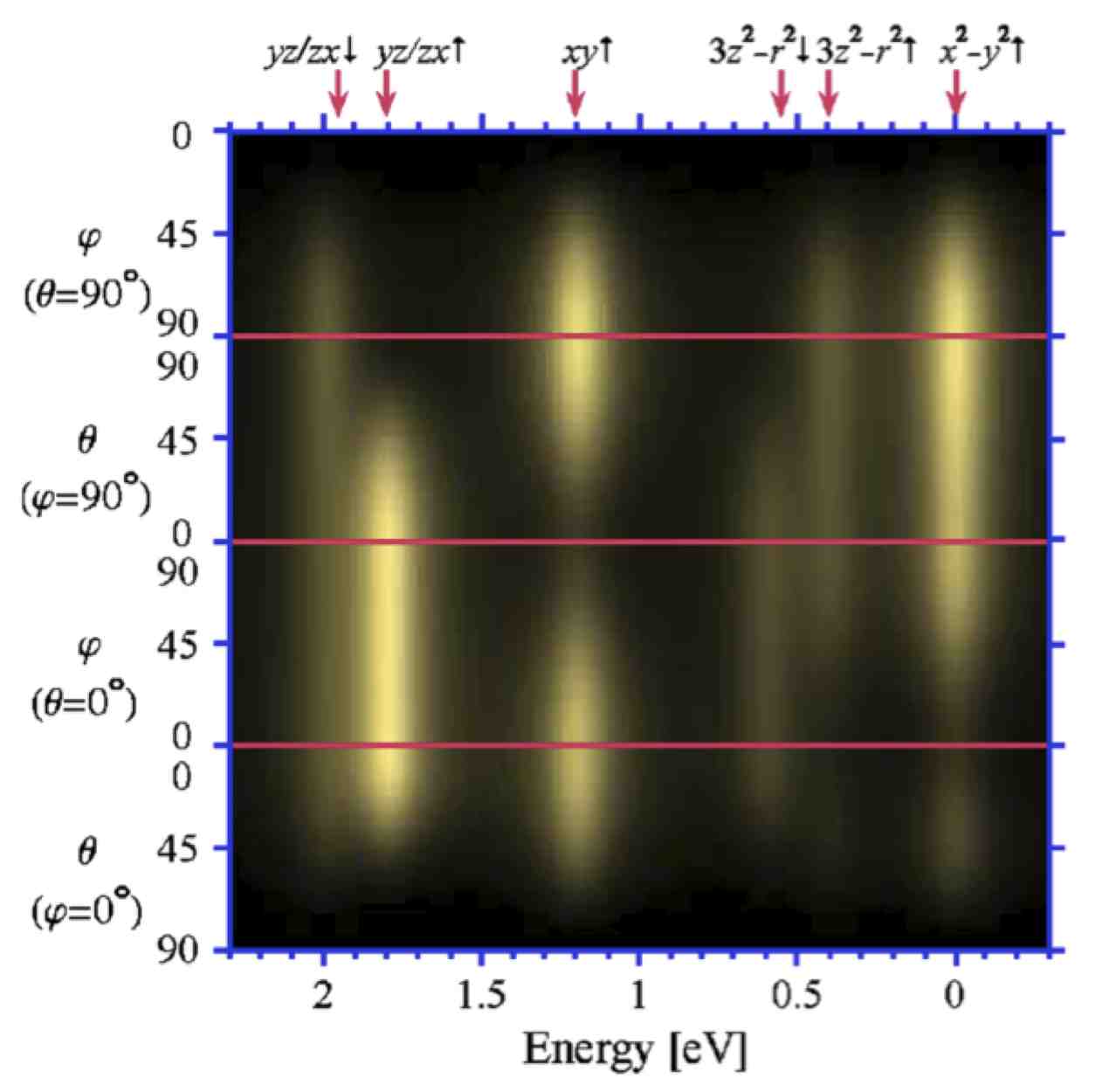}
   \end{center}
   \caption{RIXS intensity for scattering from a single Cu$^{2+}$ ion with initial state $3d_{x^2-y^2}$ and spin up along the $z$-axis~\cite{Veenendaal2006}. \label{fig:vanveenendaal2006_3}}
\end{figure}

A similar effect is observed in the cuprates where a decrease in the elastic line is observed when changing from $\sigma$ to $\pi$ polarized incoming x-rays, see NI,V$\rightarrow$ NI,H in Fig.~\ref{fig:Ghiringhelli2004_3} and $\varphi=90^\circ\rightarrow 0^\circ$ and $\theta=0^\circ$ in Fig. \ref{fig:vanveenendaal2006_3}. Note, the $\sigma$/$\pi$ polarization condition is sometimes referred to in the literature as vertical/horizontal~\cite{Ghiringhelli2004} and polarized/depolarized~\cite{Harada2000}. The small elastic line in RIXS for $\pi$-polarized incoming light combined with the reduction of the zero-loss peak due to ${\bf A}^2$ explain the popularity of the 90$^\circ$ $\pi$-polarized scattering condition.

When employing the commonly-used 90$^\circ$ $\pi$-polarized scattering condition, the scattering operator $w^{dd0}_0$ ($x=0$) is not relevant, and we have to consider the $w^{dd1}_q$ ($x=1$) operator. The angular dependence is given by $T^{1}_q=\frac{1}{2}\bm{\epsilon}_i'^*\times\bm{\epsilon}$, see Eq. (\ref{T1x}). In the absence of a core-hole spin-orbit coupling the effective transition operator is ${\underline w}^{dd1}_q=L_\mu$, where $L_\mu$ is the angular momentum with $\mu=1,0,-1=x,z,y$. Let us clarify this with the example of a divalent Cu$^{2+}$ ion in $D_{4h}$ symmetry, which is typical for the cuprates. 

The ground state of Cu$^{2+}$ in $D_{4h}$ has a hole in the $x^2-y^2$ orbital ($\mu=2$). Let us take a 90$^\circ$ $\pi$-polarized scattering condition with the incoming wavevector along the $z$ direction. In this geometry, we have $\bm{\epsilon}=\hat{\bf x}$, $\bm{\epsilon}'_1=\hat{\bf y}$, and $\bm{\epsilon}'_2=\hat{\bf z}$. Taking the outer product gives $T_{1\xi}\sim\hat{\bf z}, -\hat{\bf y}$ corresponding to $\xi=0,-1$. For $\xi=0$, the selection rules simplify to $\mu'=-\mu$, see Eq. (\ref{selectionmu}), where the minus sign is due to the fact that the orbital moment is an odd operator. Since the ground state is $\mu=2$, the final state that can be reached is $\mu'=-2$ or the $xy$ orbital. For $\bm{\epsilon}'_2=\hat{\bf z}$, we need to consider $\xi=-1$. Applying the selection rule gives $\mu'=1$ or the $zx$ orbital. From Fig. \ref{fig:vanveenendaal2006_3}, we see indeed that for $\varphi=0^\circ$ and $\theta=0^\circ$, the strongest intensity is observed for $x^2-y^2\uparrow\rightarrow xy\uparrow, yz/zx\uparrow$. Additional intensity comes from spin-dependent and higher-order ($Q=2$) terms.

At this point it is useful to make a comparison with nonresonant inelastic x-ray scattering (IXS), where $dd$ transitions are also observed~\cite{Larson2007,Haverkort2007}. Whereas in RIXS the $dd$ excitations are a result of two consecutive ${\bf p}\cdot{\bf A}$ processes, in nonresonant IXS, the local excitations are due to higher-order terms in the expansion of $e^{i{\bf q}\cdot{\bf r}}$ of the ${\bf A}^2$ term. Finite scattering amplitudes are only obtained for even operators $x=0,2,4$~\cite{Veenendaal2008}.  Therefore, the scattering from $x^2-y^2 (\mu=2)\rightarrow xy (\mu=-2)$ requires $|\Delta \mu|=4$. The transition is due to hexadecapolar transitions ($x=4$) which are generally weaker than the quadrupolar ($x=2$) transitions in nonresonant IXS~\cite{Haverkort2007}.  When the transferred momentum is in the $xy$ plane, odd $\xi$ are zero due to the angular distribution. Scattering can occur with $\xi=2$. The selection rule in Eq.~(\ref{selection}), then gives, for the cuprates, quadrupolar transitions from $x^2-y^2\rightarrow 3z^2-r^2$ ($\mu=2\rightarrow 0$)~\cite{Veenendaal2008}. This example clearly shows that, although nonresonant and resonant IXS probe the same excitations, their intensities can be strongly different due to the different selection rules involved.

 For undoped high-$T_c$ cuprates, RIXS has clearly shown the feasibility of studying the variations in crystal field among the various superconductors~\cite{Butorin1995,Kuiper1998,Duda2000a,Ghiringhelli2004,Ghiringhelli2007}. The strong polarization dependence of the spectral features aids in the peak assignment. The RIXS intensities are generally well described within a CuO$_4$ cluster model. For doped cuprates, larger clusters are necessary~\cite{Ghiringhelli2007} to properly describe the effects of nonlocal screening~\cite{Veenendaal1993a,Veenendaal1993b}. RIXS has also been consistently interpreted using small cluster models for TiO$_2$~\cite{Harada2000} and FeTiO$_3$~\cite{Agui2009}, NaV$_2$O$_5$~\cite{Veenendaal2004}, MnO~\cite{Ghiringhelli2006a,Ghiringhelli2008}, manganites~\cite{Agui2005},  CoO~\cite{Magnuson2002,Chiuzbaian2008}, NiO~\cite{Groot1998,Magnuson2002a,Matsubara2005,Chiuzbaian2005,Ghiringhelli2005,Huotari2008}. This clearly indicates a good understanding of the RIXS process at the $L$ and $M$ edges in transition-metal compounds.

A very interesting feature of RIXS at the $L$ and $M$ edges is that the large core-hole spin-orbit coupling enables spin flip processes, since spin is not a good quantum number in the intermediate state. This is reflected in the presence of spin-dependent operators ${\hat {\underline v}}_q^{x1z}$ in Eq. (\ref{operspindependent}). This opened the way for the measurement of magnons and their dispersion~\cite{Ament2009a,Braicovich2009b}, which will be discussed in Section \ref{sec:magnetic_coupling}. 

Before this a crystal-field excitation combined with a spin-flip was observed by~\textcite{Chiuzbaian2005}. In NiO, starting from the $\underline{3d}_{x^2-y^2\uparrow}\underline{3d}_{3z^2-r^2\uparrow}$($^3A_2$ in octahedral symmetry) with $S_z=1$, transitions were observed to  ${\underline 3d}_{t_{2g}\downarrow}{\underline 3d}_{e_g\uparrow}$ (with $T_1$ or $T_2$ symmetry) with $S_z=0$ final states, demonstrating the applicability of RIXS to study magnetic excitations. Following \textcite{Groot1998}, the authors employ a crystal field model for a single Ni$^{2+}$ ion and extract a superexchange constant $6J = 125 \pm 15$ meV. However, it is not entirely clear whether the superexchange for excited states is equivalent to the superexchange causing the antiferromagnetic order. In addition, exciting a combined orbital and spin excitation makes the observation of a single-magnon dispersion impossible~\cite{Ament2009a}.

\subsubsection{dd excitations with indirect RIXS}
\label{sec:dd_indirect}
The RIXS cross section at the transition-metal $K$-edge is dominated by charge-transfer excitations due to the strong core-hole potential. However, $dd$ excitations are also observed. In the pre-edge region, the $dd$ excitations, as seen for example in NiO~\cite{Huotari2008} and cuprates~\cite{Kim2004b} are a result of direct RIXS due to quadrupolar transitions. 

The $dd$ excitations observed when the incoming photon energy is tuned to the main absorption line ($1s\rightarrow 4p$), for example, in cuprates
and NiO~\cite{Veenendaal2010a} are more difficult to understand since direct transitions into the $3d$ shell are very weak. Although, in principle, these excitations can be due to hybridization with the $4p$ band~\cite{Hancock2010}, that coupling is forbidden in most symmetry directions and very weak elsewhere in the Brillouin zone (unless the inversion symmetry of the crystal is broken). Additionally, it does not reproduce the right intensities for NiO~\cite{Veenendaal2010a}. The most likely origin of the $dd$ excitations in the $1s\rightarrow 4p$ excitation region is the interaction between the excited $4p$ electron and the open $3d$ shell. 

Using the methodologies developed in Section \ref{sec:effindirect}, an effective transition operator can be obtained~\cite{Veenendaal2010a}
\begin{eqnarray}
F^x_{fg,q}=\sum_x P(\omega_{\bf k},\omega_{{\bf k}'}) c_x  \bra{f} w^{ddx}_q \ket{g}
\end{eqnarray}
where the operators related to the $1s$ and $4p$ level can be removed leading to an effective $3d$ transition operator ${\bf w}^{ddx}$. The resonance function $P(\omega_{\bf k},\omega_{{\bf k}'})$ is of the form of  Eq. (\ref{resonancefunctionPT}) or (\ref{resonancefunctionUCL}). The coefficients $c_k=2F^0-\frac{2}{15}G^1-\frac{3}{35}G^3, \frac{1}{5}G^1+\frac{3}{35}G^3, \frac{4}{35}F^2-\frac{1}{15}G^1-\frac{3}{245}G^3$ for $k=0,1,2$ and $F^0$, $F^2$, $G^1$, and $G^3$ are expressed in terms of radial integrals of the $3d$-$4p$ Coulomb interaction~\cite{Griffith1961}. 

There is a remarkable similarity between the effective spin-independent operator from  two consecutive dipolar transitions in direct $L$- and $M$-edge RIXS, see Eq. (\ref{scatspinindependent}), and $dd$ excitations created by the $3d$-$4p$ Coulomb interaction as observed with indirect RIXS. The resonance function, obviously, is different for direct and indirect RIXS and an additional constant $c_x$ enters. However, the same one-particle  tensor operator is obtained, although the $4p$-$3d$ Coulomb interaction gives an electron operator instead of a hole operator. The angular dependence is again given by  $T^1_q=[{\bf e}'^*, {\bf e}]^x_q$. 

The $dd$ excitations created change the symmetry of the valence shell. This change in symmetry has to be reflected in a change in symmetry of the excited $4p$ electron. Therefore, the $4p$ electron is no longer a spectator as is often assumed in $K$-edge RIXS. However, care must be taken with the reverse conclusion. A change in symmetry in the $4p$ state (as occurs in a $90^\circ$ $\pi$-polarized scattering condition) does not directly imply that the symmetry of the valence band has changed in the scattering process. The symmetry relation are local point group arguments that do not hold throughout the Brillouin zone, or, in other words, in the wide $4p$ band the different $4p$ states are mixed allowing excitation in one particular orbital and de-excitation from another.
 
Ishii {\it et al.}~\cite{Ishii2010} measured the polarization of the outgoing
   photons and performed a symmetry-based analysis of the $dd$
   excitations at the Cu $K$-edge of KCuF$_3$. The Cu ions are in a
   $3d^9$ configuration, where the hole has $e_g$ character in the
   ground state. Although a quantitative evaluation of the
   intensity requires a microscopic model for the scattering process,
   the authors point out that certain RIXS selection rules are in
   qualitative agreement with their data: $t_{2g}$-$e_g$ excitations appear in all scattering
   configurations that they used, while $e_g$-$e_g$ excitations only
   occur in some. No dispersion is seen within the experimental resolution ($600$ meV).
\subsubsection{Orbitons \label{sec:orbitons}}
Traditional $dd$ excitations, the subject of the previous sections, arise because of a strong crystal field splitting of the $d$ levels due to a local lowering of the symmetry of the lattice. In cases where crystal field splitting is small and superexchange dominates the interaction between orbital degrees of freedom, the orbital excitations have a completely different character: they can in principle move through the system and thus have an appreciable dispersion, in contrast to $dd$ excitations. Such dispersing orbital excitations are often referred to as orbitons or orbital waves. The experimental observation of such dispersive orbital modes is a hot topic~\cite{Tokura2000,Saitoh2001,Polli2007} -- for instance the observation of orbitons in LaMnO$_3$ with Raman scattering remains controversial~\cite{Saitoh2001,Grueninger2002,Saitoh2002}. RIXS is a promising technique for measuring orbitons because it is both directly sensitive to dipole forbidden $dd$ excitations and has enough momentum to probe their dispersion in a large part of the Brillouin zone.

\begin{figure}
\begin{center}
 \includegraphics[width=0.8\columnwidth]{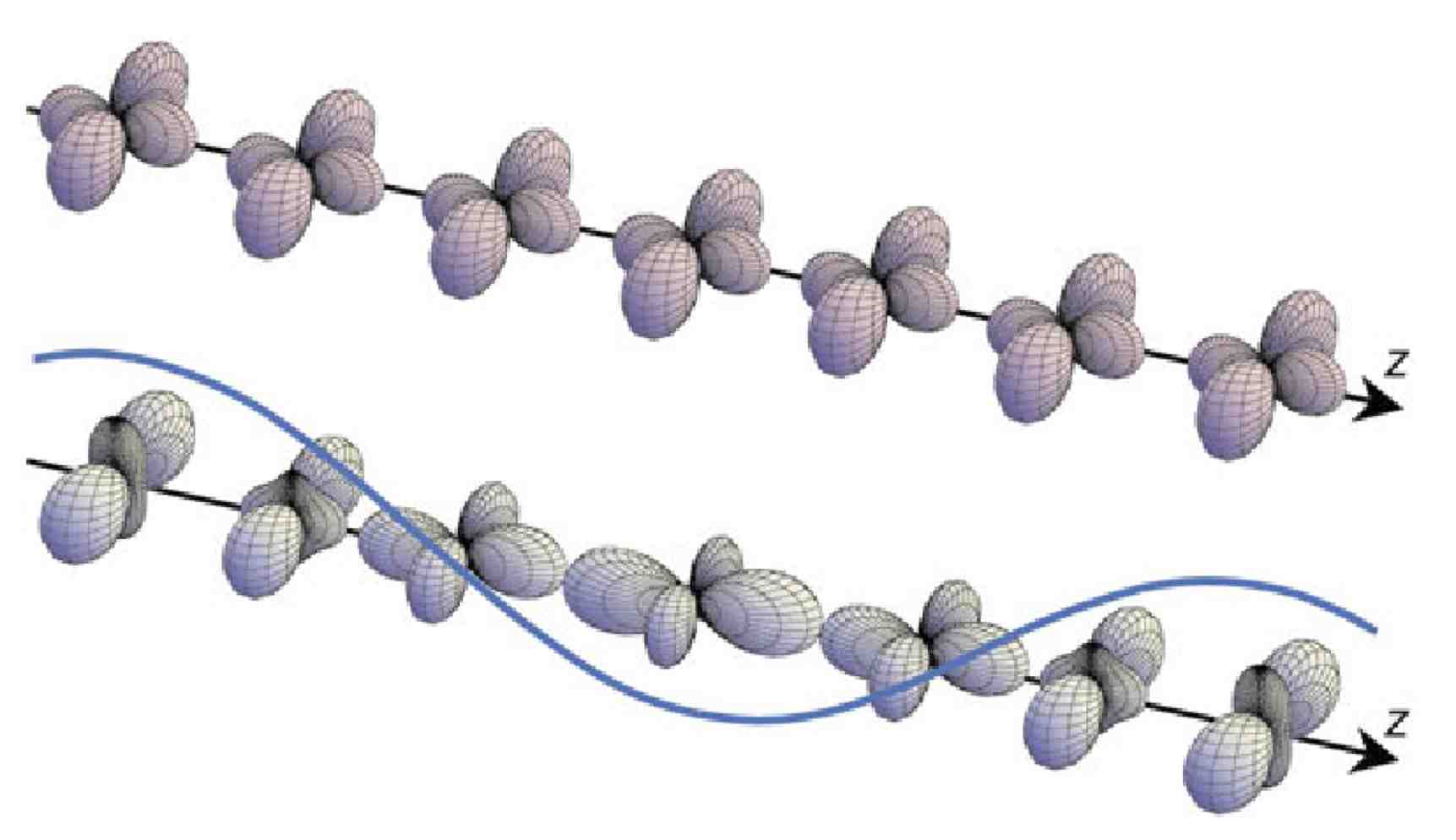}
\end{center}
\caption{Schematic illustration of the orbital wave. Top, the electron cloud in the ground state. Bottom, a snapshot of the electron cloud when an orbital wave is excited~\cite{Saitoh2001}.}
\label{fig:Saitoh2001_1}
\end{figure}

Orbitons can occur in highly symmetric lattices, where the $3d$ orbital degrees of freedom can remain active. The interaction between orbital degrees of freedom gives rise to an orbital ordered ground state. There are two types of interactions between orbital degrees of freedom
on different sites, one related to the lattice and the other to
superexchange~\cite{Kugel1982}. A cooperative Jahn-Teller lattice
distortion removes the local orbital degeneracy and stabilizes a
certain type of orbital ordering, concomitant with a lattice
distortion of the same symmetry. Superexchange is well-known to
generate magnetic interactions between spins. It relies on the
presence of virtual hopping processes that, on $180^\circ$ bonds,
promote antiferromagnetic alignment of electron spins.
In the virtual hopping processes, not only the electron's spins can be effectively interchanged, but also the orbitals that the electrons occupy. Superexchange can thus also drive
orbital ordering. The type of orbital order promoted by superexchange
is closely related to the magnetic order, be it ferro- or
antiferromagnetic, because the hopping amplitudes depend both on the spins and orbitals involved, as stated by the Goodenough-Kanamori rules. If the superexchange interaction between orbitals dominates, the couplings to the lattice can be viewed as perturbations to the orbiton dynamics, see e.g.~\cite{Khaliullin2005,Brink2001,Brink2004}.

{\it Orbiton Scattering in RIXS}
There are three different mechanisms by which photons can scatter from orbitons in RIXS. The first and simplest one is by direct RIXS: the incident X-ray photon promotes a core electron to a certain valence orbital, and an electron from a different orbital on the same site fills the core-hole. This process is discussed in Section~\ref{sec:dd_direct}. 

In indirect RIXS, there are two coupling mechanisms: the on-site shakeup mechanism~\cite{Ishihara2000} and inter-site superexchange bond modulation~\cite{Ishihara2000,Forte2008b}. In the shakeup mechansim, the potential of the core-hole and photo-excited electron locally change the orbital of one of the valence electrons.

In the superexchange bond-modulation mechanism the excitation of
orbitons is due to the core-hole potential modifying the strength of
the superexchange bonds in the intermediate state. This modification
generates orbital flips on the two sites that form the bond. This
scattering process was analyzed within the UCL
expansion~\cite{Forte2008b}, which in this case is well-controlled because the energy associated with the modulation of superexchange bonds is much smaller than the inverse core-hole life-time $\Gamma$ at TM $K$-edges. This photon-orbiton scattering mechansim is similar to the process by which bi-magnons are excited in indirect RIXS, see Section~\ref{sec:magneticindirect}.

The symmetry of the intermediate state potential determines which
on-site orbital shakeup transitions are possible. It is argued by
\textcite{Ament2009b} that multiplet effects at TM $L$-edges, like
relativistic core spin-orbit coupling, drive the rapid evolution of
the intermediate states and thus tend to wash out any particular
symmetry of the potential when the core-hole lifetime broadening
$\Gamma$ is much smaller than the multiplet energy scales. Transitions
of different symmetries are all expected to become equal and weak in strength,
except for the $A_{1(g)}$ component, which becomes
dominant. Consequently the shakeup channel gives low-energy orbital spectra that are very similar for different $L$ edges.

\paragraph{Titanates.}

Orbital excitations were investigated at the Ti $L_3$-edge RIXS in the
titanates LaTiO$_3$ and YTiO$_3$ ~\cite{Ulrich2009}. In these
materials, the orbitally active Ti ions form a cubic lattice. Every Ti
ion has a single electron in the $3d\ t_{2g}$ subshell. The
La-compound is antiferromagnetic, while YTiO$_3$ is a ferromagnet. Two
scenarios have been proposed for the orbital ground state of LaTiO$_3$:
a traditional static orbital ordered one~\cite{Pavarini2005} and a
fluctuating orbital state~\cite{Khaliullin2003}. YTiO$_3$ has an
orbital ordered ground state, where the order is being driven by either a cooperative Jahn-Teller distortion or superexchange interactions.

\begin{figure}
   \begin{center}
   \includegraphics[width=\columnwidth]{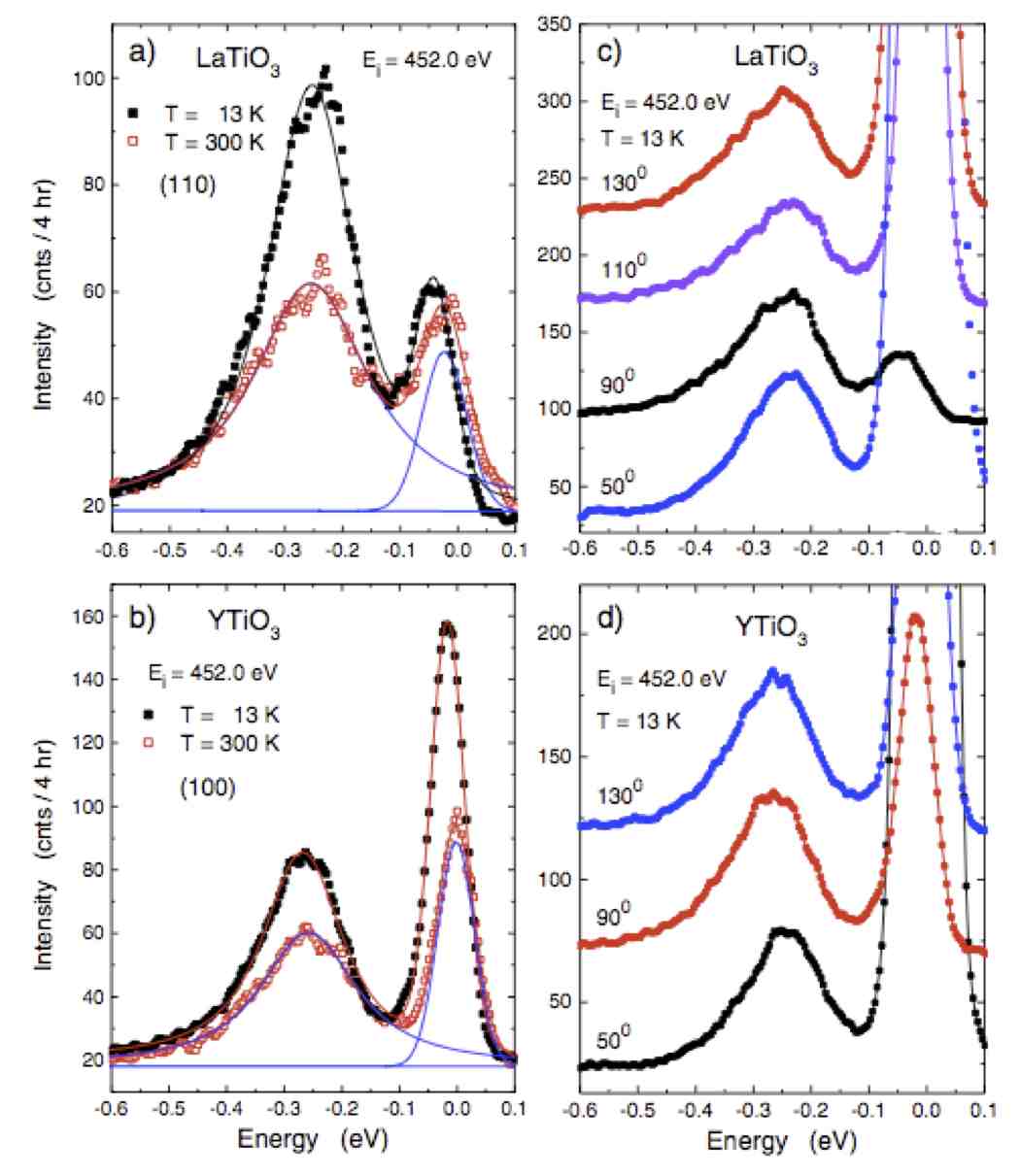}
   \end{center}
   \caption{RIXS spectra at the Ti $L_3$ edge of LaTiO$_3$ and YTiO$_3$ (\cite{Ulrich2009}). For the La compound, ${\bf q}$ is directed along the $[110]$-direction, while for the Y compound, ${\bf q}$ is along the $[100]$-direction. The incident photons were $\pi$-polarized. Figs.~(a) and (b) show the temperature dependence at scattering angle $2\theta = 90^{\circ}$, and Figs.~(c) and (d) display the dependence on scattering angle $2\theta$.\label{fig:Ulrich2009_1}}
\end{figure}

At incident energies of $452.0$ and $454.6$ eV, corresponding to
excitations into the $t_{2g}$ and $e_g$ levels,
experiments have found a loss feature at $250$ meV in both
compounds, see Fig.~\ref{fig:Ulrich2009_1} \cite{Ulrich2008,Ulrich2009}. It is attributed to orbital excitations.
Upon lowering the temperature well below the magnetic transition temperature, the orbital excitations
gain approximately $50 \%$ in spectral weight (see
Figs.~\ref{fig:Ulrich2009_1} (a) and (b)), similar to the behavior of
magnons in Raman spectra of the cuprates. This temperature dependence
is hard to explain in the scenario of a lattice driven orbital
ordering. In addition the spectral weight increases with increasing
transferred momentum in YTiO$_3$, while it decreases in LaTiO$_3$. The
authors note that this behavior is opposite to that of the magnetic
excitations, which is natural for superexchange driven orbital
ordering. The ${\bf q}$-dependence at the $t_{2g}$ edge is shown in Figs.~\ref{fig:Ulrich2009_1} (c) and (d). The orbital excitation shows little dispersion, ruling out that the 250 meV feature is a single orbiton. Comparison of the $e_g$ edge data of YTiO$_3$ to theory gives a best description of the data by single-site coupling to fluctuating orbitals. However, the theory also predicts a single orbiton peak, which is not observed in the experiment.

A detailed theoretical analysis of RIXS on YTiO$_3$ was given
by~\textcite{Ament2009b}, who considered separately single site
shake-up scattering and two-site bond-modulation processes. A
situation in which single-site RIXS processes are considered in a
superexchange driven orbital ordered ground state gives the best fit to
experiments. The $250$ meV feature is described by two-orbiton modes,
which disperse very little.

When the isotropic symmetry-component of the core-hole potential dominates over all others, the single-site RIXS processes are suppressed and two-site bond modulation processes can become visible. It is found that the two-site channel yields two-orbiton excitations, reproducing the peaks at $250$ meV and capturing the trend of increasing intensity with increasing ${\bf q}$, albeit the ${\bf q}$ dependence is too strong compared to the experimental data. From a fit of their theory to Raman and RIXS spectra, the authors obtain an orbital superexchange constant $J_{\rm orb} = 60-80$ meV in YTiO$_3$.

Calculations for a lattice distortion dominated scenario generate a two-peak feature at $250$ meV that does not capture the intensity trend correctly. Further, the authors note that this scenario cannot explain both RIXS and Raman scattering spectra with the same crystal field parameters.

\paragraph{Manganites.}

LaMnO$_3$ is the mother compound of the colossal magnetoresistance manganites. The Mn ions in LaMnO$_3$ are in a $3d^4$ configuration and
form a cubic lattice. Three electrons occupy the $t_{2g}$ orbitals and form an $S=3/2$ spin. The fourth electron occupies one of the two $e_g$ levels and is therefore orbitally active. A strong cooperative Jahn-Teller distortion stabilizes an alternating $3x^2$-$r^2$/ $3y^2$-$r^2$ orbital ordering pattern in the $ab$ plane, where spins order ferromagnetically. Superexchange therefore also contributes to the orbital ordering.

Optical Raman scattering features at around $150$ meV, ascribed to
orbitons by~\textcite{Saitoh2001}, were interpreted as multi-phonon
excitations by~\textcite{Grueninger2002}, see also \textcite{Saitoh2002}. Whether RIXS can observe orbitons in this manganite was analyzed theoretically -- low-energy RIXS data on LaMnO$_3$ has not been reported yet.

Orbital RIXS at the Mn $L$ and $K$ edges were calculated
by~\textcite{Ishihara2000} for a model Hamiltonian where orbital and
spin degrees of freedom interact via superexchange. The $L$ edge is
treated in the fast collision approximation and only single orbitons
appear in this case. In the A-type magnetically ordered ground state
phase of LaMnO$_3$, the orbiton modes have approximately ${\bf
  q}$-independent intensity.  For indirect RIXS at the Mn $K$ edge,
\textcite{Ishihara2000} consider both possible excitation
mechanisms. The shakeup process, mediated by the 3d-4p interaction,
gives the same result as for the $L$ edge, except for the polarization dependence. The inter-site process yields one- and two-orbiton excitations. The calculated single-orbiton intensity shows a strong ${\bf q}$-dependence.

\begin{figure}
   \begin{center}
   \includegraphics[width=0.7\columnwidth]{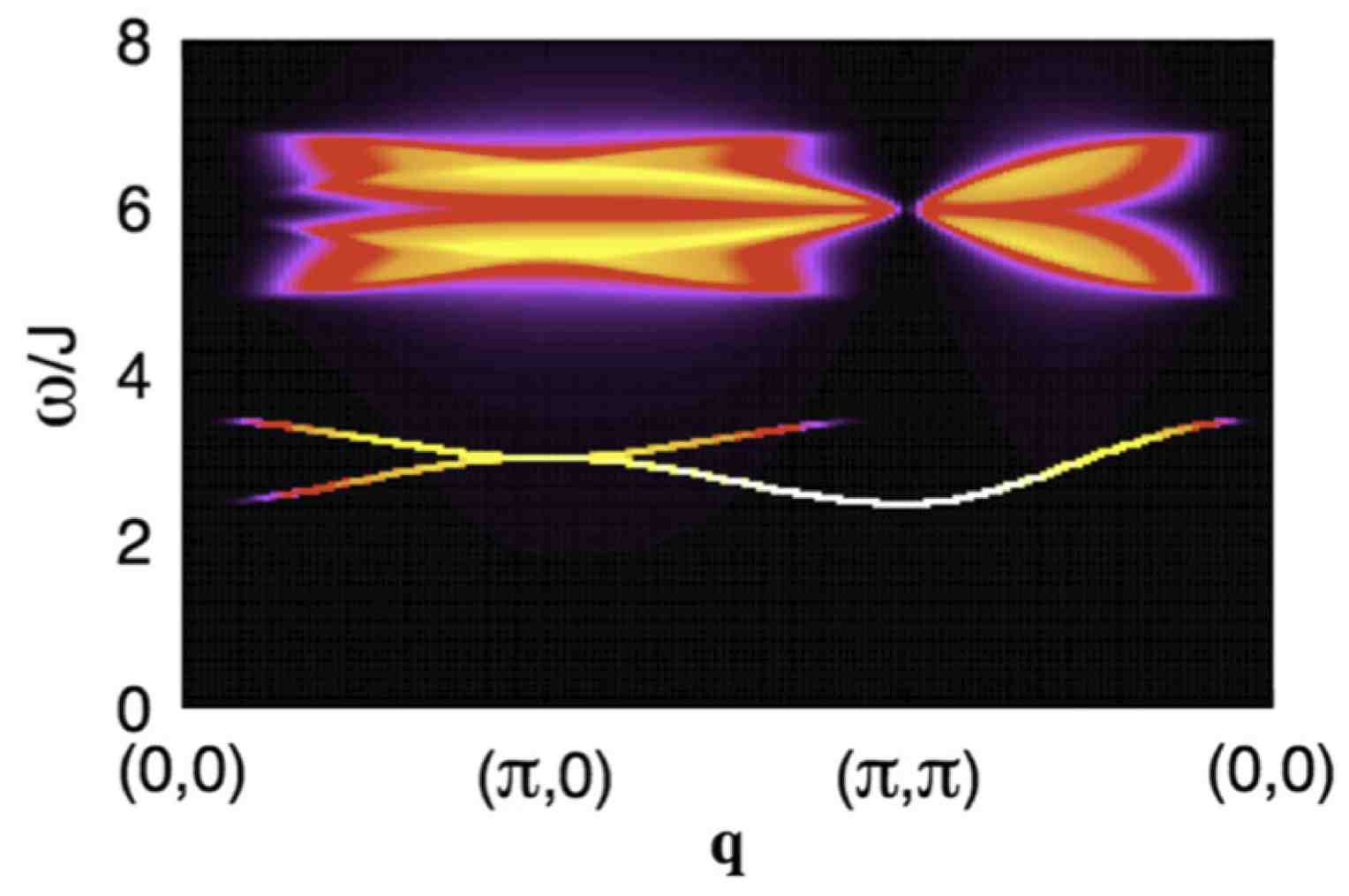}
   \end{center}
   \caption{$K$-edge RIXS spectrum of orbiton excitations in the A-type antiferromagnetic phase of a 2D $e_g$ system \cite{Forte2008b}.\label{fig:Forte2008_2}}
\end{figure}

The indirect RIXS spectrum of an orbitally active $e_g$ system such as
LaMnO$_3$ was calculated by \textcite{Forte2008b} by considering a
model Hamiltonian for the $e_g$ orbital degrees of freedom interacting
by superexchange. The authors calculate the bond modification
scattering channel microscopically by applying the UCL expansion (see
Section~\ref{sec:UCL}). The orbital RIXS spectrum, shown in
Fig.~\ref{fig:Forte2008_2}, shows both one- and two-orbiton excitation
features. The one-orbiton spectrum is sharp with a strongly ${\bf
  q}$-dependent intensity peaking at the orbital ordering vector. The
two-orbiton contributions cause broader features at somewhat higher
energy, the intensity of which is also strongly momentum
dependent. At ${\bf q} = (\pi,\pi)$ and at the $\Gamma$-point, the intensity vanishes.
\subsection{Magnetic Excitations\label{sec:magnetic}}

Measuring magnetic correlations in materials is traditionally the domain of inelastic neutron scattering. With great precision it can probe lower energy magnon dispersions and, more generally, the dynamic magnetic susceptibility $\chi({\bf q},\omega$) of spin-solids and liquids. It is a very powerful and mature technique, but it also has systematic difficulties. As a rule it requires for instance large sample volumes and experiments with materials that contain neutron absorbing elements (e.g. Cd or Gd) are difficult.

As neutrons carry a spin $\frac{1}{2}$, a projected angular momentum of $\Delta S_z=0$ or 1 can be transferred to the material in inelastic neutron scattering. The spin of the photon is often called polarization. Although the photon spin couples only very weakly to the spin of the electron, it does affect the orbital angular momentum of the electron. The spin can be changed due to the presence of additional interactions, such as the spin-orbit coupling, which mixes the electron's orbital angular momentum and its spin. It was already observed in the early days of RIXS \cite{Tsutsui1999} that as a photon has an angular momentum $L=1$, it can transfer $\Delta L_z=0, 1$ or 2 to the solid, potentially creating not only single magnon excitations with $\Delta S_z=1$, but also two-magnon excitations with $\Delta S_z=2$. Conservation of angular momentum in principle allows for such excitations to be created in photon scattering.

In the past few years the resolving power of RIXS has rapidly increased, reaching combined energy resolutions better than 0.1 eV. This has made it feasible experimentally to probe elementary magnetic excitations with both direct and indirect RIXS in a number of transition metal compounds. A large effort has been devoted to benchmarking magnetic RIXS by using the technique on paradigmatic magnetic materials such as NiO, La$_2$CuO$_4$ and Sr$_{14}$Cu$_{24}$O$_{41}$. These materials are close realizations of the elementary spin Heisenberg model in $3$, $2$ and $1$ dimensions, respectively. The successes of these studies helped to establish RIXS as an inelastic magnetic scattering technique. This made it possible to take the next step and address with RIXS open physics questions on magnetism in more complex materials, for instance in doped high temperature superconducting cuprates \cite{Hill2008,Ellis2010,Braicovich2009a,Braicovich2009b,Schlappa2009}.

From a theoretical perspective, however, it is important to note that transfer of angular momentum requires a microscopic coupling between the spin moment of the electrons in a solid and the orbital angular momentum introduced in the solid by a photon. For valence electrons the spin-orbit coupling is normally weak, but for electrons within the atomic core it can be very large, up to tens of eV's. In RIXS this strong spin-orbit coupling of the intermediate state core-hole can generate a large coupling between the orbital moment induced by a photon and the valence electron's spin moment. In what follows we will review this spin excitation mechanism in detail.

Magnetic excitations can also be probed in the $\Delta L_z=0 $, channel i.e. without transfering photon angular momentum $L$ to
valence electron spin $S$. This is a well established process in optical Raman scattering, where the electromagnetic radiation couples to two-magnon excitations (an $S_z=+1$ plus $-1$ magnon combined to a total $S_z=0$ excitation). This coupling is due to the electromagnetic field that modulates the strength of bonds between spins \cite{Elliott1963,Fleury1968,Devereaux2007}. Two-magnon scattering is also weakly allowed in optical conductivity experiments, where this $\Delta S_z=0$ magnetic mode can be excited together with a phonon. Being low energy photon techniques, optical probes have the advantage of an outstanding energy resolution, but the disadvantage of being restricted to essentially $\hbar{\bf q}=0$.

The $\Delta L_z=0 $ channel is also present in RIXS. In fact it is the dominant magnetic scattering channel if the intermediate state
core-hole is not subject to a spin-orbit interaction. This is the case at a transition metal or oxygen K-edge, where the core
hole is in a $1s$ ($L=0$) shell. In this type of RIXS experiment, as in optical Raman scattering,  the microscopic coupling between the electromagnetic radiation and the $S_z=0$ two-magnon excitation is due to a modulation of the strength of magnetic bonds. As we will review in the sections to come, the difference is that in RIXS the modulation is caused by the intermediate state core-hole modifying the magnetic exchange processes \cite{Hill2008,Brink2007,Forte2008a}. This gives rise to transition matrix elements that are very different from optical Raman scattering. Compared to optical probes the resolution of K-edge RIXS is far poorer, but it has the advantage of being able to probe the dispersion of the two-magnon excitations throughout the Brillouin zone.

This section is organized as follows. In Sec.~\ref{sec:magnetic_coupling} the various ways in which the RIXS process can excite magnetic degrees of freedom are discussed. The recent theoretical and experimental progress is discussed in parallel in Sec.~\ref{sec:magnetic1D} through \ref{sec:magnetic3D}, where we focus on magnetic materials with interactions in $1$, $2$ and $3$ dimensions. In this section we focus on purely magnetic excitations, i.e., magnetic excitations without, for example, accompanying orbital excitations as described in Sec.~\ref{sec:crystal_field_excitations}.

\subsubsection{Coupling to magnetic excitations with RIXS \label{sec:magnetic_coupling}}
As indicated above, both direct and indirect RIXS can probe magnetic excitations, be it of different kind. In direct RIXS, single spin-flip excitations can be made at the $2p,3p \rightarrow 3d$ edges of Ni and Cu because of the large spin-orbit coupling of the  $2p$ core-hole~\cite{Groot1998,Veenendaal2006,Ament2009a}.  For copper these edges split into intermediate states with total angular momentum $j=1/2$ ($L_2, M_2$ edges) and $j=3/2$ ($L_3,M_3$ edges). The edges are well-resolved in XAS spectra (particularly at the $L$ edge), and therefore in RIXS one selects either the intermediate states with $j=1/2$ or $j=3/2$. This implies that the spin and orbital angular momentum separately are no longer good quantum numbers in the intermediate state. Therefore orbital and spin orbital angular momentum can be exchanged and direct spin-flip processes can in principle be allowed in RIXS, unlike in optical spectroscopy.

For indirect RIXS at transition metal K-edges ($1s \rightarrow 4p$), the core hole couples to the spin degree of freedom by locally modifying the superexchange interactions~\cite{Brink2007,Forte2008a}. In such a process, the total spin of the valence electrons is conserved, and only excitations where at least two spins are flipped (with total $\Delta S^z = 0$) are allowed. A similar indirect process can also occur at the oxygen $K$ edge. At the copper and nickel $L$ and $M$ edges, both direct and indirect RIXS processes are allowed.

It should be noted that the RIXS process can also locally change the size of the magnetic moment. Transitions with $\Delta S = 1$ and $2$ are allowed for certain magnetic ions, for instance a transition of high spin Ni$^{2+}$ into its low spin state. The energy scale of excitations with $\Delta S \neq 0$ is set by Hund's rule coupling $J_H$, which is typically of the order of 1 eV. These excitations therefore have a much higher energy than the $\Delta S = 0$ ones, whose energy scale is usually set by the superexchange constant $J \sim 100$ meV.

When single spin-flip scattering is allowed in RIXS, one can also measure the dispersion of magnons. This might not be obvious because magnons are after all not local spin-flips but highly delocalized magnetic excitations. But one should keep in mind that the incident photon can be scattered at any equivalent site, leading to a final state that is a superposition of spin-flips at equivalent sites. Such a final state carries a non-local magnetic excitation with momentum $\hbar {\bf q}$. Just as in neutron scattering, the RIXS cross section therefore consists of a local structure factor (depending on polarization, experimental geometry and the excitation mechanism) multiplied by the appropriate spin susceptibility~\cite{Ament2009a,Haverkort2009}. For non-interacting spins, the susceptibility is uniform and featureless, but for systems with interatomic spin-spin interactions the susceptibility acquires a strong ${\bf q}$-dependence, which may be measured in momentum-resolved RIXS.

\paragraph{Direct RIXS.}
Here we give an overview of the cases in which direct RIXS can cause spin-flip excitations without involving any additional change in the orbital component of the valence electrons. Combined excitations of that kind were discussed in Sec.IV.E. Pure spin-flip scattering creates a single magnon quasiparticle and causes a change of angular momentum of the spin system $\Delta S_z=1$. Processes of this type were studied theoretically for Ni$^{2+}$ and Cu$^{2+}$ by \cite{Groot1998,Veenendaal2006,Ament2009a,Haverkort2009}. \textcite{Groot1998} showed numerically that there are no pure spin-flips present for a Cu$^{2+}$ ion when the spin is aligned along the $z$-axis and the hole occupies the $3d_{x^2-y^2}$ orbital. This can also be seen in Fig.~\ref{fig:vanveenendaal2006_3}, where the $3d_{x^2-y^2}$ initial state has spin-up along the $z$-axis, and the ion is positioned in a $D_{4h}$ crystal field. The prefactor for the effective operator that creates spin flips $S_\pm$ is $\mu^2-4$, which means that spin flips are not allowed for $\mu=\pm 2=x^2-y^2,xy$ \cite{Veenendaal2006}, as is confirmed numerically. 

In contrast, for Ni$^{2+}$ in an octahedral crystal field and spin parallel to the $z$-axis, pure spin-flip transitions with $\Delta S_z = 1,2$ are possible without altering the orbital part of the $^3A_{2}$ ground state \cite{Groot1998}. Note that the crystal field is not crucial in these observations: it only singles out the particular ground state used in the calculations.

It was subsequently shown theoretically \cite{Ament2009a} and experimentally \cite{Braicovich2009b} that a pure spin-flip transition for a Cu$^{2+}$ ion with a hole in the $3d_{x^2-y^2}$ orbital does occur if the spin is not parallel to the $z$-axis. In this case, single spin-flip excitations are a result of the $L_z S^z$ operator in the core hole spin-orbit coupling. When looking at the effective $3d$ transition operators, spin flips are a result of $S_z$. This operator has a prefactor $\mu^2-1$ and spin flips are therefore allowed for $x^2-y^2$ orbitals if the spin is in the $xy$ plane (on the other hand, they are now forbidden for $\mu=\pm 1=zx,yz$). For the cuprates these results put $L$ edge RIXS as a technique on the same footing as neutron scattering.

Using linear spin wave theory, \textcite{Haverkort2009} calculated the direct magnetic RIXS $L_{2,3}$ spectra for Cu$^{2+}$ ($S=1/2$) and Ni$^{2+}$ ($S=1$) ions embedded in $1$, $2$ and $3D$ systems. Although pure Heisenberg spin chains do not intrinsically order, N\'eel order could be stabilized by external forces like inter-chain coupling, and linear spin wave theory would then be a valid approach. The RIXS spectrum is factorized into a local transition operator and a spin susceptibility factor, following \textcite{Ament2009a}. Haverkort found that the local operator can be described in terms of fundamental X-ray absorption spectra, like the magnetic linear and circular dichroic spectra.

In principle the same holds at the Cu M edge ($\sim$75 eV). However there the photon momentum is small, so that only magnons in a very
small portion of the Brillouin zone can be probed. In general, for $3d$ transition metal ions, the $2p$ core-hole has both a spin-orbit coupling and a lifetime about an order of magnitude larger than $3p$ core-hole. Therefore, the spin-orbit split $L_2$ and $L_3$ edges are
better resolved than the $M_{2,3}$ peaks. Consequently the $L$ edges offer a better opportunity to measure magnons with direct RIXS.

\paragraph{Indirect RIXS. \label{sec:magneticindirect}}

For indirect RIXS, double spin-flip excitations can be made because the core hole locally modifies the superexchange constant $J$ \cite{Brink2007,Forte2008a}. This perturbation does not commute with the Hamiltonian of the magnetic system, and consequently magnetic excitations are created.

This is best illustrated by considering in detail a superexchange process involving two ions that both have a singly occupied orbital. If an electron hops from one ion to the other a doubly occupied site is created, at the cost of a Coulomb energy $U$. In an exchange process the second electron subsequently hops in the opposite direction. The two electrons are exchanged, and the effective amplitude for this process is, in second order perturbation theory, $J \propto t^2/U$, where $t$ is the hopping amplitude of the participating electrons. The superexchange $J$ coupling is modified by the presence of a core-hole \cite{Brink2007}.

\begin{figure}
   \begin{center}
   \includegraphics[width=1.0\columnwidth]{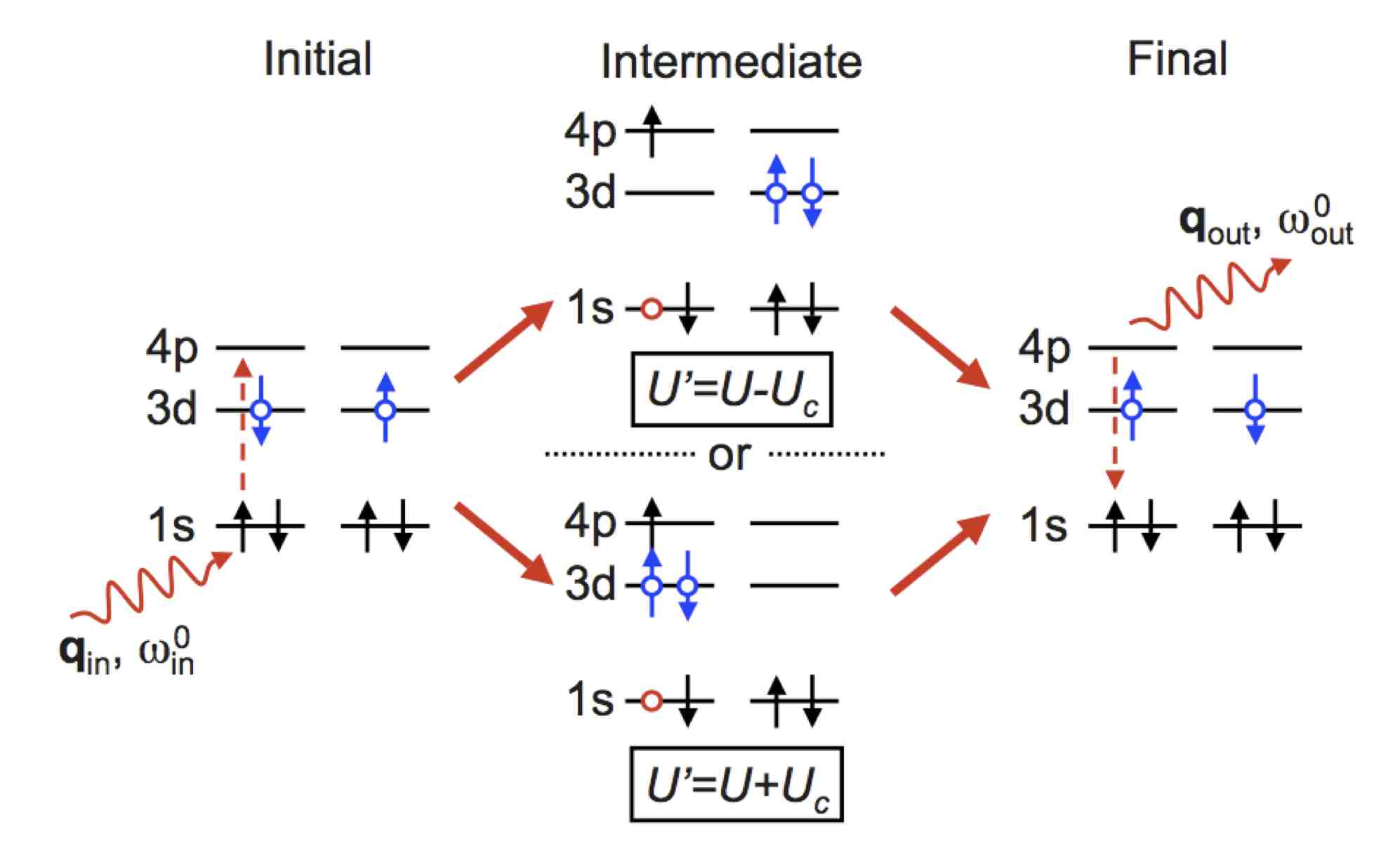}
   \end{center}
   \caption{Indirect, magnetic RIXS mechanism for the Cu$^{2+}$ $K$ edge \cite{Forte2008a}. In the intermediate state, the $3d$ holes either screen or not screen the core hole. The Coulomb repulsion $U$ is effectively replaced by $U'$, incorporating the effect of the core hole potential $U_c$. Note that the open circles represent holes, and the arrows indicate the spin direction. \label{fig:Forte2008a_1}}
\end{figure}

At the transition metal $K$ edge the photo-excited electron ends up far (some eVs) above the Fermi level. The valence electrons tend to screen the core hole, see Fig.~\ref{fig:Forte2008a_1}. Screening affects the superexchange coupling: in the intermediate state of the superexchange process, there is a doubly occupied site, and the electrons can be either both screening the core hole, or both occupying a neighboring ion. In the first case, the double-occupancy energy $U$ is decreased by the core hole potential $U_c$, while in the latter case it is increased by $U_c$. Because in this way $U$ is effectively modified, $J$ is also different at the core hole site.

RIXS experiments at the Cu $K$ edge of La$_2$CuO$_4$ and Nd$_2$CuO$_4$ showed a feature at $500$ meV, which was attributed to two-magnon modes \cite{Hill2008,Ellis2010}. This discovery triggered theoretical work: detailed calculations of the magnetic response functions within the Ultrashort Core-hole Lifetime expansion (see Sec. \ref{sec:UCL}) were initiated in \cite{Brink2007} and expanded upon in \cite{Forte2008a} by including effects of longer range magnetic exchange interactions and finite temperature, which give rise to a generally weak single-magnon contribution to the RIXS signal. \textcite{Nagao2007} computed the magnetic RIXS spectra within this theoretical framework, also taking the magnon-magnon interactions into account. These interactions modify the detailed line-shapes of two-magnon RIXS spectra and cause a transfer of spectral weight to lower energies.

At the transition metal $L$ edges, a similar mechanism occurs (in addition to the direct single spin-flip scattering discussed above) because the photo-excited electron in the $3d$ subshell frustrates the local superexchange bonds. For Cu$^{2+}$ ions, the intermediate state configuration is $3d^{10}$ and superexchange is blocked completely: locally, $J=0$ \cite{Braicovich2009a}.

Also at the oxygen $K$ edge of transition metal oxides a similar modification of $J$ is expected because the photo-excited electron is again disturbing the superexchange process. Transitions at the oxygen $K$ edge are possible because of hybridization of the O $2p$ orbitals with the transition metal ions' unoccupied orbitals.

\subsubsection{1D systems -- Chains and ladders\label{sec:magnetic1D}}

When magnetic ions are Heisenberg-coupled together in a (1D) chain, collective magnetic excitations (spinons) emerge. Two parallel spin chains can be coupled together to form ladders. If the couplings between the chains (i.e. the rungs of the ladder) are dominant, singlets will form on the rungs in the ground state. The singlet to triplet excitations on top of such a ground state are $S=1$ triplons, which disperse because of coupling along the chains. The term triplon is used to distinguish these three-fold degenerate elementary triplet excitations from magnons which are the elementary excitations of  long-range ordered antiferromagnets~\cite{Schmidt2003}.
There are many variants of these systems, for instance, `ladders' with three or more chains, but they are not yet investigated with RIXS.

Sr$_{14}$Cu$_{24}$O$_{41}$ was the first one-dimensional system that has been studied with RIXS down to energies low enough to resolve the elementary magnetic excitations \cite{Schlappa2009}. This compound has alternating layers of chains and ladders of spins. It superconducts under pressure when doped, and for that reason the system has attracted considerable interest, especially because it does not have the typical two-dimensional perovskite CuO$_2$ structure common to the high-$T_c$ superconductors. The ladder has a singlet, spin-liquid ground state that can carry spin $1$ triplon excitations. Superexchange interactions in the edge-sharing chains are weak because they are mediated by a $90^{\circ}$ bond, and the resulting excitations cannot, at present, be resolved from the elastic peak in RIXS experiments.

\begin{figure}
   \begin{center}
      \includegraphics[width=1.0\columnwidth]{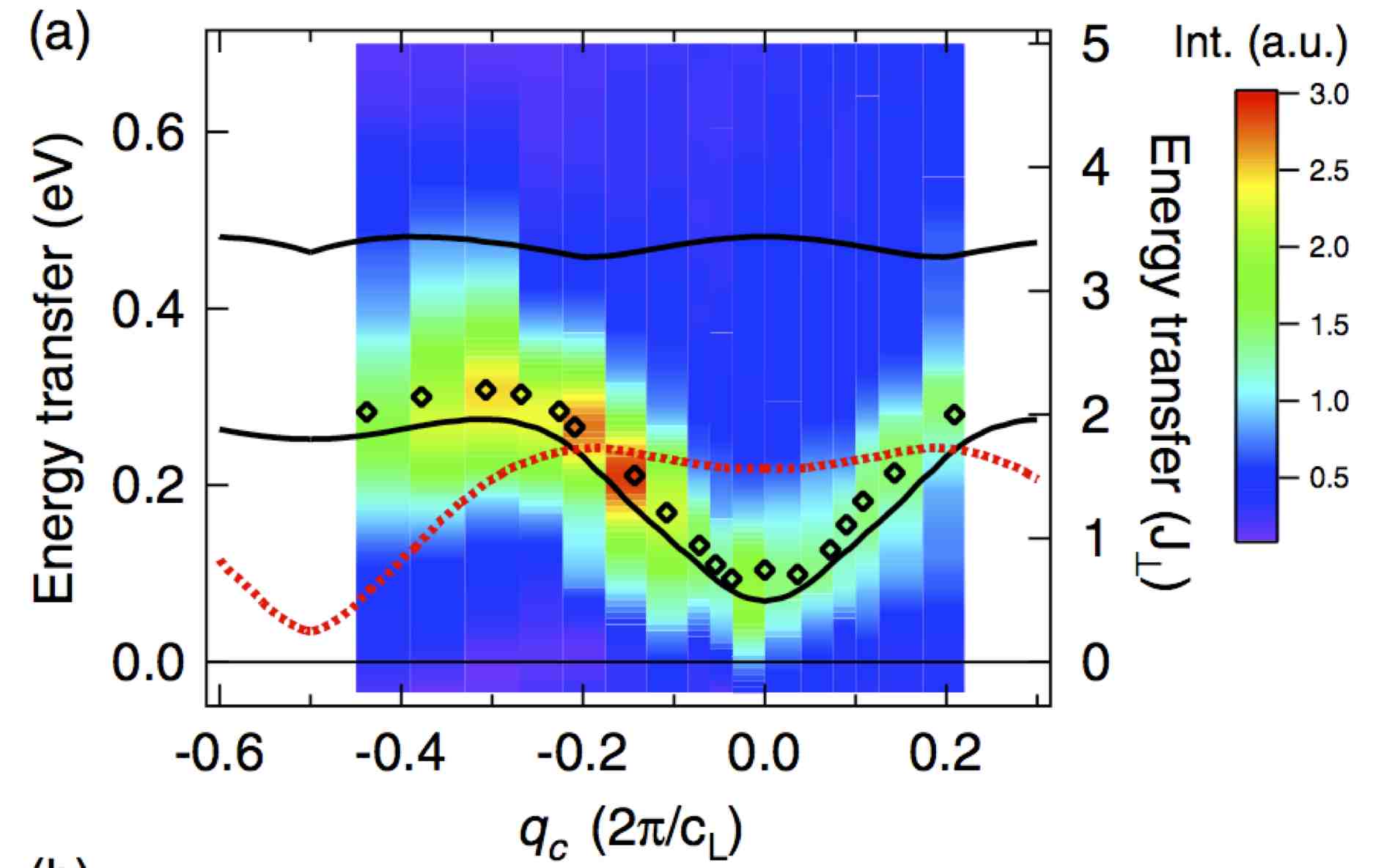}
   \end{center}
   \caption{RIXS data after subtraction of the elastic signal \cite{Schlappa2009}. The open black diamonds indicate the center of mass, obtained by fitting a Gaussian to the spectra. Theoretical one-triplon (red dashed line) and two-triplon (black full lines) dispersions are also shown. The right axis is scaled in units of J along the rungs, extracted from a single band Hubbard model.\label{fig:Schlappa2009_3a}}
\end{figure}

\textcite{Schlappa2009} investigated the Cu $L_3$ edge of Sr$_{14}$Cu$_{24}$O$_{41}$. The authors see two-triplon excitations (on the spin-liquid ladder) which are gapped by $100 \pm 30$ meV at the $\Gamma$-point, see Fig.~\ref{fig:Schlappa2009_3a}. This region of the Brillouin zone is inaccessible to inelastic neutron scattering due to lack of intensity, although it is possible to see the single triplon gap with neutrons, which is located away from the $\Gamma$-point.

To investigate spin chains, Sr$_2$CuO$_3$ is an ideal compound. It is a corner-sharing spin chain system with short-range AFM order. The superexchange constant $J$ along the chain is extremely large ($\sim 200-250$ meV), so that excitations are easy to resolve in RIXS. Further, the inter-chain coupling is about an order of magnitude smaller, ensuring 1D behavior.

\textcite{Seo2006} measured the RIXS spectrum of Sr$_2$CuO$_3$ at the Cu $1s \rightarrow 3d$ quadrupole transition, with an energy resolution of $\sim 450$ meV. The authors observe a feature attributed to dd-excitations at an energy loss of $2$ eV. They claim that this feature disperses by $\sim 0.2$ eV with a periodicity of $\pi$ (half the lattice periodicity), and attribute this dispersion to two-spinon excitations on top of the dd-excitation.

\begin{figure}
   \begin{center}
   \includegraphics[width=.775\columnwidth]{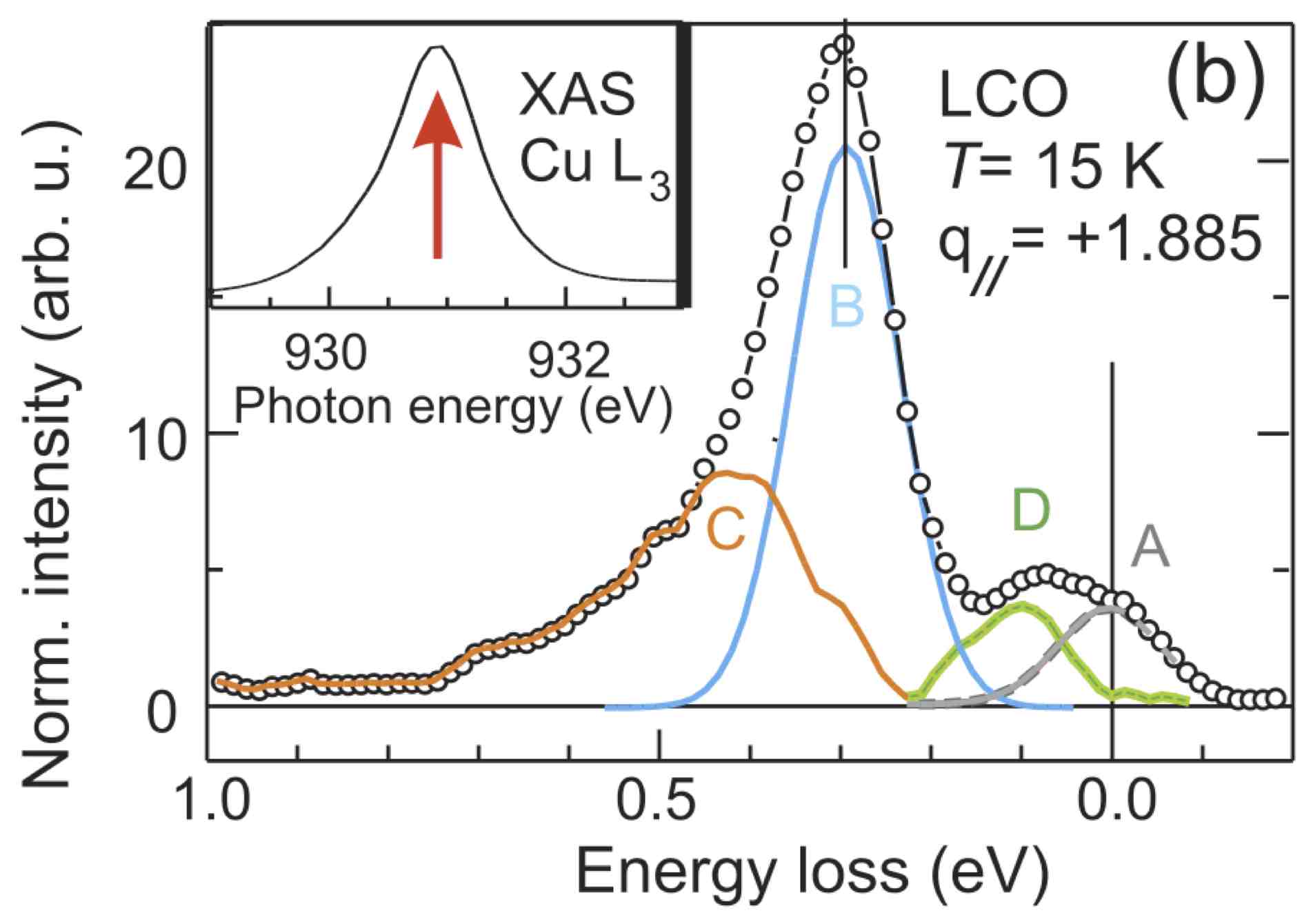}
   \end{center}
   \caption{Single magnon RIXS data at the copper $L_3$ edge of La$_2$CuO$_4$ with an energy resolution of $140$ meV. The transferred momentum (projected on the copper-oxide plane) is along the $(\pi,0)$ direction. The decomposition of the low-energy spectrum is shown, and it is found that the maximum follows the single magnon dispersion.  \cite{Braicovich2009b}. \label{fig:Braicovich2009b_1b}}
\end{figure}

\subsubsection{2D systems -- Magnons and Bimagnons}

Of the wide variety of (quasi) two-dimensional magnetic materials, so far only the two-dimensional antiferromagnetic cuprates have been studied experimentally with RIXS. 2D irdates have also been considered from a theoretical viewpoint~\cite{Ament2010b}. That the attention focusses on cuprates is not surprising, given their central role in high-$T_c$ superconductivity. The undoped cuprates are described well in terms of  2D Heisenberg antiferromagnets. When doped with either holes or electrons, the magnetic order melts and superconductivity appears. In undoped cuprates the elementary low-energy magnetic excitations are well separated from the gapped charge excitations, but upon doping these excitations mix and thus so will their RIXS response.

RIXS can create single and double spin-flip excitations, which are equivalent to single and bimagnon excitations in Heisenberg antiferromagnets. Both have been observed \cite{Hill2008,Braicovich2009b}, and are reviewed below. In addition also three-magnon scattering is predicted to contribute substantially to the magnetic spectral weight~\cite{Ament2010a}.

There are many more 2D materials of interest that so far have not been studied with RIXS, like frustrated magnets, doped cuprates, spin-orbit coupling dominated materials, etc. This is not only due to scarcity of beam-time, but also because for RIXS materials with large magnetic exchange integrals causing relatively high-energy magnetic excitations are preferable in view of the restricted energy resolution.

\paragraph{Single magnon.}

In many two-dimensional cuprates, single spin-flip excitations are allowed in direct RIXS, i.e., at the Cu $L$ and $M$ edges~\cite{Ament2009a}. For instance, La$_2$CuO$_4$, CaCuO$_2$, Sr$_2$CuO$_2$Cl$_2$, and Nd$_2$CuO$_4$ all have the copper spins (residing on the $3d_{x^2-y^2}$ orbital) oriented in the $xy$-plane. The above-mentioned materials are two-dimensional Heisenberg antiferromagnets, and in such systems a spin-flip is described in terms of the single magnon final states. Also in strong spin-orbit coupled iridates, e.g., Sr$_2$IrO$_4$, RIXS can probe the elementary magnetic excitations~\cite{Ament2010b}.

The direct RIXS cross section consists of the atomic spin-flip factor multiplied by the spin-flip susceptibility, which accounts for the collective response. The atomic factor depends on the polarization vectors and the spin orientation. \onlinecite{Haverkort2009} relates these atomic factors to fundamental x-ray absorption spectra. The susceptibility factor depends only on momentum transfer. In the antiferromagnetic cuprates, the latter factor peaks at the magnetic ordering vector ${\bf q} = (\pi,\pi)$, while going linearly to zero around the $\Gamma$-point.

\begin{figure}
   \begin{center}
   \includegraphics[width=.7\columnwidth]{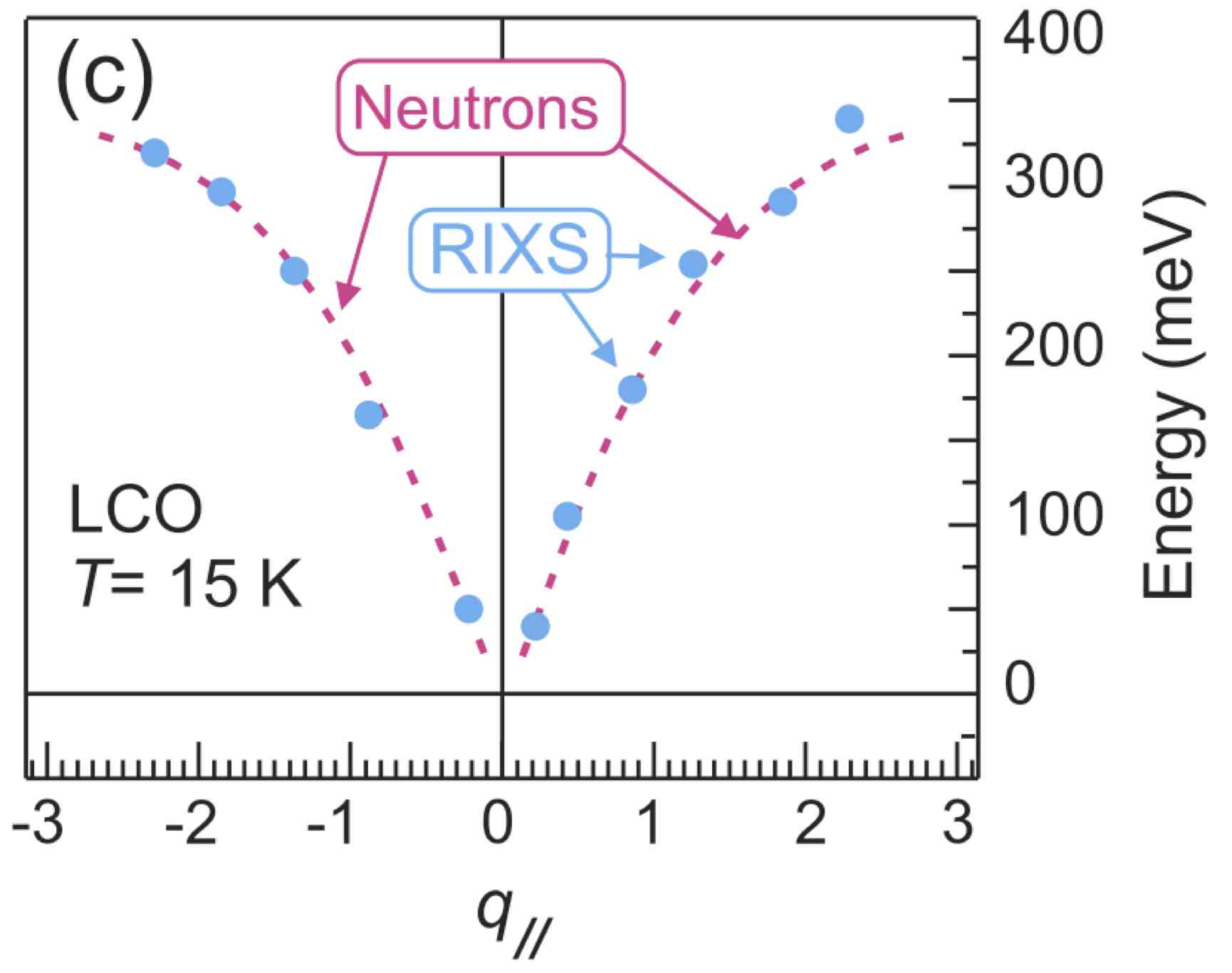}
   \end{center}
   \caption{Comparing single magnon neutron data and RIXS measurements at the copper $L_3$ edge of La$_2$CuO$_4$. Details as in Fig.~\ref{fig:Braicovich2009b_1b} \cite{Braicovich2009b}. \label{fig:Braicovich2009b_1c}}
\end{figure}

\subparagraph{Cuprates}
Single-magnon excitations were observed in RIXS experiments at the Cu $L_3$ edge on a thin film of La$_2$CuO$_4$ by \textcite{Braicovich2009b}. The authors found that the maximum of the magnetic RIXS spectrum (Fig.~\ref{fig:Braicovich2009b_1b}) follows the magnon dispersion measured with neutron scattering very closely, see Fig.~\ref{fig:Braicovich2009b_1c}. It is remarkable that RIXS allows the measurement of a magnon dispersion such a thin film sample with high accuracy.

\textcite{Guarise2010} measured the magnon dispersion in 2D antiferromagnetic insulator Sr$_2$CuCl$_2$O$_2$. The spin-waves display a large (70 meV) dispersion between the zone-boundary points ($\pi$,0) and ($\pi$/2,$\pi$/2). From an analysis of the dispersion in terms of $t$-$t'$-$t"$-$U$ Hubbard model, \textcite{Guarise2010} conclude there is a significant electronic hopping beyond nearest-neighbor Cu ions.

\textcite{Braicovich2009b} also studied the RIXS response of underdoped La$_{2-x}$Sr$_x$CuO$_4$ with a hole concentration of $x=0.08$. Due to the observed asymmetry in the cross section when changing ${\bf q}$ to $-{\bf q}$, the RIXS spectrum could be assigned predominantly to magnetic excitations, following theory~\cite{Ament2009a}. At low temperatures, two dispersing branches are observed: the high-energy one (not previously observed) follows the magnon dispersion of the undoped compound within $10 \%$, while the low-energy one coincides with neutron measurements of doped systems around $(\pi,\pi)$. The authors caution that ${\bf q} =(\pi,\pi)$ no longer has to be equivalent to $(0,0)$ since there is no long range magnetic order at this doping. The two branches merge at high temperature. \textcite{Braicovich2009b} suggest that the two branches signal phase separation on a microscopic scale. These branches are visible down to a transfered momentum of $0.44$ \AA$^{-1}$, and the authors argue that sustaining coherent excitations of such momentum, requires magnetic patches much larger than the corresponding length scale of $15$ \AA. The patch size has also been estimated with other techniques that have a much longer measuring time scale. These measurements find a smaller patch size, leading the authors to conclude that the underdoped compound is dynamically phase separated.

\begin{figure}
   \begin{center}
   \includegraphics[width=\columnwidth]{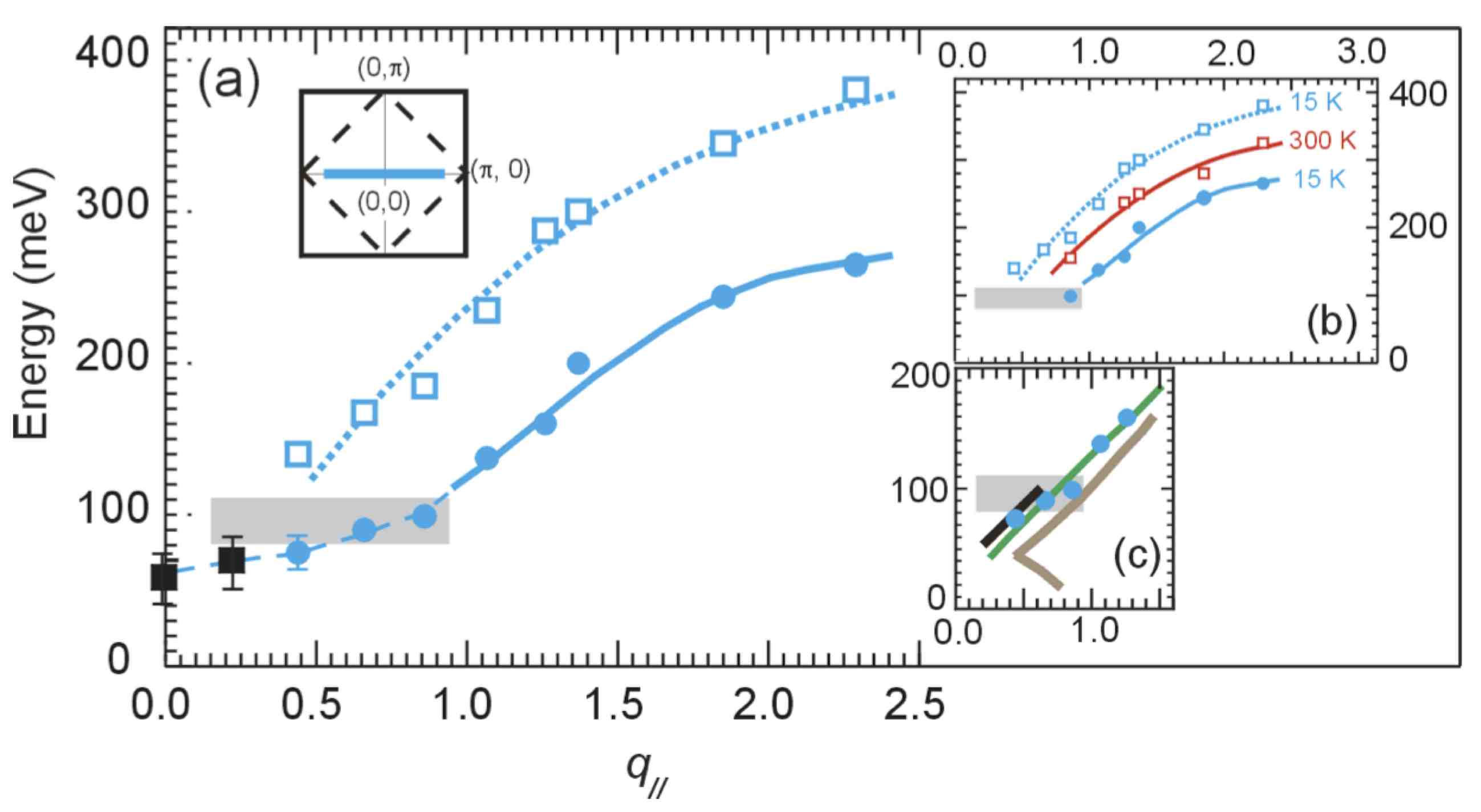}
   \end{center}
   \caption{The RIXS spectra of the underdoped cuprate La$_{1.92}$Sr$_{0.08}$CuO$_4$ shows two dispersing branches at low temperature (a), which merge at room temperature (b). The green line in Fig.~(c) displays the corresponding neutron data. The data were taken at the copper $L_3$ edge with an energy resolution of $140$ meV. The transfered momentum (projected on the copper-oxide plane) is along the $(\pi,0)$ direction \cite{Braicovich2009b}. \label{fig:Braicovich2009b_4}}
\end{figure}

\begin{figure}
   \centering
   \includegraphics[width=\columnwidth]{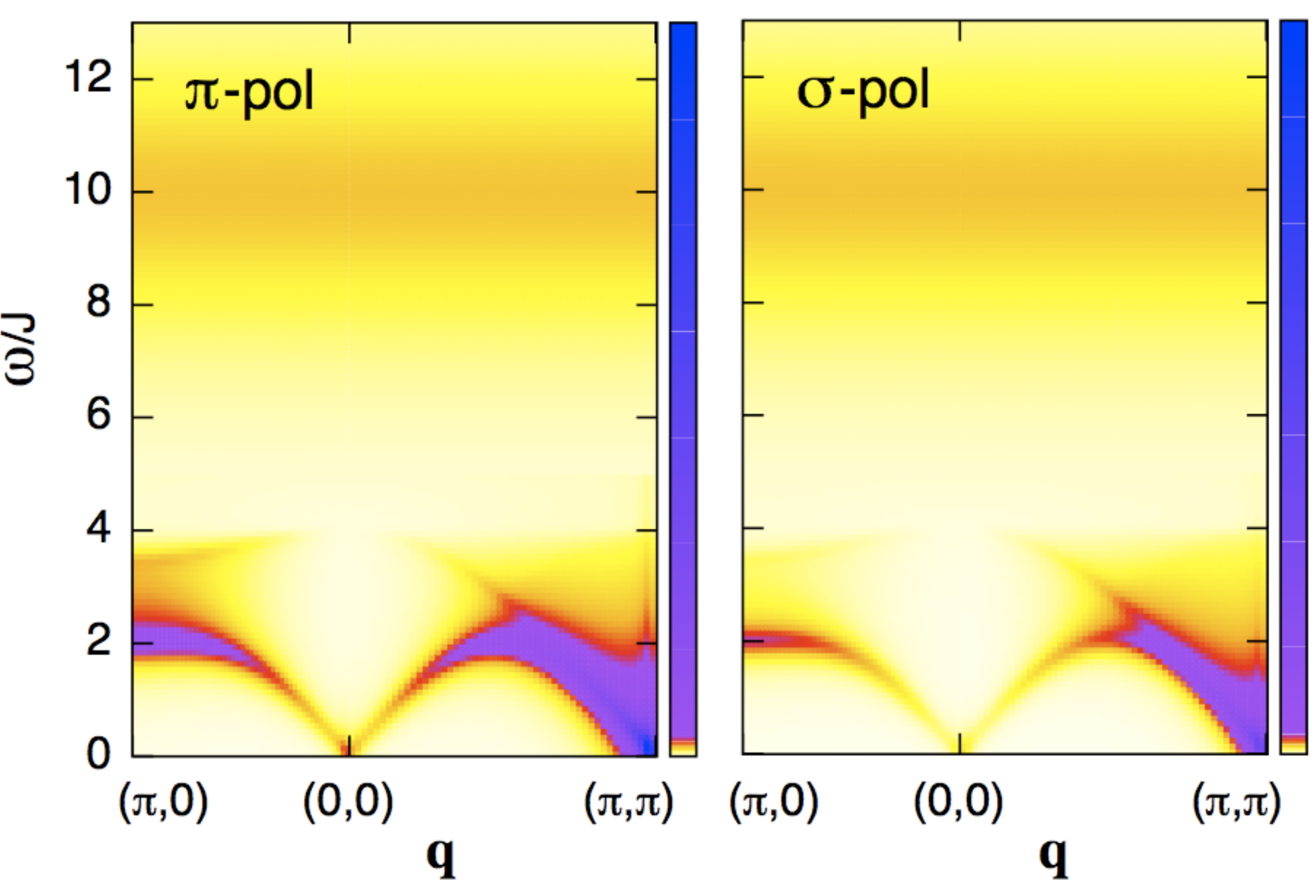}
   \caption{Theoretical RIXS spectrum of Sr$_2$IrO$_4$ at the Ir $L$-edge for $\pi$ and $\sigma$ polarization in $90^\circ$ scattering geometry, and where ${\bf q}$ is in the first Brillouin zone. The broad peak at $10J$ is due to transitions between the spin-orbital multiplets. From this peak one can determine the spin-orbit coupling $\lambda$. At low energy, strongly dispersive single- and double-magnon features are seen~\cite{Ament2010b}. \label{fig:Ament2010b_1}}
\end{figure}

\subparagraph{Iridates}
The relativistic spin-orbit coupling in Ir$^{4+}$ dominates the spin-orbital dynamics, and the ionic ground state is a Kramers doublet which can be described by a pseudo-spin $S=1/2$. In Sr$_2$IrO$_4$, the $t_{2g}$ analog of La$_2$CuO$_4$, these spin-1/2 moments couple by superexchange and yield an antiferromagnetic Heisenberg model. \textcite{Ament2010b} have calculated the Ir $L$-edge RIXS cross section for Ir$^{4+}$ ions in an octahedral field. They have theoretically shown for Sr$_2$IrO$_4$ that, next to transitions between spin-orbit multiplets, RIXS can also probe the magnon spectrum, as shown in Fig.~\ref{fig:Ament2010b_1}. At the $L_2$-edge, no intensity is expected, in contrast to the $L_3$-edge.

\paragraph{Bi-magnon.}

For direct RIXS the projected spin angular momentum {\bf of the material} may be changed by $1$ or $2$ units by a transfer of angular momentum of the photon to the electron spins in the system. For indirect RIXS such a direct transfer of angular momentum is not possible and the spin angular momentum of the material cannot be changed in the scattering process. Therefore, the leading order allowed magnetic excitations that result from indirect processes are bimagnon excitations. These have been observed with RIXS in numerous two-dimensional cuprates and are theoretically well understood.

\begin{figure}
   \begin{center}
   \includegraphics[width=0.8\columnwidth]{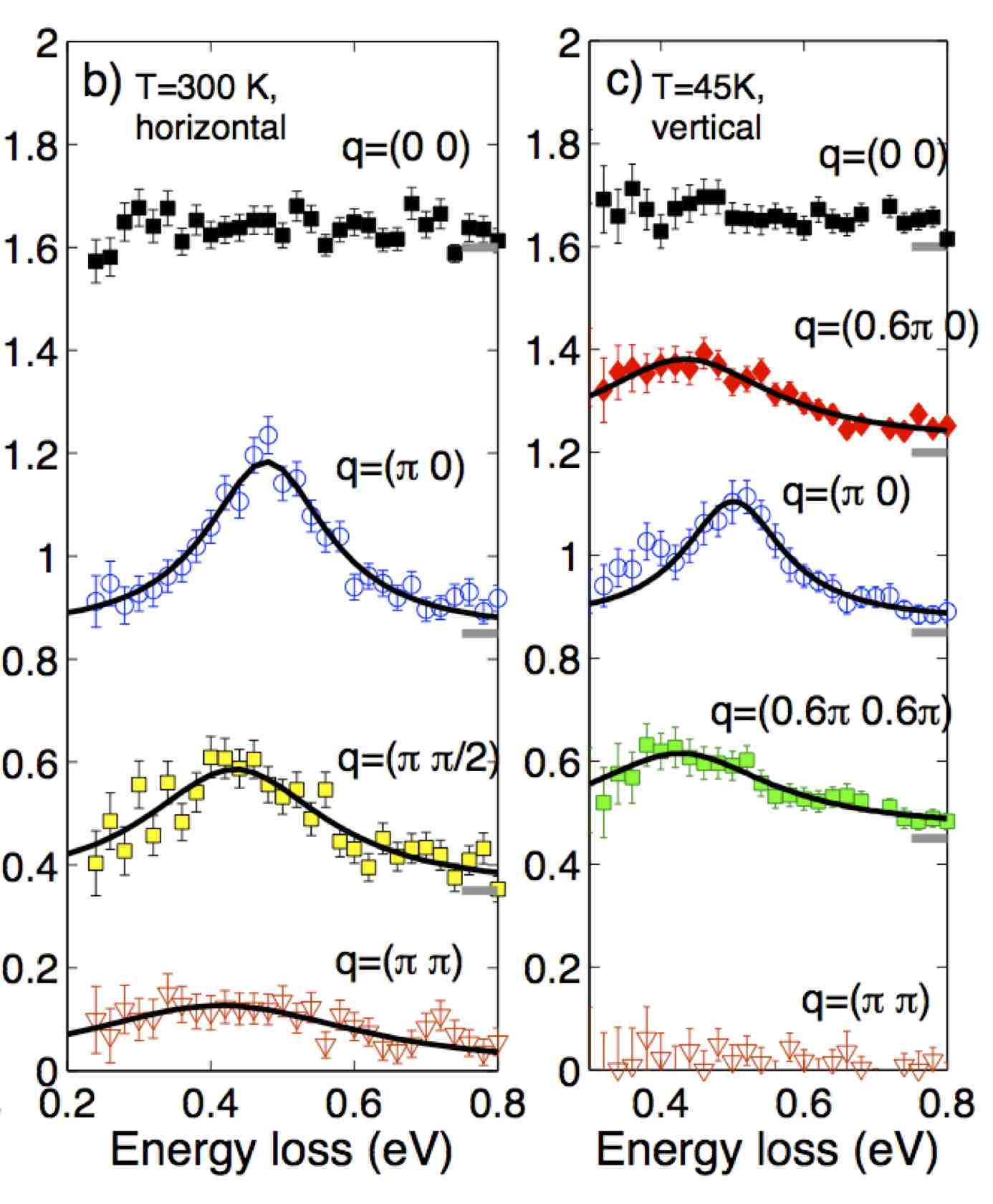}
   \end{center}
   \caption{RIXS spectra at the copper $1s \rightarrow 4p$ edge for a few points in the two-dimensional Brillouin zone of La$_2$CuO$_4$ \cite{Hill2008,Ellis2010}. Fig.~(b) is taken with $\pi$-polarization at room temperature, (c) with $\sigma$-polarization at $45$ K. The excitation energy is $8994$ eV. The solid lines are Lorentzians fitted to the data. The elastic peak has been subtracted from these data.\label{fig:Ellis2010_2bc}}
\end{figure}

The first experimental observation of the momentum dependence of bimagnon excitations in the cuprates, at the copper $K$ edge, was performed by \textcite{Hill2008}. The measurements were done at $T= 20$ K, and with an energy resolution of $120$ meV. The authors measured indirect RIXS spectra at the Cu $1s \rightarrow 4p_{\pi}$ edge on La$_2$CuO$_4$ and Nd$_2$CuO$_4$, and both materials showed a peak at momentum transfer ${\bf q} = (\pi,0)$ around an energy loss of $500$ meV, which they interpreted as a bimagnon excitation. This resonant feature disappeared quickly with hole doping, similar to bimagnon Raman scattering. Further, when the polarization was changed from parallel to the $z$-axis to perpendicular to it, the feature vanished. The authors measured at five points in the Brillouin zone, and found zero intensity at ${\bf q} = (0,0)$ and $(\pi,\pi)$, while the bimagnon DOS is highest at these points. The RIXS intensity peaked at ${\bf q} = (\pi,0)$, see Fig.~\ref{fig:Ellis2010_2bc}.

In a follow-up study, the energy and intensity of the ${\bf q}=(\pi,0)$ peak in La$_2$CuO$_4$ at the Cu $K$-edge are tracked as a function of temperature and doping, providing evidence that the peak is a two-magnon excitation \cite{Ellis2010}. Data was taken at three beamlines. At the APS 30ID beamline, the incident photons were $\pi$-polarized and the scattering geometry $\sim 90^{\circ}$. A combined energy resolution of $120$ meV (FWHM) was achieved. At the APS 9ID beamline, the incident photons were $\sigma$-polarized and the combined energy resolution $130$ meV (FWHM). Measurements at SPring-8's BL11XU beamline were done in the same geometry as at the 30ID beamline; the resolution was $440$ meV. The ${\bf q}$-dependence of the peak as found by \textcite{Hill2008} is confirmed, see Fig.~\ref{fig:Ellis2010_2bc}. The authors find that the intensity of the peak decreases with temperature, vanishing only at $T= 500$ K, far above the N\'eel temperature $T_N = 320$ K. This behavior is similar to Raman data on various cuprates, and also consistent with neutron measurements. The width of the peak was estimated to be $\sim 160$ meV, although it could not be established firmly because of the energy resolution of $120-130$ meV. The peak shift as a function of temperature is identical to Raman data on other cuprates when the temperature is renormalized to the N\'eel temperature. The authors argue that this is evidence that magnon-magnon interactions are the source of this shift. The decreasing intensity on the other hand is attributed to phonons. Doping with Sr also causes the $500$ meV peak to shift to lower energy and lose intensity rapidly following a phenomenological formula. One problem in attributing the peak to two-magnon excitations is its energy: $500$ meV where interacting magnon theory predicts $400$ meV. \textcite{Ellis2010} suggest a strongly momentum-dependent magnon-magnon interaction to explain this fact: it should be strong in the Raman case of ${\bf q} = (0,0)$ and weaker at $(\pi,0)$.

\begin{figure}
   \begin{center}
   \includegraphics[width=0.6\columnwidth]{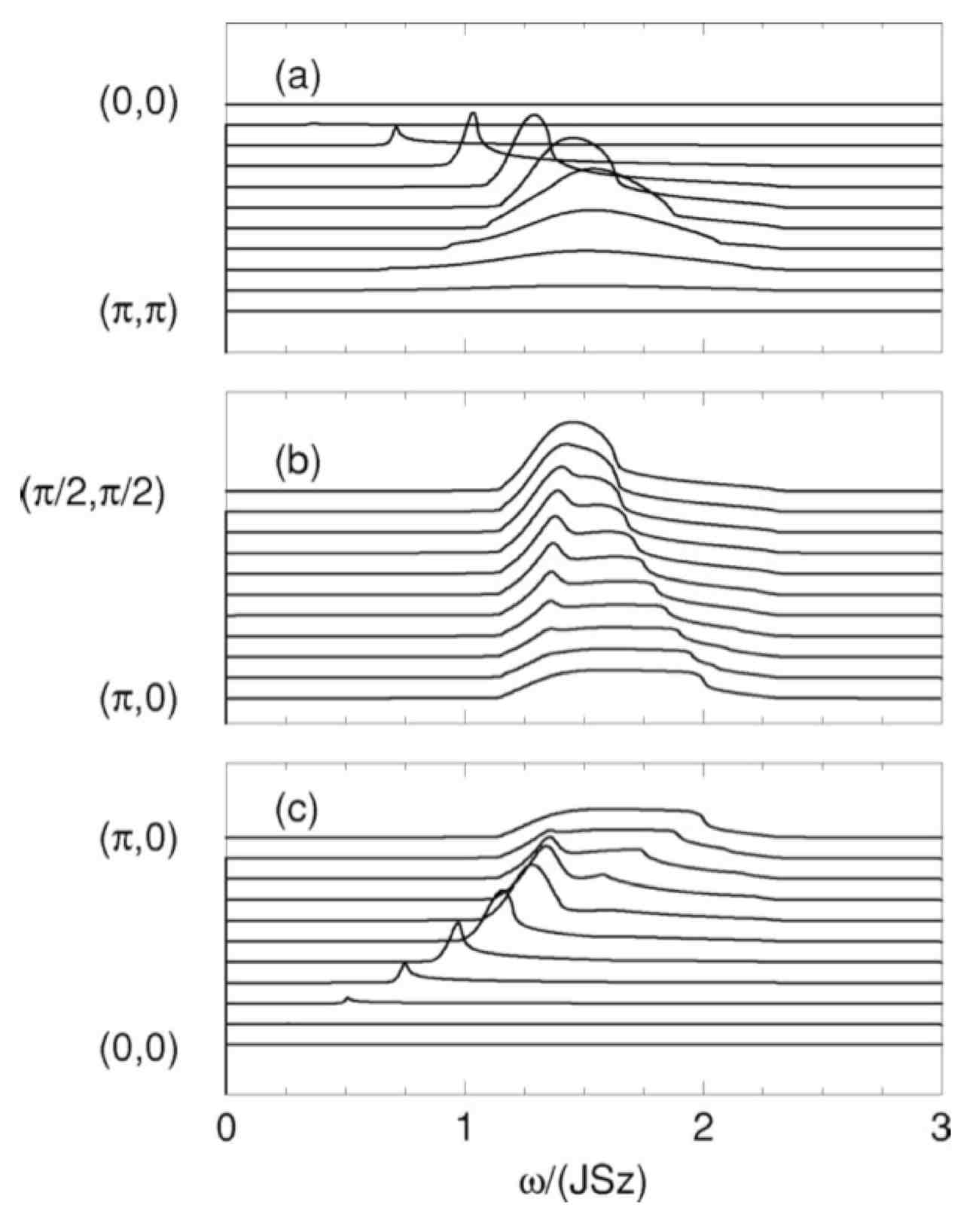}
   \end{center}
   \caption{Bimagnon RIXS spectra obtained with interacting spin wave theory for a nearest-neighbor $S=1/2$ Heisenberg model \cite{Nagao2007}. The vertical axis indicates transferred momentum projected to the CuO$_2$ plane and $\omega$ is the energy loss. For the $2D$ square lattice, $z = 4$.\label{fig:bimagnon}}
\end{figure}

From the theoretical side, it was noticed already very early on that excitations related to spin degrees of freedom, such as two-magnon Raman scattering, are present in RIXS spectra computed for small Hubbard Hamiltonian clusters for  RIXS on the $K$-edge of copper~\cite{Tsutsui1999} and oxygen~\cite{Harada2002}. \textcite{Brink2005b,Brink2007} proposed the microscopic mechanism by which the bimagnons couple to the intermediate state core-hole in RIXS --  by locally modifying the superexchange constant $J$ (see Sec.~\ref{sec:magnetic_coupling}). This approach yielded a momentum dependence of the cross section in agreement with experiment and in particular reproduced the lack of intensity at ${\bf q} = (0,0)$ and $(\pi,\pi)$. The magnetic RIXS response within this framework including magnon-magnon interactions was computed by~\textcite{Nagao2007} and is shown in Fig.~\ref{fig:bimagnon}. Magnon-magnon interactions are observed to broaden the loss features and and at the same time enhance lower-energy spectral weight with respect to the non-interacting theory of~\textcite{Brink2005b,Brink2007}.

Later theoretical work took the approach of extending bimagnon Raman scattering theory \cite{Elliott1963,Fleury1968} to finite momentum transfer \cite{Vernay2007,Donkov2007}. This approach is in essence non-resonant because it omits the core-hole (and its effect on the valence electrons) from the analysis.

\textcite{Vernay2007} consider a single-band, half-filled Hubbard model, and assume that the incoming x-ray photon excites a valence electron to a neighboring site, making it a doubly occupied one in the intermediate state. In the final state, one of the two electrons on the doubly occupied site has hopped back, emitting the outgoing photon. They classify spectra according to polarization geometries and find that the fully symmetric component transforming as $A_{1g}$ vanishes at ${\bf q} = (0,0)$ and both $A_{1g}$ and $B_{1g}$ geometries vanish at $(\pi,\pi)$. For $B_{1g}$ geometries they identified how the bi-magnon intensity disperses from the zone center (Raman) at $\sim 2.8J$ to $4J$ at $(\pi,0)$ due to the reduction of magnon-magnon interactions with increasing $q$, while the $A_{1g}$ bi-magnon intensity is strongest at $(\pi,0)$ appearing at $4J$.

\textcite{Donkov2007} take a similar approach. Starting from a half-filled single-band Hubbard model, they make a mean-field approximation and introduce an antiferromagnetic order parameter, obtaining an upper and lower Hubbard band. They also assume that, instead of making core transitions, the X-rays excite electrons from the lower Hubbard band to the upper one. These electron-hole excitations couple to magnons via the full Hubbard $U$ interaction, which was initially treated on a mean-field level. The authors find that in the $B_{1g}$ experimental geometry the intensity peaks either at ${\bf q} = (0,0)$ for small magnon-magnon interactions or at $(\pi,0)$ for large interactions. The bimagnon peak disperses to lower energies away from ${\bf q} = (0,0)$. For the $A_{1g}$ geometry, both the intensity and dispersion go up away from ${\bf q} = (0,0)$. The intensity peaks at $(\pi,0)$ and is zero at $(0,0)$. For both geometries, the authors predict a finite intensity at ${\bf q} = (\pi,\pi)$.
    
\textcite{Forte2008a} calculated the RIXS bimagnon scattering via the core-hole at finite temperatures, considering in addition longer-range superexchange interactions, comparing the theory in detail with the $K$-edge experiments of~\textcite{Hill2008}. The authors employed the Ultra-short Core-hole Life-time (UCL) expansion in the framework in which the bimagnons are created by the core-hole~\cite{Brink2007} and also calculated the sub-leading order of the UCL expansion, which leads to finite bimagnon spectral weight at ${\bf q} = (0,0)$.

At the Cu $L$ and $M$ edges, the coupling mechanism to bimagnon excitations is different, as clarified in Sec.~\ref{sec:magnetic_coupling}, but the resulting cross section is the same to first order in the UCL expansion, see Fig.~\ref{fig:bimagnon}. This comes about because, instead of modifying the superexchange constant to some new value $J'$, superexchange is locally blocked at the Cu $L$ and $M$ edges: $J' = 0$ \cite{Braicovich2009a}.

Measuring at many different transferred momenta along the ${\bf q} = (0,0) - (\pi,0)$ direction, \cite{Braicovich2009a} report bimagnon excitations at the Cu $L_3$ edge of La$_2$CuO$_4$ and CaCuO$_2$. At $T = 30$ K and at room temperature (which are both below the N\'eel temperature), they see a feature dispersing from $0$ meV at ${\bf q} = (0,0)$ to approximately $400$ meV at ${\bf q} = (2\pi/3,0)$ in both $\sigma$ and $\pi$ polarization. The authors assign it to a bimagnon excitation, which disagrees with theoretical work that predicts that a strong single magnon peak should be present as well \cite{Ament2009a}. In following experimental work, where the energy resolution was improved from $400$ meV to $140$ meV, the authors reinterpret the dispersing feature as a single magnon \cite{Braicovich2009b}. The remaining spectral weight at slightly higher energies is now interpreted as bimagnon excitations.

The effect of chemical pressure on 2D antiferromagnetic cuprates is studied by comparing the Cu $L_3$ RIXS spectra of CaCuO$_2$ and BaCuO$_2$ \cite{Bisogni2009}. By substituting Ba for Ca, the in-plane lattice parameter increases from $3.84 \pm 0.01$ \AA \ to $3.96 \pm 0.01$ \AA . This causes the hopping integrals to decrease, and the authors observe that, correspondingly, the superexchange constant in BaCuO$_2$ is $0.67 \pm 0.08$ times the CaCuO$_2$ value. They extract this ratio by comparing the position of the magnetic peak in the two compounds, which scales linearly with $J$. Putting $J$ at $120$ meV in CaCuO$_2$, they obtain a superexchange constant of $85 \pm 10$ meV for BaCuO$_2$. The spectra are normalized so as to keep the integrated weight of the dd-excitations the same in both compounds. With such a normalization, the spectral weight of the magnetic peak in the Ba compound is found to be to less than half the Ca value. This is in agreement with what one would expect for bimagnons, since the bimagnon spectral weight scales as $J^2$. The magnetic peak was therefore interpreted by the authors as a bimagnon peak, although they later changed this interpretation to single magnon \cite{Braicovich2009b}.

At the Cu $M$ edge, experiments are more difficult because of the large elastic peak, even though the experimental resolution is usually higher than at other edges. RIXS spectra taken at the Cu $M_2$ edge of CaCuO$_2$ with a resolution of $140$ meV show two low-energy features \cite{Freelon2008}. In the data $dd$-excitations are suppressed and it is uncertain what causes this effect. The authors observe a loss feature  at $390$ meV when the incoming polarization is parallel to the $c$-axis. This being a single magnon excitation can be excluded as at the $M$ edge ${\bf q} \approx {\bf 0}$ and for these momenta the single magnon energy vanishes. The authors interpret it as a bimagnon excitation.

The bimagnon excitation can in principle also be observed at the oxygen $K$ edge ($1s \rightarrow 2p$ transition). Although the cross section for the oxygen $K$ edge is relatively small, this case is interesting because the single magnon excitation is forbidden here: there is no spin-orbit coupling for $1s$ core orbitals. In this sense the situation is similar to the copper $K$ edge. \textcite{Harada2002} investigated Sr$_2$CuO$_2$Cl$_2$ at the oxygen $K$ edge with an energy resolution of $0.5$ eV. The transferred momentum is fixed by the $90^{\circ}$ scattering geometry and near-normal incidence of the x-rays.  For scattering with $\sigma$ polarization a feature is visible around $0.5$ eV which the authors interpret as a as a two-magnon excitation. A calculation for a Cu$_4$O$_{13}$ cluster reproduces the feature at 0.5 eV and its polarization dependence, but underestimates the intensity, possibly due to finite size effects.

\subsubsection{3D systems -- Magnons and Bimagnons \label{sec:magnetic3D}}

A number of 3D magnetic Mott insulators have been studied with RIXS: in particular NiO, NiCl$_2$ and CoO. The prototypical strongly correlated antiferromagnet NiO has been investigated most extensively, and we therefore start by briefly summarizing its magnetic properties.

NiO has the same crystal structure as NaCl, with Ni$^{2+}$ ions (3$d^8$ configuration) centered in an octahedron of oxygen ions, generating a crystal field of that same symmetry. In the ground state the $t_{2g}$ levels of Ni$^{2+}$ are fully occupied, with the remaining two electrons occupy the $e_g$ orbitals, with parallel spins to that the Ni$^{2+}$ ion has an $S=1$ ground state. The spin points along the $\langle1 1 $-$2\rangle$ direction. Superexchange interactions mediated by the oxygen ions give rise to four (weakly coupled) AFM subsystems, ordering at $T_N = 523$ K and described at low temperature by an $S=1$ Heisenberg model on a cubic lattice. Magnons are the elementary excitations of this Heisenberg antiferromagnet. The magnon density of states is strongly peaked around $6J$, where $6J= 114$ meV in NiO \cite{Hutchings1972}.

Atomic calculations for a Ni$^{2+}$ ion in an octahedral crystal field by \textcite{Groot1998} show that at the Ni L-edge $\Delta S_z = 1,2$ transitions are possible. 
These RIXS final states correspond to single ($\Delta S_z = 1$) and bimagnon ($\Delta S_z = 2$) excitations. 

Final states with $\Delta S_z = 2$ are interesting because they cannot be reached with neutron scattering, optical Raman scattering or indirect RIXS. In addition to the $\Delta S_z = 2$ bimagnons, also $\Delta S_z = 0$ bimagnons can be excited in NiO via the indirect RIXS process. In this case the bimagnon scattering comes about by the core-hole in the intermediate state locally disrupting the magnetic bonds, as discussed in Section \ref{sec:magneticindirect} for the cuprates.

{\it Besides the mentioned} low-energy magnon excitations -- in which only the $z$-projections of the local spin moments changes on the Ni sites, RIXS can also cause local spin transitions, i.e. a transition from a high-spin triplet ($S=1$) state of Ni$^{2+}$ to a low spin singlet $S=0$ state. Such spin state transitions are at much higher energy than the magnons as they are at an energy scale set by the atomic Hund's rule coupling $J_H \approx 1.3$ eV.

\paragraph{(Bi)-Magnons.}

The first measurement of magnetic excitations in NiO with RIXS was performed on the nickel $M$ edge by \textcite{Chiuzbaian2005}. Because of the overwhelming elastic peak, the observation of pure spin-flip transitions was impossible with their energy resolution of $130$ meV. 

In follow-up work on NiO, it proved to be possible to observe two purely magnetic excitations at $\sim 95$ and $190$ meV \cite{Ghiringhelli2009a}. These experiments had nearly the same energy resolution ($120$ meV), but profited from the fact that at the Ni $L_3$ edge the elastic peak is much smaller than at the $M$ edge.These excitations were not observed to disperse.

To establish that the peaks are of magnetic nature, the authors also measured NiCl$_2$. Like NiO, NiCl$_2$ also has Ni$^{2+}$ ions in a (slightly weaker) octahedral crystal field of Cl$^-$ ions. The superexchange constant is an order of magnitude smaller than in NiO, and therefore any magnetic peak will be buried in the elastic line, while crystal field excitations should remain almost unchanged. The two low-energy peaks seen in NiO are absent in the NiCl$_2$ spectra, indicating that they are indeed magnetic excitations. 

The authors model the RIXS spectrum using a single Ni$^{2+}$ ion in an effective exchange field. The resulting spin-flip transitions take the $S=1$ ground state into an $S=1$ final state, where the projection on the magnetization axis $\alpha$ can change with $\Delta S_{\alpha} = 1,2$. This local model describes the data reasonably well, although the single spin-flip peak deviates from the calculations. 

Measuring the dispersion of magnon and bimagnon excitations in a 3D system is challenging, but a dispersion should be present just as it is in the magnetic 1 and 2D cuprates. The single magnon dispersion is expected to be similar to the one observed with neutron scattering. The bimagnon signal (with $\Delta S_z = 2$) is found to hardly disperse in a model calculation on a 3D spin one nearest-neighbor Heisenberg model on a cubic lattice. Rather it peaks around $12J$, following the bimagnon DOS \cite{Haverkort2009}. 

Of the early transition metal oxides, RIXS spectra have been published of the 3D antiferromagnet LaTiO$_3$ and ferromagnet YTiO$_3$ with $T_N = 146$ K and $T_C = 27$ K respectively \cite{Ulrich2009}. In both compounds, the Ti$^+$ ions are in the $3d^1$ configuration: the spin $S=1/2$. Spectra with an energy resolution of $55$ meV were taken at the Ti $L_3$ edge at room temperature and at $13$ K, i.e., above and below the magnetic ordering temperatures. The authors point out that in the magnetically ordered state, the low-energy region is dominated by a single magnon peak at approximately $40$ meV for LaTiO$_3$ and $16$ meV for YTiO$_3$. The energies agree with neutron scattering measurements. This magnetic peak could only be distinguished from the elastic line in $90^{\circ}$ scattering geometry with $\pi$-polarization. At room temperature, where both compounds are paramagnetic, the peaks shift to $22$ meV for LaTiO$_3$ and zero for YTiO$_3$.

\paragraph{Local Spin Transitions.}

NiO has also been studied at the O $K$ edge, where direct RIXS spin-flips are forbidden because of the absence of spin-orbit coupling in the O $1s$ core orbital. \textcite{Duda2006} measure an excitation around $1.9$ eV, in $\sigma$-polarization only. The authors interpret this feature as a `double singlet' excitation on two neighboring Ni ions: $\ket{\uparrow\uparrow , \downarrow\downarrow} \mapsto \ket{\uparrow\downarrow , \downarrow\uparrow}$, leaving both Ni ions in an $S=0$ state. The energy of this excitation -- twice Hund's rule coupling $J_H \approx 1.3$ eV -- should be offset by $0.7$ eV to account for the hybridization energy. Note that this energy scale is much higher than the $\sim 190$ meV for the bimagnon excitation seen at the Ni $L_3$ edge \cite{Ghiringhelli2009a}. The polarization dependence of the double singlet excitation seen in experiment agrees with calculations on a Ni$_6$O$_{19}$ cluster in which the central O ion is excited by an X-ray photon. This model yields a double singlet only in $\sigma$-polarization, but not in $\pi$-polarization.

\begin{figure}
  \begin{center}
    \includegraphics[width=0.5\columnwidth]{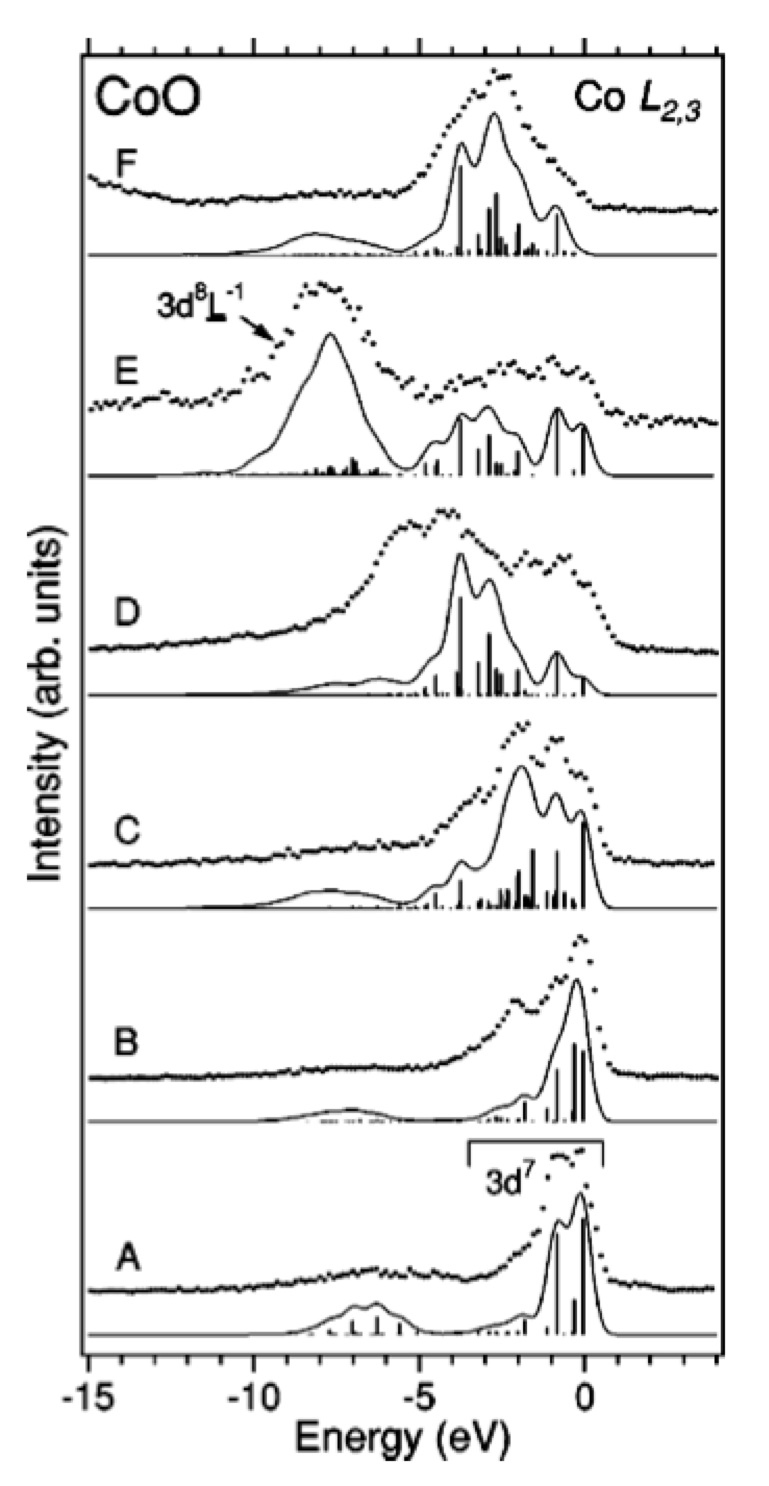}
  \end{center}
  \caption{Experimental (dots) and theoretical (solid line) RIXS spectra of CoO at the Co $L$ edges \cite{Magnuson2002}. The theoretical curve is a broadening of the vertical bars. The excitation energy runs from $778.1$ eV (A) to $785.2$ eV (E) at the $L_3$ edge, and $795.0$ eV (F) is at the $L_2$ edge. The work of \textcite{Chiuzbaian2008} includes similar spectra with a slightly better combined energy resolution. \label{fig:magnuson2002}}
\end{figure}

Going one step back in the periodic table, \textcite{Magnuson2002} and \textcite{Chiuzbaian2008} studied CoO. The Co$^{2+}$ ions are in a $3d^7$ configuration and in the ground state assume spin $S=3/2$. A cubic crystal field ($10Dq \approx 0.5$ eV) combined with Hund's rule selects the configuration $t_{2g}^5\ e_g^2$, and the orbital ground state has $T_{1g}$ symmetry with substantial charge-transfer character. The spectrum consist of several spin-quartets and -doublets within $\sim 3$ eV \cite{Magnuson2002}.

\textcite{Chiuzbaian2008} measured Co $M_{2,3}$ and Co $L_3$ RIXS, while \textcite{Magnuson2002} measured the Co $L_3$ edge only, see Fig.~\ref{fig:magnuson2002}. The lowest energy excited states (below $\sim 3$ eV) are dominated by crystal field states derived from atomic spin-quartet states, while above $\sim 3$ eV, states derived from spin-doublets dominate the spectrum \cite{Chiuzbaian2008}. On the theory side, both authors simulate the RIXS and XAS spectra within a single impurity Anderson model including multiplet effects. \textcite{Chiuzbaian2008} performed a calculation on a CoO$_6$ cluster including $3d$ spin-orbit coupling but without exchange field. Including exchange ($9J = 0.3$ eV) improves agreement between theoretical and experimental RIXS spectra, as shown by \textcite{Magnuson2002}, but reduces the agreement for the Co $L_3$ XAS spectra. All experiments were performed at room temperature, which is far above the N\'eel temperature $T_N \approx 288$ K.

\section{Summary and outlook \label{Summary}}
In this review, we have surveyed Resonant Inelastic X-ray Scattering with a focus on the creation of elementary excitations. Recent years have seen a tremendous improvement in instrumentation for the sources, for the beamlines and for the spectrometers. As a result, there has been a tremendous advance in the technique. Whereas early work primarily dealt with excitations on the scale of electron volts, such as $dd$ and charge-transfer excitations, more recent experiments have resolved low-energy features such as magnons.

It would be interesting to see if this can be pushed down to electronic pairing gaps, such as the superconducting and charge-density wave gap, or more exotic ones, such as pseudogaps in cuprates. RIXS provides an alternative and complementary look at the charge dynamics of a material. In the coming years, the experimental RIXS work in the $L$ and $M$ edge regions likely will be focused on establishing itself as a viable technique to measure elementary excitations at low energies, and investigate in detail the momentum dependent elementary charge response, {\em viz} magnons, orbitons and phonons, of correlated and uncorrelated materials. At the $K$ edge, the majority of the work has been performed on cuprates and this research is likely to continue. However, more-detailed extensions to other transition-metal compounds, such as manganites, pnictides, and ferroelectric materials, can undoubtedly be expected. Furthermore, the work on other strongly correlated systems, such as the edges of the rare-earths, is currently extremely limited and is expected to grown. In addition, with the advent of new synchrotron sources, there will be a push to improving the experimental resolution towards the 10 meV level. As has been demonstrated many times to date, each improvement in resolution brings with it the discovery of new details in the excitation spectrum of these materials. In addition to searching for electronic gaps, mentioned above, it would also be of particular interest to challenge theory to reproduce the finest of details seen in these response functions with this improved resolution.

Apart from bulk samples, it would also be interesting to make a comparison of lower-dimensional systems, such as thin films, interfaces, and nanostructured materials, which have not been covered in this review. Each of these are exciting areas in which new excitations and new physics is to be anticipated. RIXS is likely the only tool that can probe these systems over the relevant energy and momentum scales. Another fruitful area of growth for the technique will be in the study of materials where other spectroscopies may not be able to provide information on elementary valence excitations. Examples here include complex materials where the chemical selectivity aids in the analysis, or materials in extreme environments, such as high pressure or strong electric/magnetic fields, that are difficult to access with techniques such as neutron scattering and photoemission.

Concurrent with the experimental advances, theory has made great progress in analyzing the RIXS cross section.
Determining the charge dynamics in strongly correlated systems is a deep and complex problem. In elucidating this problem, it is very helpful to have effective methods that establish the correlation function that RIXS measures under certain conditions, and its relation to other spectroscopic techniques. Additional theoretical methods can aid in providing a deeper understanding of the charge dynamics which is relevant in understanding the optical and conductive properties of materials. Here advancements in the theory would include techniques which do not suffer from finite size effects, are able to address many intertwined degrees of freedom, and have as few approximations as possible.

The core-hole present in RIXS represents in many cases a strong local perturbation to the electronic system. As a result, dynamical mean-field theory (DMFT) may be an important technique to apply to this scattering process as it can treat local dynamical many-body effects exactly. DMFT (and its cluster extensions) is well suited to include both local and non-local electronic correlations, and for certain model Hamiltonians this can be done exactly. As a result the multi-particle correlation functions, including multi-particle vertex renormalizations, can be formulated to calculate various spectroscopies, such as RIXS, as well as x-ray absorption and emission, and core level photoelectron spectroscopies. Progress has been made for non-resonant inelastic x-ray scattering across a metal-insulator transition~\cite{Devereaux2003a,Devereaux2003b,Freericks2003} and in a CDW phase \cite{Matveev2009}.
Moreover recent extensions to treat resonant Raman scattering with the Hubbard bands~\cite{Shvaika2004,Shvaika2005}
point towards a future applicability for RIXS.
One of the key ingredients will be the development of a formalism within DMFT to calculate the core-hole propagator. It is to be hoped that such development lies in the near future.

\begin{figure}
\begin{center}
\includegraphics[width=\columnwidth]{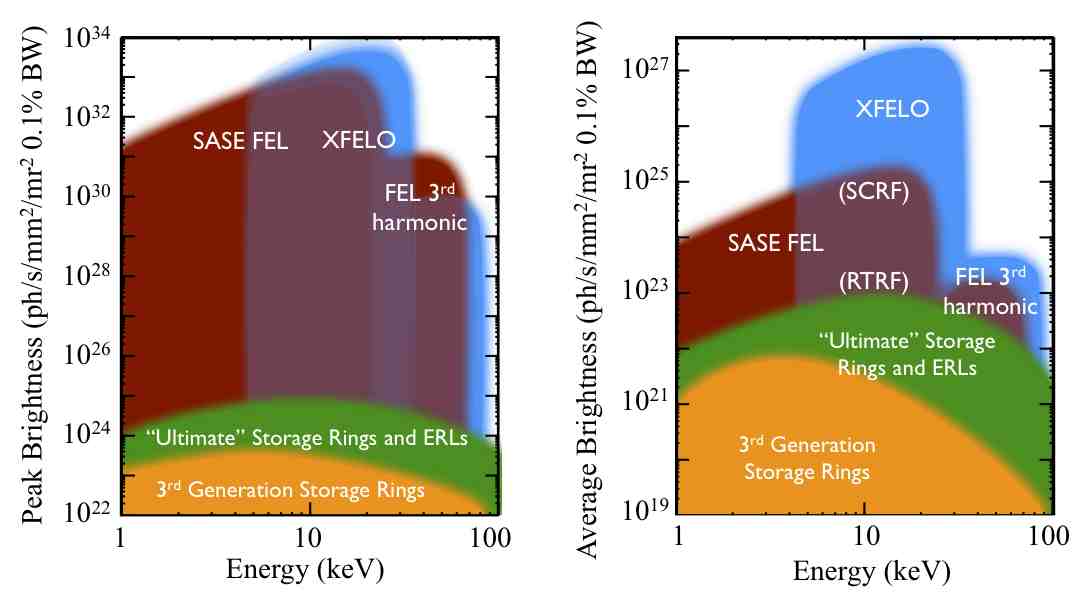}
\end{center}
\caption{Left: Peak brightness and Right: Average brightness of
  present and future x-ray sources.  Present third generation
  synchrotron light sources average brightnesses are in the $10^{20}$
  $phot/s/0.1\%BW/mm^2/mrad^2$ range. A brilliance enhancement by two
  orders of magnitude is envisaged for {\it ultimate} x-ray storage
  ring-based light sources and energy recovery linacs (ERLs). Free
  electron lasers (FELs) based on self-amplification of spontaneous
  emission (SASE) are pulsed and can enhance the peak brightness by
  $\sim 8$ orders of magnitude~(see \textcite{Saldin2000} and
  references therein). Feasibility studies show that X-Ray FEL
  Oscillator can greatly enhance the {\it average} x-ray
  brightness~\cite{Kim2008}. Courtesy of K.-J. Kim and R. Lindberg \label{fig:BrightnessGraph}}
\end{figure}

One of the newest developments is the extension of RIXS into the time-domain.
The very recent availability of ultra-short and ultra-bright x-ray pulses generated by free electron lasers such as FLASH at DESY and the Linac Coherent Light Source (LCLS) at SLAC National Accelerator Laboratory offers unique opportunities for investigating the temporal dynamics of electronic states using time-resolved x-ray scattering and spectroscopies. Pulses of x-ray laser light from LCLS are many orders of magnitude brighter and/or several orders of magnitude shorter than what can be produced by any other x-ray source available now or in the near future, see Fig.~\ref{fig:BrightnessGraph}. These characteristics will enable frontier new science in areas that include discovering and probing new states of matter, understanding and following chemical reactions and energy conversion processes in real time, imaging non-crystalline biological materials at atomic resolution, and imaging chemical and structural properties of materials on the nanoscale. In addition, the strong coherence of these sources open the door to performing coherent stimulated RIXS \cite{Harbola2009}.

From a theoretical point, concepts developed for RIXS are close to pump-probe experiments. In fact, a number of groups
have advocated the concept of viewing RIXS as an effective pump-probe experiment, where the absorption is the pump and the emission is the probe. The core-hole lifetime is the effective clock.
These time-dependent measurements require the inclusion of non-equilbrium dynamics in spectroscopy. Although current theoretical frameworks describe charge and spin dynamics in the discussion of RIXS, after excitation, the total energy of the system is always conserved. When studying the time domain, the system will dissipate energy by coupling to the surroundings. The inclusion of nonequilbrium effects in x-ray spectroscopy is still largely an unexplored field.

Other source developments may provide additional opportunities for inelastic x-ray scattering techniques. Recent feasibility studies~\cite{Kim2008} for X-ray free-electron oscillators, first proposed by~\textcite{Colella1984}, show the possibility of producing x-ray beams with a bandwidth of 2 meV in the 10 keV region. Related possiblities include so-called seeded-FELs, which also produce transform limited photon pulses at high repetition rates. The presence of such sources would come a long way towards solving the major experimental RIXS issue of a limited photon flux, holding out the promise of many orders of magnitude (up to $10^6$) gain in incident flux. This is a tantalizing prospect allowing one to imagine studying single magnons in small exchange systems, full dispersion curves, superconducting gaps, etc.

Summarizing, with the current steady progress in instrumentation, combined with the increasing theoretical expertise, the use of resonant inelastic x-ray scattering techniques is becoming increasingly prevalent. The future for RIXS as a characterization tool able to probe elementary excitations in real materials under real conditions looks bright.
\section{Acknowledgments}


We would like to express our thanks to current and former members of our groups for their cooperation, daily discussions and many useful comments in the field of RIXS:
Ken Ahn,
Cheng-Chien Chen,
Maria Daghofer,
Fiona Forte,
Stephane Grenier,
Liviu Hozoi,
Steven Johnston,
Young-June Kim,
Antoine Klauser,
Stefanos Kourtis,
Pasquale Marra,
Brian Moritz,
Adam Sorini,
Francois Vernay, and
Krzysztof Wohlfeld.
We also would like to take this opportunity to acknowledge our colleagues and collaborators in the field:
Peter Abbamonte,
Ercan Alp,
Arun Bansil,
Bernardo Barbiellini,
Valentina Bisogni,
Lucio Braicovich,
Nick Brookes,
Diego Casa,
Guillaume Chabot-Couture,
Ralph Claessen,
Steve Cramer,
Arthur Fedro,
Byron Freelon, 
Jim Freericks,
Jochen Geck,
Giacomo Ghiringhelli,
Thomas Gog,
Martin Greven,
Marco Grioni,
Frank de Groot,
Jason Hancock,
Zahid Hasan,
Maurits Haverkort,
Di-Jing Huang,
Zahid Hussain,
Sumio Ishihara,
Mark Jarrell,
Bernhard Keimer,
Giniyat Khaliullin,
B.J. Kim,
Jung-Ho Kim,
Akio Kotani,
Gerrit van der Laan,
Ben Larson,
Xiaosong Liu,
Sadamichi Maekawa,
Bob Markiewicz,
Alexander Moewes,
Juana Moreno,
Marco Moretti Sala,
Tatsuya Nagao,
John Rehr,
George Sawatzky,
Thorsten Schmitt,
Winfried Sch\"ulke,
Z.X. Shen,
Yuri Shvyd'ko,
Hidenori Takagi,
Carsten Timm,
Takami Tohyama,
G\"otz Uhrig,
Mary Upton, and
Jan Zaanen.

The work of LJPA and JvdB is supported by the Dutch Science Foundation FOM. This research benefited from the RIXS collaboration supported by the Computational Materials Science Network (CMSN) program of the Division of Materials Science and Engineering, U.S. Department of Energy, Grant No. DE-FG02-08ER46540. MvV was  supported by  U.S. DOE under contract DE-FG02-03ER46097. T. P. D. was supported by the Office of Science of the U.S. Department of Energy under Contract No. DE-AC02-76SF00515. The U.S. DOE, Office of Science, Office of Basic Energy Sciences supported work at Argonne National Laboratory under contract DE-AC02-06CH11357 and work at Brookhaven National Laboratory under Contract No. DE-AC02-98CH10886.

\pagebreak

\appendix
\section{Spherical Tensors\label{sec:spehricaltensors}}
The use of spherical tensors is prevalent in x-ray spectroscopy  resulting from expansions around the sites where the absorption takes place. The angular dependencies are typically expressed in terms of spherical harmonics $Y_{lm} (\theta_{{\hat {\bf r}}},\varphi_{{\hat {\bf r}}})$. Since the angles $\theta_{{\hat {\bf r}}}$ and $\varphi_{{\hat {\bf r}}}$ determine the direction of the unit vector ${\hat {\bf r}}={\bf r}/r$, we can also use the shorthand $Y_{lm} ({\hat {\bf r}})$ for the spherical harmonic in the direction of ${{\hat {\bf r}}}$. The spherical harmonics for a particular $l$ form a tensor ${\bf C}^l ({\hat {\bf r}})$ with normalized components $C^l_m ({\hat {\bf r}})=\sqrt{4\pi/[l]}Y_{lm} ({\hat {\bf r}})$ using the shorthand $[l_1\cdots l_n]=(2l_1+1)\cdots (2l_n+1)$. The rank of the tensor is $l$ and its components are $m=l,l-1,\cdots,-l$. Note that the components of a vector form the tensor $r{\bf C}^1 ({\hat {\bf r}})$.

It is often useful to couple tensors together to from a new tensor. This can be done via the tensor coupling
\begin{equation}
[{\bf T}^{l'},{\bf V}^l]^x_q =
\sum_{mm'}
w^{ll'x}_{mm'q} (T^{l'}_{m'})^\dagger V^l_{m},
\end{equation}
where the tensors ${\bf V}^l$ and ${\bf T}^{l'}$ of rank $l$ and $l'$, respectively, are coupled to a new tensor of rank $x$. The coefficients $w^{ll'x}_{mm'q}$ are defined as
\begin{equation}
w^{ll'x}_{mm'q} = (-1)^{l'-m'}
\left (
\begin{array}{ccc}
l' & x& l \\
-m' & q & m
\end{array}
\right )
n_{l'lx}^{-1}.
\label{wllx2}
\end{equation}
the term between parentheses is a $3j$ symbol, which is directly related to the Clebsch-Gordan coefficients \cite{Varshalovich1988,Yutsis1988,Brink1962}.
We use normalized tensor products that have the advantage that more nicely defined products are obtained when dealing with physical properties \cite{Thole1994a,Thole1994b}. This is achieved by defining the normalization constants
\begin{equation*}
		n_{ll'x} = \frac{(l+l')!}{(l+l'-x)!x!}\sqrt{\frac{ (L-2l)!(L-2l')!(L-2 x)!}{(L+1)!}},
\end{equation*}
with $L=l+l'+x$. Important examples of tensor products are the inner product $[{\bf T}^l,{\bf V}^l]^0_0={\bf T}^l\cdot {\bf V}^l$ and the outer product $[{\bf T}^1,{\bf V}^1]={\bf T}^1\times {\bf V}^1$.  Without normalization factors, these tensor products would contain square roots.

An identity used in this paper is the recoupling of the product of two inner products to the inner product of two tensors: 
\begin{equation}
({\bf a}^l \cdot {\bf b}^l) ({\bf A}^{l'} \cdot {\bf B}^{l'})=
\sum_{x=|l-l'|}^{l+l'} [x] n_{ll'x}^2 [ {\bf a}^l ,{\bf A}^{l'}]^x \cdot [ {\bf b}^l ,{\bf B}^{l'}]^x
\label{recoupling}
\end{equation}
This recoupling allows us to combine similar quantities to a total tensor of rank $x$.
\pagebreak
\section{Table of symbols}

\begin{table}[!h]
\vspace{0.5cm}
\centering\begin{tabular}{c|l|l}
\hline 
Symbol & \multicolumn{1}{c}{Significance} & \\
\hline
$\ket{g}$ & Ground state of the system & \\
$\ket{n}$ & Intermediate state of RIXS process & \\
$\ket{f}$ & Final state of the system & \\

$E_g$ & Ground state energy of the system & s\\
$E_n$ & Intermediate state energy of the system & s\\
$E_f$ & Final state energy of the system & s\\

$\ket{{\tt g}} = \ket{g;{\bf k}{\boldsymbol \epsilon}}$ & ground state of system plus incident photon & \\
$\ket{{\tt f}} = \ket{f;{\bf k}'{\boldsymbol \epsilon}'}$ &  final state of system plus scattered photon & \\

$E_{\tt g}$ & $= E_g + \hbar \omega_{\bf k}$ & s \\
$E_{\tt f}$ & $= E_f + \hbar \omega_{{\bf k}'}$ & s \\

$ \Gamma_n,\Gamma$ & Inverse lifetime of intermediate state & s \\

\hline

$\hbar {\bf k}$ / $\hbar {\bf k}'$& momentum of incident/scattered photon & v\\
$\hbar {\bf q}$ & transferred momentum $\hbar{\bf k}' - \hbar{\bf k}$ & v\\
${\boldsymbol \kappa}$ & reduced wave vector ${\boldsymbol \kappa} =
{\bf q} -n{\bf G}$ & v\\
${\bf G}$ & Reciprocal lattice vector & v\\

$\hbar \omega_{\bf k}$ / $\hbar \omega_{{\bf k}'}$ & Incident/scattered photon energy, ${\bf k} = k \hat{{\bf k}}$ & s\\
$\hbar \omega$ & Transferred energy $\hbar \omega_{\bf k}- \hbar
\omega_{{\bf k}'}$ & s \\
$\boldsymbol \epsilon$ / ${\boldsymbol \epsilon}'$ & Polarization of
incident/scattered photon & v \\
$\sigma$ / $\pi$ & Polarization vector is perpendicular ($\sigma$) &
\\
& or parallel ($\pi$) to the scattering plane &  \\
$2\theta$ & Scattering angle (between ${\bf k}$ and ${\bf k}'$) & s \\
R & Radius of curvature of grating & s \\

\hline

$I$ & Double-differential RIXS cross section
& s \\
$w$ &  Transition rate & s\\
$d\Omega$ & Solid angle of scattered photons & s \\
${\cal F}_{fg}$ & RIXS scattering amplitude & s\\
${\bf F}^x$ & Fundamental RIXS scattering amplitude & t \\
$F_{q'q}$  & RIXS scattering amplitude & t\\
			& in components of $r_q$ &  \\
${\bf T}^x$ & Angular and polarization dependence of ${\bf F}^x$ & t \\
$P$ &  Resonance function & s\\
$W^x$ &  Effective transition operator & o\\

\hline

$x$ & rank of tensor of fundamental  & s\\
& spectra and operators & \\
$\hat{w}^{clL}_M$ & Core to valence level orbital & o\\
& transition operator & \\
$\bar{E}$ & Average energy & s \\
$\underline{\hat{w}}^{llL}_M$ & d-d orbital transition operator & o\\
$\underline{\hat{V}}^{llz}_q, \hat{v}^{xyz}_q$ & d-d spin-orbital transition operators & o\\
$P_{1s,4p}$ & 1s $\rightarrow$ 4p dipole transition amplitude & s\\
$n_{ll'x}$ & Normalization constant & s \\
$p_{lL}(k)$ & $= [lL] n^2_{1lL} k^l /[l]!!$ & s\\
$[x]$ & $= 2x+1$ & s \\
$[x_1 x_2 \dots x_n]$ & $=(2x_1+1) (2x_2+1) \dots (2x_n+1)$ & s \\
$[ , ]^x$ & Tensor product, see Appendix \ref{sec:spehricaltensors} &
t \\
$Y_{lm}$ & Spherical harmonics & s \\
$R_{nl}$ & Radial wave functions & s \\
$j_l$ & Spherical Bessel functions & s \\
${\bf r}^{(l)}$ & $= r^l \hat{{\bf r}}^{(l)}$, $\hat{r}^{(l)}_m =
\sqrt{4\pi /[l]}
Y_{lm} (\hat{{\bf r}})$ & t \\
$\begin{pmatrix} A&B&C\\a&b&c \end{pmatrix}$ & $3j$ symbol & s \\

\hline
\end{tabular}
\end{table}

\begin{table}
\vspace{0.5cm}
\centering\begin{tabular}{c|l|l}
  
\hline

${\bf r}$, ${\bf p}$ & Electron position, momentum & v o\\
t & Time & s \\
${\bf R}$ & Ion position & v \\
${\bf A}$ & Electromagnetic vector potential & v o\\
$\phi$ & Electric Potential & s o \\
${\bf E}$, ${\bf B}$ & Electric, magnetic field & v o\\
${\boldsymbol \sigma}$ & Vector of Pauli matrices: ${\boldsymbol
  \sigma} = (\sigma_x, \sigma_y, \sigma_z)$ & v o \\
$\rho$ & Charge density & s o \\

\hline

$H$ & Hamiltonian & o\\
$G(z)$ & Green's function & o \\
$z_{\bf k}$ & $= E_g + \hbar \omega_{\bf k} + i \Gamma$ & s\\
$a^{{\dag}}$ / $a^{\phantom{\dag}}$ & Photon creation/annihilation operator & o\\
${\cal D}$ & Electromagnetic transition operator & o\\
${\bf D}$ & Electromagnetic dipole operator & v o\\
$v_F$ & Fermi velocity & s\\

$\Delta$ & Charge transfer energy & s \\
$t,t',t''$ & $1^{\rm st}$, $2^{\rm nd}$ and $3^{\rm rd}$ neighbor & \\
& hopping amplitudes & s \\
$U, U_{3d,3d}$ & On-site d-d Coulomb energy & s \\
$U_c, U_{1s,3d}, Q$ & Core hole-valence electron Coulomb energy & s \\
$J$ & Superexchange constant & s\\
$S_{\bf q} (\omega)$ & Dynamic structure factor & s \\
$3d^n$ & 3d levels contain $n$ electrons & \\
$\underline{L}$ & Hole at the ligand site & \\

\hline

$m$ & Mass of electron & s\\
-$e$ & Charge of electron & s\\
$c$ & Speed of light in vacuum & s\\
$\hbar$ & Planck's constant & s\\
$a_0$ & Bohr radius & s\\
$\alpha$ & Fine structure constant $e^2/4\pi \epsilon_0 \hbar c$ & s \\
$r_e$ & Classical electron radius $e^2 / (4\pi \epsilon_0 mc^2)$ & s\\
Z & Atomic number & s\\
$\epsilon_0$ &  Vacuum permittivity  & s\\
${\cal V}$ & Volume & s \\
$N$ & Number of electrons & s \\

\hline
\end{tabular}
\caption{Table of symbols used in this article, the physical quantity they denote and if they are used as a scalar (s), vector (v), tensor (t) or operator (o).}
\label{tabdist}
\end{table}

\pagebreak

\pagebreak
\bibliographystyle{apsrmp}

\bibliography{rmp}

\end{document}